\pdfoutput=1

\documentclass[11pt,twoside,a4paper,cmspaper,final,collab]{cms-tdr}
\def\svnVersion{265421}\def\cmsCernNo{2013-037}\def\cmsCernDate{\today}\def\cmsMessage{Published in the Journal of Instrumentation as \href{http://dx.doi.org/10.1088/1748-0221/9/10/P10009}{\doi{10.1088/1748-0221/9/10/P10009.}}}

\begin{document}\cmsNoteHeader{TRK-11-001}

\hyphenation{had-ron-i-za-tion}
\hyphenation{cal-or-i-me-ter}
\hyphenation{de-vices}
\RCS$Revision: 264908 $
\RCS$HeadURL: svn+ssh://svn.cern.ch/reps/tdr2/papers/TRK-11-001/trunk/TRK-11-001.tex $
\RCS$Id: TRK-11-001.tex 264908 2014-10-23 09:33:47Z tomalini $
\setcounter{secnumdepth}{4}
\newcommand{\subsubsubsection}[1]{\paragraph{#1}}
\newcommand{\Kzero}{\PKz\xspace}
\newcommand{\mumsq}{\ensuremath{\,\mu\text{m}^{2}}\xspace}
\newcommand{\metre}{\ensuremath{\,\text{m}}\xspace}
\newcommand{\metresq}{\ensuremath{\,\text{m}^{2}}\xspace}
\newcommand{\mrad}{\ensuremath{\,\text{mrad}}\xspace}
\providecommand{\rms}{\textsc{rms}\xspace}
\newcommand{\ifANnote}[1]{\ifthenelse{\boolean{cms@an}}{#1}{}}
\newcommand{\ifPAPER}[1]{\ifthenelse{\NOT\boolean{cms@an}}{#1}{}}
\newcommand{\PIXELAV}{\textsc{Pixelav}\xspace}
\newcommand{\zi}{\ensuremath{z^T_i}}
\newcommand{\zk}{\ensuremath{z^V_k}}
\newcommand{\zkprime}{\ensuremath{z^V_{k'}}}

\cmsNoteHeader{TRK-11-001} % This is over-written in the CMS environment: useful as preprint no. for export versions
\title{Description and performance of track and primary-vertex reconstruction with the CMS tracker}

\date{\today}

\abstract{
  A description is provided of the software algorithms developed for the CMS tracker both for reconstructing
  charged-particle trajectories in proton-proton interactions and for using
  the resulting tracks to estimate the positions of the LHC luminous region and individual primary-interaction
  vertices. Despite the very hostile environment at the LHC, the performance obtained with these algorithms is found to
  be excellent. For \ttbar events under typical 2011 pileup conditions,
  the average track-reconstruction efficiency for promptly-produced charged particles with transverse momenta of
  $\pt > 0.9$\GeV is 94\% for pseudorapidities of $\abs{\eta} < 0.9$ and 85\% for $0.9 < \abs{\eta} < 2.5$. The
  inefficiency is caused mainly by hadrons that undergo nuclear interactions in the tracker material. For isolated muons, the
  corresponding efficiencies are essentially 100\%.
  For isolated muons of $\pt = 100$\GeV emitted at $\abs{\eta} < 1.4$,
  the resolutions are approximately  2.8\% in \pt, and respectively, 10\mum and 30\mum in the transverse and longitudinal
  impact parameters.
  The position resolution achieved for reconstructed primary vertices that correspond to interesting pp
  collisions is 10--12\mum in each of the three spatial dimensions. The tracking and vertexing software is fast and flexible,
  and easily adaptable to other functions, such as fast tracking for the trigger, or dedicated
  tracking for electrons that takes into account bremsstrahlung.
}

\hypersetup{%
pdfauthor={CMS Collaboration},%
pdftitle={Description and performance of track and primary-vertex reconstruction with the CMS tracker},%
pdfsubject={CMS},%
pdfkeywords={CMS, software, tracker, tracking, vertexing}}

\maketitle %maketitle comes after all the front information has been supplied

\tableofcontents
\newpage

\section{Introduction}
\label{sec:intro}

At an instantaneous luminosity of $10^{34}$\percms, typical of that expected at the Large Hadron Collider (LHC), with the proton bunches crossing at intervals of 25\unit{ns},
the Compact Muon Solenoid (CMS) tracker is expected to be traversed by about 1000 charged particles at each bunch crossing, produced by
an average of more than twenty proton--proton (pp) interactions. These multiple interactions
are known as \textit{pileup}, to which prior or later bunch crossings can also contribute because of the
finite time resolution of the detector.
Reconstructing tracks in such a high-occupancy environment is immensely challenging. It is difficult to
attain high track-finding efficiency, while keeping the fraction of \textit{fake} tracks small. Fake tracks
are falsely reconstructed tracks that may be formed from a combination of unrelated hits or from a genuine
particle trajectory that is badly reconstructed through the inclusion of spurious hits.
In addition, the tracking software must run sufficiently fast to be used not only for offline event
reconstruction (of ${\approx}10^9$ events per year), but also for the CMS High-Level Trigger (HLT),
which processes events at rates of up to 100~kHz.

The scientific goals of CMS \cite{Bayatian:2006zz,Ball:2007zza} place demanding requirements on the
performance of the tracking system. Searches for high-mass dilepton resonances, for example, require
good momentum resolution for transverse momenta $\pt$ of up to 1\TeV. At the same time,
efficient reconstruction of tracks with very low \pt of order 100\MeV is needed for studies of hadron
production rates and to obtain optimum jet energy resolution with particle-flow techniques
\cite{CMS_PAS_PFT-09-001}. In addition, it is essential to resolve nearby tracks, such as those
from 3-prong $\tau$-lepton decays. Furthermore, excellent impact parameter resolution is needed for
a precise measurement of the positions of primary pp interaction vertices as well as for identifying b-quark jets
\cite{Chatrchyan:2012jua}.

While the CMS tracker \cite{:2008zzk} was designed with the above requirements in mind, the track-finding
algorithms must fully exploit its capabilities, so as to deliver the desired performance. The goal of this
paper is to describe the algorithms used to achieve this and show the level of performance attained.
The focus here is purely on pp collisions, with heavy ion collisions being beyond the scope of this document.
Section~\ref{sec:cmsTracker} introduces the CMS tracker; and Section~\ref{sec:localReco} describes the
reconstruction of the \textit{hits} created by charged particles crossing the tracker's sensitive layers.
The algorithms used to reconstruct tracks from these hits are explained in Section~\ref{sec:trackReco}; and
the performance obtained in terms of track-finding efficiency, proportion of fake tracks and track parameter resolution is presented in
Section~\ref{sec:trackPerformance}. Primary vertices from pp collisions are distributed over a
luminous region known as the beam spot. Reconstruction of the beam spot
and of the primary vertex positions is described in Section~\ref{sec:beamSpotAndPV}. This is intimately connected with tracking,
since on the one hand, the beam spot and primary vertices are found using reconstructed tracks, and on the other hand, an approximate knowledge
of their positions is needed before track finding can begin.
All results shown in this paper are based on
pp collision data collected or events simulated at a centre-of-mass energy of $\sqrt{s}=7$\TeV in 2011.
The simulated events include a full simulation of the CMS detector response based on \GEANTfour
\cite{Agostinelli:2002hh}. All events are reconstructed using software from the same period.
The track-reconstruction algorithms have been steadily evolving since then, but still have a similar
design now.

The CMS detector \cite{:2008zzk} was commissioned initially using cosmic ray muons and subsequently
using data from the first LHC running period.
Results obtained using cosmic rays in 2008 \cite{craft_paper} are extensively documented
in several publications pertaining to the pixel detector~\cite{craft_pixel_paper}, strip
detector~\cite{craft_strip_paper}, tracker alignment~\cite{craft_alignment_paper}, and magnetic
field~\cite{craft_bfield_paper}, and are of particular relevance to the present paper.
Results from the commissioning of the tracker using pp collisions in 2010 are presented in
\cite{TRK10_001}.

\section{The CMS tracker \label{sec:det}}
\label{sec:cmsTracker}

The CMS collaboration uses a right-handed coordinate system, with the origin at the
centre of the detector, the $x$-axis pointing to the centre of the
LHC ring, the $y$-axis pointing up (perpendicular to the plane of the LHC ring), and with
the $z$-axis along the anticlockwise-beam direction. The polar angle
$\theta$ is defined relative to the positive $z$-axis and the azimuthal
angle $\phi$ is defined relative to the $x$-axis in the $x$-$y$ plane.
Particle pseudorapidity $\eta$ is defined as $-\ln[\tan(\theta/2)]$.

\begin{figure}[hbtp]
  \centering
    \includegraphics[width=1.00\textwidth]{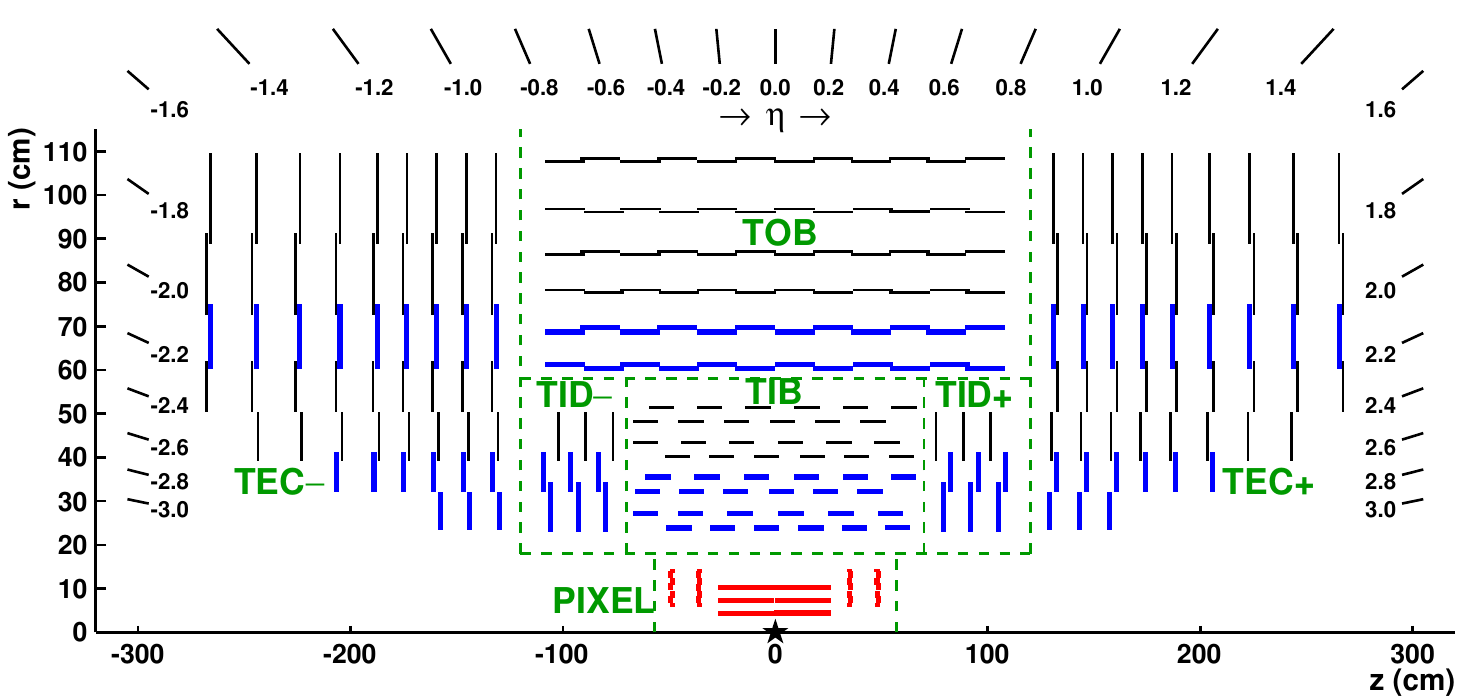}
    \caption{Schematic cross section through the CMS tracker in the $r$-$z$ plane. In this view,
    the tracker is symmetric about the horizontal line $r=0$, so only the top half is shown here.
    The centre of the tracker, corresponding to the approximate position of the pp collision point,
    is indicated by a star. Green dashed lines help the reader understand which modules belong to each of
    the named tracker subsystems. Strip tracker modules that provide 2-D hits are shown by thin, black
    lines, while those permitting the reconstruction of hit positions in 3-D are shown by thick, blue lines.
    The latter actually each consist of two back-to-back strip modules, in which one module is rotated
    through a `stereo' angle. The pixel modules, shown by the red lines, also provide 3-D hits. Within a given
    layer, each module is shifted slightly in $r$ or $z$ with respect to its neighbouring modules,
    which allows them to overlap, thereby avoiding gaps in the acceptance.}
    \label{fig:TrackerLayout}
\end{figure}

The CMS tracker \cite{:2008zzk} occupies a cylindrical volume 5.8\metre in length and 2.5\metre in diameter,
with its axis closely aligned to the LHC beam line. The tracker is
immersed in a co-axial magnetic field of 3.8\unit{T} provided by the CMS solenoid.
A schematic drawing of the CMS tracker is shown in
Fig.~\ref{fig:TrackerLayout}.
The tracker comprises a large silicon strip tracker with a small silicon pixel tracker inside it. In the
central pseudorapidity region, the pixel tracker consists of three co-axial barrel layers at radii between
4.4\cm and 10.2\cm and the strip tracker consists of ten co-axial barrel layers extending outwards to a
radius of 110\cm. Both subdetectors are completed by endcaps on either side of the barrel,
each consisting of two disks in the pixel tracker, and three small plus nine large disks in the
strip tracker. The endcaps extend the acceptance of the tracker up to a pseudorapidity of $\abs{\eta}<2.5$.

The pixel detector consists of cylindrical barrel layers at radii of 4.4, 7.3
and 10.2\cm, and two pairs of endcap disks at $z= {\pm}34.5$ and $\pm$46.5\cm.
It provides three-dimensional (3-D) position measurements of the hits
arising from the interaction of charged particles with its sensors.
The hit position resolution is approximately 10\mum in the transverse coordinate and 20--40\mum in the
longitudinal coordinate, while the third coordinate is given by the sensor plane position.
In total, its 1440 modules cover an area of about 1\metresq and have 66 million pixels.

The strip tracker has 15\,148 silicon modules, which in total cover
an active area of about 198\metresq and have 9.3 million strips.
It is composed of four
subsystems. The Tracker Inner Barrel (TIB) and Disks (TID)
cover $r < 55$\cm and $\abs{z} < 118$\cm, and are composed of four barrel layers,
supplemented by three disks at each end.
These provide position measurements in $r\phi$ with a resolution of approximately 13--38\mum.
The Tracker Outer Barrel (TOB) covers $r > 55$\cm and $\abs{z} < 118$\cm and consists
of six barrel layers providing position measurements in $r\phi$ with a resolution of approximately 18--47\mum.
The Tracker EndCaps (TEC) cover the region $124 < \abs{z} < 282$\cm. Each TEC
is composed of nine disks, each containing up to seven concentric rings of silicon strip modules, yielding
a range of resolutions similar to that of the TOB.

To refer to the individual layers/disks within a subsystem, we use a numbering convention
whereby the barrel layer number increases with its radius and the endcap disk number increases
with its $\abs{z}$-coordinate. When referring to individual rings within an endcap disk, the ring
number increases with the radius of the ring.

The modules of the pixel detector use silicon of 285\mum thickness, and achieve resolutions that are
roughly the same in $r\phi$ as in $z$, because of the chosen pixel cell size of $100\times 150$\mumsq
in $r\phi\times z$.
The modules in the TIB, TID and inner four TEC rings use silicon that is 320\mum thick, while those in
the TOB and the outer three TEC rings use silicon of 500\mum thickness.
In the barrel, the silicon strips usually run parallel to
the beam axis and have a pitch (\ie, the distance between neighbouring strips) that varies from 80\mum in the inner TIB layers to 183\mum in the inner
TOB layers. The endcap disks use wedge-shaped sensors with radial strips, whose pitch varies from
81\mum at small radii to 205\mum at large radii.

The modules in the innermost two layers of both the TIB and the TOB,
as well as the modules in rings 1 and 2 of the TID, and 1, 2 and 5 of the TEC, carry a second
strip detector module, which is mounted back-to-back to the first and rotated in the plane
of the module by a `stereo' angle of 100\mrad. The hits from these two modules, known as `$r\phi$' and
`stereo hits', can be combined into \textit{matched hits} that provide
a measurement of the second coordinate ($z$ in the barrel and $r$ on the disks). The achieved
single-point resolution of this measurement is an order of magnitude worse than in $r\phi$.

The principal characteristics of the tracker are summarized in Table~\ref{tab:TrackerGeom}.

\begin{figure}
\centering
\includegraphics[width=0.45\textwidth]{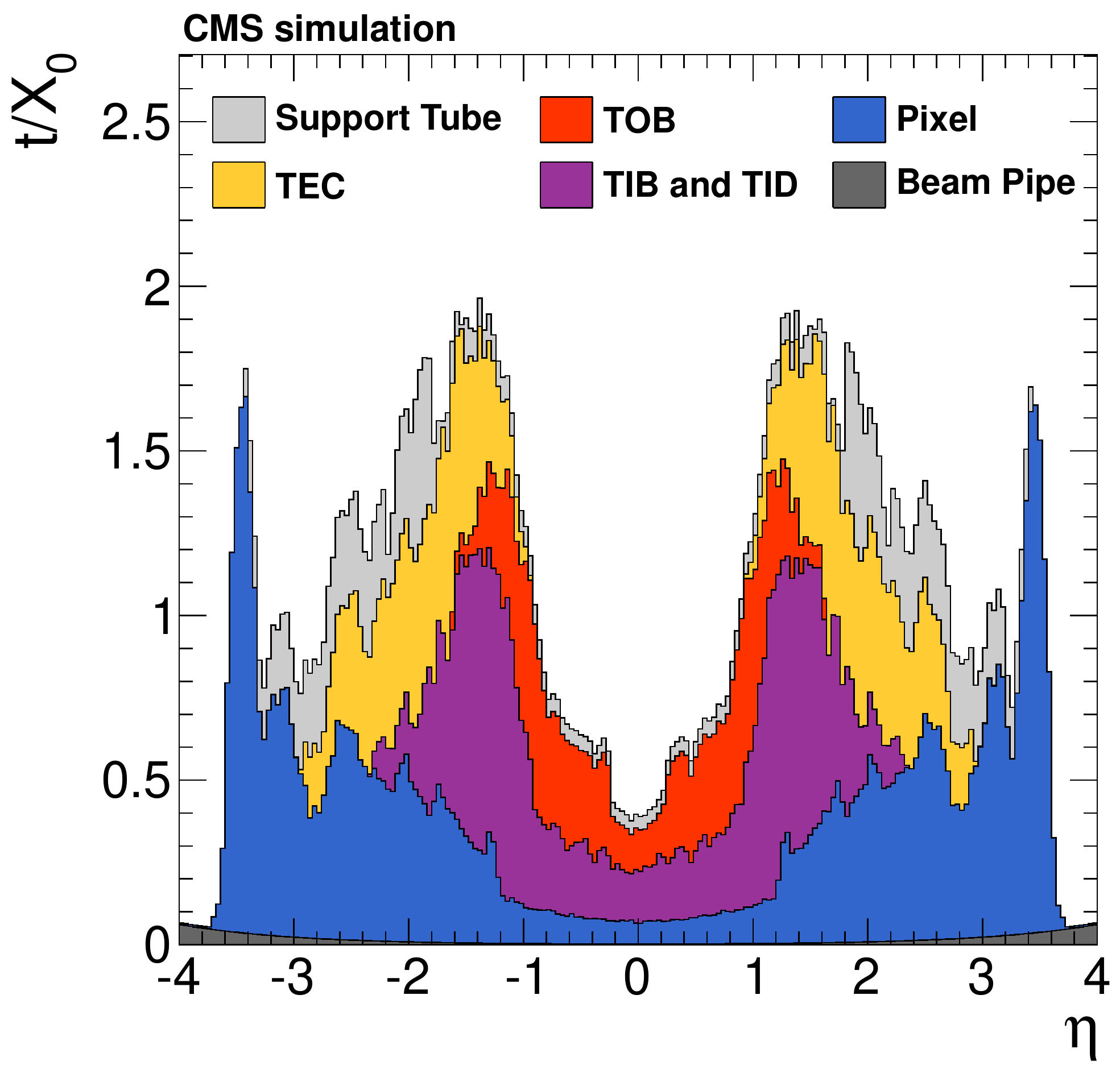}
\includegraphics[width=0.45\textwidth]{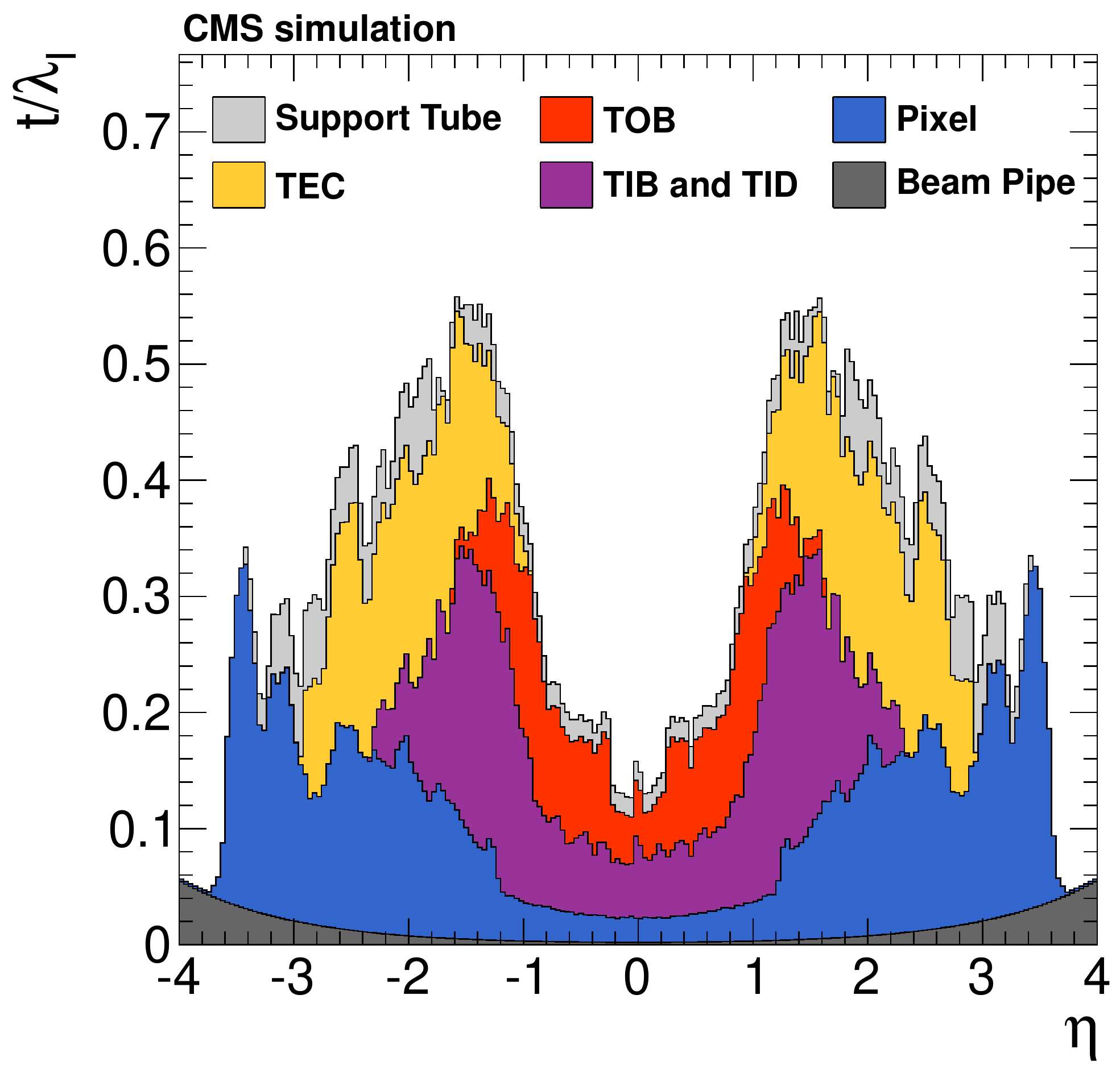}
\caption{Total thickness $t$ of the tracker material traversed by a particle produced at the
nominal interaction point, as a function of pseudorapidity $\eta$, expressed in units of radiation
length $X_0$ (left) and nuclear interaction length $\lambda_I$ (right). The contribution to the total
material budget of each of the subsystems that comprise the CMS tracker is shown, together
with contributions from the beam pipe and from the support tube that surrounds the tracker.
\label{fig:MaterialBudget}}
\end{figure}

Figure~\ref{fig:MaterialBudget} shows the material budget of the
CMS tracker, both in units of radiation lengths and nuclear interaction lengths, as estimated
from simulation. The simulation describes the tracker material budget with an
accuracy better than 10\% \cite{CMS_PAS_TRK-10-003}, as was established by measuring the distribution of
reconstructed nuclear interactions and photon conversions in the tracker.

\begin{table}[htbp]
\centering
\topcaption{\label{tab:TrackerGeom} A summary of the principal characteristics of the various tracker
subsystems. The number of disks corresponds to that in a single endcap. The location specifies
the region in $r$ ($z$) occupied by each barrel (endcap) subsystem.}
\begin{tabular}{llll}
\hline
Tracker subsystem   &   Layers   &   Pitch   &   Location   \\
\hline
    Pixel tracker barrel                  &       3 cylindrical     &    $100\times 150$\mumsq    &   $4.4 < r < 10.2$\cm   \\
    Strip tracker inner barrel (TIB)      &       4 cylindrical     &    80--120\mum              &   $20 < r < 55$\cm    \\
    Strip tracker outer barrel (TOB)      &       6 cylindrical     &    122--183\mum             &   $55 < r < 116$\cm   \\
\hline
    Pixel tracker endcap                  &       2 disks           &    $100\times 150$\mumsq    &   $34.5 < \abs{z} < 46.5$\cm     \\
    Strip tracker inner disks (TID)       &       3 disks           &    100--141\mum             &   $58 < \abs{z} < 124$\cm \\
    Strip tracker endcap (TEC)            &       9 disks           &    97--184\mum              &   $124 < \abs{z} < 282$\cm \\
\hline
\end{tabular}
\end{table}

\section{Reconstruction of hits in the pixel and strip tracker}
\label{sec:localReco}

The first step of the reconstruction process is referred to as \textit{local reconstruction}. It consists of the clustering of \textit{zero-suppressed} signals above specified thresholds in
pixel and strip channels into hits,
and then estimating the cluster positions and their uncertainties defined in a local orthogonal coordinate system $(u,v)$ in
the plane of each sensor.
A pixel sensor consists of $100\times 150\mum^2$ pixels with the $u$-axis oriented parallel to the shorter pixel edge.
In the strip sensors, the $u$-axis is chosen perpendicular to the central strip in each sensor (which in the TEC is not parallel to the
other strips in the same sensor).

\subsection{Hit reconstruction in the pixel detector}
\label{sec:localPixelReco}

In the data acquisition system of the pixel detector~\cite{Kotlinski:2006jz}, zero-suppression is performed in the readout chips of the sensors~\cite{Kastli:2005jj},
with adjustable thresholds for each pixel.
This pixel readout threshold is set to a single-pixel threshold corresponding to an equivalent charge of 3200 electrons.
Offline, pixel clusters are formed from adjacent pixels, including both side-by-side and corner-by-corner
adjacent cells. Each cluster must have a minimum charge equivalent to 4000 electrons. For comparison, a minimum ionizing particle deposits usually around 21000 electrons.
Miscalibration of residual charge caused by pixel-to-pixel differences of the charge injection capacitors, which are used to calibrate the pixel gain,
are extracted from laboratory measurements and included in the Monte Carlo (MC) simulation.

Two algorithms are used to determine the position of pixel clusters.
A fast algorithm  (described in Section~\ref{sec:firstPassPixel}) is used during track seeding and pattern recognition, and a more precise algorithm (Section~\ref{sec:templatesPixel}), based on cluster shapes, is used in the final track fit.

\subsubsection{First-pass hit reconstruction}
\label{sec:firstPassPixel}

The position of a pixel cluster along the transverse ($u$) and longitudinal ($v$) directions on the sensor is obtained
as follows. The procedure is described only for the case of the $u$ coordinate, but is identical for
the $v$ coordinate.

The cluster is projected onto the $u$-axis by summing the charge collected in pixels with the same
$u$-coordinate \cite{Cucciarelli:687475}. The result is referred to as a \textit{projected cluster}.
For projected clusters that are only one pixel large, the $u$-position is given by the
centre of that pixel, corrected for the Lorentz drift of the collected charge in the CMS magnetic field.
For larger projected clusters, the hit position $u_{\text{hit}}$ is determined using the relative charge in the
two pixels at each end of the projected cluster:
\begin{linenomath}
\begin{equation}
\label{eqn:pixelhit}
u_{\text{hit}} = u_{\text{geom}} + \frac{Q^u_{\text{last}} - Q^u_{\text{first}}}{2(Q^u_{\text{last}} + Q^u_{\text{first}})}
\left|W^u - W^u_{\text{inner}}\right| - \frac{L_u}{2},
\end{equation}
\end{linenomath}
where $Q_{\text{first}}$ and $Q_{\text{last}}$ are the charges collected in the first and last pixel of the
projected cluster, respectively; $u_{\text{geom}}$ is the position of the geometrical centre of the
projected cluster; and the parameter $L_u/2 = D\tan \Theta^u_L/2$ is the Lorentz shift along the $u$-axis, where $\Theta^u_L$ is the
Lorentz angle in this direction, and $D$ is the sensor thickness.
For the pixel barrel, the Lorentz shift is approximately $59\mum$.
The parameter $W^u_{\mathrm{inner}}$ is the geometrical width of the projected cluster, excluding its first and last
pixels. It is zero if the width of the projected cluster is less than three pixels.
The \textit{charge width} $W^u$ is defined as the width expected for the deposited charge, as estimated from the angle of the track with respect to the sensor, and equals
\begin{linenomath}
\begin{equation}
W^u = D \left| \tan \left ( \alpha^u - \pi/2 \right ) + \tan\Theta^u_L \right|,
\end{equation}
\end{linenomath}
where the angle $\alpha^u$ is the impact angle of the track relative to the plane of the sensor, measured
after projecting the track into the plane perpendicular to the $v$-axis. If no track is available,
$\alpha^u$ is calculated assuming that the particle producing the hit moved in a straight line from the
centre of the CMS detector.

The motivation for Eq.~(\ref{eqn:pixelhit}) is that the charge deposited by the traversing particle is expected to only partially cover the two pixels at each end of the projected cluster.
The quantity $W^u - W^u_{\text{inner}}$, which is expected to have a value between zero and twice the pixel pitch, (a modified version of Eq.~(\ref{eqn:pixelhit}) is used for any hits
that do not meet this
expectation), provides an estimate
of the total extension of charge into these two outermost pixels, while the relative
charge deposited in these two pixels provides a way to deduce how this total distance is shared between them.
The distance that the charge extends into each of the two pixels can thereby be deduced.
This gives the position of the two edges of the charge distribution, and the mean value of these edges, corrected for the Lorentz drift, equals the position of the cluster.

\subsubsection{Template-based hit reconstruction}
\label{sec:templatesPixel}

The high level of radiation exposure of the pixel detector can affect significantly the collection of charge
by the pixels during the detector's useful life.
This degrades particularly the performance of the standard hit reconstruction algorithm, sketched in the
previous section, as this algorithm only uses the end pixels of projected clusters when determining
hit positions.
The reconstructed positions of hits can be biased by up to $50\mum$ in highly irradiated sensors, and the
hit position resolution can be severely degraded.
In the template-based reconstruction algorithm, the observed distribution of the cluster charge is compared to expected projected distributions,
 called \textit{templates}, to estimate the positions of hits~\cite{CMS_NOTE_2007_033}.

The templates are generated based on a large number of simulated particles traversing
pixel modules, which are modelled using the detailed \PIXELAV simulation~\cite{Swartz:2003ch,Chiochia:2004qh, Swartz:2005vp}.
Since the \PIXELAV program can describe the behaviour of irradiated sensors, new templates can be generated over the life of the detector to maintain the
performance of the hit reconstruction.
To allow the template-based algorithm to be applied to tracks crossing the silicon at various angles,
different sets of templates are generated for several ranges of the angle between the particle trajectory and
the sensor.
Working in each dimension independently, each pixel is subdivided into nine bins along the $u$ (or $v$) axis, where
each bin has a width of one-eighth of the size of a pixel and the end bins are centred on the pixel boundaries.
The $u$ (or $v$) coordinate of the point of interception of the particle trajectory and the pixel (defined as the
position at which the track crosses the plane that lies halfway between the front and back faces of the sensor)
is used to assign the interception point to one of the nine bins, $j$, indicating its location within the pixel.
The charge profile of the cluster produced by each particle is projected into an array that is 13
pixels long along the $u$ axis (or 23 pixels long along the $v$ axis) and centred on the intercepted pixel.
The resulting charge in each element $i$ of this array is recorded.
Only clusters with a charge below some specified angle-dependent maximum, determined from simulation, are used, as the charge distributions can be distorted by the significant ionization caused by energetic delta rays.
This procedure provides an accurate determination of the projected cluster distributions, determined by effects of geometry, charge drift, trapping, and charge induction.
In each dimension, the mean charge $S_{i,j}$ in bin $(i,j)$, averaged over all the particles, is then
determined.
In addition, the \rms charge distributions for the two projected pixels at the two ends
of the cluster are extracted, as are the charge in the projected pixel that has the
highest charge within the cluster, and the cluster charge, both averaged over all tracks.

The charge distribution of a reconstructed cluster, projected onto either the $u$ or $v$ axis,
can be described in terms of a charge $P_i$ in each pixel $i$ of the cluster. This can be compared to the
expected charge distributions $S_{i,j}$ stored in the templates, so as to determine the bin $j$
where the particle is likely to have crossed the sensor, and hence the best estimate of the
reconstructed hit position. This is accomplished by minimizing a $\chi^2$
function for several or all of the bins:
\begin{align}
\chi^2 (j) & =  \sum_i \left ( \frac {P_i - N_j S_{i,j} }
{\Delta P_i } \right )^2,
\label{eq:fulleq}
\\
\intertext{with}
N_j  & =  \sum_i  \frac {P_i}{ \left (\Delta P_i  \right )^2 }
\Big / \sum_i  \frac {S_{i,j}}{ \left (\Delta P_i  \right )^2 }.
\end{align}
In this expression, $\Delta P_i$ is the expected \rms of a charge $P_i$ from the \PIXELAV
simulation and $N_j$ represents a normalization factor between the observed cluster charge and the
template.
While a sum over all the template bins yields an absolute minimum, different strategies can be used to optimize the performance of the algorithm as a function of allowed CPU time.
As described in Section~\ref{sec:TrackFit}, this  $\chi^2$ is also used to reject outliers
during track fitting, in particular pixel hits on a track that are incompatible with
the distribution expected for the reconstructed track angle.

A simplified estimate of the position of a hit is performed for cluster projections consisting of a single pixel by correcting the
 position of the hit for bias from
 Lorentz drift and possible radiation damage.
The bias is defined by the average residual of all single-pixel clusters, as detailed below..

For cluster projections consisting of multiple pixels, the estimate of the hit position is further
refined. The charge template expected for a track crossing the pixel at an arbitrary position $r$, near
the best $j$ bin is approximated by the expression
$(1-r)S_{i,j-1} + r S_{i,j+1}$. Substituting this expression in place of $S_{i,j}$ in Eq.~(\ref{eq:fulleq}), and
minimizing $\chi^2$ with respect to $r$, yields an improved estimate of the hit position.

Finally, the above-mentioned hit reconstruction algorithm is
 applied to the same \PIXELAV MC samples originally used to generate the templates.
Since the true hit position is known, any bias in the reconstructed hit position can be determined
and accounted for when the algorithm is run on collision data.
In addition, the \rms of the difference
between the reconstructed and true hit position is used to define the uncertainty in
the position of a reconstructed hit.

\subsection{Hit reconstruction in the strip detector}
\label{sec:striphit}

The data acquisition system of the strip detector \cite{ttdr} runs algorithms on off-detector electronics
(namely, on the modules of the \textit{front-end driver} (FED)~\cite{FED}) to subtract pedestals (the baseline signal
level when no particle is present) and common mode noise (event-by-event
fluctuations in the baseline within each tracker readout chip), and to perform zero-suppression.
Zero-suppression accepts a strip if its charge
exceeds the expected channel noise by at least a factor of five, or if both the strip and one of its neighbours have
a charge exceeding twice the channel noise.
As a result, information for only a small fraction of the channels in any given event is retained for offline
storage.

Offline, clusters are seeded by any channel passing zero-suppression that has a charge at least a factor of three greater than the corresponding
channel noise~\cite{Bayatian:2006zz}. Neighbouring strips are added to each seed, if their strip charge
is more than twice the strip noise. A cluster is kept if its total charge is a factor five larger than the cluster noise,
defined as $\sigma_{\text{cluster}} = \sqrt{\scriptstyle{ \sum_i \sigma_i^2}}$, where $\sigma_i$ is the noise for strip $i$,
and the sum runs over all the strips in the cluster.

The position of the hit corresponding to each cluster is determined from the charge-weighted average of
its strip positions, corrected by approximately 10\mum  (20\mum) in the TIB (TOB)
to account for the Lorentz drift. One additional correction is made to compensate
for the fact that charge generated near the back-plane of the sensitive volume of the thicker silicon
sensors is inefficiently collected. This inefficiency shifts the cluster barycentre along the direction
perpendicular to the sensor
plane by approximately 10\mum in the 500\mum thick silicon, while its effect is negligible in the
320\mum thick silicon.
The inefficient charge collection from the sensor backplane is caused by the narrow time window during
which the APV25 readout chip~\cite{French:2001xb} integrates the collected charge,
and whose purpose is to reduce background from out-of-time hits.

The uncertainty in the hit position is usually parametrized as a function of the expected
width of the cluster obtained from the track angle (\ie, the `charge width' defined in
Section~\ref{sec:firstPassPixel}). However, in rare cases, when the observed width of a cluster exceeds the expected width by at least a factor of 3.5, and is incompatible with it, the uncertainty in the position is then
set to the `binary resolution', namely, the width of the cluster divided by $\sqrt{12}$. This broadening of the cluster is caused by capacitive coupling between the strips or energetic delta rays.

\subsection{Hit efficiency}
\label{s:hit_efficiency}

The hit efficiency is the probability to find a cluster in a given silicon sensor that has been traversed by a charged particle.

\begin{figure}[t]
  \centering
    \includegraphics[width=0.48\linewidth]{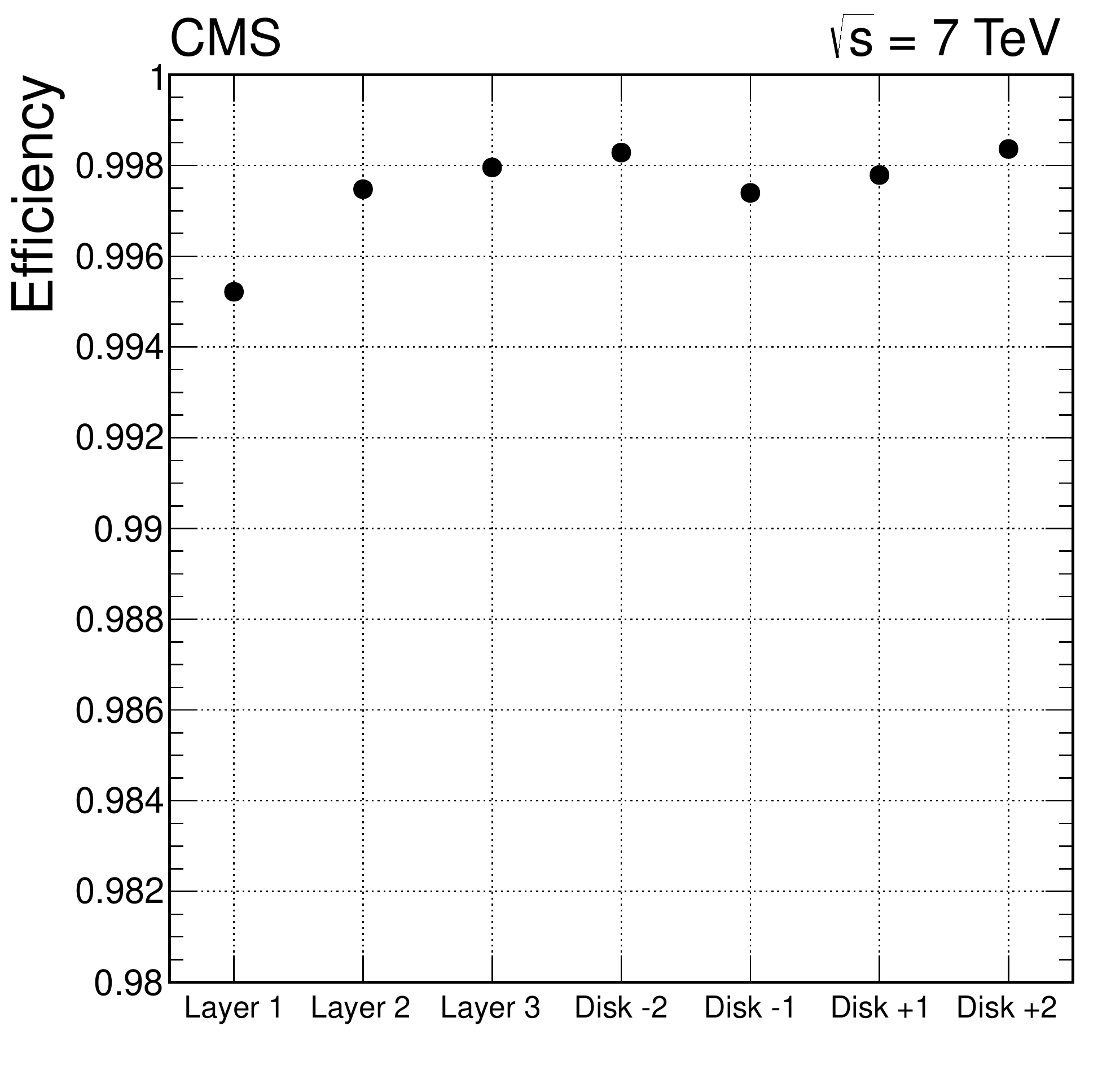}
    \includegraphics[width=0.48\linewidth]{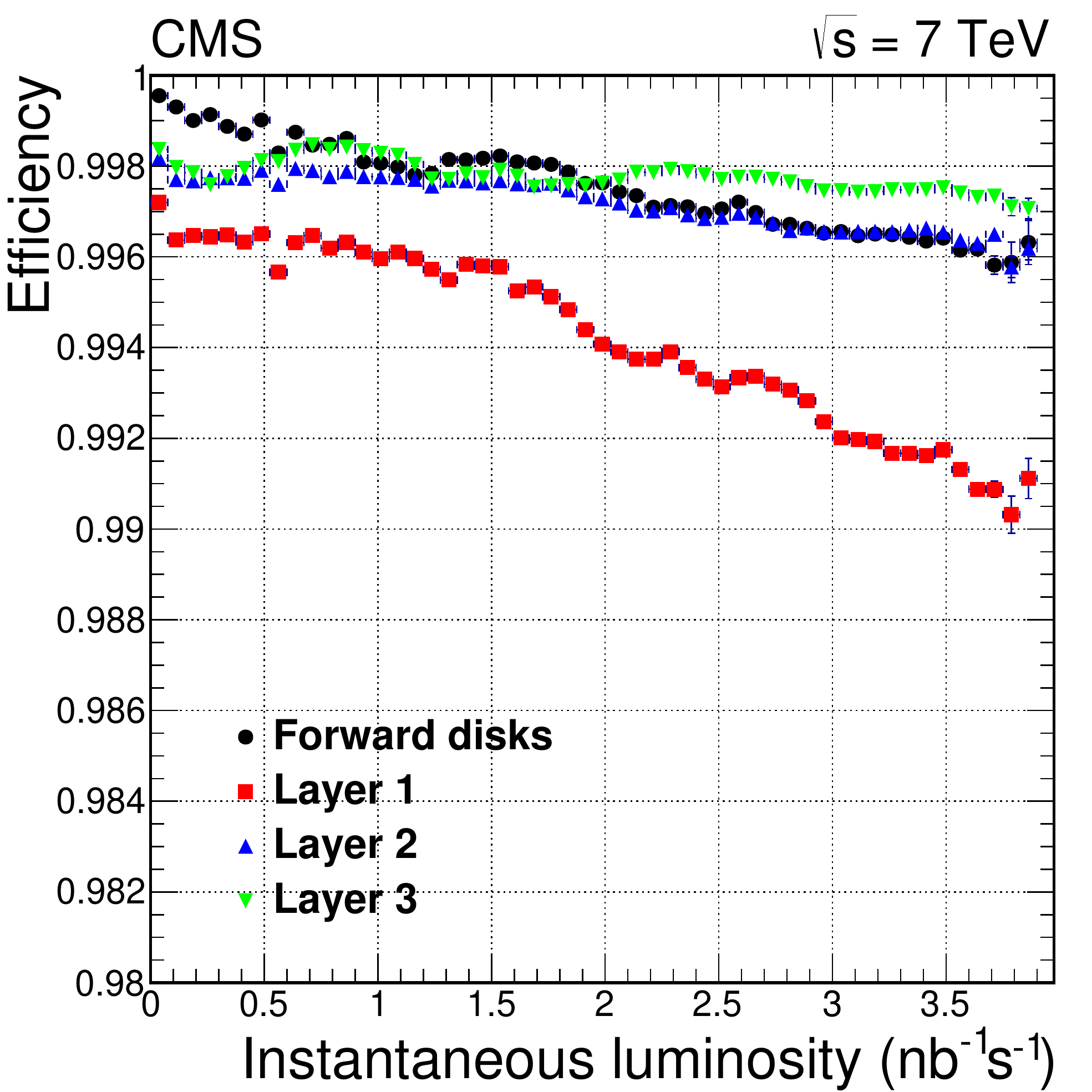}
    \caption{The average hit efficiency for layers or disks in the pixel detector excluding defective modules  (left), and
the average hit efficiency as a function of instantaneous luminosity (right). The peak luminosity ranged from 1 to  $4\,\mathrm{nb^{-1}s^{-1}}$ during the data taking.}
    \label{fig:hitEffPixel}

\end{figure}

In the pixel detector, the efficiency is measured using isolated tracks originating from the primary vertex.
The \pt is required to be $>$1\GeV,
 and the tracks are required to be reconstructed with a minimum of 11~hits measured in the strip detector.
Hits from the pixel layer under study are not removed when the tracks are reconstructed. To minimize any ensuing bias,
all tracks are required to have hits in the other two pixel layers, ensuring thereby
that they would be found even without using the studied layer.
A restrictive selection is set on the impact parameter to reduce false tracks and tracks from secondary interactions.
To avoid inactive regions and to allow for residual misalignment, track trajectories passing near the edges
of the sensors or their readout chips are excluded. Specifically, they must not pass within
0.6\mm (1.0--1.5\mm) of a sensor edge in the pixel endcap (barrel) or within 0.6\mm of the edge of a pixel
readout chip.
The efficiency is determined from the fraction of tracks to which either a hit is associated in the layer under study, or if it is found within $500\mum$
of the predicted position of the track. Given the high track density, only tracks that have no additional trajectories within 5\mm are considered so as to reduce
false track-to-cluster association.
The average efficiency for reconstructing hits is $>$99\%, as shown in Fig.~\ref{fig:hitEffPixel}(left),
when excluding the 2.4\% of the pixel modules known to be defective.
The hit efficiency depends on the instantaneous luminosity and on the trigger rate, as shown in Fig.~\ref{fig:hitEffPixel}(right).
The systematic uncertainty in these measurements is estimated to be 0.2\%.
Several sources of loss have been identified.
First, the limited size of the internal buffer of the readout chips cause a dynamic inefficiency that increases with the instantaneous luminosity and with the trigger rate.
Single-event upsets temporarily cause loss of information at a negligible rate of approximately two readout chips per hour.
Finally, readout errors signalled by the FED modules depend on the rate of beam induced background.

\begin{figure}[t]
  \centering
    \includegraphics[width=0.5\linewidth]{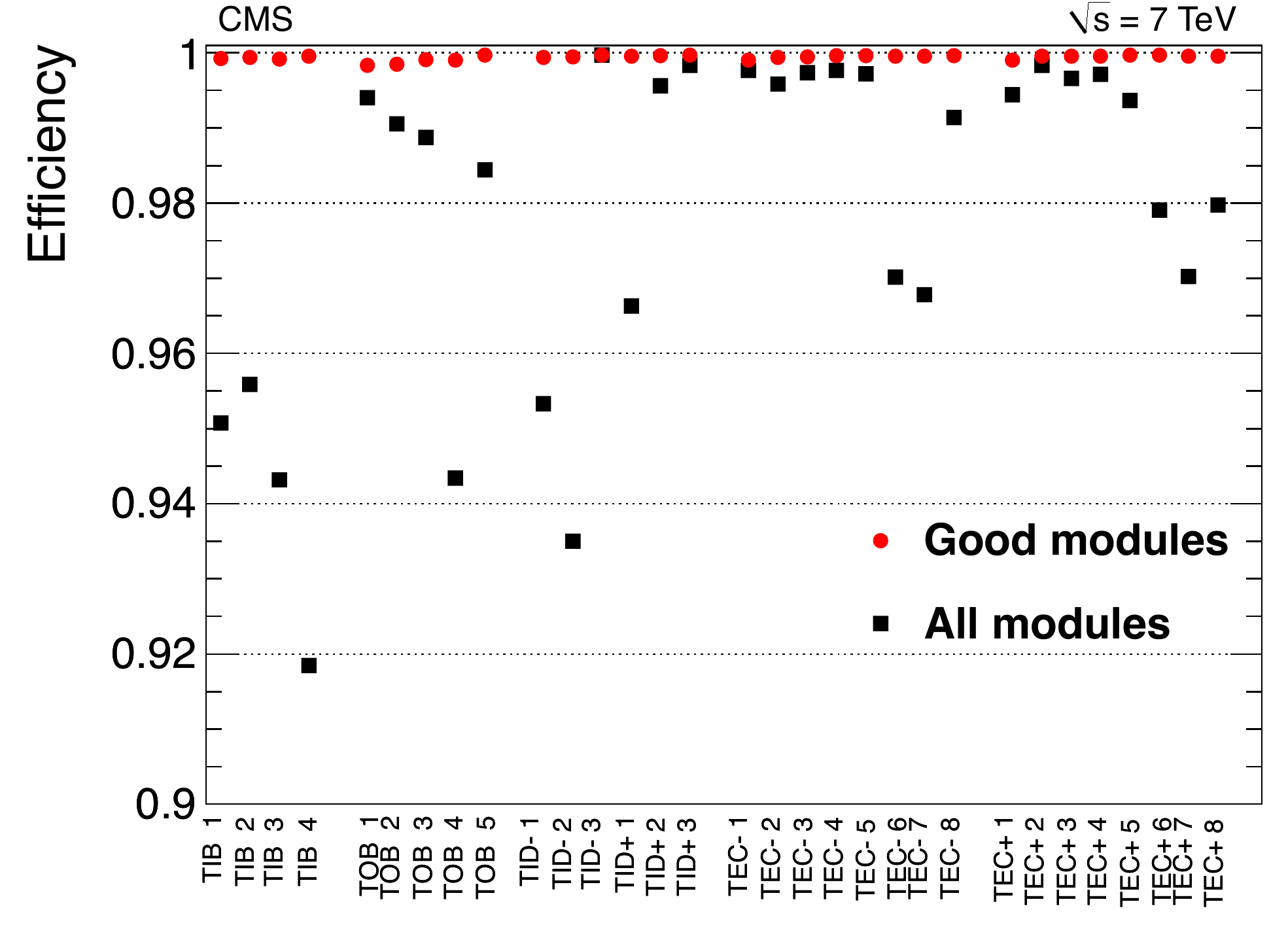}
    \caption{Average hit efficiency for layers or disks in the strip tracker.
The black squares show the hit efficiency in all modules, and the red dots for modules included in the readout.}
    \label{fig:hitEffStrip}
\end{figure}

The efficiency in the strip tracker is measured using tracks that have a minimum of eight hits in the pixel and strip detectors. Where two hits are found in one of the closely-spaced double layers, which consist
of $r\phi$ and stereo modules, both hits are counted separately.
The efficiency in any given layer is determined using only the subset of tracks that have at least one hit in
subsequent layers, further away from the beam spot.
This requirement ensures that the particle traverses the layer under study, but
also means that the efficiency cannot be measured in the outermost layers of the TOB
(layer~6) and the TEC (layer~9).
To avoid inactive regions and to take account of any residual misalignment, tracks that cross a module
within five standard deviations from the sensor's edges, based on the uncertainty in the extrapolated track trajectory,
are excluded from consideration.
The efficiency is determined from the fraction of traversing tracks with a hit anywhere within the non-excluded region of a traversed module.
In the strip tracker, 2.3\% of the modules are excluded because of short circuits of the high voltage,
communication problems with the front-end electronics, or other faults.
Once the defective modules are excluded from the measurement, the overall hit efficiency is 99.8\%, as shown in Fig.~\ref{fig:hitEffStrip}.
This number is compatible with the 0.2\% fraction of defective channels observed during the construction of the strip tracker.

All defective components of the tracker are taken into account, both in the MC simulation of the detector and in the reconstruction of tracks.

\subsection{Hit resolution}

The hit resolution in the pixel and strip barrel sensors has been studied by measuring residuals, defined by the difference between the measured and the expected hit
position as predicted by the fitted track. Each trajectory is refitted excluding the
hit under study in order to minimize biases of the procedure.

\begin{figure}[t]
  \centering
\includegraphics[width=0.5\linewidth]{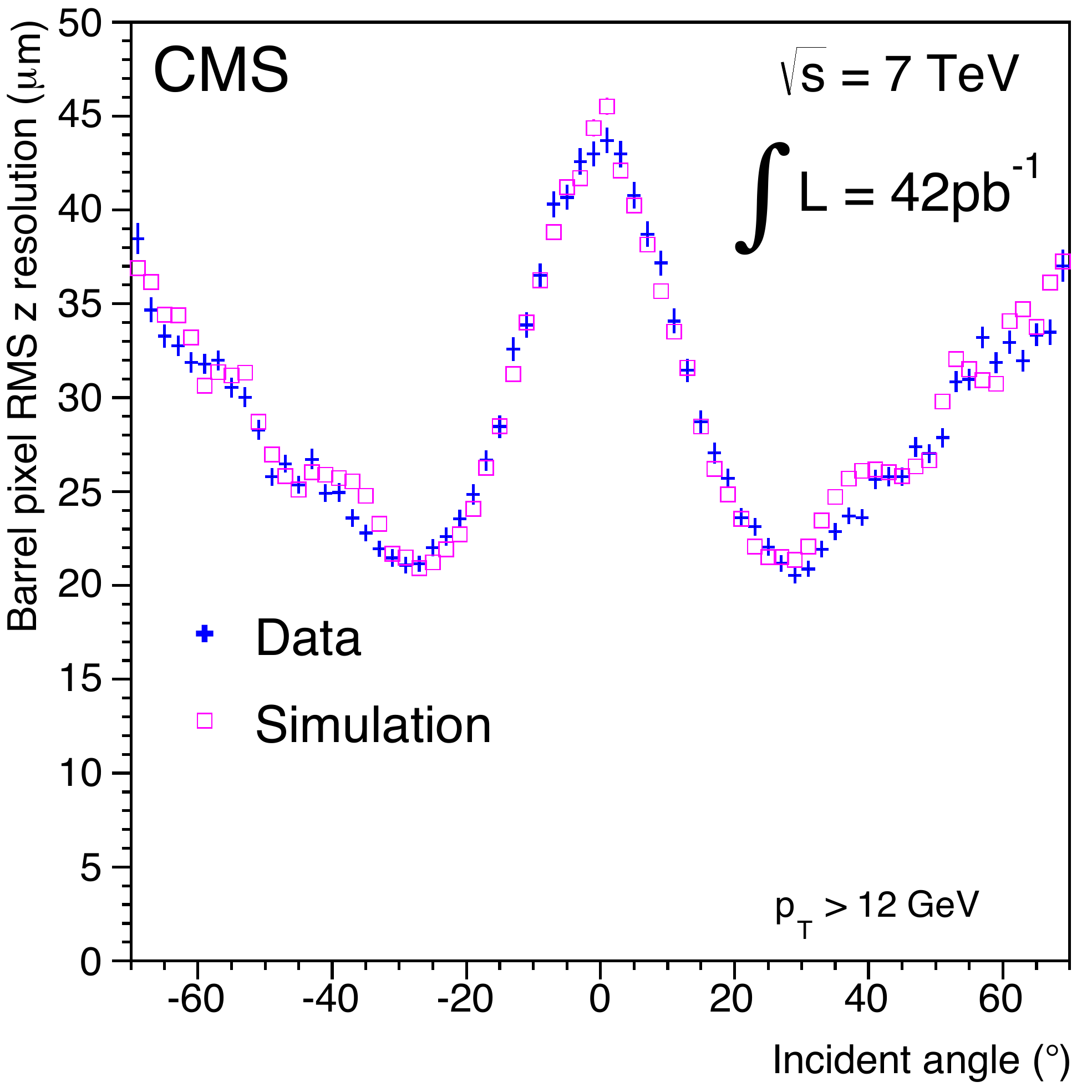}
\caption{Resolution in the longitudinal ($z$)  coordinate of hits in the barrel section of the pixel detector, shown
as a function of the incident angle of the track, which is defined as $90^\circ-\theta$, and equals the angle of the
track relative to the normal to the plane of the sensor. Data are compared with MC simulation for tracks with $\pt>12$\GeV.
}
  \label{fig:pixResZ}
\end{figure}

The resolution of the pixel detector is measured from the RMS width of the hit residual distribution in the
middle of the three barrel layers, using only tracks with $\pt > 12$\GeV, for which multiple scattering
between the layers does not affect the measurement. The expected hit position in the middle layer, as
determined from the track trajectory, has an uncertainty that is dominated by the resolution of the hits
assigned to the track in the first and third barrel layers. Assuming that the three barrel layers all have
the same hit resolution $\sigma_\text{hit}$ and because they are approximately equally spaced in radius from
the z-axis of CMS, then this uncertainty is given by $\sigma_\text{hit}/\sqrt{2}$. Adding this in quadrature
with the uncertainty $\sigma_\text{hit}$ in the measured position of the hit in the middle layer, demonstrates
that the RMS width of the residual distribution is given by $\sigma_\text{hit}\sqrt{3/2}$. The measured hit
resolution $\sigma_\text{hit}$ in the $r\phi$ coordinate, as derived using this formula, is 9.4\mum.
The resolution in the longitudinal direction is shown in Fig.~\ref{fig:pixResZ}, and found to agree within
1\mum with MC simulation.
The longitudinal resolution depends on the angle of the track relative to the sensor.
For longer clusters, sharing of charge among pixels improves the resolution, with optimal resolution reached for interception angles of $\pm 30^\circ$.

Because of multiple scattering, the uncertainty in track position in the strip detector is usually much larger than the inherent resolution;
consequently, individual residuals of hits are not sensitive to the resolution.
However, the difference in a track's residuals for two closely spaced modules can be
measured with much greater precision.
Any offset in a track's position caused by multiple scattering will
be largely common to both modules.
A technique based on tracks passing through overlapping modules from the same tracker layer is
employed to compare the difference in residuals for the two measurements in the overlapping modules~\cite{tifPaper}.
The difference in hit positions ($\Delta x_\text{hit}$) can be compared to the difference in
predicted positions ($\Delta x_\text{pred}$) derived from the track trajectory, and their difference, fitted to a Gaussian function, provides a hit resolution convoluted with the uncertainty from
the trajectory propagation. The bias from translational misalignment between modules affects only
the mean of the Gaussian distribution, and not its \rms width.
As the two overlapping modules are expected to have the
same resolution, the resolution of a single sensor is determined by dividing this \rms width by $\sqrt{2}$.

Only tracks of \textit{high purity} (defined in Section~\ref{sec:TrackSelection})
are used for the above-described study. To reduce the uncertainty from multiple
 Coulomb scattering, the track momenta are required to be $>$10\GeV.
The $\chi^2$ probability of the track fit is required to be $>$0.1\%,
 and the tracks are required to be reconstructed using a minimum of six hits in the strip detector.
Tracks in the overlapping barrel modules are analysed only when the residual rotational
 misalignment is less than 5\mum.
Remaining uncertainties from multiple scattering and rotational
 misalignment for the overlapping modules are included as systematic
 uncertainties of the measurement.

Sensor resolution depends strongly on the size of the cluster and on the pitch of the sensor.
The resolutions for the strip detector are shown in Table~\ref{tab:sstHitRes}, where they are compared to the predictions from MC simulation.
The resolution varies not only as a function of the cluster width, but also as a function of pseudorapidity, as the energy deposited by a charged particle
in the silicon depends on the angle at which it crosses the sensor plane. The resolution is worse in simulation than in data, implying the need for additional tuning
of the MC simulation. The results in the table are valid only for tracks with momenta $>$10\GeV. At lower momenta, the simulations indicate
that the resolution in hit position improves, but this is not important for tracking performance, as
the resolution of the track parameters for low-momentum tracks is dominated by the multiple scattering and by not  the hit resolution.

\begin{table}[bt]
\centering
\topcaption{\label{tab:sstHitRes} %.
A comparison of hit resolution in the barrel strip detector as measured in data with the
corresponding prediction from simulation, for track momenta $>$10\GeV.
The resolution is given as function of both the barrel layer and the width
of the cluster in strips. Since the resolution is observed to vary with $\phi$ and $\eta$,
a range of resolution values is quoted in each case.}
\begin{tabular}{ccccccc}
\hline
Sensor & Pitch &  & \multicolumn{4}{c}{Resolution [$\mu \mathrm{m}$] vs. width of cluster [strips]}\\ \cline{4-7}
layer & $(\mu \mathrm{m})$ & & width=1 & =2 & =3 & =4\\
\hline
\multirow{2}{*}{TIB 1--2} & \multirow{2}{*}{ 80} &
Data	& 11.7--19.1	& 10.9--17.9	& 10.1--18.1	&		\\
&& MC   & 14.5--20.5	& 15.0--19.8	& 14.0--20.6	&		\\
\hline
\multirow{2}{*}{TIB 3--4} & \multirow{2}{*}{ 120} &
Data	&  20.9--29.5	& 21.8--28.8	& 20.8--29.2	&		\\
&& MC   & 26.8--30.4	& 27.6--30.8	& 27.9--32.5	&		\\
\hline
\multirow{2}{*}{TOB 1--4} & \multirow{2}{*}{ 183} &
Data	&		& 23.4--40.0	& 32.3--42.3	& 16.9--28.5	\\
&& MC   &		& 42.5--50.5	& 43.0--48.6	& 18.8--35.2	\\
\hline
\multirow{2}{*}{TOB 5--6} & \multirow{2}{*}{ 122} &
Data	&		&  		& 18.4--26.6	& 11.8--19.4	\\
&& MC   &		&		& 26.1--29.5	& 17.8--21.6	\\
\hline
\end{tabular}
\end{table}

\section{Track reconstruction}
\label{sec:trackReco}

Track reconstruction refers to the process of using the hits, obtained from the local reconstruction
described in Section~\ref{sec:localReco}, to obtain estimates for the momentum and position
parameters of the charged particles responsible for the hits (tracks).  As part of this process, a
translation between the local coordinate system of the hits and the global coordinate system of the
track is necessary.  This translation takes into account discrepancies between the assumed and
actual location and surface deformation of detector elements as found through the alignment
process~\cite{Chatrchyan:2014wfa}.  In addition, the uncertainty in the detector element location is
added to the intrinsic uncertainty in the local hit position.

Reconstructing the trajectories of charged particles is a computationally challenging
task.  An overview of the difficulties and solutions can be found in review
articles~\cite{Regler:1996yw,Mankel:2004yv,Strandlie:2010zz}.
The tracking software at CMS is commonly referred to as the Combinatorial Track Finder (CTF),
which is an adaptation of the combinatorial Kalman filter~\cite{Billoir:1989mh,Billoir:1990we,Mankel:1997dy}, which in turn
is an extension of the Kalman filter~\cite{Fruhwirth:1987fm} to allow pattern recognition and
track fitting to occur in the same framework.
The collection of reconstructed tracks is produced by multiple passes (iterations) of the CTF track
reconstruction sequence, in a process called \textit{iterative tracking}.
The basic idea of iterative tracking is that the initial iterations search for tracks that
are easiest to find (\eg, of relatively large \pt, and produced near the interaction region).
After each iteration, hits associated with tracks are removed, thereby reducing the
combinatorial complexity, and simplifying subsequent iterations in a search for more difficult
classes of tracks (\eg, low-\pt, or greatly displaced tracks).
The presented results reflect the status of the software in use from May through August, 2011, which is
applied in a series of six iterations of the track reconstruction algorithm.  Later versions of the
software retain the same basic structure but with different iterations and tuned values for the
configurable parameters to adapt to the higher pileup conditions.  Iteration 0, the
source of most reconstructed tracks, is designed for prompt tracks (originating near the pp interaction point)
with $\pt>0.8$\GeV that
have three pixel hits.  Iteration 1 is used to recover prompt tracks that have only two pixel hits.
Iteration 2 is configured to find low-\pt prompt tracks.  Iterations 3--5
are intended to find tracks that originate outside the beam spot (luminous region of the pp
collisions) and to recover tracks not found in
the previous iterations.  At the beginning of each iteration, hits associated with high-purity
tracks (defined in Section~\ref{sec:TrackSelection}) found in previous iterations are excluded
from consideration (masked).

Each iteration proceeds in four steps:
\begin{itemize}
\item{Seed generation provides initial track candidates found using only a few (2 or 3) hits.
A seed defines the initial estimate of the trajectory parameters and their uncertainties.}
\item{Track finding is based on a Kalman filter. It extrapolates
the seed trajectories along the expected flight path of a charged particle, searching for additional hits
that can be assigned to the track candidate.}
\item{The track-fitting module is used to provide the best possible estimate of the parameters
    of each trajectory by means of a Kalman filter and smoother.}
\item{Track selection sets quality flags, and discards tracks that fail certain specified criteria.}
\end{itemize}

The main differences between the six iterations lie in the configuration of the seed generation and the final
track selection.

\subsection{Seed generation}
\label{sec:SeedGeneration}

The seeds define the starting trajectory parameters and associated uncertainties of potential
tracks.  In the quasi-uniform magnetic field of the tracker, charged particles follow helical paths and
therefore five parameters are needed to define a
trajectory.  Extraction of these five parameters requires either three 3-D hits, or two 3-D hits and a
constraint on the origin of the trajectory based on the assumption that the particle originated near the
beam spot.  (A `3-D hit' is defined to be any hit that provides a 3-D position measurement).
To limit the number of hit combinations, seeds are required to satisfy
certain weak restrictions, for
example, on their minimum \pt and their consistency with originating from the pp
interaction region.

In principle, it is possible to construct seeds in the outermost regions of the tracker, where the track density is
smallest, and then construct track candidates by searching inwards from the seeds for additional hits
at smaller distances from the beam-line.  However, there are several reasons why an alternative approach, of
constructing seeds in the inner part of the tracker and building the track candidates outwards, has been chosen
instead.

\begin{figure}[hbtp]
  \centering
    \includegraphics[width=0.9\textwidth]{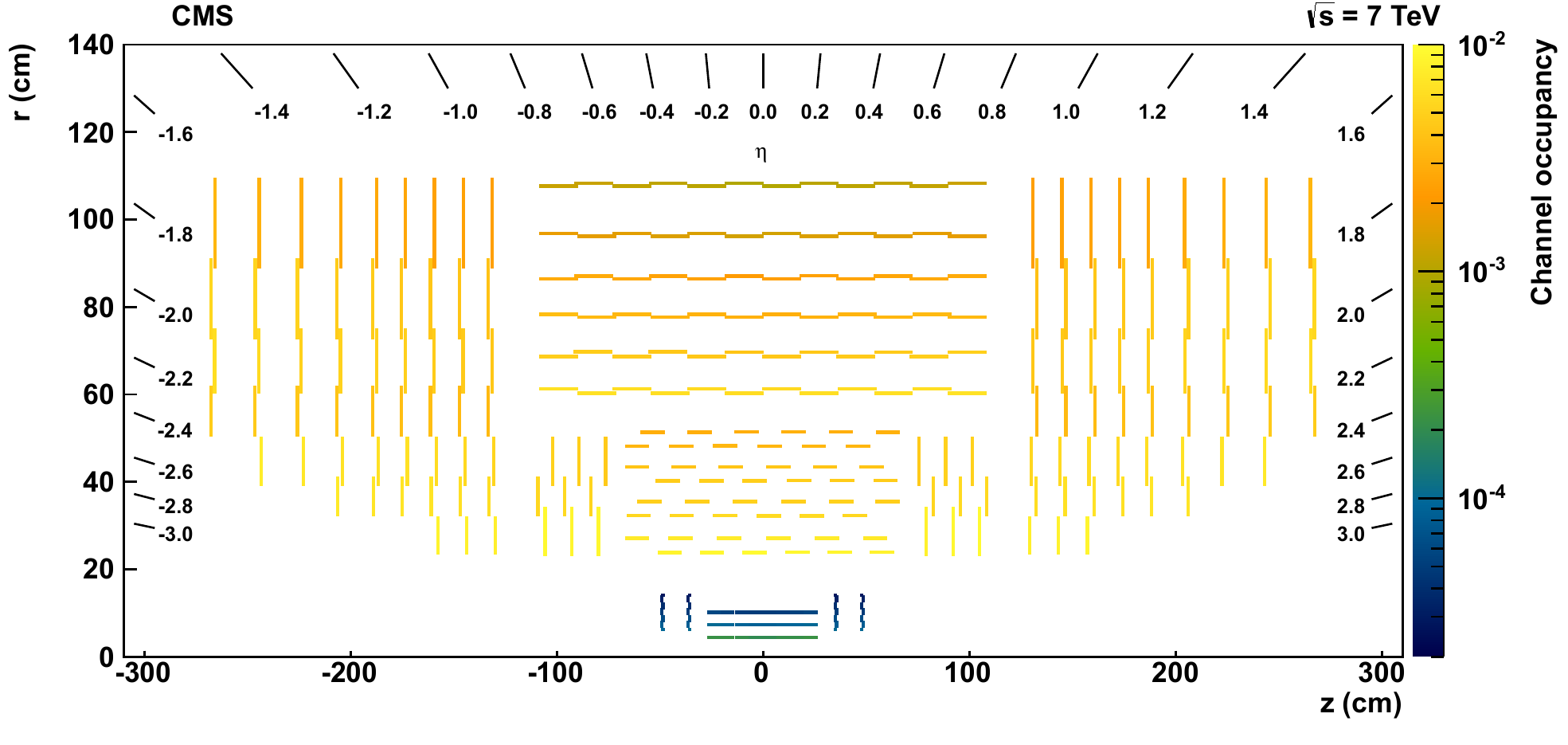}
    \caption{Channel occupancy (labelled by the scale on the right) for CMS silicon detectors in events taken
with unbiased triggers with an average of nine pp interactions per beam crossing, displayed as a function of $\eta$,
$r$, and $z$.}
    \label{fig:TrackerOccupancy}
\end{figure}

\begin{figure}[hbtp]
  \centering
  \includegraphics[width=0.6\textwidth]{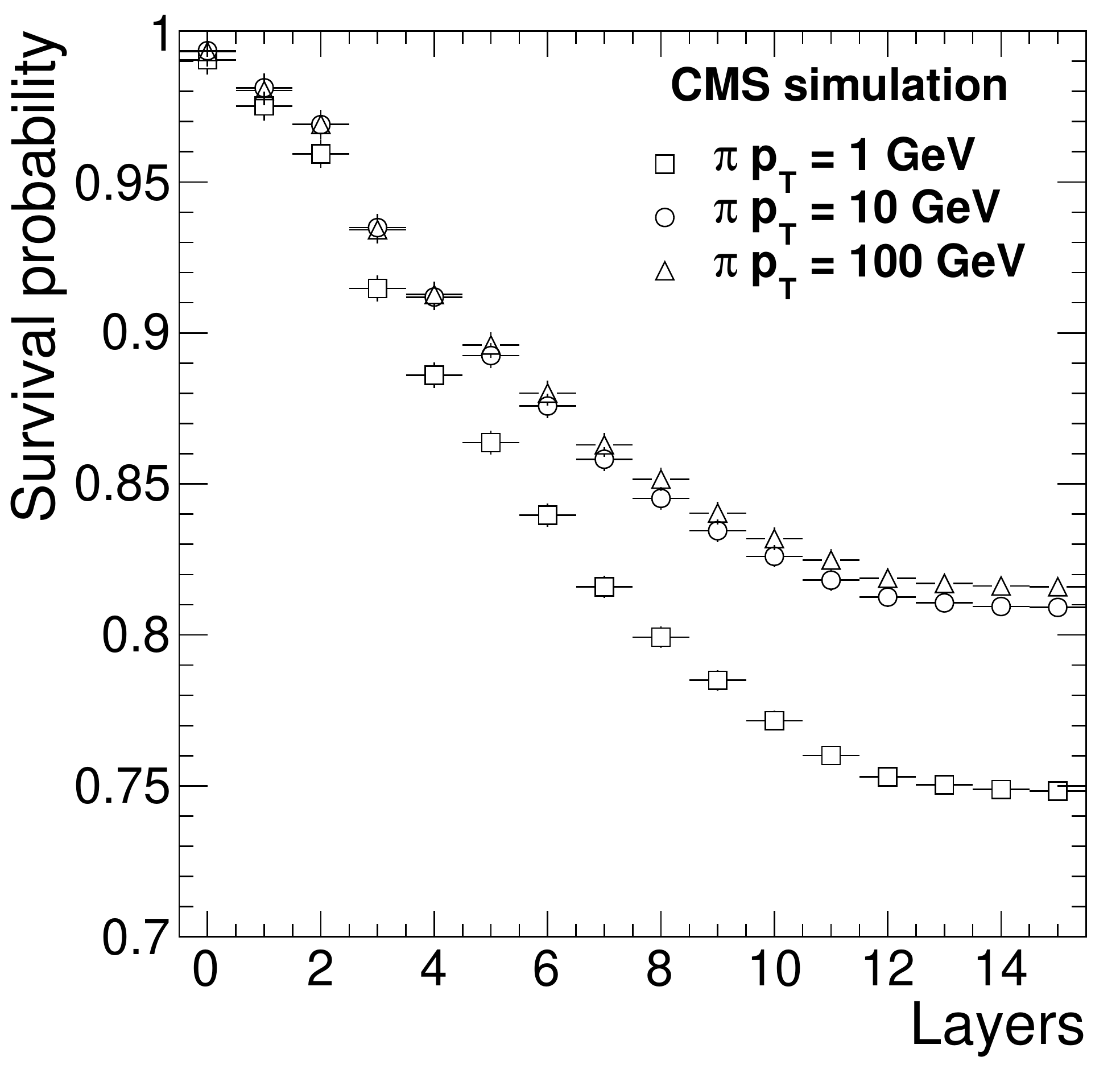}
  \caption {Fraction of pions produced with $\abs{\eta}<2.5$ that do not undergo a nuclear interaction in the
tracker volume, as a function of the number of traversed layers.}
\label{fig:PionSurvival}
\end{figure}

First, although the track density is much higher in the inner region of the tracker, the high granularity
of the pixel detector ensures that the channel occupancy (fraction of channels that are hit) of the inner pixel layer is much lower than
that of the outer strip layer. This can be seen in Fig.~\ref{fig:TrackerOccupancy},
which shows the mean channel occupancy in strip and pixel sensors
in data collected with a `zero-bias' trigger, (which took events from randomly selected non-empty LHC bunch
crossings).  This data had a mean of about nine pp interactions per bunch crossing.
The channel occupancy
is 0.002--0.02\% in the pixel detector and 0.1--0.8\% in the strip detector.
Second, the pixel
layers produce 3-D spatial measurements, which provide more constraints and
better estimates of trajectory parameters.  Finally, generating seeds in the inner tracker leads to
a higher efficiency for reconstructing tracks.  Although most high-\pt muons traverse the
entire tracker, a significant fraction of the produced pions interact inelastically in the
tracker~(Fig.~\ref{fig:PionSurvival}).  In addition, many electrons lose a significant
fraction of their energy to bremsstrahlung radiation in the tracker.  Therefore, to ensure high
efficiency, track finding begins with trajectory seeds created in the inner region of the tracker.
This also facilitates reconstruction of low-momentum tracks that are deflected by the strong magnetic field
before reaching the outer part of the tracker.

Seed generation requires information on the position of the centre of the reconstructed beam spot,
obtained prior to track finding using the method described in Section~\ref{sec:beamspot}. It
also requires the locations of primary vertices in the event, including
those from pileup events.
This information is obtained by running a very fast track and vertex reconstruction algorithm,
described in Section~\ref{sec:pixeltrackvertex}, that uses only hits from the pixel detector.
The tracks and primary vertices found with this algorithm are known as \textit{pixel tracks} and
\textit{pixel vertices}, respectively.

The seed generation algorithm is controlled by two main sets of parameters: \textit{seeding layers} and
\textit{tracking regions}.  The seeding layers are pairs or triplets of detector layers in which hits are searched
for.  The tracking regions specify the limits on the acceptable track parameters, including the
minimum \pt, and the maximum transverse and longitudinal distances of closest approach to the assumed production point of the particle, taken to be located either at the centre of the reconstructed beam spot or at a
pixel vertex.
If the seeding layers correspond to pairs of detector layers, then seeds are constructed using one hit in each layer.
A hit pair is accepted as a seed if the corresponding track parameters are
consistent with the requirements of the tracking region.  If the seeding layers correspond to triplets of
detector layers, then, after pairs of hits are found in the two inner layers of each triplet, a
search is performed in the outer detector layer for another hit.  If the track parameters derived from the
three hits are compatible with the tracking region requirements, the seed is accepted.  It is
also possible to check if the hits associated with the seed have the expected charge distribution from the
track parameters: a particle that enters the detector at a grazing angle will have a larger
cluster size than a particle that enters the detector at a normal angle.  Requiring the reconstructed
charge distribution to match the expected charge distribution can remove many fake seeds.

In simulated $\ttbar$ events at $\sqrt{s} = 7$\TeV, more than 85\% of the charged particles produced
within the geometrical acceptance of the tracker $(\abs{\eta}<2.5)$ cross three pixel layers and
can therefore be reconstructed starting from trajectory seeds
obtained from triplets of pixel hits.  Nevertheless, other trajectory seeds are also needed, partially to
compensate for inefficiencies in the pixel detector (from gaps in coverage, non-functioning modules, and saturation
of the readout), and partially to reconstruct particles not produced directly at the pp
collision point (decay products of strange hadrons, electrons from photon conversions, and particles from
nuclear interactions).  To improve the speed and quality of the seeding algorithm, only 3-D space
points are used, either from a pixel hit or a \textit{matched} strip hit.  Matched strip hits are obtained
from the closely-spaced double strip layers, which are composed of two sensors mounted back-to-back,
one providing an $r\phi$ view and one providing a stereo view (rotated by 100\mrad relative to the other,
in the plane of the sensor).  The `$r\phi$' and `stereo hits' in such a layer are combined into a
matched hit, which provides a 3-D position measurement.  Table~\ref{tab:IterativeSeeds} shows the seeding requirements for
each of the six tracking iterations. The seeding layers listed in this table are defined as follows:

\begin{itemize}
\item{Pixel triplets are seeds produced from three pixel hits.
    These seeds are used to find most of the tracks corresponding to promptly produced charged
    particles.  The three precise 3-D space points provide seeds of high quality and with well-measured
    starting trajectories.  A mild constraint on the compatibility of these trajectories with the centre of the beam spot is employed, to remove seeds inconsistent with promptly produced particles.  Also, the
    charge distribution of each pixel hit is
    required to be compatible with that expected for the crossing angle of the seed trajectory and the
    corresponding sensor.}
\item{Mixed pairs with vertex constraint are seeds that use two hits and a third
    space-point given by the location of a pixel vertex. If more than one pixel vertex is found in
    an event, which often happens because of pileup, all are considered in turn. The pixel vertices are required to pass
    quality criteria; the most important is that a vertex must contain at least four
    pixel tracks.  The two hits used for these seeds can be provided by the pixel tracker, or by the two inner
    rings of the three inner TEC layers, where the TEC layers are used to increase coverage in the
    very forward regions.}
\item{Mixed triplets are seeds produced from three hits formed from a combination of pixel
    hits and matched strip hits. Each triplet contains between one and three pixel hits and $< 3$
    strip hits.  This iteration is implemented for finding displaced tracks and prompt tracks that do not
    have three hits in the pixel detector.
    The beam spot related constraint is less restrictive, providing higher efficiency for finding
    tracks arising from decays of hadrons containing s, c, or b quarks, photon conversions, and
    nuclear interactions.}
\item{Strip pairs are seeds constructed using two matched hits from the strip detector.
    Iteration 4 uses the two inner TIB layers and rings 1--2 of the TID/TEC, which are the same
    strip layers used in Iteration 3. In Iteration 5, hits from the two inner TOB layers and ring 5
    of the TEC are used for seeds.  These two iterations have even weaker constraints on the
    compatibility of the seed trajectory with the centre of the beam spot than has
    Iteration 3, and they do not require pixel hits.  These iterations are therefore useful for
    finding tracks produced outside of the pixel detector volume or tracks that do not leave hits in the
    pixel detector.}
\end{itemize}

\begin{table}[htbp]
\centering
  \topcaption{\label{tab:IterativeSeeds} The configuration of the track seeding for each of the six
    iterative tracking steps. Shown are the layers used to seed the tracks, as well as the requirements on the minimum \pt and
    the maximum transverse ($d_0$) and longitudinal ($z_0$) impact parameters relative to the centre of the beam spot.
    The Gaussian standard deviation corresponding to the length of the beam spot along the $z$-direction is $\sigma$.  The asterisk symbol
    indicates that the longitudinal impact parameter is calculated relative to a pixel vertex
    instead of to the centre of the beam spot.}
\begin{tabular}{cclll}
\hline
Iteration & Seeding layers & \multicolumn{1}{c}{\pt (\GeV)} & \multicolumn{1}{c}{$d_0$ (cm)} & \multicolumn{1}{c}{$\abs{z_0}$} \\
\hline
    0     & Pixel triplets		&      $>$0.8         &    $<$0.2      &   ${<}3\sigma$   \\
    1     & Mixed pairs with vertex&   $>$0.6         &    $<$0.2      &   ${<}0.2\cm^*$  \\
    2     & Pixel triplets		&      $>$0.075       &    $<$0.2      &   ${<} 3.3\sigma$  \\
    3     & Mixed triplets   	&      $>$0.35        &    $<$1.2      &   ${<}10$\cm     \\
    4     & TIB 1+2 \& TID/TEC ring 1+2 &$>$0.5       &    $<$2.0      &   ${<}10$\cm     \\
    5     & TOB 1+2 \& TEC ring 5 &    $>$0.6         &    $<$5.0      &   ${<}30$\cm     \\
\hline
\end{tabular}
\end{table}

\subsection{Track finding}
\label{sec:TrackFinding}

The track-finding module of the CTF algorithm is based on the Kalman filter
method~\cite{Fruhwirth:1987fm,Billoir:1989mh,Billoir:1990we,Mankel:1997dy}.
The filter
begins with a coarse estimate of the track parameters provided by the trajectory seed, and then
builds track candidates by adding hits from successive detector layers, updating the
parameters at each layer.  The information needed at each layer includes the location and
uncertainty of the detected hits, as well as the amount of material crossed, which is used to estimate
the effects of multiple Coulomb scattering and energy loss.  The track finding is implemented in
the four steps listed below.

The first step (navigation) uses the parameters of the track candidate, evaluated at the current layer,
to determine which adjacent layers of the detector can be intersected through an extrapolation of the
trajectory, taking into account the current uncertainty in that trajectory.
The navigation service can be configured to propagate along or opposite to the momentum vector, and
uses a fast \textit{analytical propagator} to find the intercepted layers.  The analytical propagator assumes a
uniform magnetic field, and does not include effects of multiple Coulomb scattering or energy
loss. With these assumptions, the track trajectory is a perfect helix, and the propagator can therefore
extrapolate the trajectory from one layer to the next using rapid analytical calculations.
In the barrel, the cylindrical geometry makes navigation particularly easy, since the extrapolated
trajectory can only intercept the layer adjacent to the current one.  In the endcap and
barrel-endcap transition regions, navigation is more complex, as the crossing from one layer
does not uniquely define the next one.

The second step involves a search for compatible silicon modules in the layers returned by the
navigation step.  A module is considered compatible with the trajectory if the position at which
the trajectory intercepts the module surface is no more than some given number (currently three) of
standard deviations outside the module boundary.  The propagation of the trajectory parameters, and of the
corresponding uncertainties, to the sensor surface involves mathematical operations and routines that are
generally quite time-consuming~\cite{Strandlie:927379}.  Hence, the code responsible for searching for
compatible modules has been optimized to limit the number of sensors that are
considered, while preserving an efficiency of $>$99\% in finding the relevant
sensors. A complication is that the design of the CMS tracker is such that sensors
often slightly overlap their neighbours,
meaning that a particle can cross two sensors in the same layer. This possibility is accommodated by
dividing the compatible modules in each layer into groups of mutually exclusive modules,
defined such that if a particle passes through one member of a group, it is not physically possible for
it to pass through a second member of the same group. Any two modules that have some overlap are not mutually
exclusive, and are therefore assigned to different groups. This feature is used in the third and fourth
steps of the track finding, described next.

The third step forms groups of hits, each of which is defined by the collection of all the hits from one of the
module groups.  A configurable parameter provides the possibility of adding a ghost hit
to represent the possibility that the particle failed to produce a hit in the module group, for example,
as a result of module inefficiency.  The hit positions and uncertainties are refined using the trajectory
direction on the sensor surface, to calculate more accurately the Lorentz drift of the ionization-charge
carriers inside the silicon bulk.  A $\chi^2$ test is used to check which of the hits are
compatible with the extrapolated trajectory. The current (configurable) requirement is
$\chi^2<30$ for one degree of freedom (dof).  The $\chi^2$ calculation takes into account both the hit and trajectory uncertainties.
In the endcap regions and the barrel-endcap transition regions, the extrapolation distances and the
amount of material traversed are generally greater,
with correspondingly larger uncertainties in the trajectory, and the probability
of finding spurious hits compatible with the track tends therefore to be greater.

The fourth and last step is to update the trajectories. From each of the original track candidates,
new track candidates are formed by adding exactly one of the compatible hits from each module grouping
(where this hit may be a ghost hit).
As the modules in a given group are mutually exclusive,
it would not be expected that a track would have more than one hit contributing from each group.
The trajectory parameters for each new candidate are then
updated at the location of the module surface, by combining the information from the added hits with the extrapolated
trajectory of the original track candidate.

For the above second, third, and fourth steps of the procedure, a more accurate \textit{material propagator}
is used when extrapolating the track trajectory, which includes the effect of the material in the tracker.  This differs from the method of the
simple analytical propagator, in that it increases the uncertainty in the trajectory parameters according
to the predicted \rms scattering angle in the tracker material. It also adjusts the momentum of
the trajectory by the predicted mean energy loss of the Bethe--Bloch equation. Since all
detector material is assumed to be concentrated in the detector layers, the track propagates along a simple helix
between the layers, allowing the material propagator to extrapolate the track analytically.  The ghost
hits include the effect of material without providing position information to the propagator.

All resulting track candidates found at each layer are then propagated to the next compatible
layers, and the procedure is repeated until a termination condition is satisfied.  However, to avoid a rapid
increase in the number of candidates, only a limited number (default is 5) of the candidates
are retained at each step, with the best candidates chosen based on the normalized $\chi^2$ and a bonus given for
each valid hit, and a penalty for each ghost hit.  The standard termination conditions are if a
track reaches the end of the tracker or contains too many missing hits (limit is $N_\text{lost}$),
or if its \pt drops below a user specified value.
The number of missing hits on a track is equal
to the number of ghost hits, except that hits not found due to attributable known detector
conditions, for example, if a detector module is turned off, are not counted.  The building of a
trajectory can also be terminated when the uncertainty in its parameters falls below a given
threshold or the number of hits is above a threshold; these kinds of termination conditions tend to
be used only in the high-level trigger (HLT), where the required accuracy on track parameters
is often reached after 5 or 6 hits are added to the track candidate, and the continuation of the
track building would correspond to a waste of CPU time.

When the search for hits in the outward direction reveals a minimum number of valid hits
($N_\text{rebuild}$), an inwards search is initiated for
additional hits. Otherwise, the track candidate remains as formed.
The inwards search starts by taking all of the hits assigned to the track, excluding those belonging
to the track seed, and using them to fit the track trajectory.  In case this exclusion of the seeding hits leaves
fewer than $N_\text{rebuild}$ hits to fit, some of the seeding hits are also used
(taking first the outer contributions) so as to obtain at least $N_\text{rebuild}$ hits.
Then, as in the outward track building, the
trajectory is propagated inwards through the seeding layers and then further, until the inner
edge of the tracker is reached or too many ghost hits are found.  There are three reasons for
this inward search.  First, additional hits can be found in the seeding layers (for example,
from overlapping sensors). Second, hits can be found in layers closer to the interaction region
than the seeding layers. Third, when strip layers are used in seeding, matched hits are used to
increase computational speed and reduce the combinations of hits available for seeding.  However,
some $r\phi$ or stereo hits are not part of any matched hit.  While these hits are
not available during seeding, they can be found during the inward track building process.
The effect of the inward search is an increase in the mean number of hits per track by 0.15,
(\ie, a 1\% increase relative to a total of $\approx$14 hits), which translates to a better
signal-to-background ratio, impact parameter resolution, and \pt resolution, with maximum improvements of
2\%, 1\%, and 0.5\%, respectively.

The track of a single charged particle can be reconstructed more than once, either starting from
different seeds, or when a given seed develops into more than one track candidate.  To remedy this
feature, a trajectory cleaner is applied after all the track candidates in a given iteration
have been found.  The trajectory cleaner calculates the fraction of shared hits between two track
candidates:
$f_\text{shared} = \frac{N^\text{hits}_\text{shared}}{\min(N^\text{hits}_1,N^\text{hits}_2)}$ where $N^\text{hits}_1$
and $N^\text{hits}_2$ are, respectively, the number of hits used in forming the first (second) track candidate.  If this fraction
exceeds the (configurable) value of 19\% (determined empirically), the trajectory cleaner removes the track with the
fewest hits; if both tracks have the same number of hits, the track with the largest
$\chi^2$ value is discarded.  The procedure is repeated iteratively on all pairs of
track candidates.  The same algorithm is applied when tracks from the six iterations are
combined into a single track collection.

The requirements applied during the track-finding stage are shown in Table~\ref{tab:IterativeFinding}
for each tracking iteration.  In addition to the requirement on $N_\text{lost}$,
the completed track candidates must also pass requirements on the minimum number of hits $(N_\text{hits})$
and minimum track \pt.
The minimum \pt requirements have very little effect, as they are
weaker than those applied to the seeds, given in Table~\ref{tab:IterativeSeeds}.  Since the later iterations
do not have strong requirements that the tracks originate close to the centre of the beam spot, the probability of
random hits forming tracks increases, which leads to more fake tracks and greater usage of CPU time.  To
compensate for this tendency, the criteria for the minimum number of hits, and maximum number of lost hits,
are tightened in the later iterations.

\begin{table}[htbp]
\centering
\topcaption{\label{tab:IterativeFinding} Selection requirements applied to track candidates
during the six iterative steps of track finding, the minimum \pt,
the minimum number of hits $N_\text{hits}$, and the maximum number of missing hits
$N_\text{lost}$. Also shown is the minimum number of hits needed to be found in the outward
track building step to trigger the inward track building step $N_\text{rebuild}$,
although candidates failing this requirement are not rejected.}
\begin{tabular}{ccccc}
\hline
Iteration & \pt (\GeVns) & $N_\text{hits}$ & $N_\text{lost}$ & $N_\text{rebuild}$ \\
\hline
    0      &       0.3     &    3    & 	 1    &   5    \\
    1      &       0.3     &    3    & 	 1    &   5    \\
    2      &       0.1     &    3    & 	 1    &   5    \\
    3      &       0.1     &    4    & 	 0    &   5    \\
    4      &       0.1     &    7    & 	 0    &   5    \\
    5      &       0.1     &    7    & 	 0    &   4    \\
\hline
\end{tabular}
\end{table}

\subsection{Track fitting}
\label{sec:TrackFit}

For each trajectory, the track-finding stage yields a collection of hits and an estimate of
the track parameters. However, the full information about the trajectory is only available at the final hit of the
trajectory (when all hits are known). Furthermore, the estimate can be biased by constraints, such as a beam spot constraint
applied to the trajectory during the seeding stage. The trajectory is therefore refitted using a
Kalman filter and smoother.

The Kalman filter is initialized at the location of the innermost hit, with the trajectory estimate obtained
by performing a Kalman filter fit to the innermost hits (typically four) on the track.
The corresponding covariance matrix is scaled up by a large factor (10 for the last iteration and 100 for the
other iterations) in order to
limit the bias.  The fit then proceeds in an iterative way through the full list of hits, from the inside
outwards, updating the track trajectory estimate sequentially with each hit. For each
valid hit, the estimated hit position uncertainty is reevaluated
using the current values of the track parameters. In the case of pixel hits, the estimated hit
position is also reevaluated.
This first filter is followed by the smoothing stage, whereby a second filter is initialized with the
result of the first one (except for the covariance matrix, which is scaled by a large factor),
and is run backward towards the beam-line.
The track parameters at the surface associated with any of its hits, can then be obtained from the
weighted average of the track parameters of these two filters, evaluated on this same surface, as
one filter uses information from all the hits found before, and the other uses information from all the hits
found after the surface.  This provides the optimal track parameters at any point, including the
innermost and outermost hit on the track, which are used to extrapolate the trajectory to the
interaction region and to the calorimeter and muon detectors, respectively.  A configurable parameter
determines whether the silicon strip matched hits are used as is or split into their component
$r\phi$ and stereo hits.  For the standard offline reconstruction, the split hits are used to improve
the track resolution, while for the HLT, the matched hits are used to improve speed.

To obtain the best precision, this filtering and smoothing procedure uses a \textit{Runge--Kutta propagator}
to extrapolate the trajectory from one hit to the next.
This not only takes into account the effect of material, but it also accommodates an inhomogeneous
magnetic field. The latter means that the particle may not move along a perfect helix, and its equations
of motion in the magnetic field must therefore be solved numerically. To do so, the Runge--Kutta
propagator divides the distance to be extrapolated into many small
steps. It extrapolates the track trajectory over each of these steps in turn, using
a well-known mathematical technique for solving first-order differential equations, called the fourth-order
Runge--Kutta method, so called because it is accurate to fourth order in the step size. The optimal step size is chosen
automatically, according to how non-linear the problem is. This automatic determination of step size
employs the method \cite{Cash:1990:VOR:79505.79507},
which is based on how well the fourth and fifth order Runge--Kutta predictions agree with
each other. Use of the Runge--Kutta propagator is most important in the region $\abs{\eta} > 1$, where
the magnetic field inhomogeneities are greatest. For example, in this region, tracks
fitted using the simple material propagator are biased by up to 1\% for particles with $\pt = 10\GeV$.
This bias is almost completely eliminated when using the Runge--Kutta propagator. To assure an accurate
extrapolation of the track trajectory, the Runge--Kutta propagator uses a detailed map of the magnetic
field, which was measured before LHC collisions to a precision of $<0.01\%$.

Estimates of the track trajectory at any other points, such as the point of closest approach to the beam-line,
can be obtained by extrapolating the trajectory evaluated at the nearest hit to that very point. This extrapolation
also uses the Runge--Kutta propagator.

After filtering and smoothing, a search is made for spurious hits (outliers), incorrectly
associated to the track.  Such hits can be related to an otherwise well-defined track,
\eg, from $\delta$-rays, or unrelated, such as hits from nearby tracks or electronic noise.
Two methods are used to find outliers.  One uses the
measured residual between a hit and the track to reject hits whose $\chi^2$ compatibility with the track
exceeds a
configurable threshold (20 for Iterations 0--4 and 30 for Iteration 5).  While a $\chi^2$ requirement
of 30 on each hit is already applied during track finding, the outlier rejection criterion provides a more
powerful restriction as it uses information from the full fit~\cite{Fruhwirth:1987fm}.  The other method
calculates a probability that a pixel hit is consistent with the track, taking into account the
charge distribution of the pixel hit, which generally comprises several pixel channels.  This probability
corresponds to the $\chi^2$ defined in Eq.~(\ref{eq:fulleq}). After removing the outlier,
the track is again filtered and smoothed and another check for outliers is made.  This continues
until no more outliers are found.  In cases where removing an outlier results in two consecutive ghost
hits, the track is terminated and the remaining outer hits discarded (although not used, a configurable parameter is available
to allow the track fitting to continue).  If a track is found to have less than three hits after outlier
rejection or for the track fitting to fail, the track is discarded (although not used, a configurable parameter
is available to return the original track).

The default value of 20 for the $\chi^2$ requirement is chosen to reject a significant fraction of
outliers, while removing few genuine hits.  With this value, approximately 20\% of the spurious outliers
are removed from tracks reconstructed in high-density dijet events, whereas ${<} 0.2\%$ of
the good hits are removed.

\subsection{Track selection}
\label{sec:TrackSelection}

In a typical LHC event containing jets, the track-finding procedure described above yields a
significant fraction of fake tracks, where a fake track is defined as a reconstructed track not
associated with a charged
particle, as defined in Section~\ref{sec:trackPerformance}.  The fake
rate (fraction of reconstructed tracks that are fake) can be reduced substantially through
quality requirements. Tracks are selected on the basis of the number of layers that have hits, whether
their fit yielded a good $\chi^2/\mathrm{dof}$, and how compatible they are with originating from
a primary interaction vertex.  If several primary vertices are present in the event,
as often happens due to pileup, all are considered. To optimize the performance, several
requirements are imposed as a function of the track $\eta$ and \pt, and on the number of layers
$(N_\text{layers})$ with an
assigned hit (where a layer with both $r\phi$ and stereo strip modules is counted as a single layer). The selection criteria are as follows.

\begin{itemize}
\itemsep 0pt
\item A requirement on the minimum number of layers in which the track has at least one associated hit. This differs from
selections based on the number of hits on the track, because more than one hit in a given layer can be assigned to a track,
as in the case of layers with overlapping sensors or double-sided layers in which two sensors are mounted back-to-back.
\item A requirement on the minimum number of layers in which the track has an associated 3-D hit (\ie, in the pixel
tracker or matched hits in the strip tracker).
\item A requirement on the maximum number of layers intercepted by the track containing no assigned hits, not
counting those layers inside its innermost hit or outside its outermost hit, nor those layers where no hit was expected because
the module was known to be malfunctioning.
\item $\chi^2/\mathrm{dof} < \alpha_0 N_\text{layers}$.
\item $\abs{d_0^\mathrm{BS}}/\delta d_0 < \left( \alpha_3 N_\text{layers}\right)^\beta$.
\item $\abs{z_0^\mathrm{PV}}/\delta z_0 < \left( \alpha_4 N_\text{layers}\right)^\beta$.
\item $\abs{d_0^\mathrm{BS}}/\sigma_{d_0}(\pt) < \left( \alpha_1 N_\text{layers}\right)^\beta$.
\item $\abs{z_0^\mathrm{PV}}/\sigma_{z_0}(\pt,\eta) < \left( \alpha_2 N_\text{layers}\right)^\beta$.
\end{itemize}

The parameters $\alpha_i$ and $\beta$ are configurable constants. The track's impact parameters
are $d_0^\mathrm{BS}$ and $z_0^\mathrm{PV}$, where $d_0^\mathrm{BS}$ is the distance from the centre of the
beam spot in the plane transverse to the beam-line and $z_0^\mathrm{PV}$ is the distance along the beam-line from the closest
pixel vertex. These
pixel vertices, described in Section~\ref{sec:pixeltrackvertex}, are required to have at least three
pixel tracks and if no pixel vertices meet this requirement, then $z_0^\mathrm{PV}$ is required to be within 3$\sigma$ of
the $z$-position of the centre of the beam spot, where $\sigma$ is the Gaussian standard deviation corresponding to the length of the beam spot in the $z$-direction.
The above selection criteria include requirements on the transverse $\abs{d_0^\mathrm{BS}}/\delta d_0$ and
longitudinal $\abs{z_0^\mathrm{PV}}/\delta z_0$ impact parameter significances of the track, where
the impact parameter uncertainties, $\delta d_0$ and $\delta z_0$, are calculated from the covariance
matrix of the fitted track trajectory.  A second pair of requirements is also imposed on these
significances, but calculated differently, with
the uncertainties in the impact parameters being parametrized in terms of \pt and polar angle of the track:
$\sigma(d_0) = \sigma(z_0\sin\theta) = a \oplus \frac{b}{\pt}$, where $\oplus$ represents the
sum in quadrature and $a$ and $b$ are
 parameters. Their nominal values are $a=30\mum$ and $b=10\mum\GeV$, but
 $b$ increases to $100\mum\GeV$ for the
\textit{loose} and \textit{tight} selection criteria used (and defined below) in Iterations 0 and 1.

The fraction of fake tracks decreases roughly exponentially as a function of the
number of layers in which the track has associated hits: $dN_{\rm fake}/dN_\text{layers} \sim \exp(-\omega N_\text{layers}) $, with
$\omega$ in the range 0.9--1.3 depending on the \pt of the track.  As a consequence, weaker selection criteria can be applied for
tracks having many hit layers, which is the reason for the chosen selection criteria.
For tracks with hits in at least 10 layers, the selection requirements on $\chi^2$ and impact parameters
are found to reject no tracks. However, the criteria become far more stringent for tracks with relatively
few hit layers.

The above quality criteria were initially optimized as a function of track \pt and $N_\text{layers}$,
so as to maximize the quality $Q(\rho) = {s}/{\sqrt{s+\rho b}}$, where $s$ is the
number of selected genuine (non-fake) tracks, $b$ is the number of selected fake tracks and $\rho\simeq 10$
inflates the importance of the fake tracks to achieve low fake rates (below 1\% for \PYTHIA QCD events with
$\hat{p}_{\mathrm{T}}$ of the two outgoing partons in the range 170--230\GeV).  As data taking conditions have evolved,
the parameters have been adjusted to maintain high efficiency and low fake rate.

The track selection criteria for each iteration are given in Table~\ref{tab:TrackSelection}.
The \textit{loose} criteria denote the minimum requirements for a track to be kept in the general track collection.
The \textit{tight} and \textit{high-purity} criteria provide progressively more stringent requirements, which reduce the
efficiency and fake rate.
In general,
high-purity tracks are used for scientific analysis, although in cases where efficiency is essential and purity is not a major
concern, the loose tracks can be used.  The criteria for the initial tracking iterations emphasise compatibility
with originating from a primary vertex as a means of assuring quality, while the criteria used for the later iterations
rely on other measures of track quality such as fit $\chi^2$ and the number of hits,
ensuring thereby that they are still useful for selecting displaced tracks.  This matches the seeding and
track-finding requirements shown in Tables~\ref{tab:IterativeSeeds}--\ref{tab:IterativeFinding}, and is aligned with the
goals for the six iterations.

After the track selection is complete, the tracks found by each of the six iterations are
merged into a single collection.

\begin{table}[htbp]
\centering
\topcaption{\label{tab:TrackSelection} Parameter values used in selecting tracks reconstructed by each
of the six iterative tracking steps.  The first table shows the three requirements on the number
of layers that contain hits assigned to tracks and the parameter $\alpha_0$
that controls selection criteria based on $\chi^2/\mathrm{dof}$. The second table shows the parameters $\alpha_i$ and $\beta$
that define compatibility of impact parameters with the interaction point.  Each parameter has
three entries, corresponding to the loose (L), tight (T), and high-purity (H) selection requirements.
Iterations 2 and 3 use two paths that emphasise track quality (Trk) or primary-vertex compatibility (Vtx).
A track produced by these iterations is retained if it passes either of these criteria.
}

\begin{tabular}{c|ccc|ccc|ccc|ccc}
\hline
\multicolumn{1}{c|}{\multirow{2}{*}{Iteration}} & \multicolumn{3}{c|}{Min layers} & \multicolumn{3}{c|}{Min 3-D layers} & \multicolumn{3}{c|}{Max lost layers} & \multicolumn{3}{c}{$\alpha_0$} \\\cline{2-13}
          & \multicolumn{1}{c}{~~L~~} &\multicolumn{1}{c}{~~T~~} &\multicolumn{1}{c|}{~~H~~} & \multicolumn{1}{c}{~~L~~} &\multicolumn{1}{c}{~~T~~} &\multicolumn{1}{c}{~~H~~}
& \multicolumn{1}{c}{~~L~~} &\multicolumn{1}{c}{~~T~~} &\multicolumn{1}{c|}{~~H~~} & \multicolumn{1}{c}{~~L~~} &\multicolumn{1}{c}{~~T~~} &\multicolumn{1}{c}{~~H~~}
\\
\hline
    0 \& 1 &  0 & 3 & 4  &  0 & 3 & 4   &  $\infty$ & 2 & 2  & 2.0 & 0.9 & 0.9     \\
    2 Trk  &  4 & 5 & 5  &  0 & 3 & 3   &  $\infty$ & 1 & 1  & 0.9 & 0.7 & 0.5     \\
    2 Vtx  &  3 & 3 & 3  &  0 & 3 & 3   &  $\infty$ & 1 & 1  & 2.0 & 0.9 & 0.9     \\
    3 Trk  &  4 & 5 & 5  &  2 & 3 & 4   &  1 & 1 & 1         & 0.9 & 0.7 & 0.5    \\
    3 Vtx  &  3 & 3 & 3  &  2 & 3 & 3   &  1 & 1 & 1         & 2.0 & 0.9 & 0.9     \\
    4      &  5 & 5 & 6  &  3 & 3 & 3   &  1 & 0 & 0         & 0.6 & 0.4 & 0.3    \\
    5      &  6 & 6 & 6  &  2 & 2 & 2   &  1 & 0 & 0         & 0.6 & 0.35 & 0.25   \\
\hline
\end{tabular}

\vspace{10pt}

\begin{tabular}{@{~}c@{~}@{~}c@{~}|ccc|ccc|ccc|ccc}
\hline
\multicolumn{1}{c}{\multirow{2}{*}{Iteration}} & \multicolumn{1}{c|}{\multirow{2}{*}{$\beta$}} & \multicolumn{3}{c|}{$\alpha_1$} & \multicolumn{3}{c|}{$\alpha_2$} & \multicolumn{3}{c|}{$\alpha_3$} & \multicolumn{3}{c}{$\alpha_4$} \\\cline{3-14}
        &         & \multicolumn{1}{c}{~~L~~} &\multicolumn{1}{c}{~~T~~} &\multicolumn{1}{c|}{~~H~~} & \multicolumn{1}{c}{~~L~~} &\multicolumn{1}{c}{~~T~~} &\multicolumn{1}{c|}{~~H~~}
& \multicolumn{1}{c}{~~L~~} &\multicolumn{1}{c}{~~T~~} &\multicolumn{1}{c|}{~~H~~} & \multicolumn{1}{c}{~~L~~} &\multicolumn{1}{c}{~~T~~} &\multicolumn{1}{c}{~~H~~} \\
\hline
    0 \& 1 &  4  & 0.55 & 0.30 & 0.30   &  0.65 & 0.35 & 0.35  &  0.55 & 0.40 & 0.40   &  0.45 & 0.40 & 0.40   \\
    2 Trk  &  4  & 1.50 & 1.00 & 0.90   &  1.50 & 1.00 & 0.90  &  1.50 & 1.00 & 0.90   &  1.50 & 1.00 & 0.90   \\
    2 Vtx  &  3  & 1.20 & 0.95 & 0.85   &  1.20 & 0.90 & 0.80  &  1.30 & 1.00 & 0.90   &  1.30 & 1.00 & 0.90   \\
    3 Trk  &  4  & 1.80 & 1.10 & 1.00   &  1.80 & 1.10 & 1.00  &  1.80 & 1.10 & 1.00   &  1.80 & 1.10 & 1.00   \\
    3 Vtx  &  3  & 1.20 & 1.00 & 0.90   &  1.20 & 1.00 & 0.90  &  1.30 & 1.10 & 1.00   &  1.30 & 1.10 & 1.00   \\
    4      &  3  & 1.50 & 1.20 & 1.00   &  1.50 & 1.20 & 1.00  &  1.50 & 1.20 & 1.00   &  1.50 & 1.20 & 1.00   \\
    5      &  3  & 1.80 & 1.30 & 1.20   &  1.50 & 1.20 & 1.10  &  1.80 & 1.30 & 1.20   &  1.50 & 1.20 & 1.10   \\
\hline
\end{tabular}
\end{table}

\subsection{Specialized tracking}

The track reconstruction described above produces the main track collection used by the CMS collaboration.
However, variants of this software are also used for more specialized purposes, as described in
this section.

\subsubsection{Electron track reconstruction}
\label{sec:GSF}

Electrons, being charged particles, can be reconstructed through the standard track reconstruction.
However, as electrons lose energy primarily through bremsstrahlung, rather than ionization, large energy
losses are common.
For example, about 35\% of electrons radiate more than 70\% of their initial energy before reaching
the electromagnetic calorimeter (ECAL)
that surrounds the tracker. The energy loss
distribution is highly non-Gaussian, and therefore the standard Kalman filter, which is optimal when all
variables have Gaussian uncertainties, is not appropriate. As a result, the
efficiency and resolution of the standard tracking are not particularly good for electrons and therefore
electron candidates are reconstructed using a combination of two techniques that make use of information,
not only from the tracker, but also from the ECAL.  As this is a subject beyond the scope of this paper,
only a brief description of these methods is given.

The first method \cite{Baffioni:2006cd} starts by searching for clusters
of energy in the ECAL.  The curvature of electrons in the strong CMS magnetic field means that
bremsstrahlung photons emitted by the electrons will, in general, strike the ECAL at $\eta$ values similar
to that of the electron, but at different azimuthal coordinates $(\phi)$. To recover this radiated energy,
ECAL \textit{superclusters} are formed, by merging clusters of similar $\eta$ over some range of $\phi$. The
knowledge of the energy and position of each supercluster, and the assumption that the electron originated near the centre of the beam spot,
constrains the trajectory of the electron through the tracker (aside from a two-fold
ambiguity introduced by its unknown charge). Tracker seeds compatible with this trajectory are sought
in the pixel tracker (and also in the TEC to improve efficiency in the forward region).

The second method \cite{CMS_PAS_PFT-10-003}
takes the standard track collection (excluding tracks found by Iteration 5, as described in Table~\ref{tab:IterativeSeeds})
and attempts to identify a subset of these tracks that are
compatible with being electrons. Electrons that suffer only little bremsstrahlung loss can be identified by searching
for tracks extrapolated to the ECAL that pass close to an ECAL cluster. Electrons that suffer large
bremsstrahlung loss can be identified by the fact that the fitted track will often have poor $\chi^2$ or few associated hits.
The track seeds originally used to generate these electron-like tracks are retained.

The seed collections obtained by using these two methods are merged, and used to initiate electron track finding. This procedure
is similar to that used in standard tracking, except that the $\chi^2$ threshold, used by the Kalman filter to decide whether a
hit is compatible with a trajectory, is weakened from 30 to 2000. This is to accommodate tracks that deviate from their
expected trajectory because of bremsstrahlung. In addition, the penalties assigned to track candidates for passing through
a tracker layer without being assigned a hit are adjusted. This is necessary because bremsstrahlung photons can
convert into $\Pep\Pem$ pairs with the track-finding algorithm
incorrectly forming a track by combining hits from the primary electron with one of the conversion electrons.

To obtain the best parameter estimates, the final track fit is performed using a
modified version of the Kalman filter, called the Gaussian Sum Filter (GSF) \cite{Adam:2003kg}.
In essence, the fractional energy loss of an electron, as it traverses material of a given thickness, is expected
to have a distribution described by the Bethe--Heitler formula. This distribution is non-Gaussian,
making it unsuitable for use in a conventional Kalman filter algorithm. The GSF technique solves this by
approximating the Bethe--Heitler energy-loss distribution as the sum of several Gaussian functions, whose means, widths, and
relative amplitudes are chosen so as to optimize this approximation.
The parameters of these Gaussian energy-loss functions are determined only once.
Each track trajectory is also represented by a mixture of several `trajectory components', where each trajectory component has
helix parameters with Gaussian uncertainties, and a `weight' corresponding to the probability that it correctly describes
the true path of the particle. Initially, a track trajectory is described by only a single such trajectory component, derived from
the track seed.
When propagating a trajectory component through a layer of material in the tracker, the
estimated mean energy of the trajectory component is reduced and its uncertainty increased, according to the
mean and width of each Gaussian component of the energy-loss distribution applied independently, in turn, to the original trajectory component.
Thus after passing through the a layer of material, each original trajectory component gives rise to several new trajectory components,
each one obtained using one of the Gaussian energy-loss functions. The weight of each new trajectory component is given by
the product of the weight of the original trajectory component and the weight of the corresponding Gaussian component of the energy-loss distribution.
To avoid an exponential explosion in the number of
trajectory components being followed, as the track candidate is propagated through successive tracker layers, the less probable
trajectory components are dropped or merged (by grouping together similar trajectory components), so as to limit their number to 12.
Each trajectory component will also be updated by the Kalman filter if an additional hit is assigned to it when passing through a layer.
When this happens, the weight of the trajectory component is further adjusted according to its compatibility with the hit.

The GSF fit provides estimates of the track parameters, whose uncertainties are described not by
a single Gaussian distribution, but instead by the sum of several Gaussian distributions, each corresponding to the uncertainty
on one of the trajectory components that make up the track.
For each parameter, the mode of this distribution is used as it
is found to provide the best estimates of the parameters.

The performance of the GSF electron tracking has been studied both with simulations \cite{Adam:2003kg}
and with data \cite{CMS_PAS_EGM-10-004}, with good agreement observed between the two.

\subsubsection{Track reconstruction in the high-level trigger}

The CMS high-level trigger (HLT) \cite{LHCC_2007_021} uses a processor farm running C++ software to achieve
large reductions in data rate. The HLT filters events selected at rates of up to 100\unit{kHz} using the
{Level-1} (hardware) trigger. Whereas {Level-1} uses information only from the CMS calorimeters and
muon detectors, the HLT is also able to capture information from the tracker, thereby adding the powerful
tool of track reconstruction to the HLT\@.  Some examples of how this improves the HLT performance are
listed below.

\begin{itemize}
\item
Requiring muon candidates that are reconstructed in the muon detectors to be confirmed
through the presence of a corresponding track in the tracker greatly reduces the false reconstruction rate and substantially
improves momentum resolution.
\item
Energy clusters found in the electromagnetic calorimeters can be identified as electrons or
photons through the presence of a track of appropriate momentum pointing to the cluster.
\item
The background rejection rate for lepton triggers can be enhanced by requiring leptons to be isolated. One method
of doing this is to use a veto on the presence of (too many) tracks in a cone around the lepton direction.
\item
Triggering on jets produced by b~quarks can be done by counting the number of tracks in a jet that have
transverse impact parameters statistically incompatible with the track originating from the
beam-line.
\item Triggers on $\tau$ decays $\tau \to \ell \nu_\ell \nu_\tau$, where $\ell =\Pe$ or $\mu$, can
be extended to $\tau \to h \nu_\tau$ decays, where $h$ represents one or more charged hadrons, by
reconstructing a narrow, isolated jet using tracks in combination with calorimeter information.
\end{itemize}

The HLT uses track reconstruction software that is identical to that used for offline reconstruction, but
it must run much faster. This is achieved by using a modified configuration of the track reconstruction.

Tracks can be reconstructed from triplets of hits found using only the pixel tracker,
as documented in Sections~\ref{sec:SeedGeneration} and \ref{sec:pixeltrackvertex}. This is extremely fast, and can be used with great effect
in the reconstruction of the primary-vertex position in the HLT, described in Section~\ref{sec:pixeltrackvertex}.

Tracks can also be reconstructed in the HLT using hits from both the pixel and strip detectors. Such
tracks have superior momentum resolution and a lower probability of being fake. However, this requires much more CPU time than just reconstructing pixel tracks, since the strip tracker does not provide the precise 3-D hits of the pixel tracker, and suffers from a higher hit occupancy. This can be mitigated using
some or all of the following techniques (the details vary significantly, depending on the type of
trigger).

\begin{itemize}
\item
Rather than trying to reconstruct all tracks in the event, \textit{regional} track reconstruction can be
performed instead, where the software is used to reconstruct tracks lying within a specified $\eta$-$\phi$ region
around some object of interest (which might be a muon, electron, or jet candidate reconstructed using the
calorimeters or muon detectors). This saves CPU time, and is accomplished by using
regional seeding. This method differs from the track seeding described in Section~\ref{sec:SeedGeneration},
in that it only forms seeds from combinations of hits that are consistent with a track heading
into the desired $\eta$-$\phi$ region. Another important ingredient of regional tracking concerns
the extraction of hits. As discussed in Section~\ref{sec:striphit}, hits are reconstructed after
unpacking the original data blocks produced by the FED readout boards. Significant time is saved
by unpacking only the data from those FED units that read out tracker modules within the
region of interest~\cite{Wingham:2008zz}.  This is not used in the offline reconstruction as the
track reconstruction searches the entire $\eta$-$\phi$ region and therefore needs all hits.
\item
Further gains in speed can be made by performing just a single iteration in the iterative tracking,
such that only seeds made from pairs of pixel hits are considered, where these hits are compatible with a
track originating within a few millimetres of a primary pixel vertex. Furthermore, the HLT uses a higher
\pt requirement when forming the seeds (usually $>$1\GeV) than is used for offline reconstruction.
These stringent requirements on track impact parameter and \pt reduce the number of seeds, and thereby the amount of
time spent building track candidates.
\item
Track finding can differ from that described in Section~\ref{sec:TrackFinding}, in that it can rely on
partial track reconstruction. With this technique, the building of each track candidate is stopped
once a specific condition is met, for example, a given minimum number of hits (typically eight), or a certain
precision requirement on the track parameters.  As a consequence,
the hits in the outermost layers of the tracker tend not to be used. While such partially
reconstructed tracks will have slightly poorer momentum resolution and higher fake rates than fully reconstructed
tracks, they also take less CPU time to construct.
\item
Other changes in the tracking configuration can further enhance the speed of reconstruction. For example, when building
track candidates from a given seed, the offline track reconstruction retains at most the best five
partially reconstructed candidates for extrapolating to the next layer. Changing this configurable
parameter to retain fewer candidates can save CPU time.
\end{itemize}

Pixel tracking and other aspects of track reconstruction absorb about 20\% of the total HLT CPU
time.  This is kept low by performing track reconstruction only when necessary, and only after other
requirements have been satisfied, so as to reduce the rate at which tracking must be performed.
Track reconstruction is employed in a variety of ways to satisfy different needs in the HLT\@.
Examples of track reconstruction at the HLT include seeds originating in the muon detector, tracking
in a specific $\eta$-$\phi$ region defined by a jet, and searching for tracks over the full
detector.  Even the most comprehensive (and slowest) track reconstruction configuration at the HLT
is more than ten times faster than the offline reconstruction of tracks in events representative of
data taken in 2011 (\ttbar + 10 pileup events).

\section{Track reconstruction performance}
\label{sec:trackPerformance}

In this section, the performance of the CTF tracking algorithm is evaluated
in terms of tracking efficiency and fake rate, track parameter resolutions, and the CPU time
required for processing collision events. Two different categories of
simulated samples are used: isolated particles and pp collision events. Comparing the results helps
one understand both the performance of the tracking for isolated particles and to what extent it is
degraded in a high hit occupancy environment.

Simulated events offer the possibility of detailed studies of track reconstruction, such as the way
characteristics of the tracker and the design of the track reconstruction algorithms influence its
performance over a wide range of particle momenta and rapidities, and how much its performance
depends on the type of charged particle being reconstructed, and on whether this particle is isolated or
not. The performance in simulation can be compared with that in data in certain regions of phase space
to verify that the results from simulation are realistic. The CMS collaboration demonstrated previously that its
simulation describes the momentum resolution of muons from \JPsi decay to an accuracy of better than
5\% \cite{CMS_PAS_TRK-10-004}; and does similarly well in describing the dimuon mass resolution of muons from Z boson
decay \cite{Chatrchyan:2013mxa}. The transverse and longitudinal impact parameters of tracks reconstructed in typical
multijet events agree in data and simulation to better than 10\% \cite{CMS_PAS_TRK-10-005}. The CMS collaboration also showed that the
tracking efficiency for particles from \JPsi and charmed hadron decays is simulated with a precision
 better than 5\% \cite{CMS_PAS_TRK-10-002}. A similar comparison for the higher-momentum muons from Z boson decay will
be presented in Section~\ref{subSec:TagAndProbe} of the present work.

The isolated particle samples that are used here consist of simple events with just a single generated muon, pion or electron,
although secondary particles may also be present due to interactions with the detector material.
The single particles are generated with a flat distribution in pseudorapidity inside the tracker acceptance $\abs{\eta} < 2.5$.
Their transverse momenta are either fixed to
1, 10 or 100\GeV, or are generated according to a flat distribution in $\ln(\pt)$.
The former set of particles with fixed momenta is used for studying the tracking performance
as a function of $\eta$, while the latter is used to quantify the performance as a function of \pt.

For pp collisions, simulated inclusive \ttbar events are used,
either with or without superimposed pileup events.
The average number of pileup collisions per LHC bunch crossing depends on the instantaneous
luminosity of the machine and on the period of data-taking over which the luminosity is averaged.
For the sake of simplicity, the number of pileup interactions superimposed on each
simulated \ttbar event is randomly generated from a Poisson distribution with mean equal to 8.
This amount of pileup corresponds roughly to what was delivered by the LHC, when averaged over the whole 2011 running period.
The \ttbar events, and also the minimum-bias events used for the pileup, are generated with
the \PYTHIA6 program~\cite{Pythia6}.

Simulated particles are paired to reconstructed tracks
for evaluating tracking efficiency, fake rate, and other quantities discussed in this section.
A simulated particle is \emph{associated} with a
reconstructed track if at least 75\% of the hits assigned to the reconstructed track originate from the simulated particle. The association of simulated hits with reconstructed hits is possible because
the simulation software records the particles responsible for the signal in each channel of the tracker.
Strip and pixel response to electronic noise is also recorded.
Reconstructed tracks that are not associated with a simulated particle are referred to as \textit{fake} tracks.

Results for tracking efficiencies and for fake rates are presented in Section~\ref{subSec:EfficiencyFakeRate}.
While the latter is evaluated only using simulated samples,
the former is also measured in data as described in Section~\ref{subSec:TagAndProbe}.
The resolution obtained for track parameters is discussed in Section~\ref{subSec:TrackParResolution}. Unless
indicated otherwise, all results pertaining to the performance are obtained using the set of `high-purity' tracks
defined in Section~\ref{sec:TrackSelection}. Finally, Section~\ref{sec:CPU} provides estimates of
the CPU time required for different components of track reconstruction.

\subsection{Tracking efficiency and fake rate}
\label{subSec:EfficiencyFakeRate}

For simulated samples, the tracking efficiency is defined as the fraction of simulated charged particles
that can be associated with corresponding reconstructed tracks,
where the association criterion is the one described at the beginning of this section.
This definition of efficiency depends not only on the quality of the track-finding
algorithm, but also upon the intrinsic properties of the tracker, such as its geometrical acceptance
and material content.
Using the same association criterion as used for the efficiency,
the fake rate is defined as the fraction of reconstructed tracks that are not
associated with any simulated particle.
This quantity represents the probability that a reconstructed
track is either a combination of unrelated hits or a genuine
particle trajectory that is badly reconstructed through the inclusion of spurious hits.  The efficiency and
fake rate  presented in this section are given as a function of \pt and $\eta$ of the
simulated particle and reconstructed track, respectively.
The efficiency is obtained for simulated particles generated
within $\abs{\eta}<2.5$,
with a production
point $<$3\cm and $<$30\cm from the centre of the beam spot for $r$ and $\abs{z}$, respectively. These criteria
select fairly prompt particles. We also require $\pt > 0.9$\GeV, for the study of efficiency
as a function of $\eta$, or $\pt > 0.1$\GeV for studying efficiency over the entire \pt spectrum.
Since the `high-purity' requirement described in Section~\ref{sec:TrackSelection} is the default track selection
for the majority of analyses in CMS, unless otherwise stated efficiency and fake rate are measured and presented here
using only the subset of reconstructed tracks that are identified as `high-purity'.

\subsubsection{Results from simulation of isolated particles}
\label{sec:PerfEffAndFakesSingleMC}

This section presents the performance of the CTF tracking software in reconstructing
trajectories of particles in events containing just a single muon, a pion or an electron.

\begin{figure}[hbtp]
  \centering
    \includegraphics[width=0.45\textwidth]{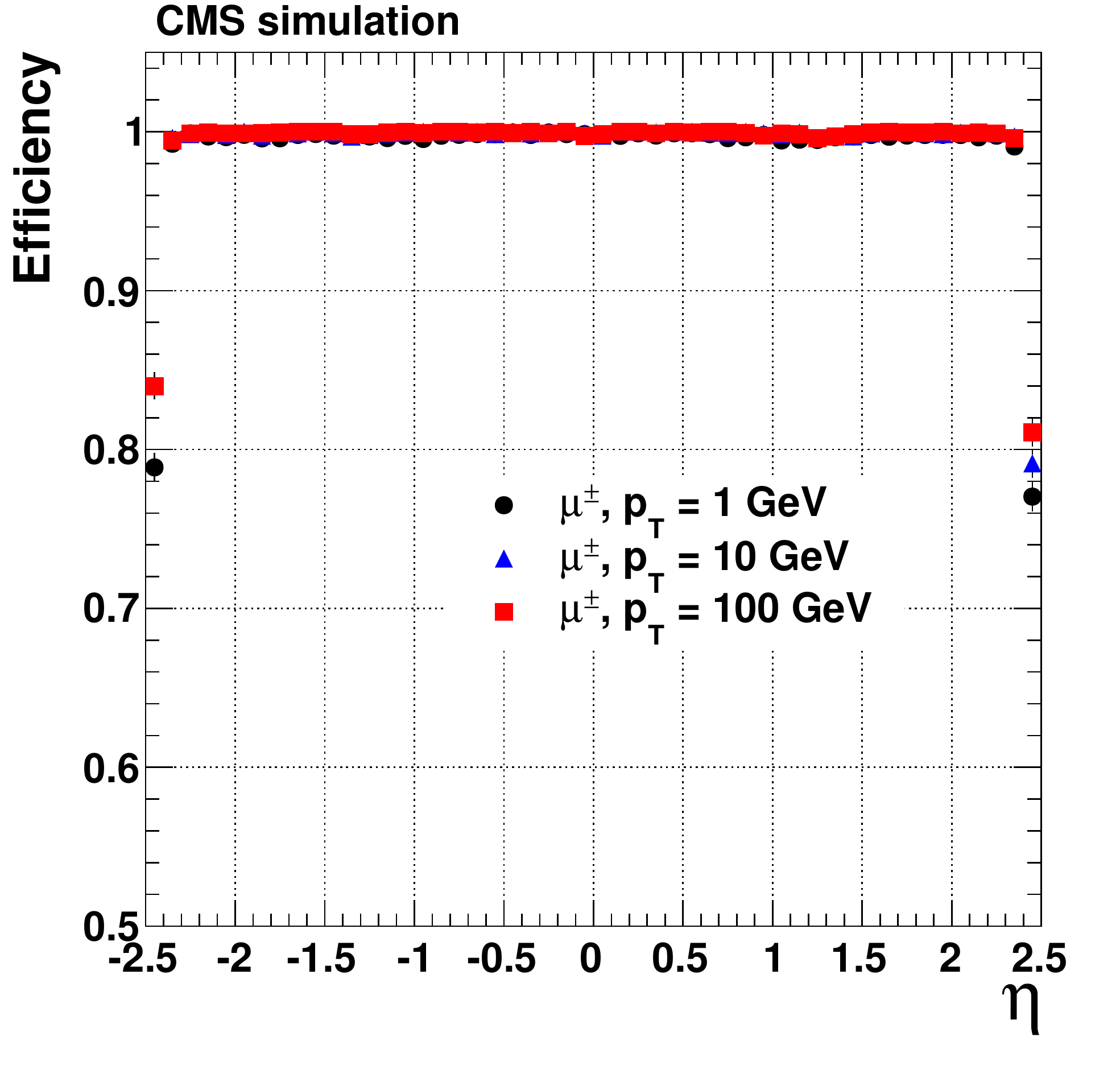}
    \includegraphics[width=0.45\textwidth]{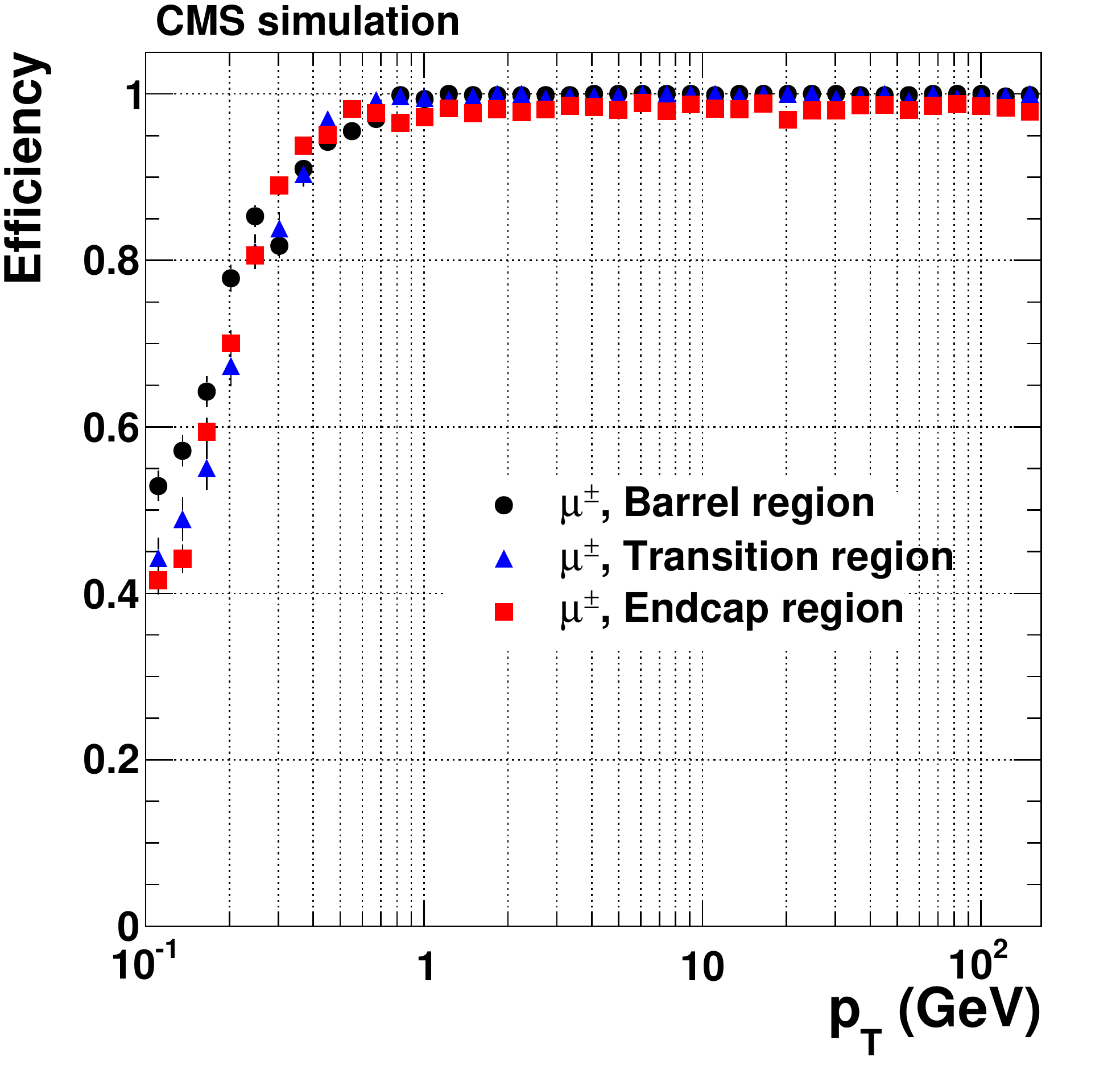}
    \includegraphics[width=0.45\textwidth]{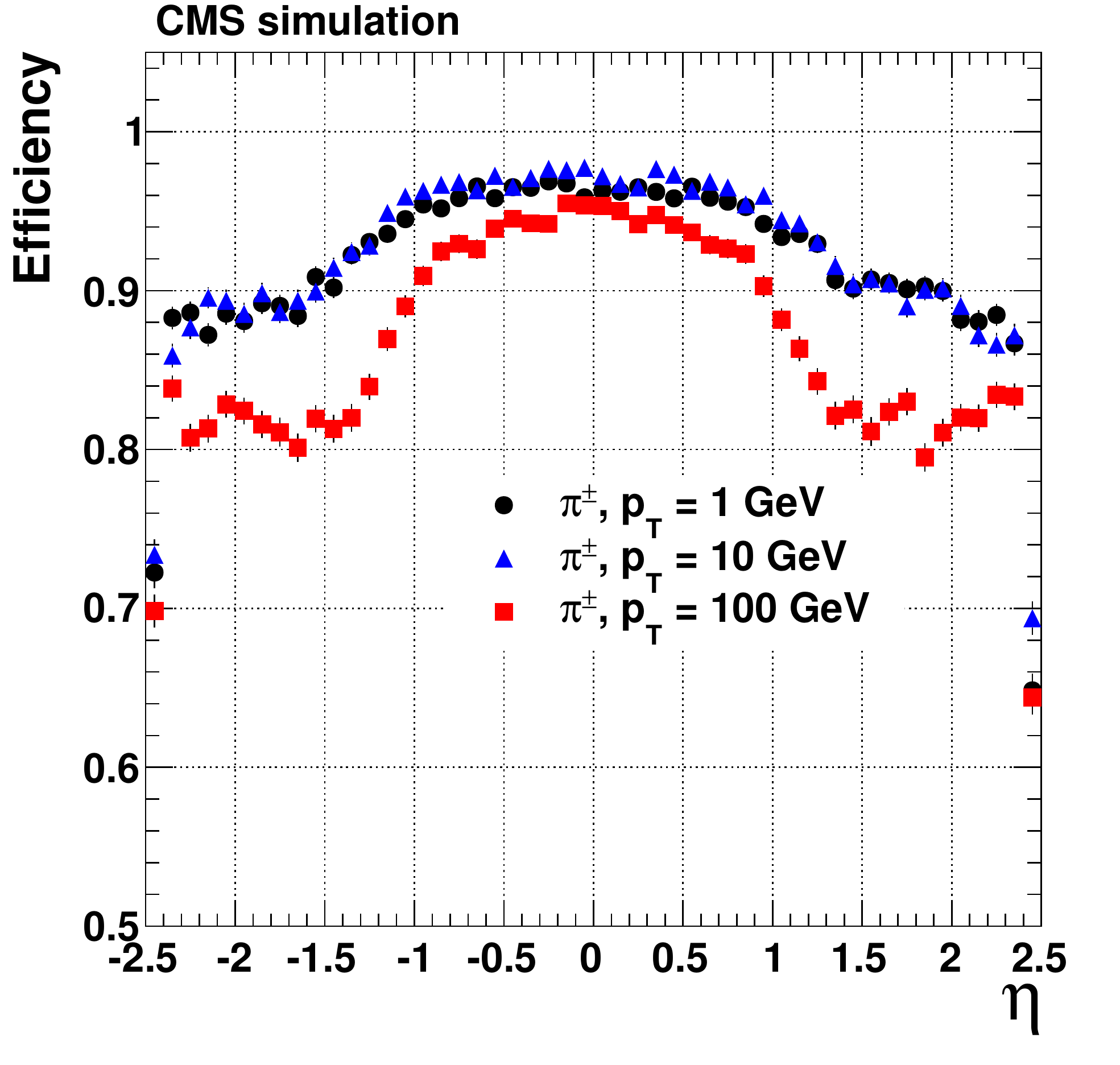}
    \includegraphics[width=0.45\textwidth]{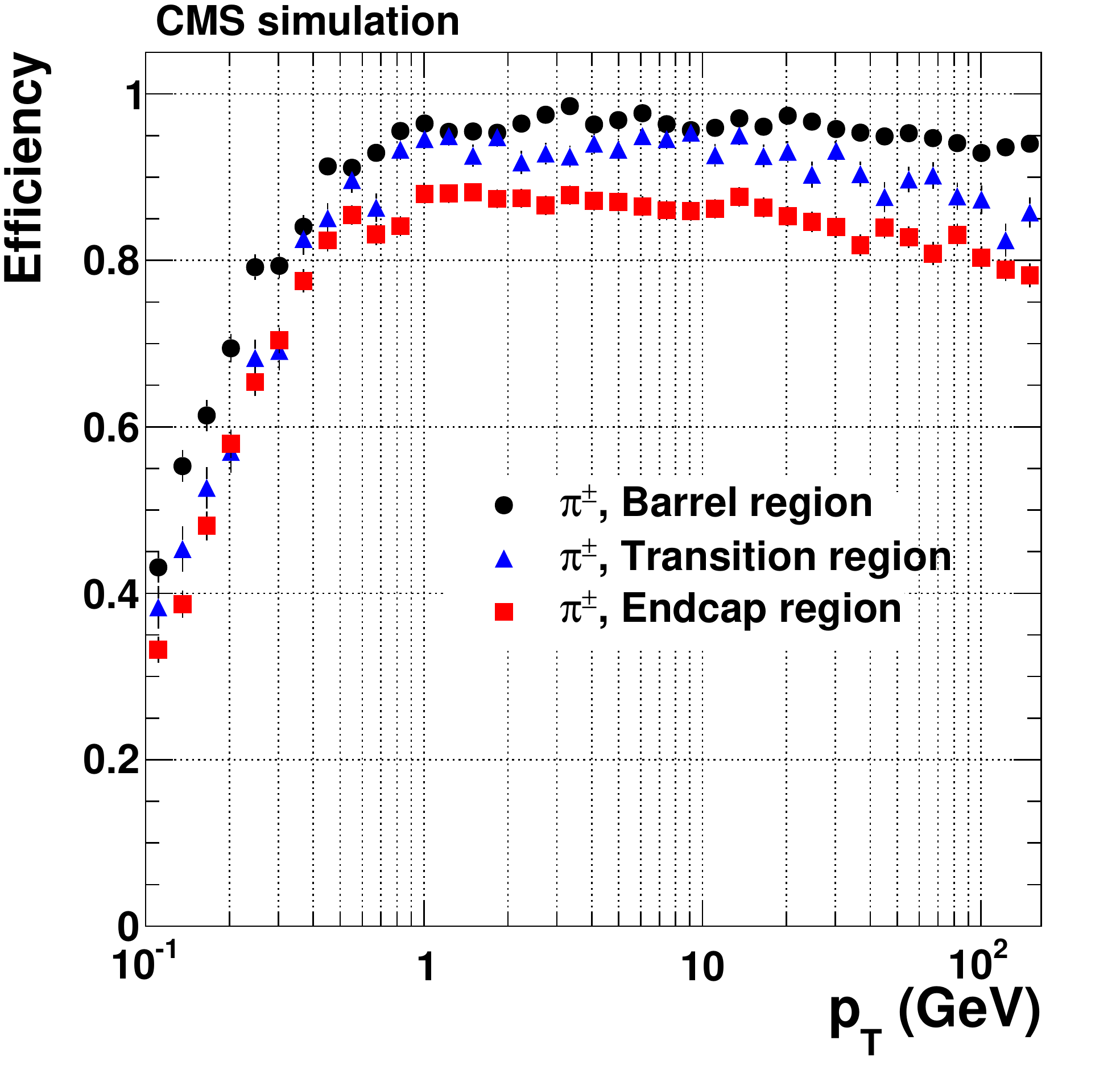}
    \includegraphics[width=0.45\textwidth]{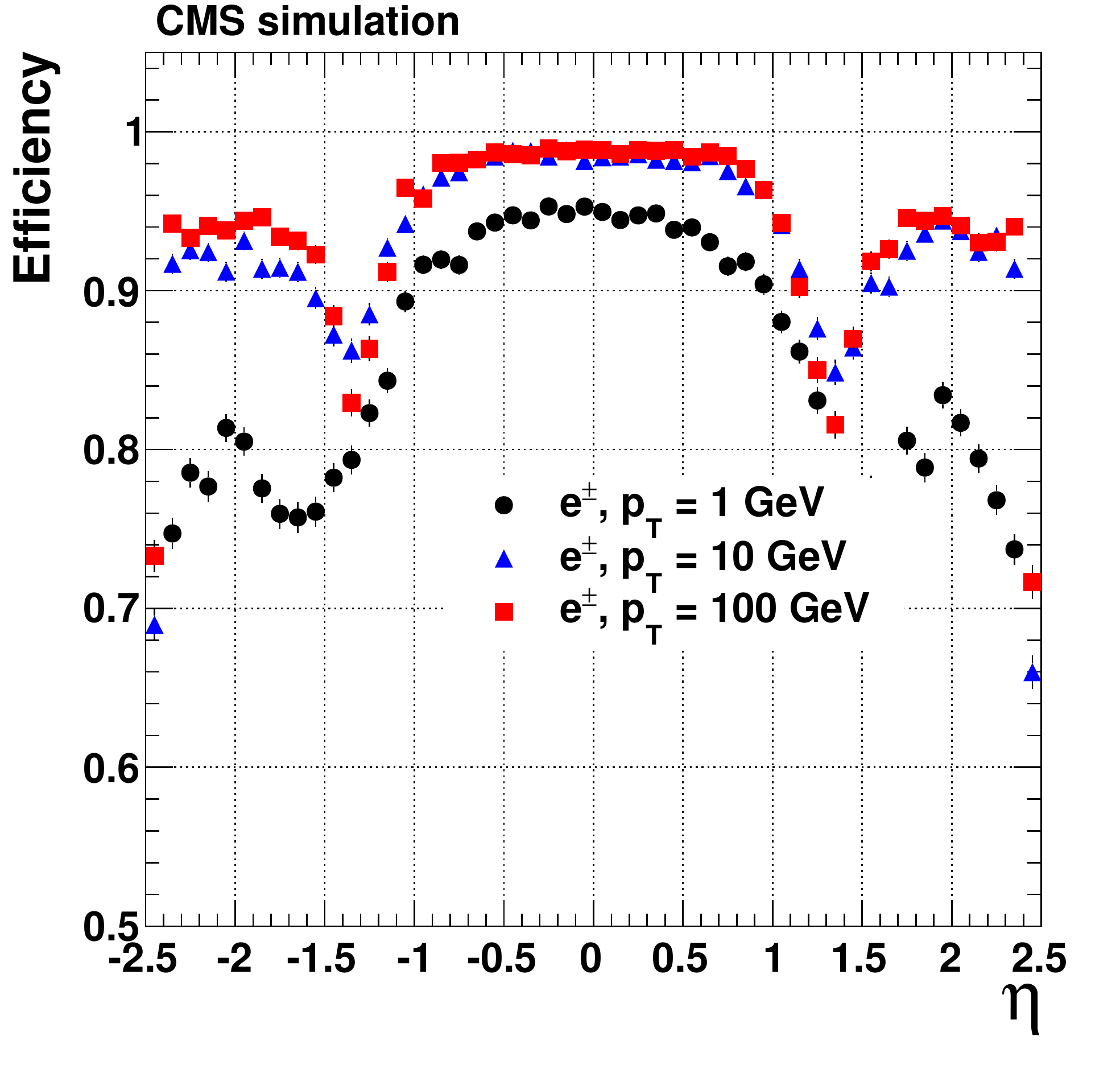}
    \includegraphics[width=0.45\textwidth]{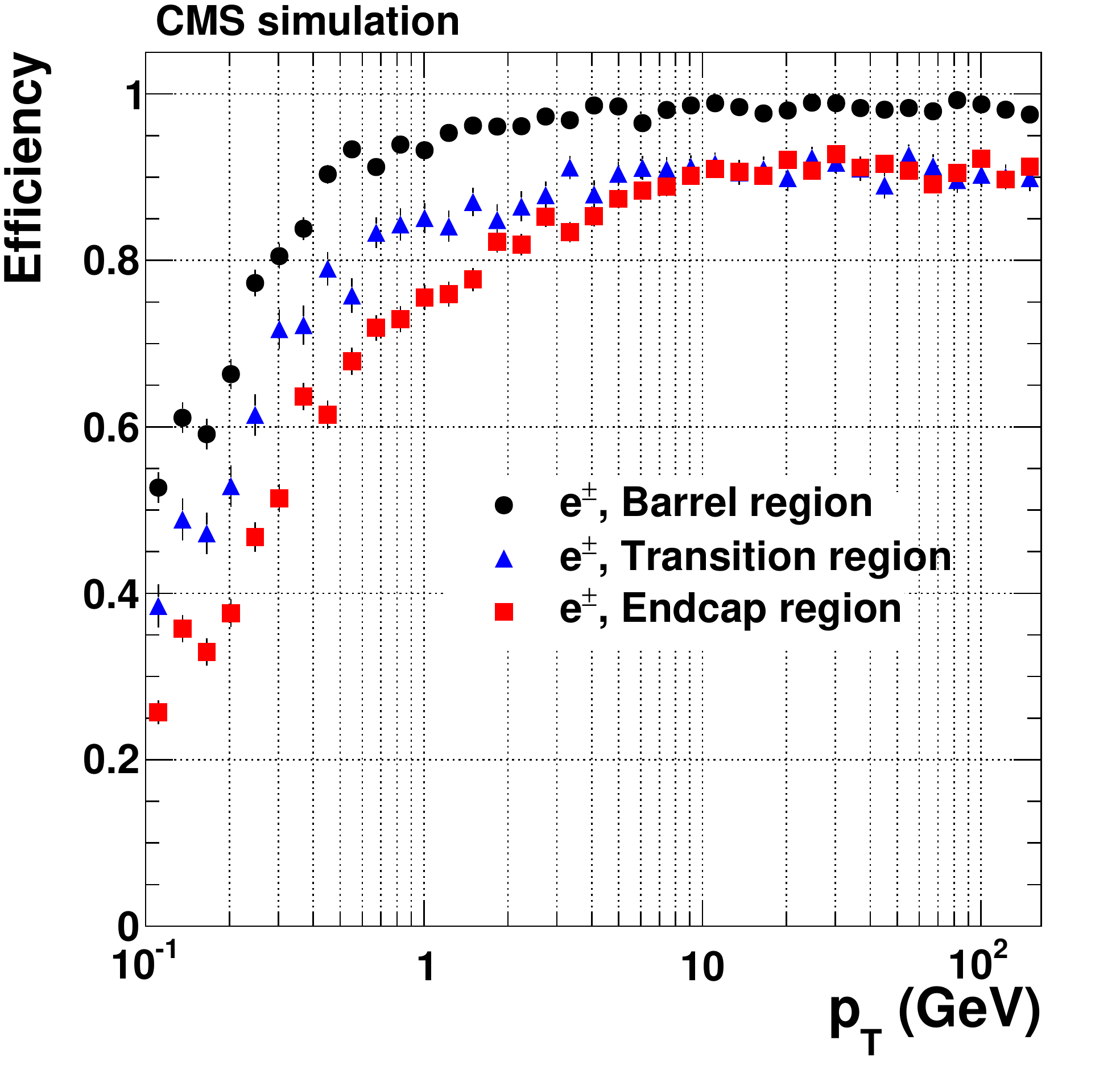}
      \caption {
    Track reconstruction \textit{efficiencies} for \textit{single, isolated muons} (top), \textit{pions} (middle) and \textit{electrons} (bottom)
    passing the \textit{high-purity} quality requirements.
    Results are shown as a function of $\eta$ (left), for $\pt = 1$, 10, and 100\GeV. They are
    also shown as a function of \pt (right), for the barrel, transition, and endcap regions, which are defined by the
    $\eta$ intervals of 0--0.9, 0.9--1.4 and 1.4--2.5, respectively.
      }
    \label{fig:SingleTrackEfficiencyMC}
\end{figure}

Muons are reconstructed better than any other charged particle in the tracker, as they mainly interact with the silicon detector
through ionization of the medium and, unlike electrons, their energy loss through bremsstrahlung is negligible.
Muons therefore tend to cross the entire volume of the tracking system,
producing detectable hits in several sensitive layers of the apparatus. Finally, muon trajectories are
altered almost exclusively by Coulomb scattering and energy loss, whose effects are straightforward to
include within the formalism of Kalman filter.
For isolated muons with  $1 < \pt < 100\GeV$, the  tracking efficiency is
$>$99\% over the full $\eta$-range of tracker acceptance, and does
not depend on \pt~(Fig.~\ref{fig:SingleTrackEfficiencyMC}, top). The fake rate is completely
negligible.

\begin{figure}[hbtp]
  \centering
    \includegraphics[width=0.45\textwidth]{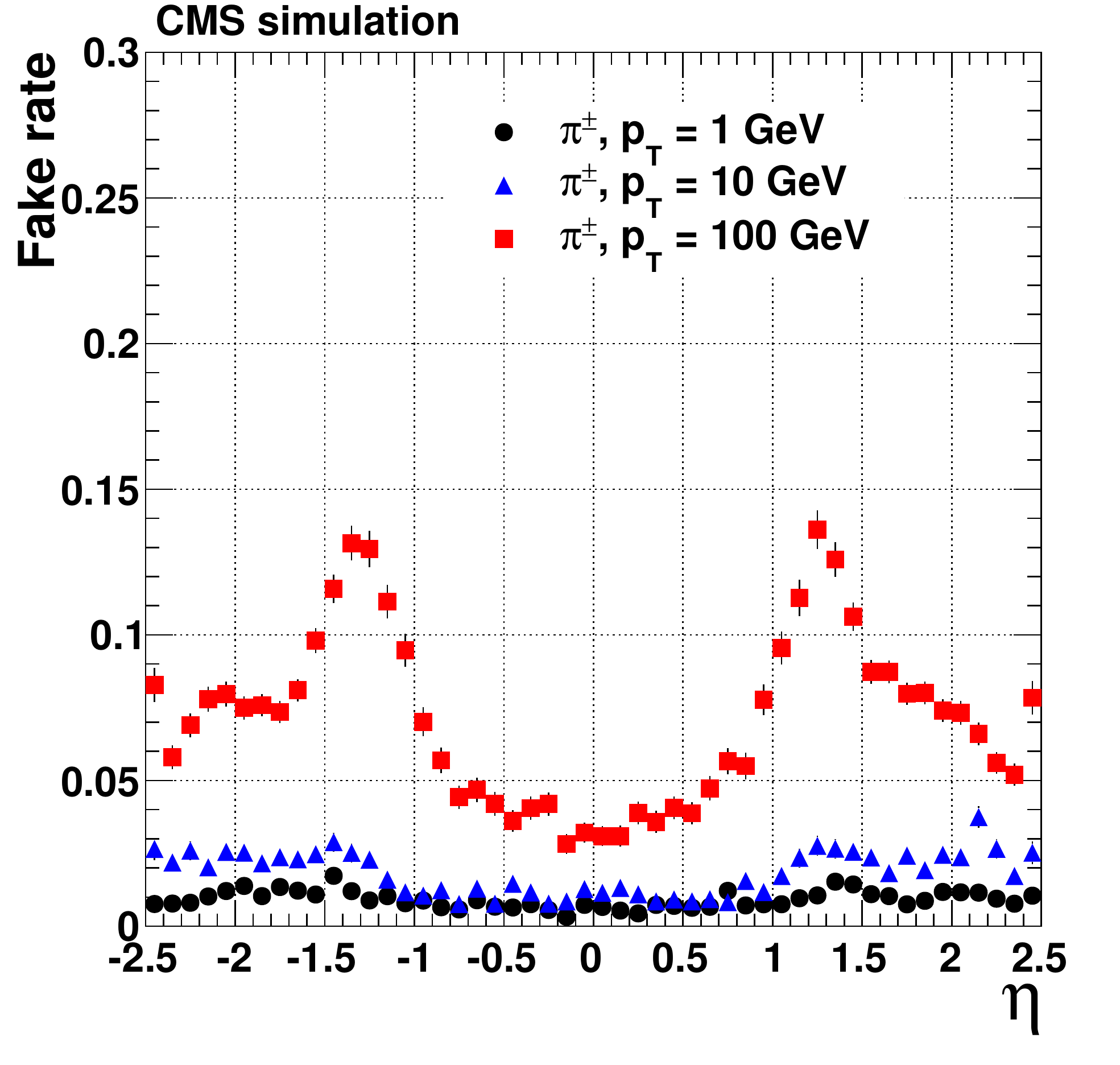}
    \includegraphics[width=0.45\textwidth]{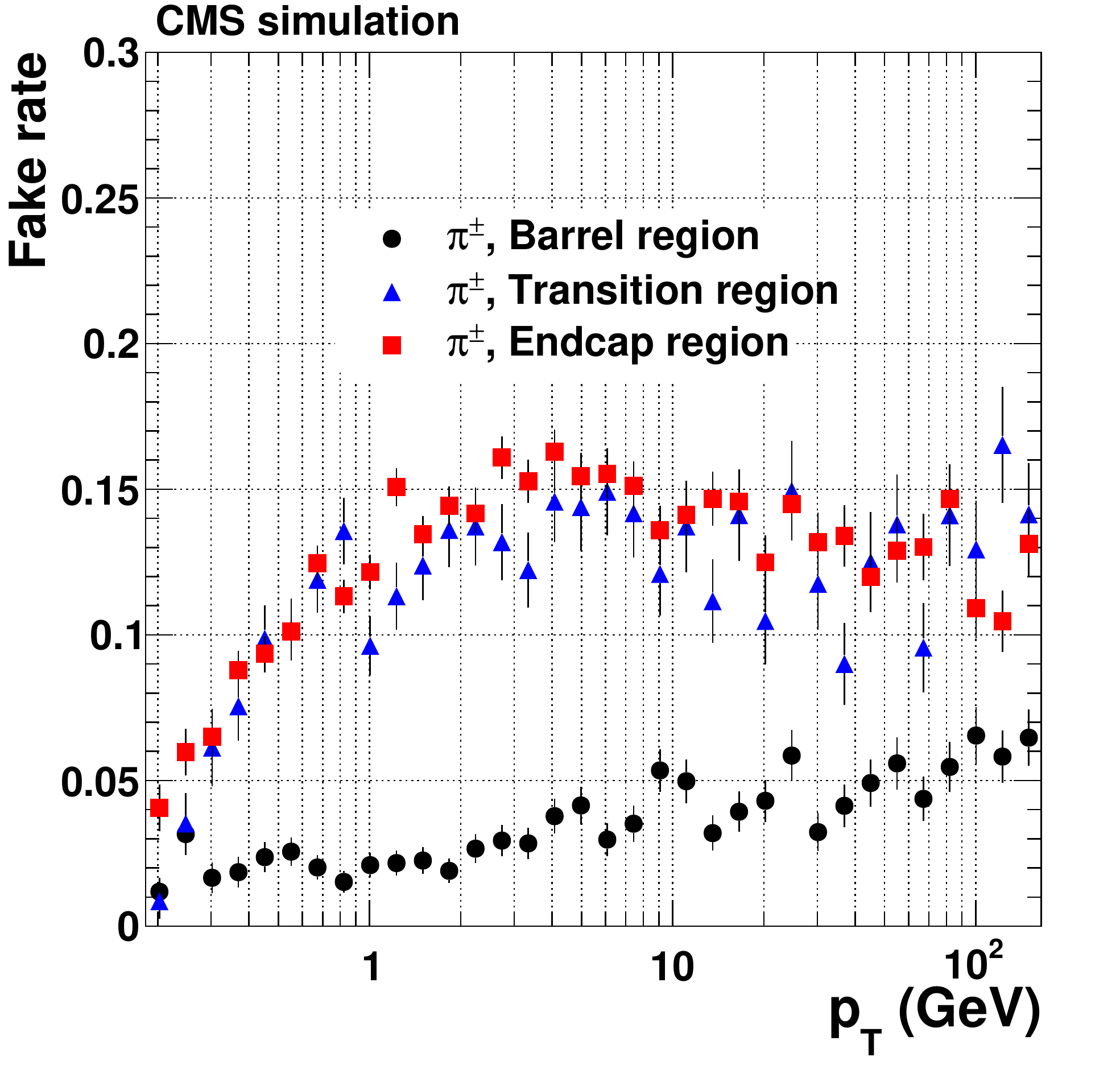}
    \includegraphics[width=0.45\textwidth]{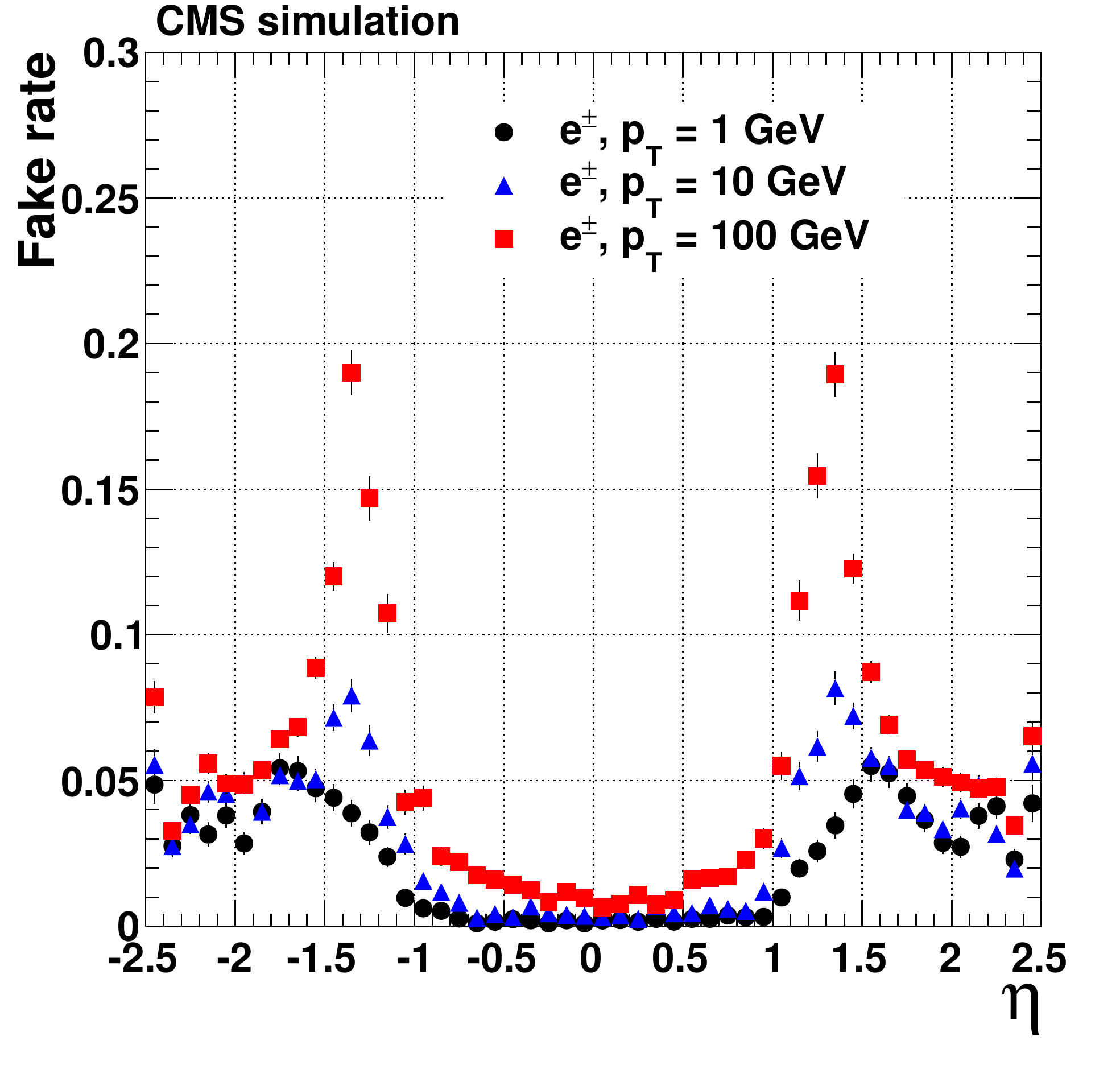}
    \includegraphics[width=0.45\textwidth]{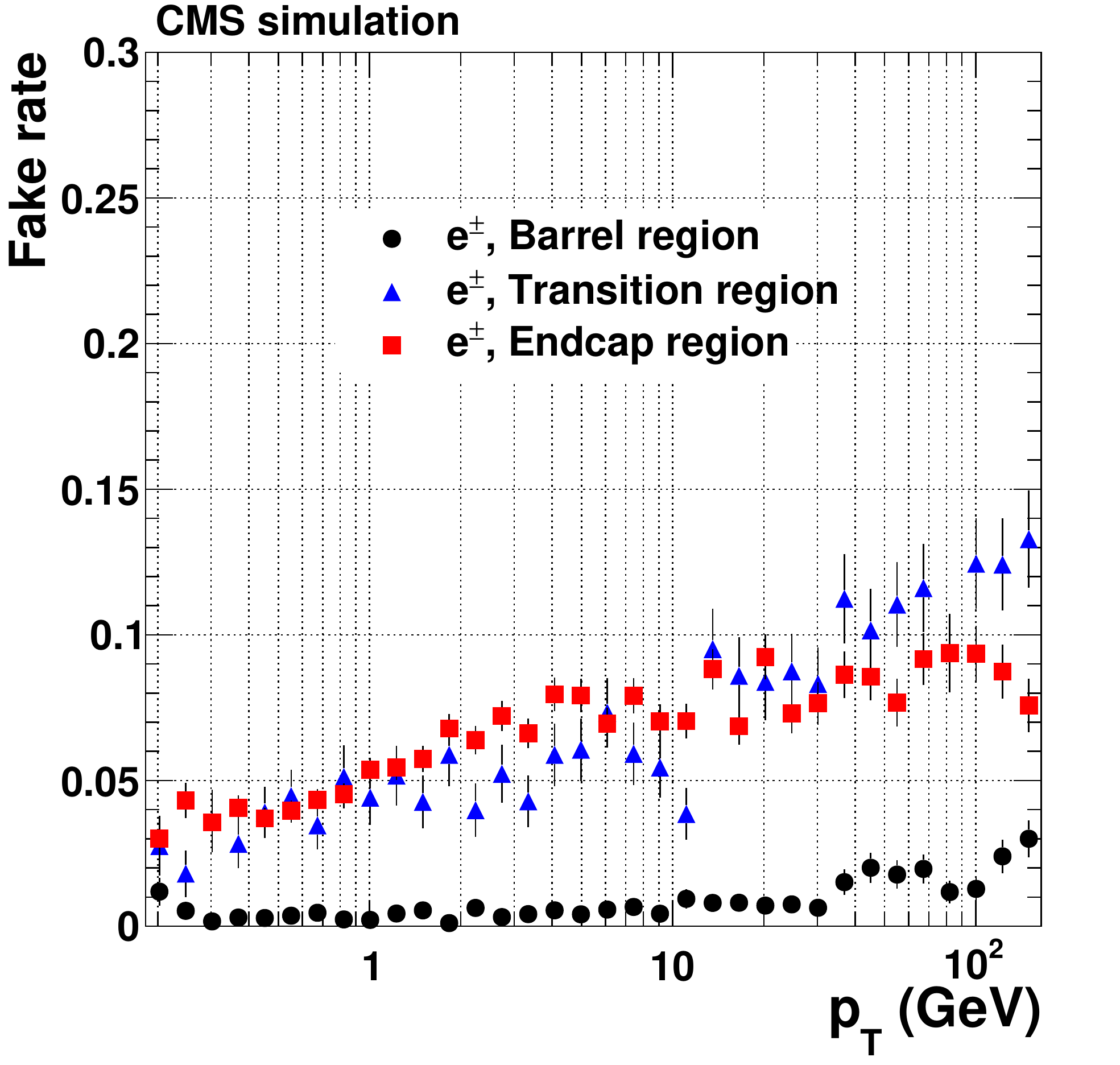}
      \caption {
    Tracking \textit{fake rate} for \textit{single, isolated pions} (top) and \textit{electrons} (bottom)
    passing \textit{high-purity} quality requirements.
    Results are shown, as a function of the reconstructed $\eta$ (left),
    for generated $\pt = 1$, 10, and 100\GeV. Results are
    also shown as a function of the reconstructed \pt (right), for the barrel, transition, and endcap regions,
    which are defined by the
    $\eta$ intervals of 0--0.9, 0.9--1.4 and 1.4--2.5, respectively. The results for \pt are obtained
    using particles generated with a flat distribution in $\ln\pt$. NB The measured fake rate depends
    strongly on the \pt distribution of the generated particles, since, for example, if no particles are
    generated in a given \pt range, most tracks reconstructed in that range must necessarily be fake.
    The generated particles used to make the plots of fake rate versus $\eta$ have a different \pt spectrum
    to those used to make the plots of fake rate versus \pt, therefore the measured fake rates in these
    two sets of plots are not directly comparable.
      }
    \label{fig:SingleTrackFakeRateMC}

\end{figure}

Charged pions, as muons, undergo multiple scattering and energy loss through ionization as they cross the tracker volume.
However, like all hadrons, pions are also subject to elastic and inelastic nuclear interactions.
The elastic nuclear interactions
introduce long tails in the distribution of the scattering angle, well beyond expectations from Coulomb
scattering. The current implementation of the track-finding algorithm assumes a track trajectory modelled by the
material propagator described in Section~\ref{sec:TrackFinding}. This takes into account Coulomb scattering, but neglects
elastic nuclear interactions. As a result, the formation of a track can be interrupted if a hadron undergoes a large-angle
elastic nuclear scattering.
Hence, a hadron can be reconstructed as a single track with fewer hits, or as two separate tracks, or it may not be found at all.
A loss of hits also degrades the precision with which the parameters of the trajectory can be estimated
(Section~\ref{subSec:TrackParResolution}).
Inelastic nuclear interactions are the main source of  tracking inefficiency for hadrons,  particularly in those regions of the
tracker with large material content.
Depending on $\eta$, up to 20\% of the simulated pions are not reconstructed~(Fig.~\ref{fig:SingleTrackEfficiencyMC}, middle).
This effect is most significant for hadrons with $\pt\lesssim 700\MeV$, because of the larger
cross sections for nuclear interactions at low energies~\cite{PDG2012}.
The tracking efficiency is also affected, along with the fake rate (Fig.~\ref{fig:SingleTrackFakeRateMC}, top), by
the secondary particles produced in inelastic processes.
This is because the products of nuclear interactions are often emitted with trajectories approximately
aligned to that of the traversing pion, particularly for large pion momenta.
As a result, it is common for the trajectory builder to combine hits
of the incoming pion with those of a secondary particle into a single track.
The degradation in efficiency and the increase in fake rate are correlated, as expected, and the loss in performance
is greatest for highest momentum pions.  In general, the merging of separate
trajectories during reconstruction is more common in the region of the barrel to endcap transition and in the endcap
regions of the tracker, as these regions contain large amounts of material. In the transition region, the proportion of
fake tracks is also high because the distances between successive hits on each track are longer,
particularly when passing from a hit in the barrel to a hit in an endcap detector.
These longer distances result in correspondingly larger uncertainties in the track trajectory extrapolation that is
performed during track building. This makes it more probable that spurious hits, such as those from secondary particles,
will be incorrectly assigned to the track. Although the extrapolation uncertainties would be equally large
for muons, the fake rate remains very small for muons, as they rarely produce secondary particles.
While the fake rate is generally  $<$2--3\% for tracks reconstructed in the sample of single pions with $\pt = 1$ or 10\GeV,
in a sample of single pions with a \pt of 100\GeV, the fake rate peaks at $\approx$15\% for $\abs{\eta} \approx 1.3$.

Electrons lose a large fraction of their energy
via bremsstrahlung radiation before they reach the outer
layers of the silicon tracker. Such radiation has an impact on the reconstruction of electrons, similar to that of
inelastic nuclear interactions on the reconstruction of charged hadrons.
First, if an electron loses most
of its energy before reaching the outer layer of the tracker, the number of hits assigned to the track
can be reduced significantly. Second, if a radiated photon converts to an electron-positron pair or
induces an electromagnetic shower,
the track finder can assign a mixture of hits from the primary electron and
from the secondary particles to a single track. This reduces tracking efficiency, increases fake rate, and
is the principal source of misidentification of charge for electrons. The efficiency and fake rate of the CTF algorithm for reconstructing electrons are shown in
Fig.~\ref{fig:SingleTrackEfficiencyMC} (bottom) and Fig.~\ref{fig:SingleTrackFakeRateMC} (bottom), respectively.
In the barrel, the efficiency for electrons exceeds 90\% for $\pt > 0.4$\GeV, and the fake rate is very small.
However, the performance is significantly worse in the endcap and barrel-endcap transition regions, because of the larger
amount of material and the correspondingly greater chance of an electron to produce an electromagnetic shower within the tracker
volume. The fake rate is particularly high in the sample of electrons with $\pt = 100$\GeV, since any secondary particle they
produce will tend to be emitted tangentially to the direction of the original electron, with the consequence that
the tracking algorithm tends to reconstruct the primary and the secondary particle as a single track.
It is important to note that, in practice, CMS achieves considerably better performance for electron reconstruction,
by using the dedicated GSF algorithm~\cite{CMS_PAS_EGM-10-004}, described in Section~\ref{sec:GSF}, rather
than the standard CTF algorithm.

\subsubsection{Results from simulated pp collision events}
\label{sec:PerfEffAndFakesCollisionMC}

This section presents the performance of the CTF tracking software for reconstructing
trajectories of non-isolated charged particles generated in simulated LHC
collisions. Compared to the results shown in the previous section for isolated
particles, the tracking performance discussed in this section is affected by an additional important feature
of LHC events: the large number of hits produced in the tracker at each LHC bunch crossing.
These hits originate from the hundreds of primary particles and their interactions selected by the CMS triggers.
Their number is increased  by the combined effects of low-energy particles spiralling in the CMS magnetic
field (``loopers''), and particles produced in temporally overlapping pileup collisions.
In the following, we give examples of the kind of difficulties encountered by
the tracking algorithm during the reconstruction of these events.

\begin{itemize}
\item Many particles can be emitted within highly collimated jets and the hits they produce are closer
  to each other than the typical uncertainty in the position of extrapolated trajectories at the sensors.
  In such situations, the trajectory builder is
  unable to assign unambiguously the hits to the corresponding trajectories. For example, hits
  corresponding to two distinct charged particles can be mixed into one or two reconstructed tracks that
  do not describe accurately either of the trajectories of the two particles.

\item Trajectories of nearby particles can be separated sufficiently in the outer layers of the
  tracker so as to be  correctly identified by the track-finding module. Nevertheless, their hits
  in the innermost layers can be so near to each other that the reconstruction algorithm often
  assigns incorrectly the hits to the relevant trajectories. Particularly
  in the innermost pixel layer particles can be so close to each other that their
  ionization signals can merge into a single cluster. In this
  case, even if the individual trajectories are reconstructed, and their momenta are
  well measured, the resolutions in their impact parameter are degraded by the formation of this merged cluster.

\item Many of the low-\pt particles from the underlying event of the hard collision, or from the other pileup
collisions, have such a low transverse momentum that they cannot escape the volume of the tracker, but
instead spiral in the magnetic field, producing many hits in the detector, increasing the complexity of the track-finding
task. Even when these circulating particles are not close to each other at their production vertex, their
large number of hits increases the probability of having uncorrelated hits
accepted as legitimate trajectories, and thereby generate reconstructed fake track.

\end{itemize}

\begin{figure}[hbtp]
  \centering
    \includegraphics[width=0.45\textwidth]{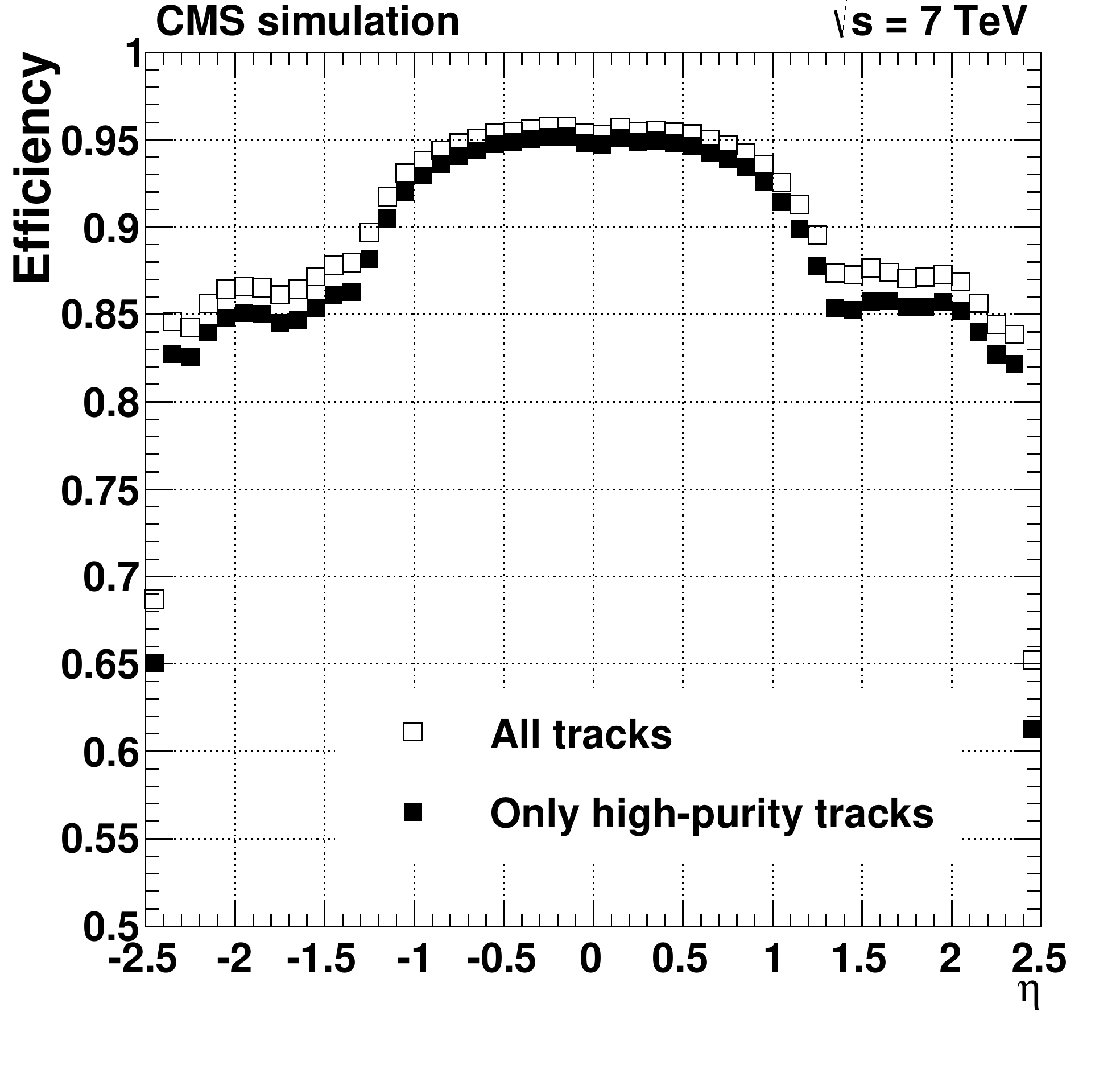}
    \includegraphics[width=0.45\textwidth]{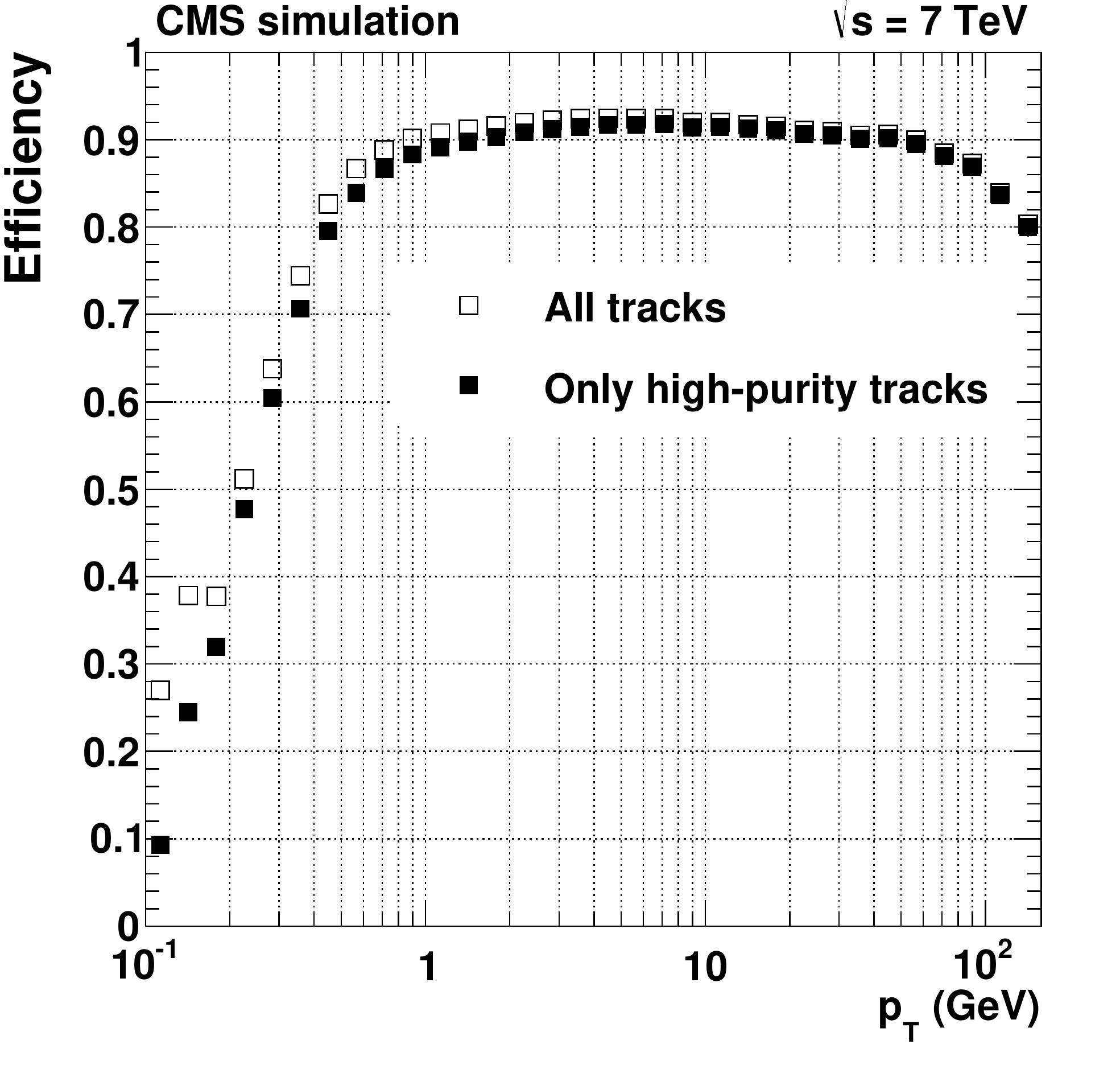} \\
    \includegraphics[width=0.45\textwidth]{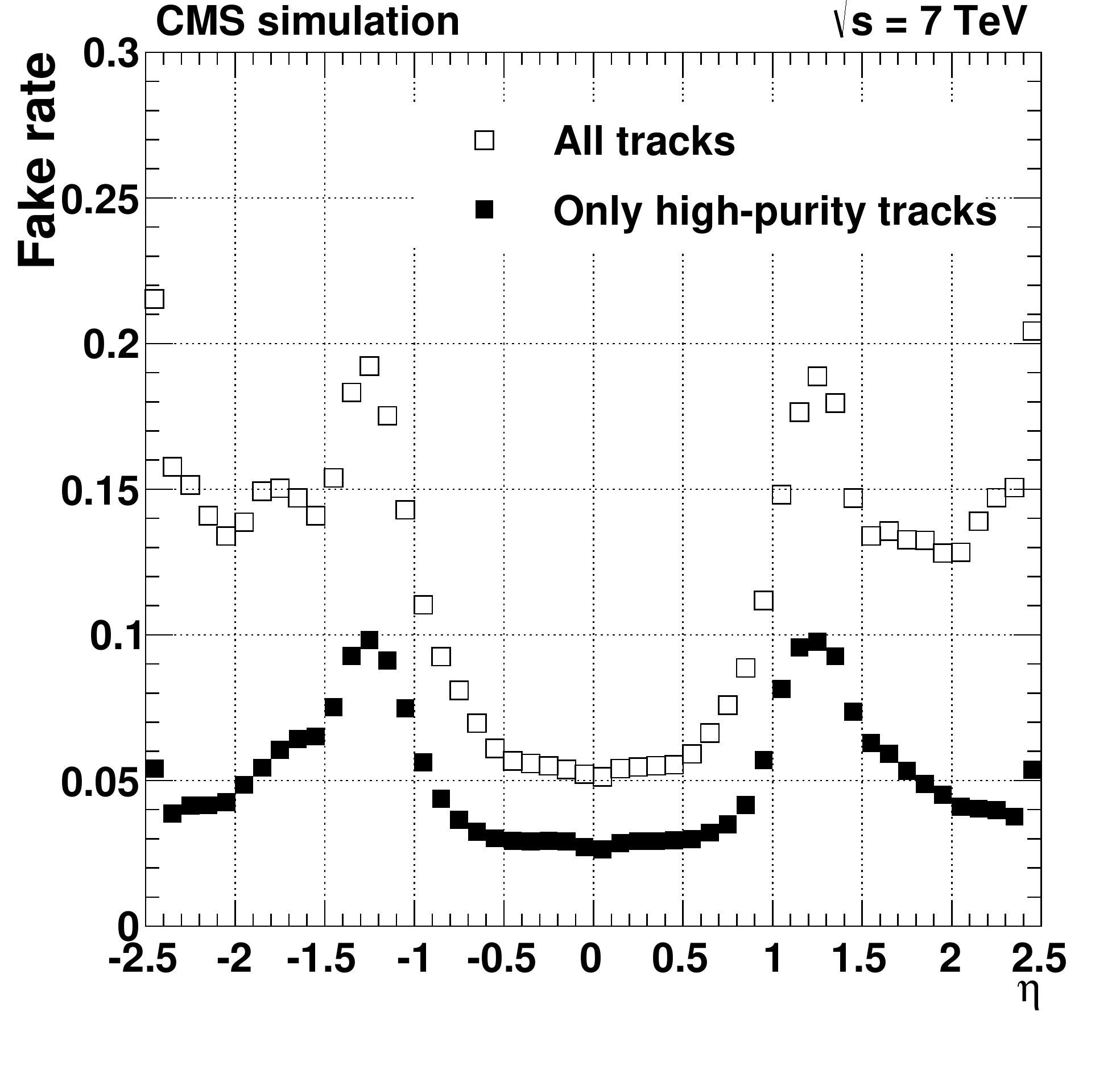}
    \includegraphics[width=0.45\textwidth]{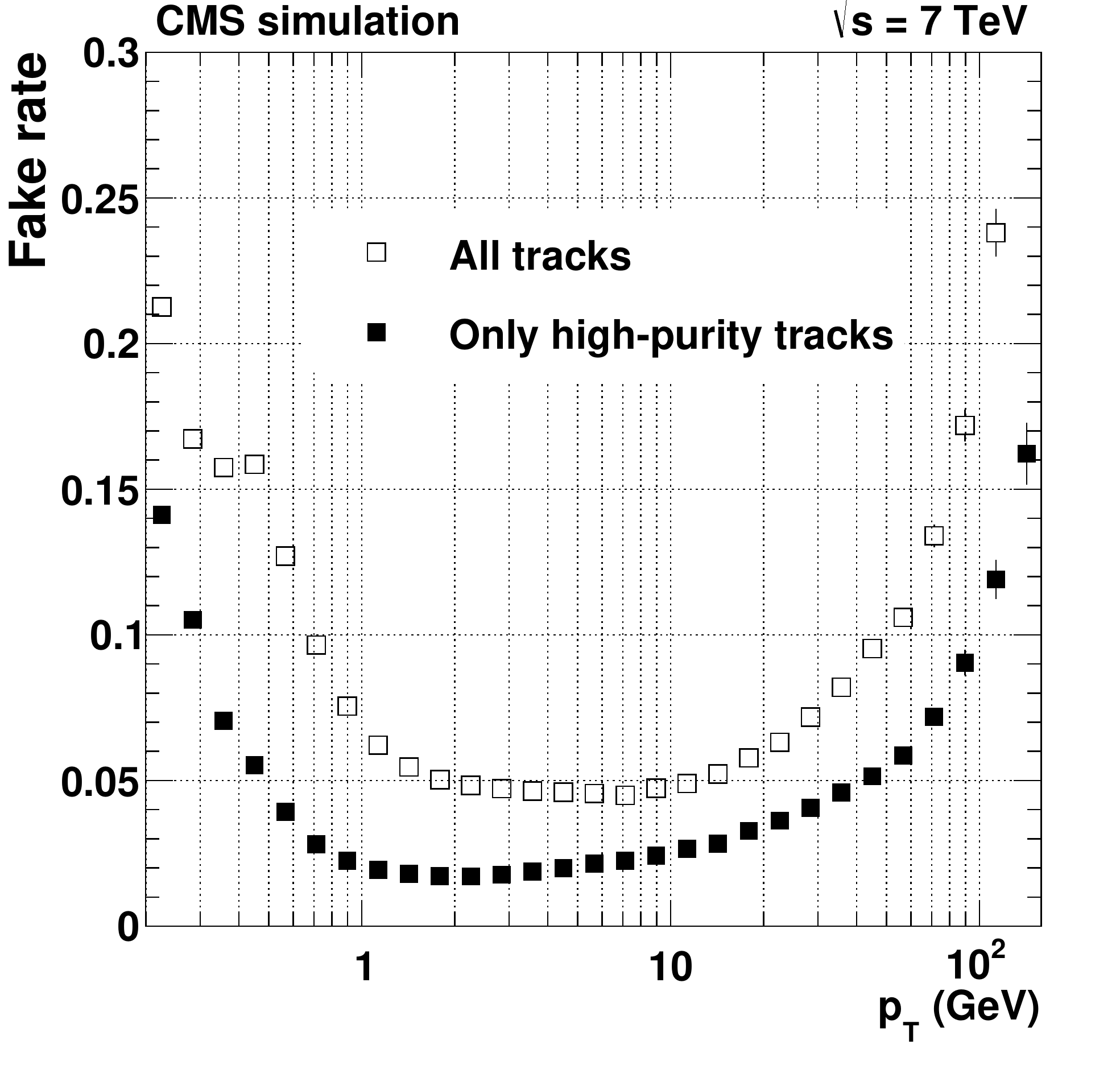} \\
  \caption {
    Tracking \textit{efficiency} (top) and \textit{fake rate} (bottom) for simulated \textit{$t\bar t$ events} that include
    superimposed pileup collisions.
    The number of pileup interactions superimposed on each simulated event is generated randomly from a Poisson distribution with mean
    value of 8.
    Plots are for all reconstructed tracks, and also for the subset of tracks passing \textit{high-purity} requirements.
    The efficiency and fake rate plots are plotted for $\abs{\eta}<2.5$, and the efficiency
    for charged particles refers to those generated  less than 3\cm (30\cm) from the centre of the beam spot in $r$ ($z$) directions.
    The efficiency as a function of $\eta$ is for generated particles with $\pt>0.9$\GeV.
    \label{fig:TTbarEfficiencyFakeRateAllvsHP}}
\end{figure}

Since most charged particles produced in LHC collisions
are hadrons, all the sources of inefficiency discussed for single, isolated pions similarly affect
the \ttbar results. The efficiency for reconstructing charged particles in \ttbar events, which is shown in
Fig.~\ref{fig:TTbarEfficiencyFakeRateAllvsHP} (top), closely resembles the reconstruction efficiency for
isolated pions shown in Fig.~\ref{fig:SingleTrackEfficiencyMC} (middle). The similarity between the two
indicates that tracking efficiency is not strongly degraded by particle multiplicity in
typical \ttbar events.
The tracking efficiency as a function of the \pt is approximately constant
for $1 < \pt < 80\GeV$, but, at small \pt, the efficiency decreases quickly
(Fig.~\ref{fig:TTbarEfficiencyFakeRateAllvsHP}, top-right) for several reasons.

\begin{itemize}
\item The pion-nucleus cross section increases rapidly for pions of energies below 0.7\GeV.

\item Track selection criteria~(see Section~\ref{sec:TrackSelection}) are much more stringent for trajectories
  of small momentum, as they correspond to the main source of fake tracks.

\item When estimating the \rms scattering angle and mean energy loss in the detector material, the trajectory
  propagator assumes that all particles have a pion mass, since the pion is the most
  common particle produced in LHC collisions. While this assumption is good for
  relativistic particles, it breaks down at low energies when particle masses become more important.

\end{itemize}

For the considered \ttbar events, charged particles with \pt larger than 80\GeV are mostly produced inside the core
of collimated jets. The inability of the trajectory builder to cope fully with regions of the tracker characterised
by extremely high-density of particles is reflected in the drop in tracking efficiency for
large \pt values.

The fake rate, shown in Fig.~\ref{fig:TTbarEfficiencyFakeRateAllvsHP} (bottom), has a similar
dependence on $\eta$ as that observed for isolated pions in Fig.~\ref{fig:SingleTrackFakeRateMC}.
However, the fake rate has a very different dependence on \pt for the two cases. In the pp
collisions, the fake rate increases for \pt
values $<$1\GeV. This is because the smaller the \pt of an initial trajectory seed, the larger the search
windows that must be used (because of multiple scattering) when searching for additional hits to form
the corresponding track
candidates. This increases the probability to assign wrong hits to a track.
The fake rate also increases at large \pt, as was the case for the single-pion samples,
partially because of the production of secondary particles in nuclear interactions, and partially
because comparatively few high-\pt particles are produced in pp collisions.

The distributions in efficiency and fake rate in Fig.~\ref{fig:TTbarEfficiencyFakeRateAllvsHP} are generated for two sets of
reconstructed tracks: all the tracks produced using the default tracking software, and only those tracks that pass
the \textit{high-purity} requirements. For a 1--2\% reduction in efficiency, the quality requirement reduces the fake rate over
the entire \pt range by more than a factor of two.

\begin{figure}[hbtp]
  \centering
    \includegraphics[width=0.45\textwidth]{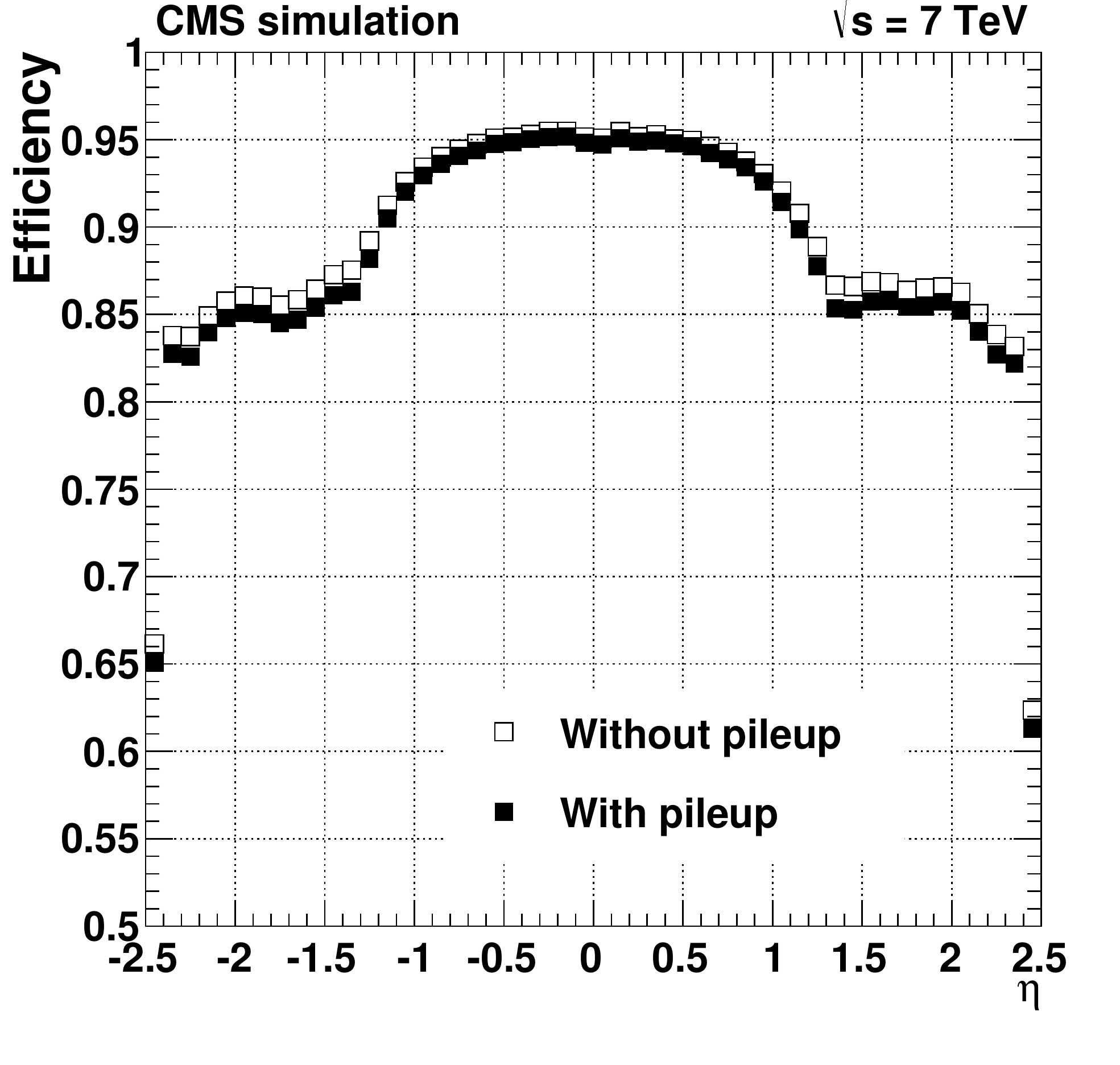}
    \includegraphics[width=0.45\textwidth]{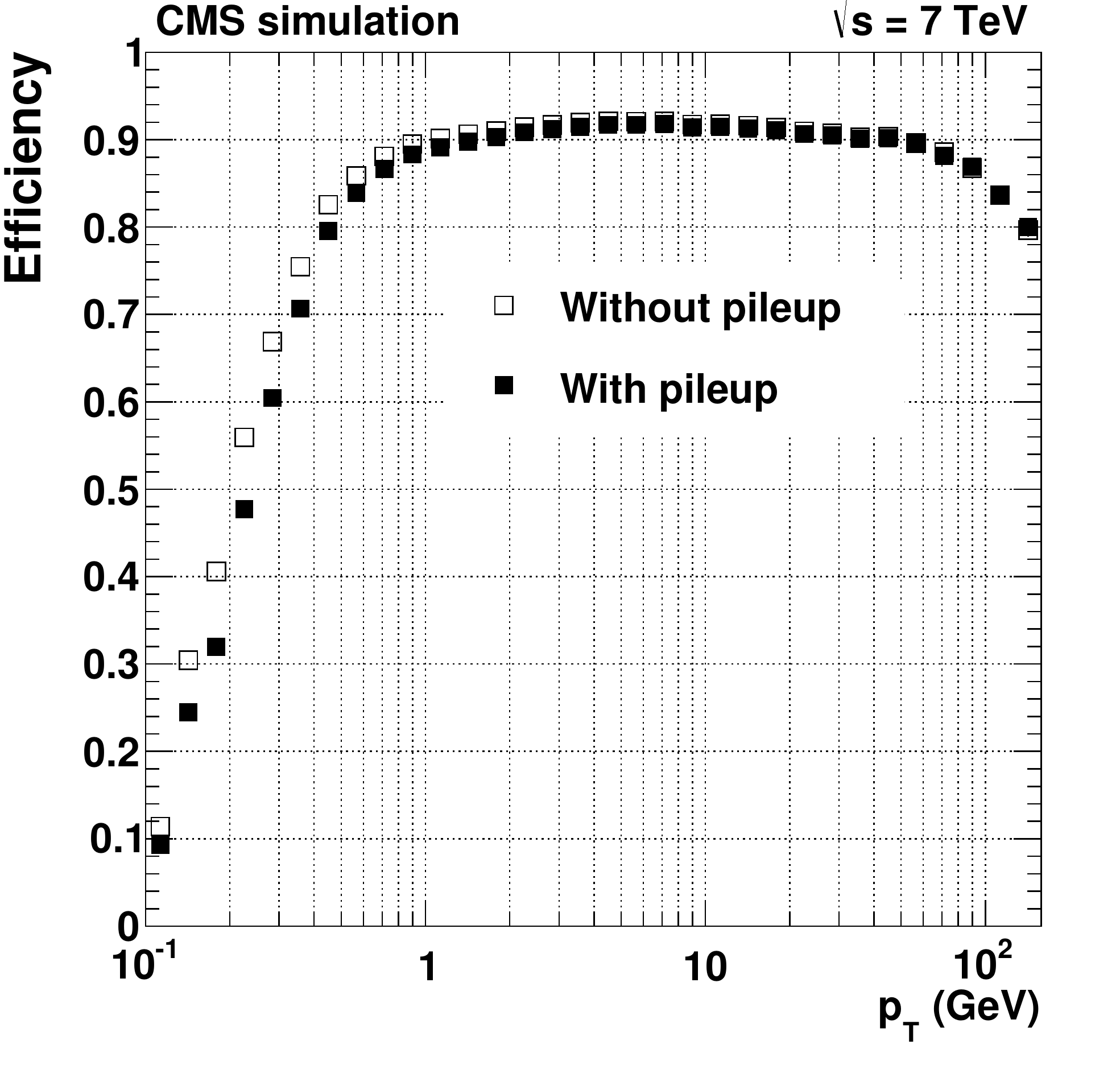} \\
    \includegraphics[width=0.45\textwidth]{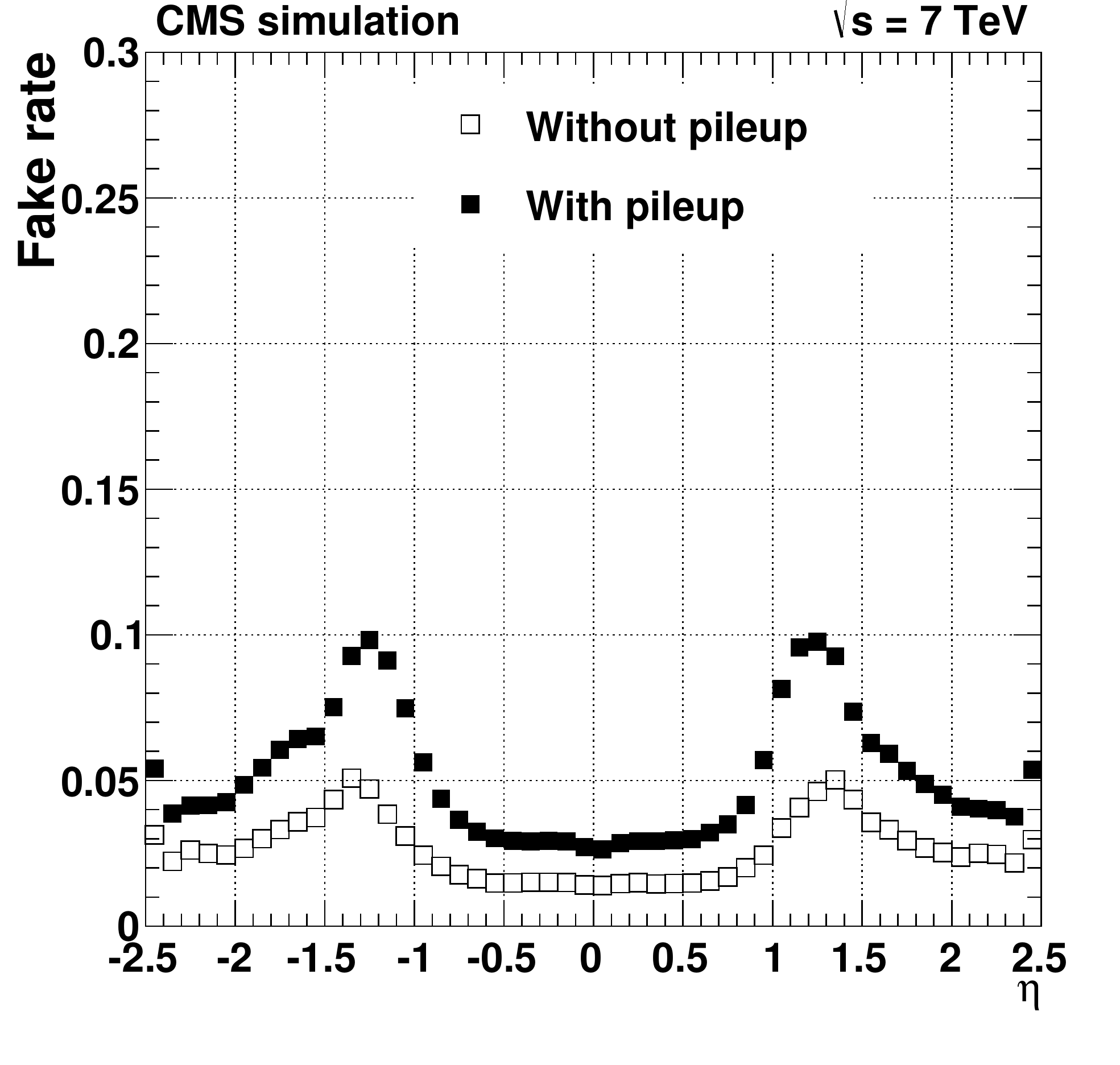}
    \includegraphics[width=0.45\textwidth]{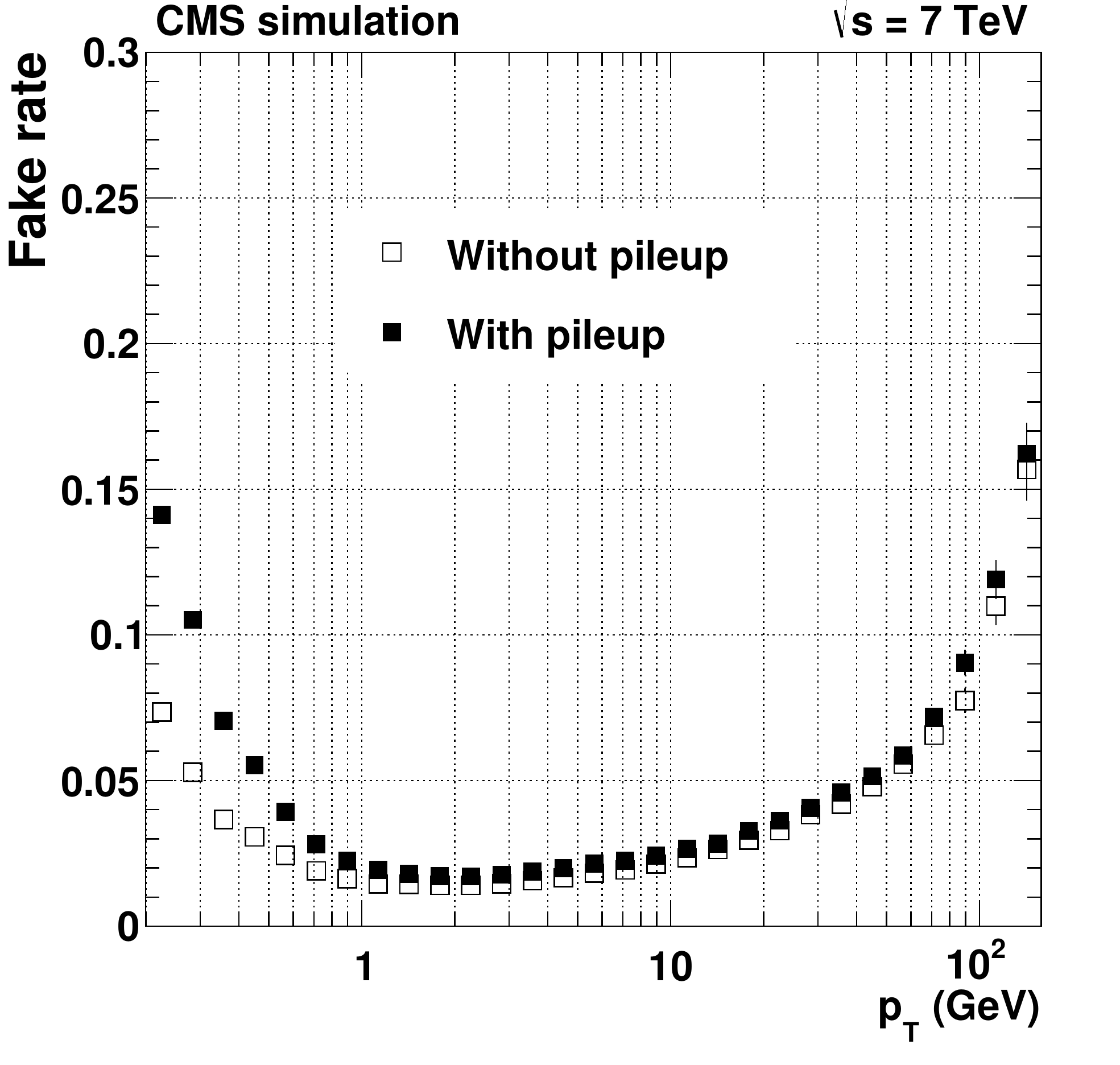} \\
  \caption {
    Tracking \textit{efficiency} (top) and \textit{fake rate} (bottom) for \textit{$t\bar t$ events} simulated with and without superimposed pileup
    collisions.
    The number of pileup interactions superimposed on each simulated event is generated randomly from a Poisson distribution with mean
    value of 8.
    Plots are produced for the subset of tracks passing the \textit{high-purity} quality requirements.
    The efficiency and fake rate plots cover $\abs{\eta}<2.5$.  The efficiency results are for charged particles produced less
    than 3\cm (30\cm) from the centre of the beam spot in $r$ ($z$) directions.
    The efficiency as a function of $\eta$ is for generated particles with $\pt>0.9$\GeV.
  \label{fig:TTbarEfficiencyFakeRateWithVsWithoutPU}}
\end{figure}

Figure~\ref{fig:TTbarEfficiencyFakeRateWithVsWithoutPU} shows the efficiency and fake rate plots for \ttbar events
simulated either with or without superimposed pileup interactions. After applying the quality requirement,
the presence of pileup significantly degrades the efficiency and fake rate only for tracks with a $\pt < 1\GeV$.

\begin{figure}[hbtp]
  \centering
   \includegraphics[width=0.6\textwidth]{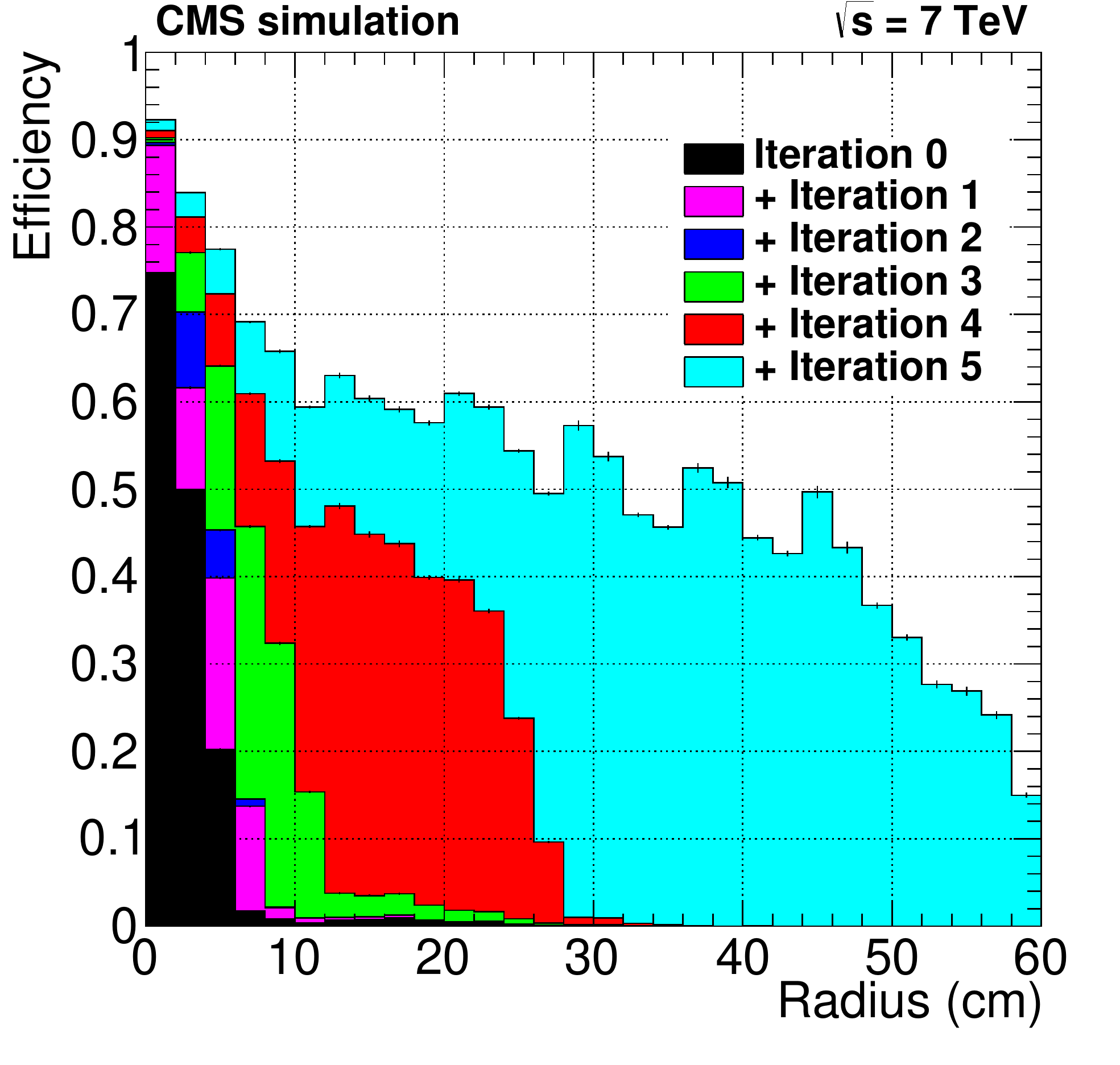}
    \caption{Cumulative contributions to the overall tracking performance from the six iterations in track reconstruction.
      The tracking efficiency for simulated \ttbar events is shown as a function of
      transverse distance ($r$) from the beam axis to the production point of each particle, for tracks with
      $\pt > 0.9$\GeV and $\abs{\eta}<2.5$, transverse (longitudinal) impact parameter $<$60 (30)\cm.
      The reconstructed tracks are required to pass the \textit{high-purity} quality requirements.}
    \label{fig:TTbarEffVsR}
\end{figure}

The CMS tracker is capable of reconstructing highly displaced tracks, such as pions from \Kzero decay, or particles
produced in nuclear interactions and photon conversions. This is very useful
for studies of B~physics, photon reconstruction, and for improving energy resolution for particle-flow reconstruction \cite{CMS_PAS_PFT-09-001}.
This capability also makes it possible to search for signatures of new phenomena, such as new long-lived particles
that decay with displaced tracks. Reconstruction of displaced tracks is carried out in
Iterations~3--5 of the 6-step iterative tracking scheme described in Section~\ref{sec:trackReco}. Charged
particles originating outside the pixel detector can also be reconstructed.
The efficiency for reconstructing this kind of charged particle
as a function of the radius of its point of production is shown in Fig.~\ref{fig:TTbarEffVsR} for \ttbar
events.

\subsubsection{Efficiency estimated from data}
\label{subSec:TagAndProbe}

A ``tag-and-probe'' method~\cite{CMS:2011aa,Chatrchyan:2012xi} allows an extraction of muon tracking
efficiency directly from
decays of known resonances.
For example, $\cPZ\to\Pgmp\Pgmm$ candidates are reconstructed using pairs of oppositely charged muons
identified in the muon chambers. Each $\cPZ$ candidate must have
one \textit{tag} muon, meaning that it is reconstructed in both the tracker and muon chambers, and one
\textit{probe} muon, meaning that it is reconstructed just in the muon chambers, with no requirement
on the tracker.
The invariant mass of each $\mu^+\mu^-$ candidate is required to be within the 50--130\GeV range, around the
91\GeV mass of the Z boson~\cite{PDG2012}.

For both data and simulated events, the tracking efficiency can be estimated as the fraction of the probe muons in $\cPZ\to\Pgmp\Pgmm$
events that can be associated with
a track reconstructed in the tracker. A correction must be made for the fact that some of the probe muons
are not genuine. This correction is obtained by fitting the dilepton mass spectrum in order to
subtract the non-resonant background, since only genuine dimuons will contribute to the resonant peak.
This must be done separately for $\Pgmp\Pgmm$ candidates in which the probe is associated (or not) with
a track in the tracker.

\begin{figure}[ht!]
  \centering
    \includegraphics[width=0.4\textwidth]{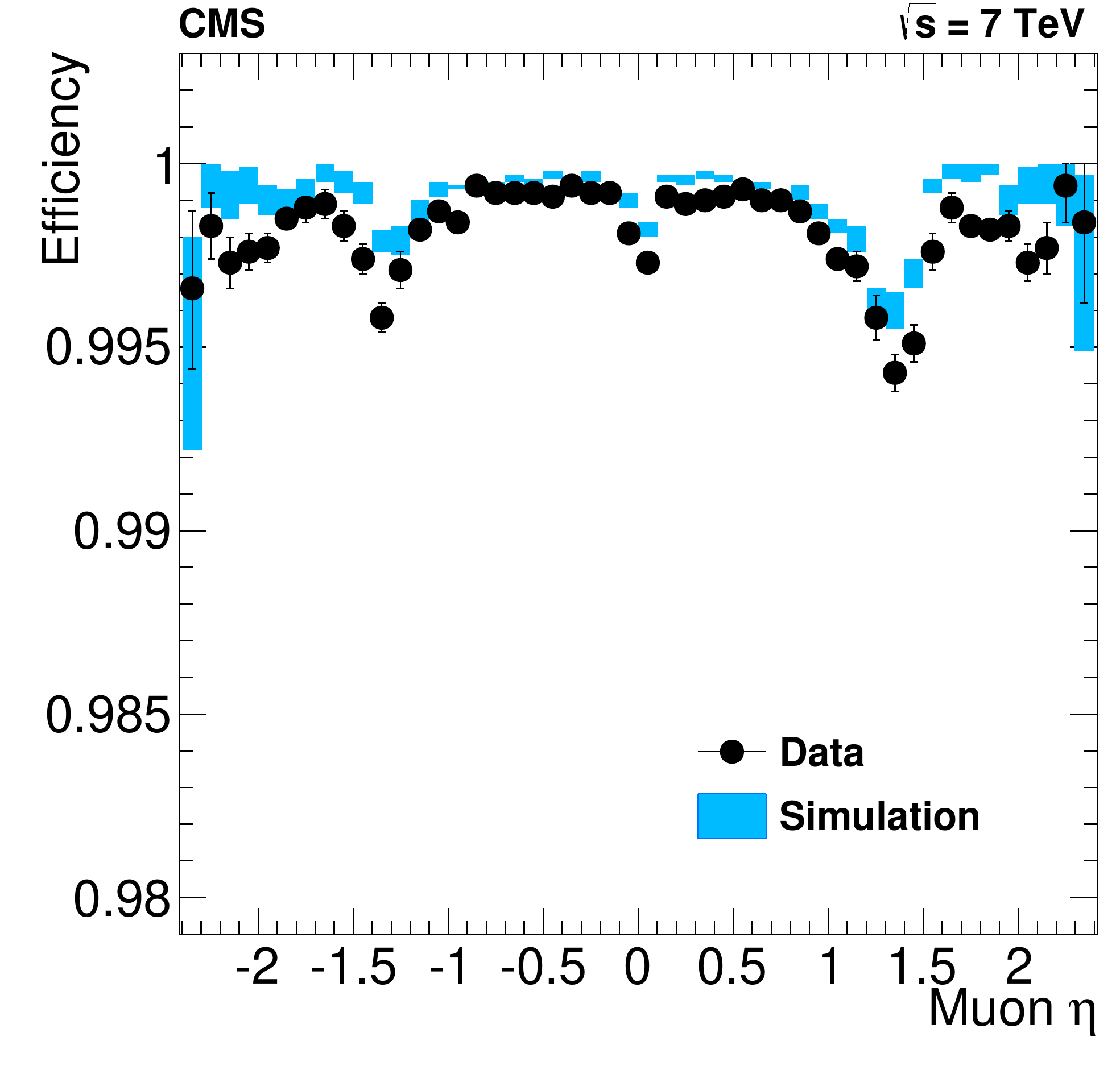}
    \includegraphics[width=0.4\textwidth]{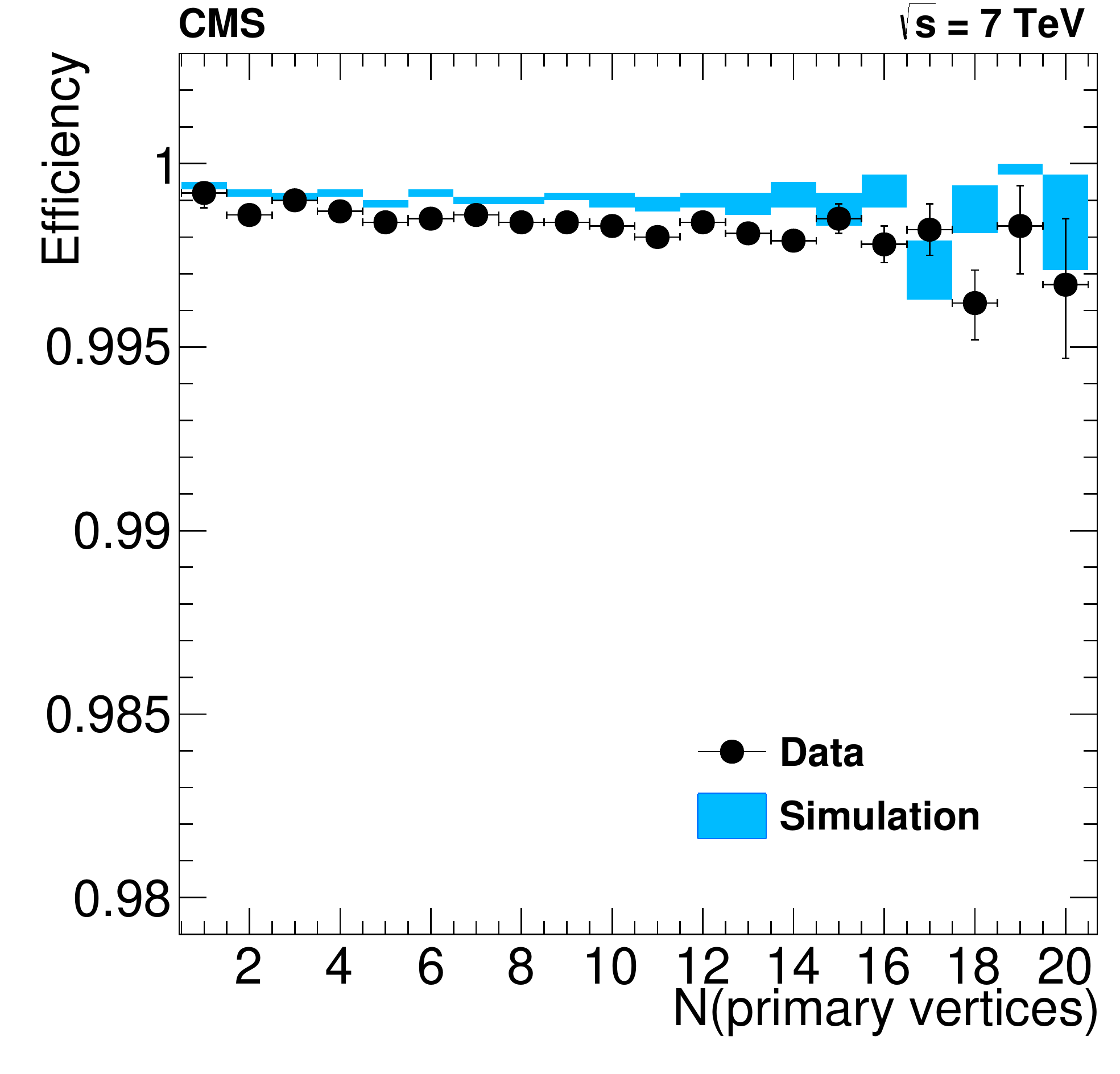}
  \caption{Tracking efficiency measured with a tag-and-probe technique,
           for muons from $\cPZ$ decays, as a function of
           the muon $\eta$ (left) and the number of reconstructed primary vertices in the event (right)
           for data (black dots) and simulation (blue bands). }
  \label{f:eff_meas}
\end{figure}

The results of fits using the tag-and-probe method are shown for data and simulation in Fig.~\ref{f:eff_meas}
as a function of the $\eta$ of the probe, as well as the number of reconstructed primary vertices in the event.
The measured tracking efficiency is $>$99\% in both data and simulation.
The data displays a $\lesssim$0.3\%  drop in tracking efficiency with increasing pileup, which is not reproduced in
the simulation. This may originate from the dynamic (pileup dependent) inefficiency of the pixel detector,
discussed in Section~\ref{s:hit_efficiency}, which is not modelled in the simulation.
The structure in the tracking efficiency when shown as a function of
$\eta$ is caused by inactive modules
and residual misalignment of the tracker. As the figure shows, these detector conditions are well reproduced in simulation.

\subsection{Resolution in the track parameters}
\label{subSec:TrackParResolution}

In the context of the reconstruction software of CMS,
the five parameters used to describe a track are: $d_0$, $z_0$, $\phi$, $\cot \theta$, and the
\pt of the track, defined at the point of closest approach of the
track to the assumed beam axis. This point is called the \textit{impact point},
with global coordinates ($x_0$, $y_0$, $z_0$). Thus $d_0$ and $z_0$ define the
coordinates of the impact point in the radial and $z$ directions ($d_0 = -y_0 \cos \phi+
x_0 \sin \phi$). The azimuthal and polar angles of the momentum vector of the track are denoted by $\phi$  and $\theta$, respectively.

The resolution in the parameters is studied using simulated events, and estimated from \textit{track residuals},
which are defined as the differences between the reconstructed track parameters and the
corresponding parameters of the generated particles. For each of the five track parameters, the
resolution is plotted as a function of the $\eta$ or \pt of the simulated charged particle.
In every bin of $\eta$ or \pt, the distribution in track residuals defines the resolution as the
half-width of the interval that satisfies both of the following requirements.
\begin{itemize}
\item The width contains 68\% of all entries (including underflows and overflows) in the distribution of
the residuals.
\item The interval is centred on the most probable value (mode) of the residuals, where this value is
taken from the peak of a double-tailed Crystal Ball function~\cite{Oreglia:1980cs} fitted to the residuals.
The function must provide different parametrizations of the tails on the left and right sides of the
residuals distribution as, especially for electrons, the distribution can be very asymmetric.
\end{itemize}

For all resolution plots, we also provide a second measure of the resolution, defined such that the
interval contains 90\% of the track residuals. This quantifies the impact of the extreme values, whereas
the resolutions for the 68\% intervals represent the core of the distribution.

\subsubsection{Results from simulation of isolated particles}
\label{sec:PerfResolutionsMCsingle}

Muons do not undergo strong interactions, and
therefore they tend to traverse the entire volume of the tracker, so
the hits on their trajectories provide a long lever arm for reconstruction.

\begin{figure}[hbtp]
  \centering
    \includegraphics[width=0.45\textwidth]{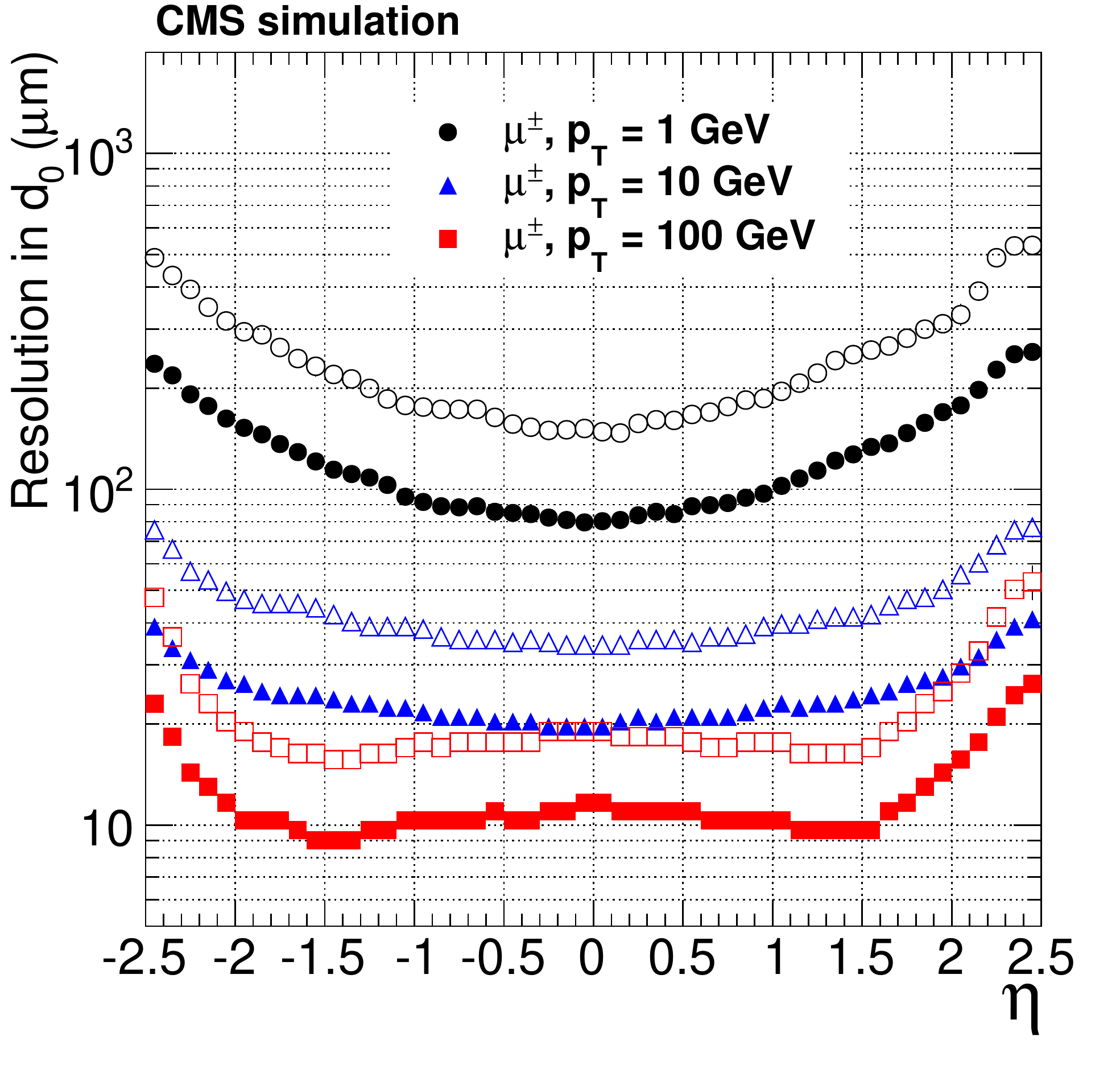}
    \includegraphics[width=0.45\textwidth]{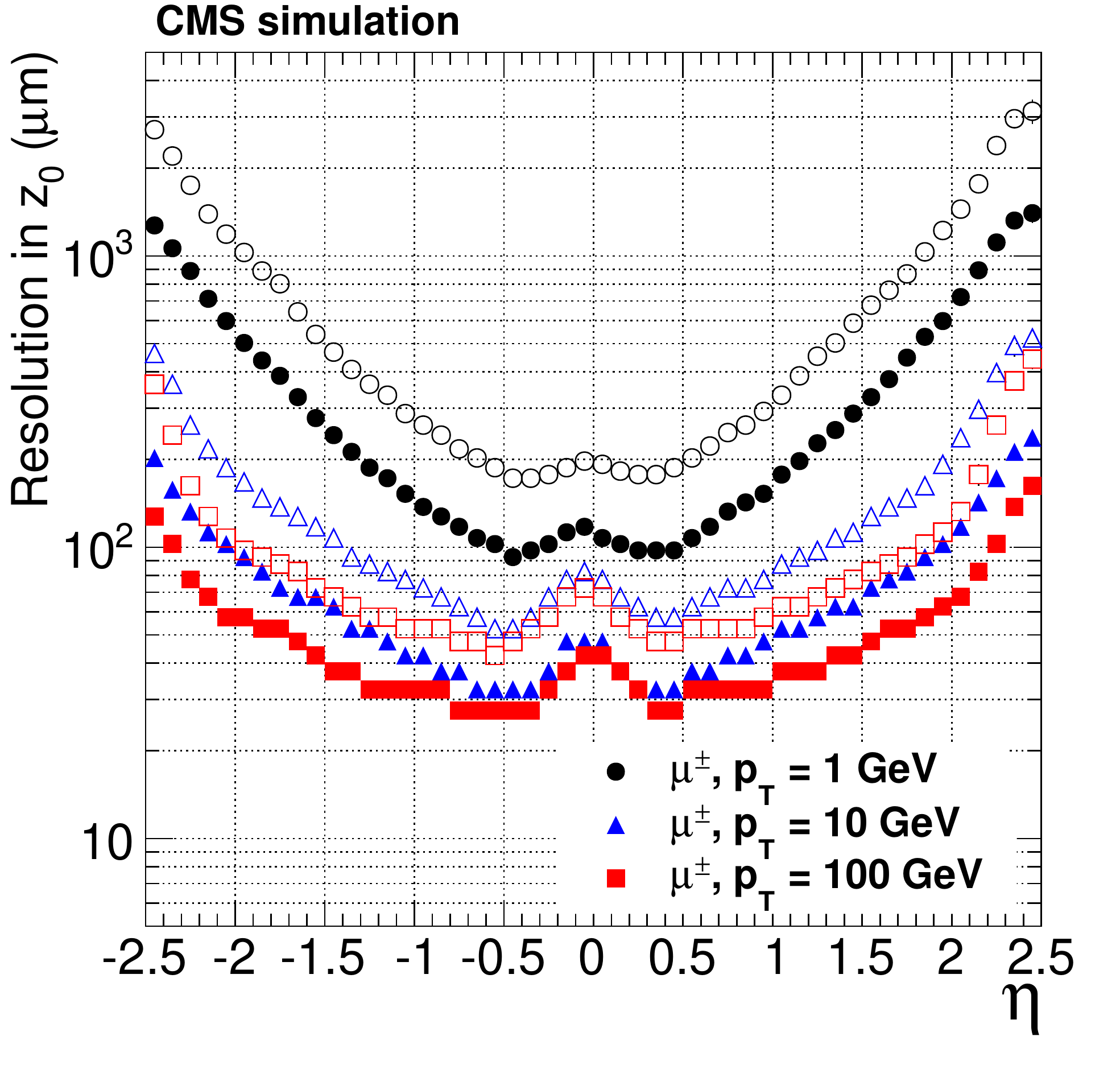}
    \includegraphics[width=0.45\textwidth]{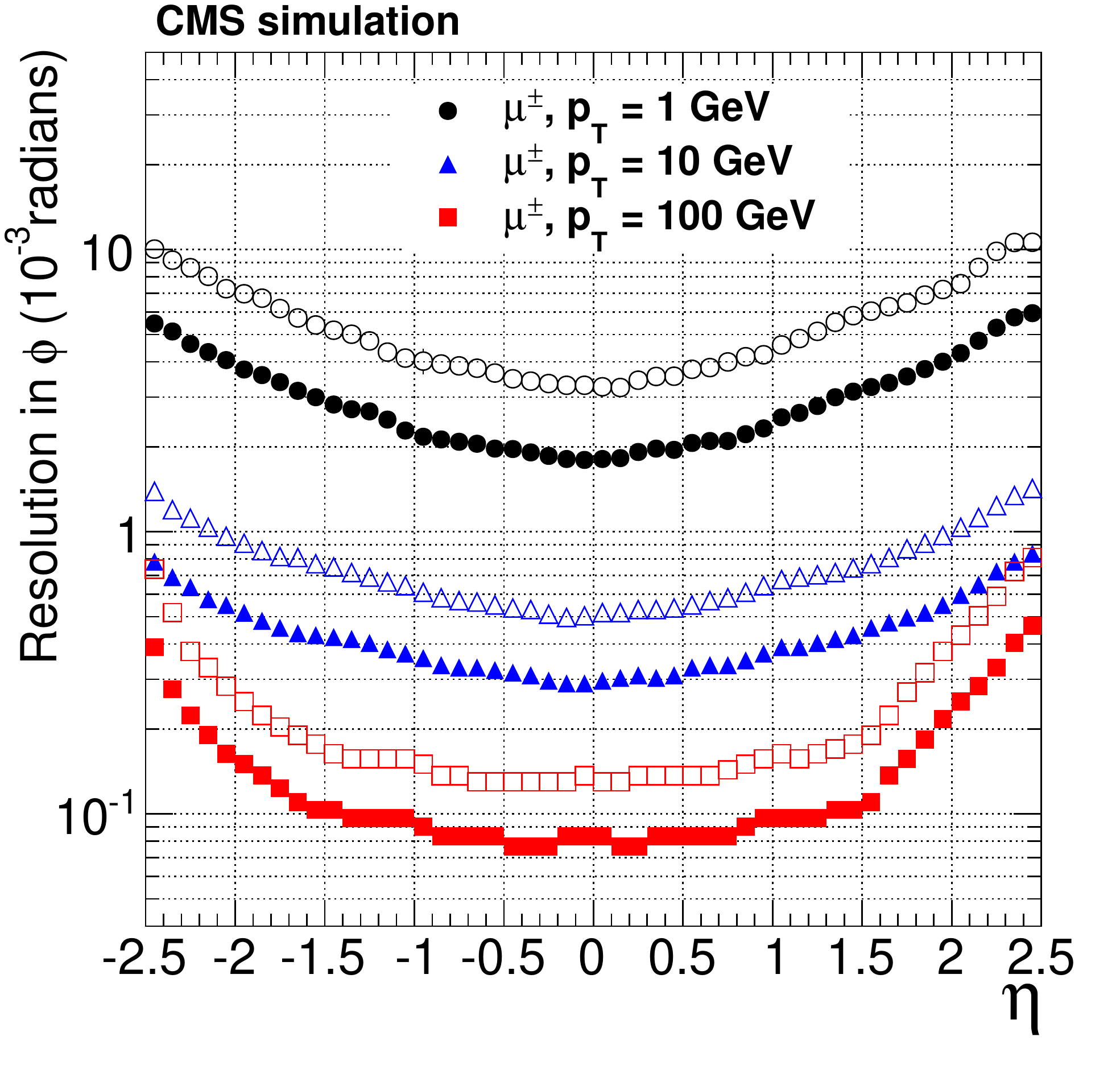}
    \includegraphics[width=0.45\textwidth]{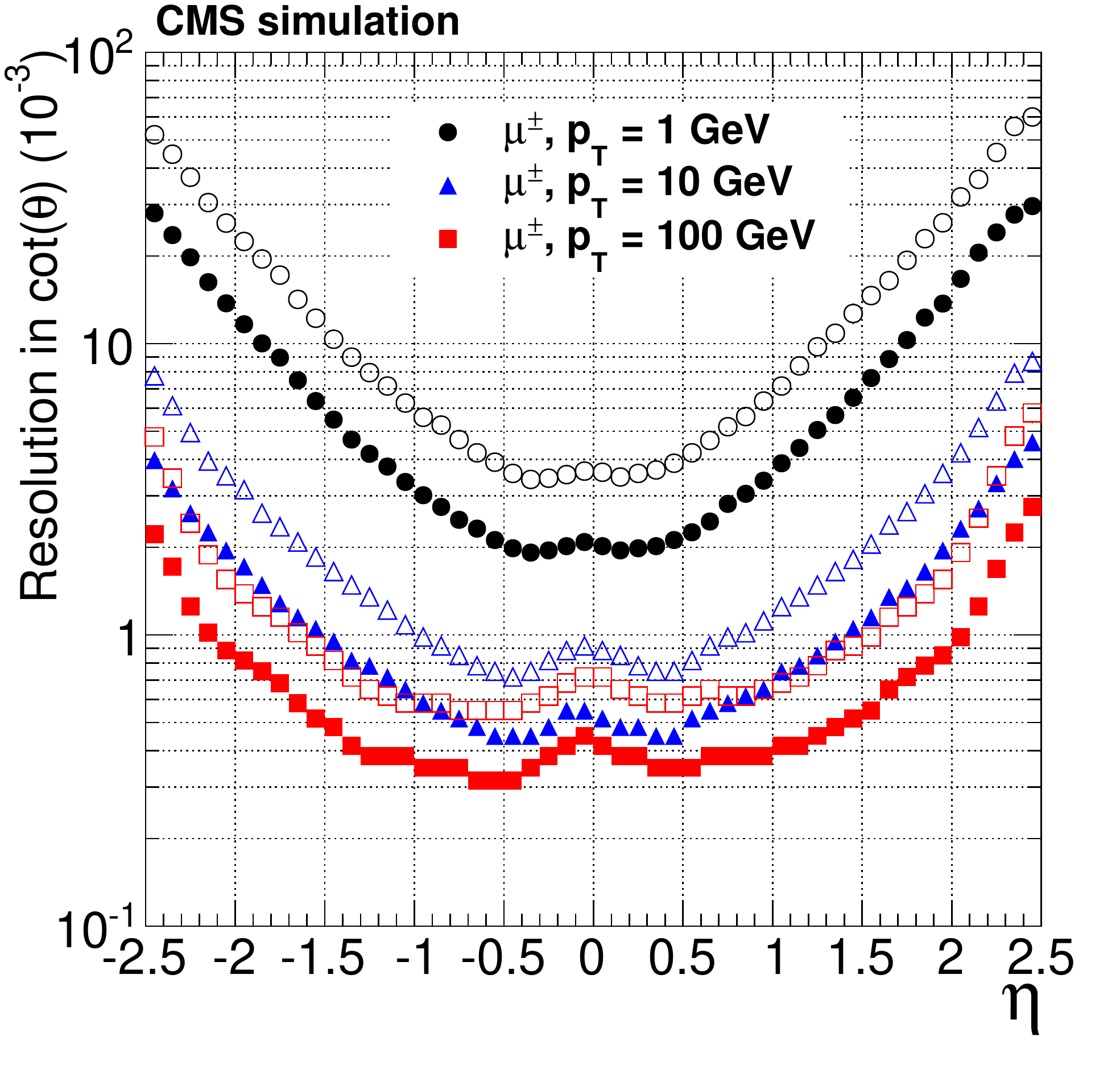}
    \includegraphics[width=0.45\textwidth]{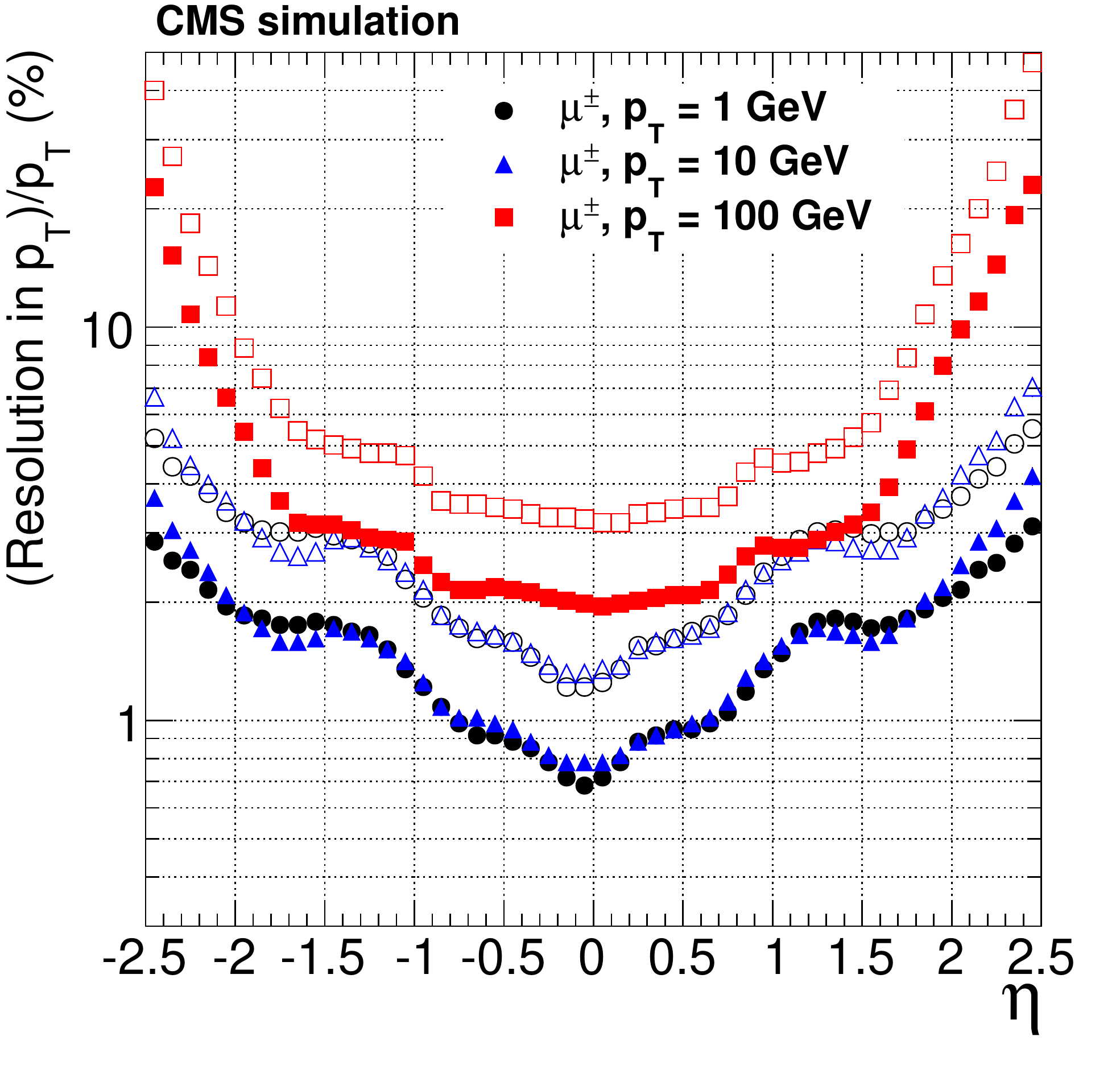}
  \caption {
    Resolution, \textit{as a function of pseudorapidity}, in the five track parameters for \textit{single, isolated muons} with
    $\pt  = 1$, 10, and 100\GeV. From top to bottom and left
    to right:  transverse and longitudinal impact parameters, $\phi$, $\cot \theta$
    and transverse momentum.
    For each bin in $\eta$, the solid (open) symbols
    correspond to the half-width for 68\% (90\%) intervals centered on the
    mode of the distribution in residuals, as described in  the text.
 }
    \label{fig:resolutionsVsEtaMuMC}
\end{figure}

\begin{figure}[hbtp]
  \centering
    \includegraphics[width=0.45\textwidth]{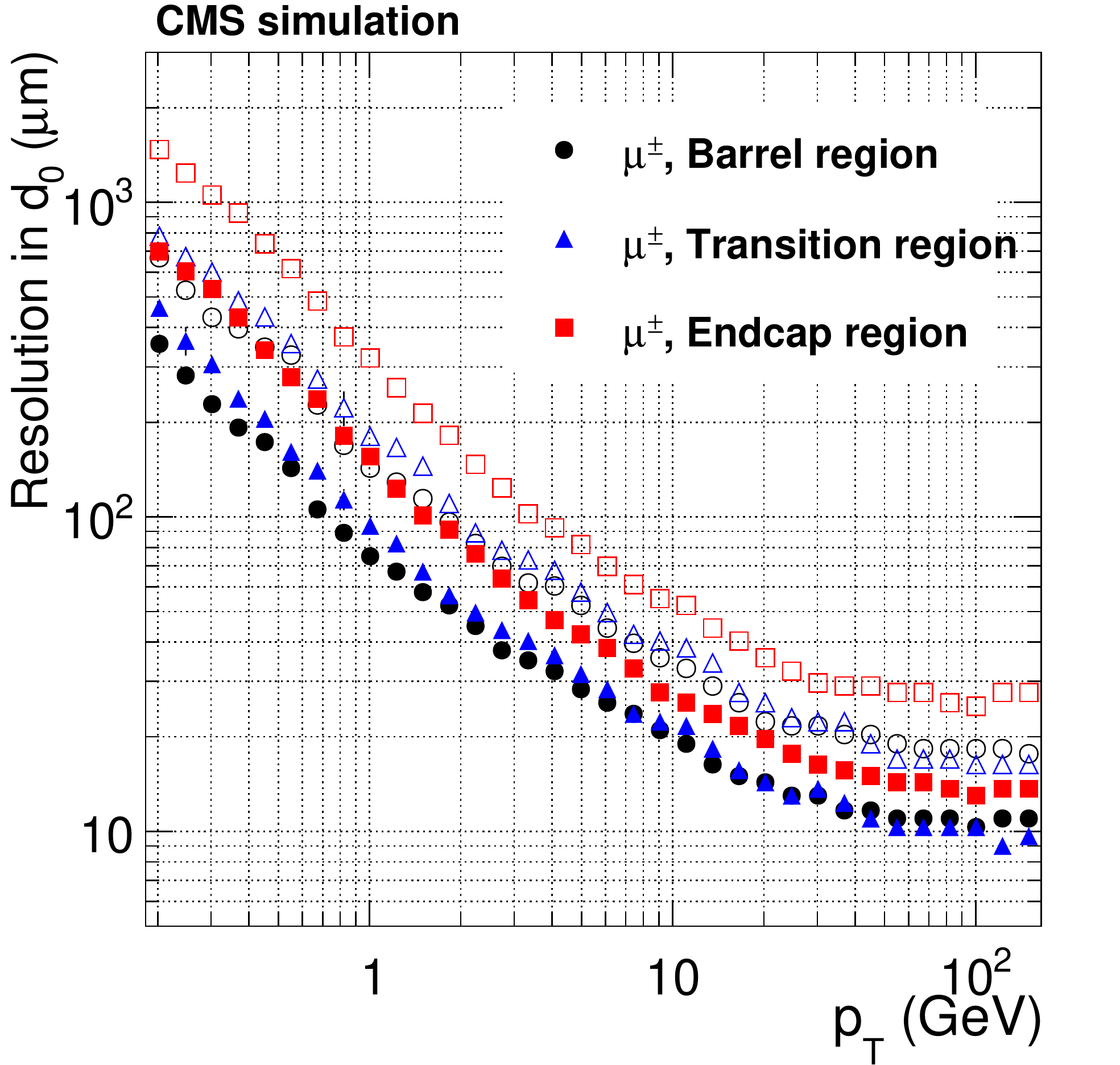}
    \includegraphics[width=0.45\textwidth]{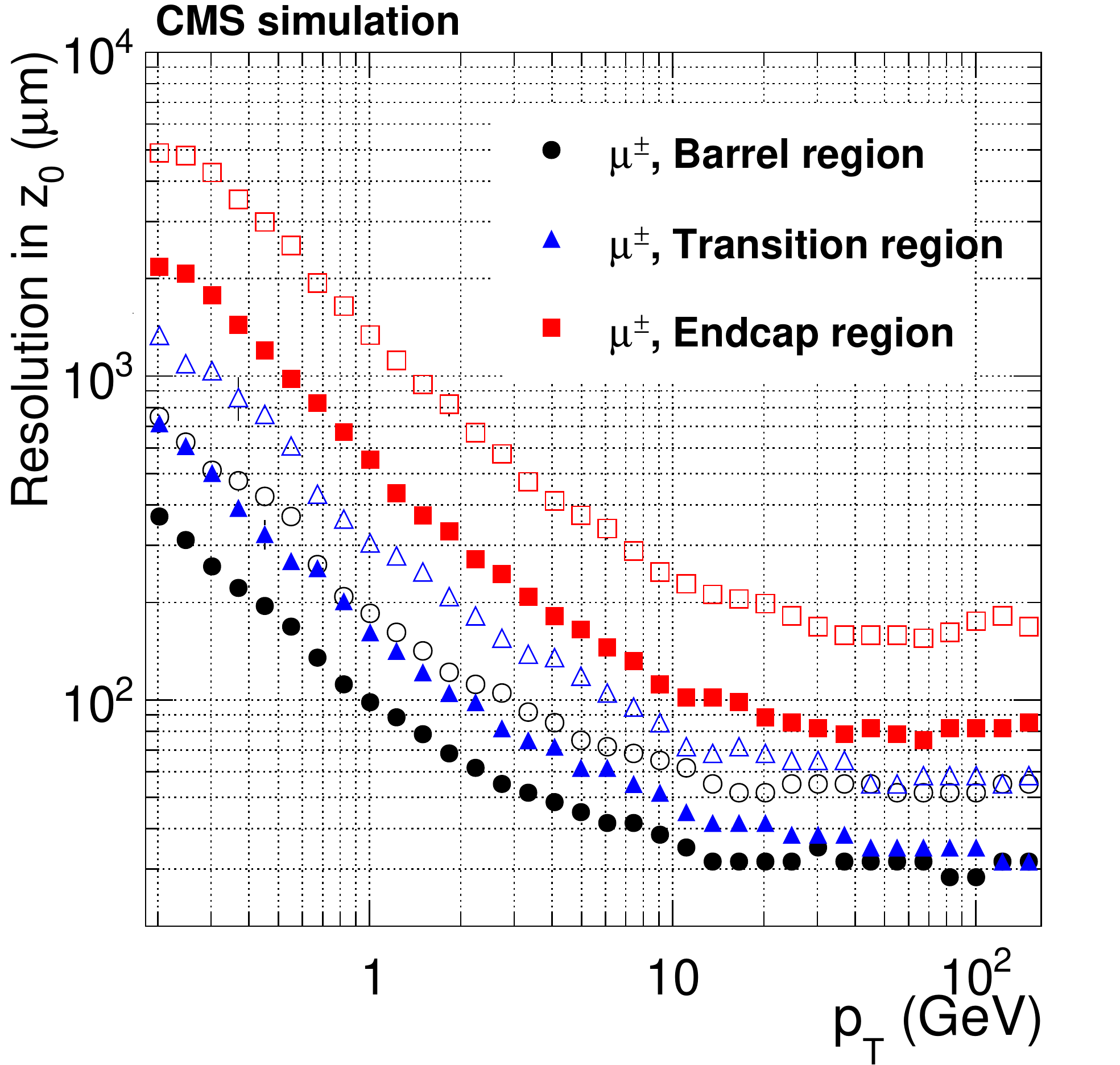}
    \includegraphics[width=0.45\textwidth]{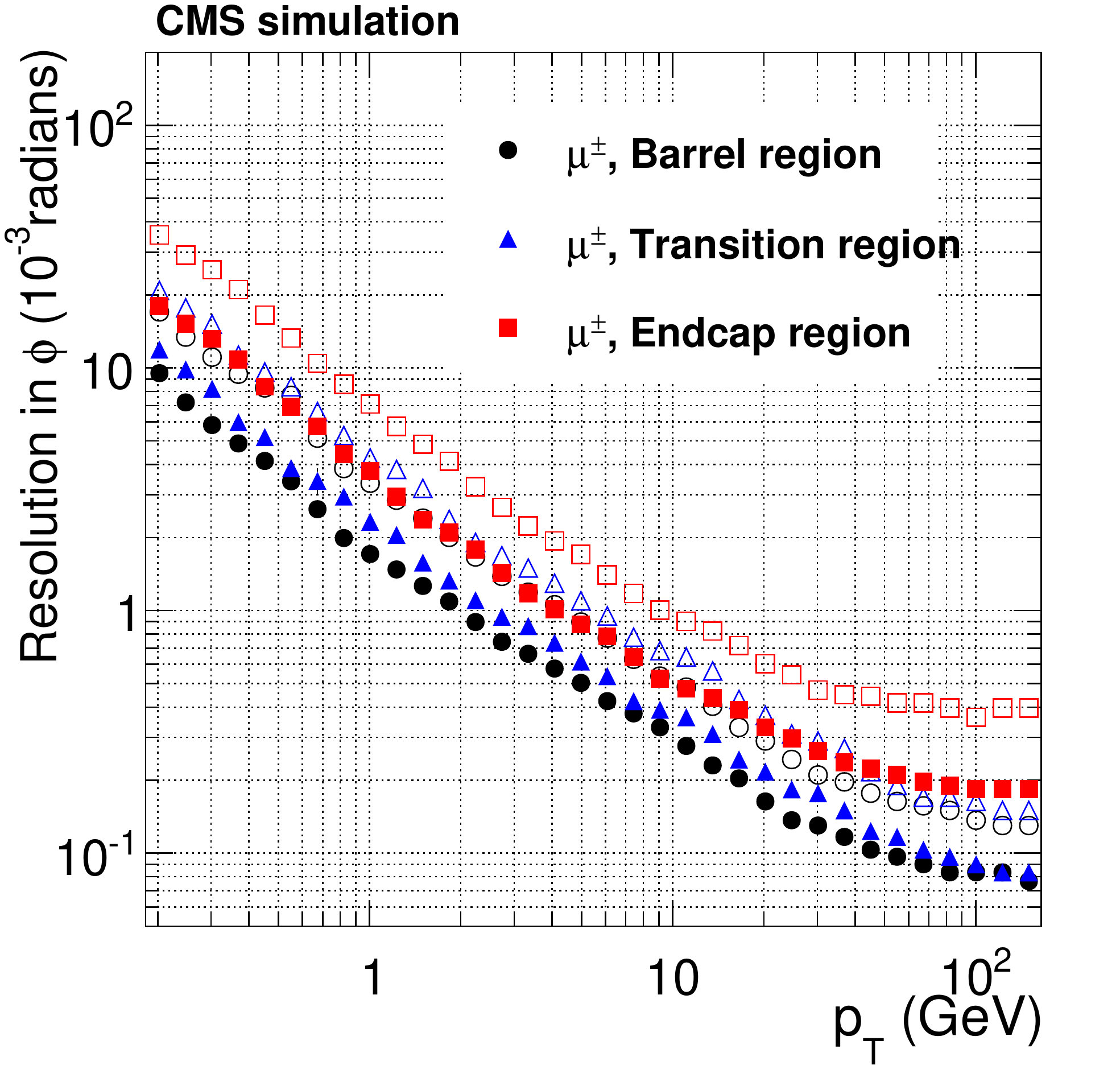}
    \includegraphics[width=0.45\textwidth]{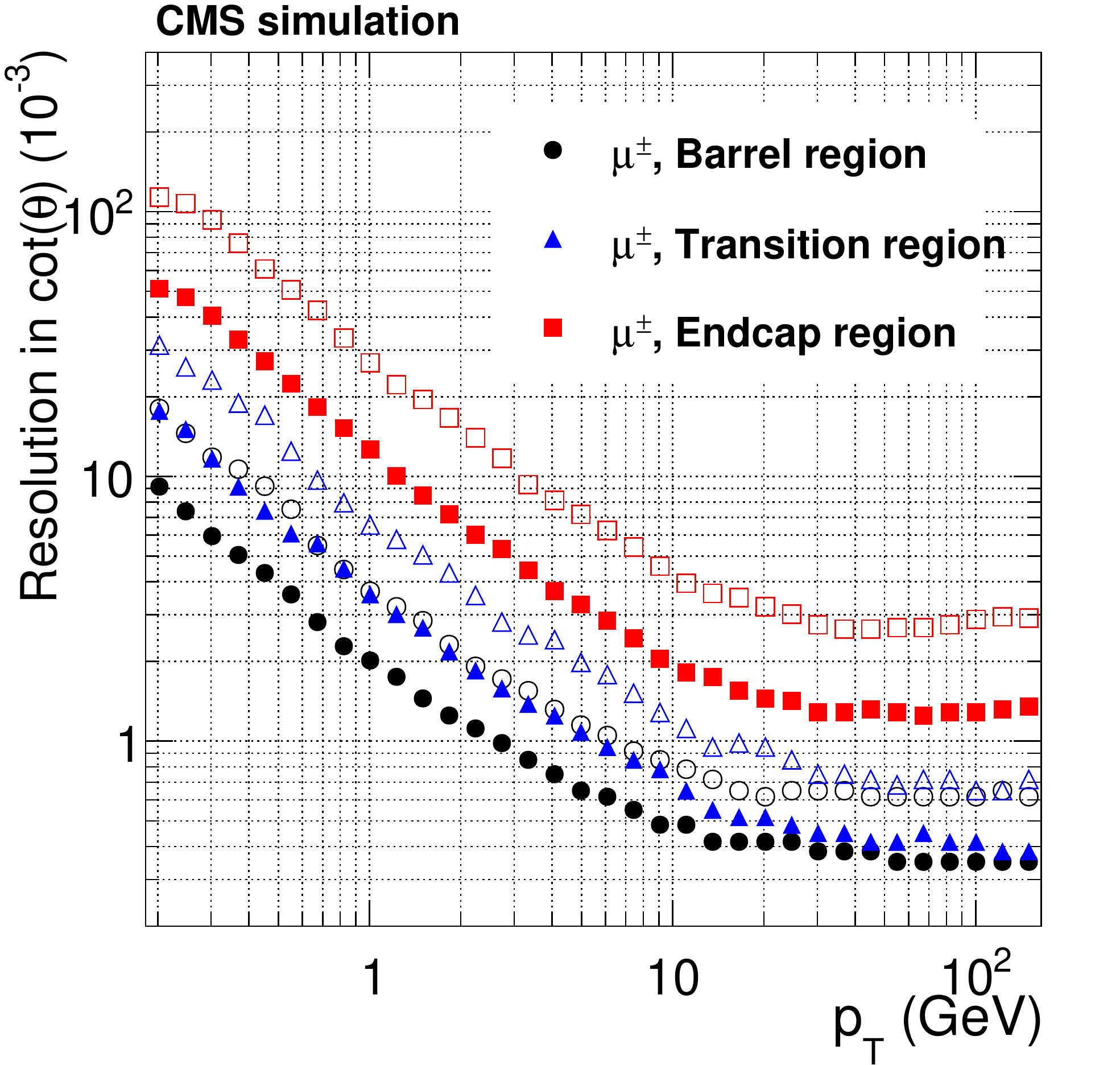}
    \includegraphics[width=0.45\textwidth]{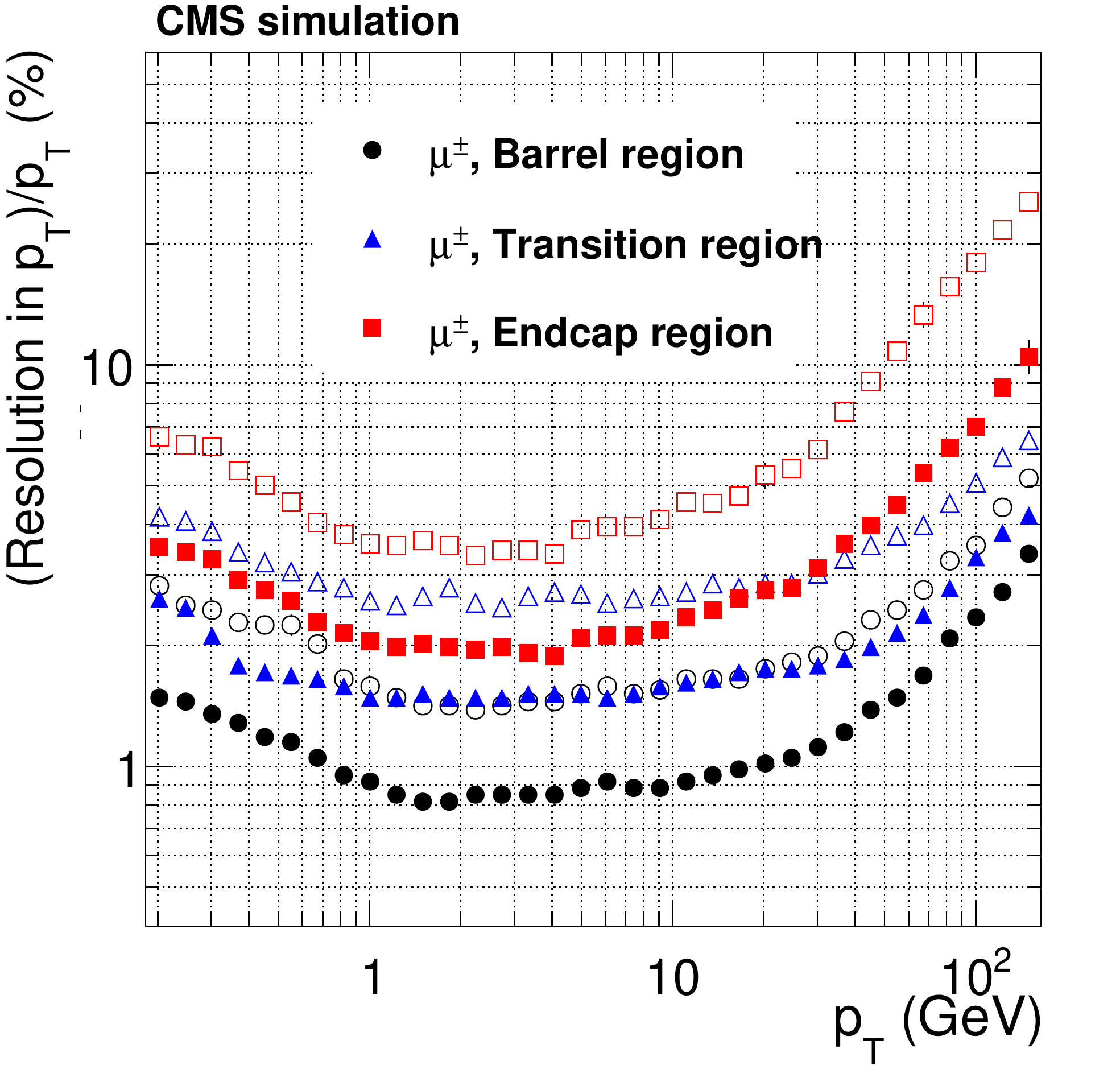}
  \caption {
      Resolution, \textit{as a function of \pt}, in the five track parameters for \textit{single, isolated muons} in
      the barrel, transition, and endcap regions, defined by
      $\eta$ intervals of 0--0.9, 0.9--1.4 and 1.4--2.5, respectively.
      From top to bottom and left to right: transverse and longitudinal impact parameters, $\phi$, $\cot \theta$
      and \pt.
      For each bin in \pt, the solid (open) symbols
      correspond to the half-width for 68\% (90\%) intervals centered on the
      mode of the distribution in residuals, as described in  the text.
 }
    \label{fig:resolutionsVsPtMuMC}
\end{figure}

Figure~\ref{fig:resolutionsVsEtaMuMC} shows the dependence on $\eta$ of the resolution for the
five track parameters, for isolated muons with $\pt = 1,$ 10, and 100\GeV.
The same resolutions, but as a function of \pt, are shown in Fig.~\ref{fig:resolutionsVsPtMuMC}.
The resolutions in both the impact parameters and the angular parameters generally deteriorate
for larger values of $\abs{\eta}$ because the extrapolation from the innermost
hit to the beam axis, where the parameters are calculated, becomes larger.
The resolutions in the transverse and longitudinal impact parameters $d_0$ and $z_0$ are shown in
the first two plots of Figs.~\ref{fig:resolutionsVsEtaMuMC} and \ref{fig:resolutionsVsPtMuMC}.
At high momentum, the impact parameter resolution
is dominated by the position resolution of the innermost hit in the pixel detector, while at
lower momenta, the resolution is degraded progressively by multiple scattering. The
improvement in $z_0$ resolution as $\abs{\eta}$ increases to 0.4 can be attributed to the
beneficial effect of charge sharing in the estimation of position of pixel
clusters~(see Fig.~\ref{fig:pixResZ}); in the barrel, as the crossing angle for the tracks in
the pixel layers increases, the clusters broaden, distributing thereby the signal over more than one
pixel, and improving the resolution in position.
The resolutions in the $\phi$ and $\cot \theta$ parameters, shown in the middle two
panels of Figs.~\ref{fig:resolutionsVsEtaMuMC} and \ref{fig:resolutionsVsPtMuMC},
have distributions in resolutions similar to those found for $d_0$ and $z_0$, respectively, for likewise reasons.
However, as the contribution of the strip tracker to the
measurement of $\phi$ and $\theta$ is important, the influence of charge sharing in the pixel
tracker is smaller.
As a function of $\eta$, the resolutions in the four track parameters $d_0$, $z_0$, $\phi$, and $\theta$,
are not exactly symmetric around $\eta = 0$. This effect is not caused by the tracker geometry, but is rather due to the noisy and dead channels
of pixel and strip modules, whose defective components are taken into account in simulation to reproduce the condition of the real detector.
The resolution in \pt is shown in the bottom panel of
Figs.~\ref{fig:resolutionsVsEtaMuMC} and \ref{fig:resolutionsVsPtMuMC}. At high transverse momentum (100\GeV),
the resolution is $\approx$2--3\% up to $\abs{\eta}=1.6$, but deteriorates at
higher $\abs{\eta}$ values, because of the shorter lever arm of these tracks in the $x$-$y$ plane of the tracker.
The degradation at $\abs{\eta} \approx 1.0$ and beyond is due to the gap between the barrel and
the endcap disks~(Fig.~\ref{fig:TrackerLayout}), and due to the inferior hit resolution of the last hits of
the track measured in TEC ring~7 compared to the hit resolution in TOB layers~5 and
6~(Table~\ref{tab:sstHitRes}).
At a transverse momentum of 100\GeV, the material in the tracker accounts for between 20 and
30\% of the transverse momentum resolution; at lower momenta, the resolution is dominated by
multiple scattering and its value reflects the amount of material traversed by the track.
The relative precision in \pt is measured to be best for tracks with $\pt \approx 3\GeV$.

\begin{figure}[hbtp]
  \centering
    \includegraphics[width=0.45\textwidth]{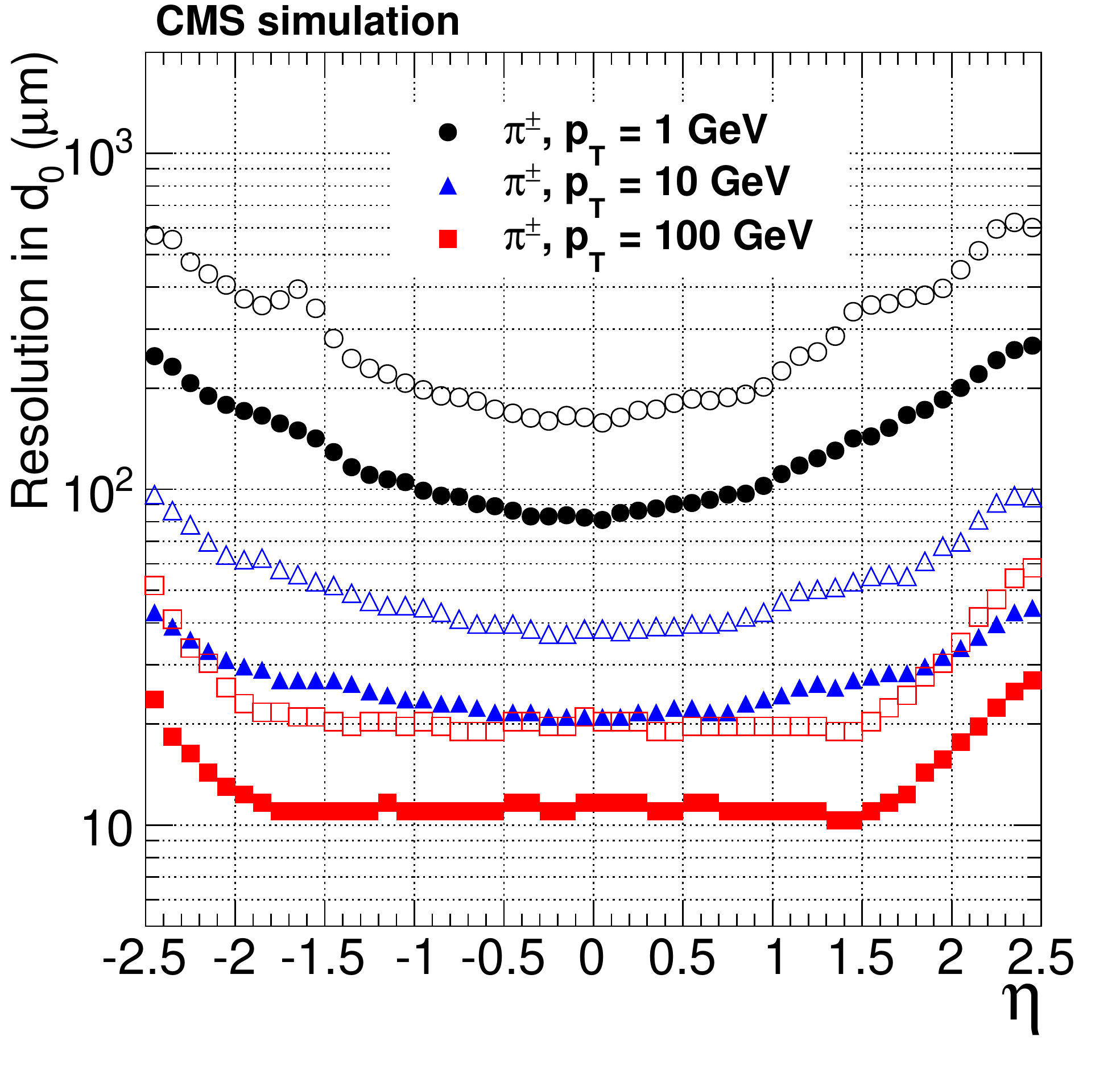}
    \includegraphics[width=0.45\textwidth]{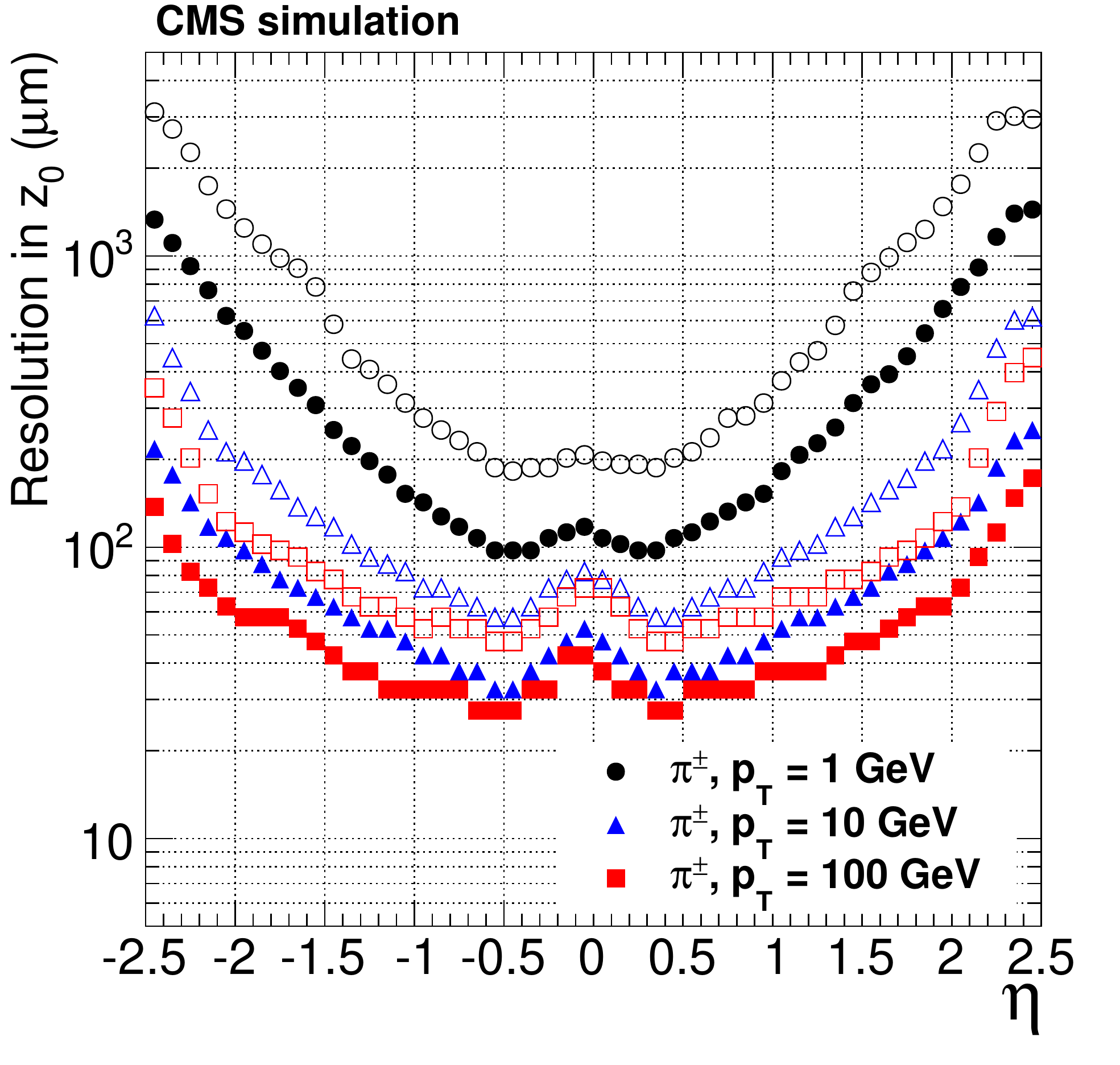}
    \includegraphics[width=0.45\textwidth]{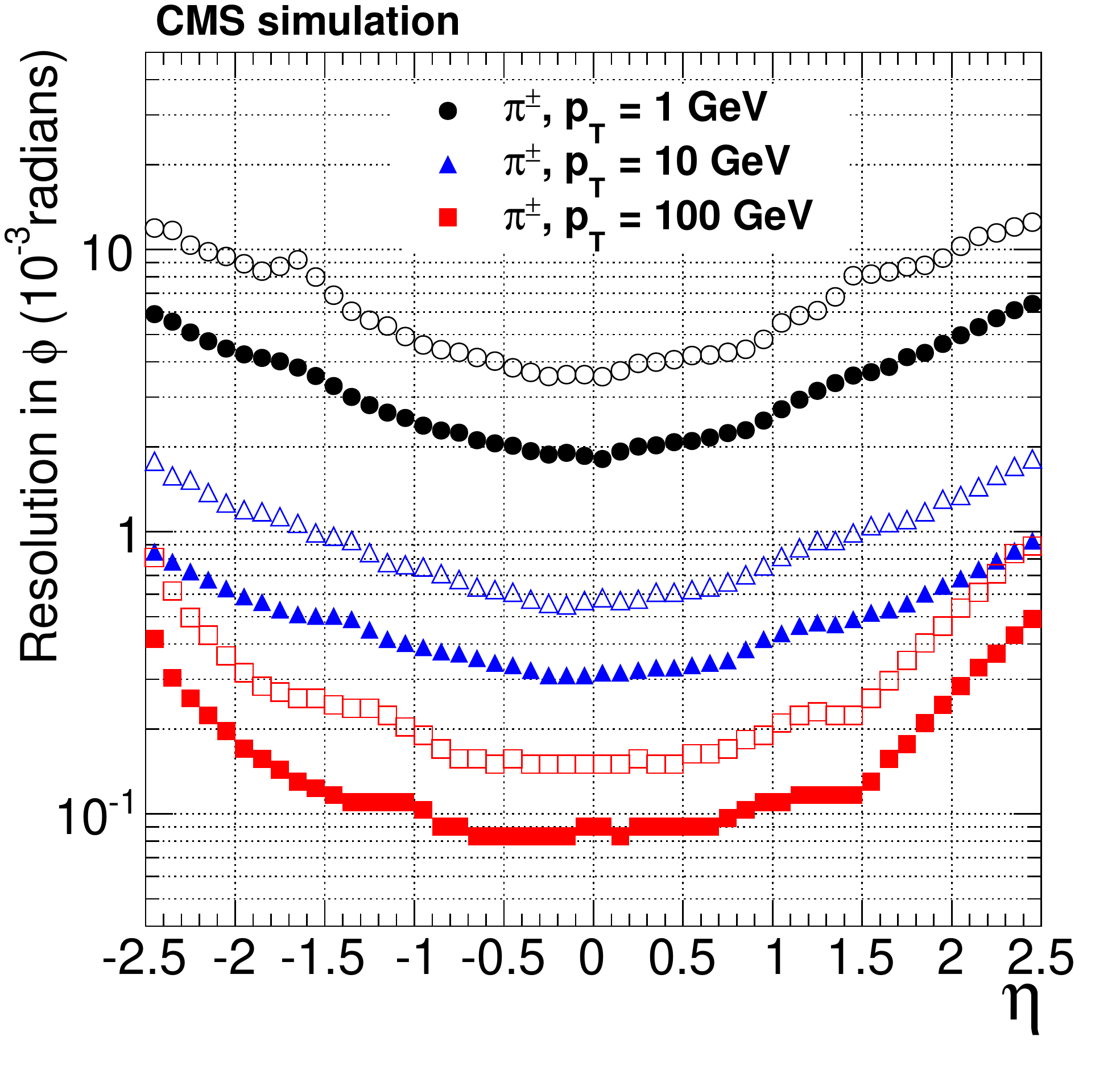}
    \includegraphics[width=0.45\textwidth]{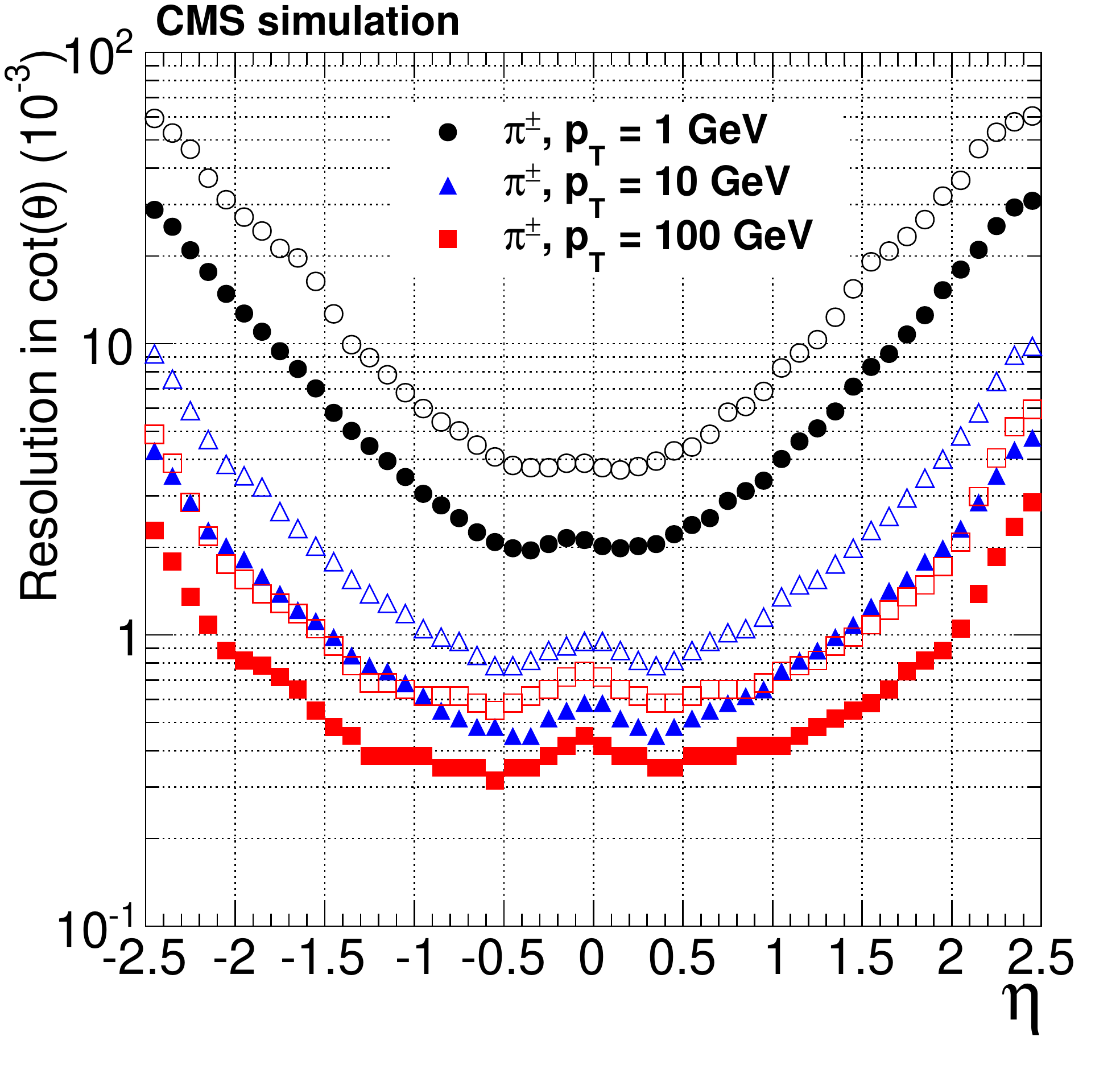}
    \includegraphics[width=0.45\textwidth]{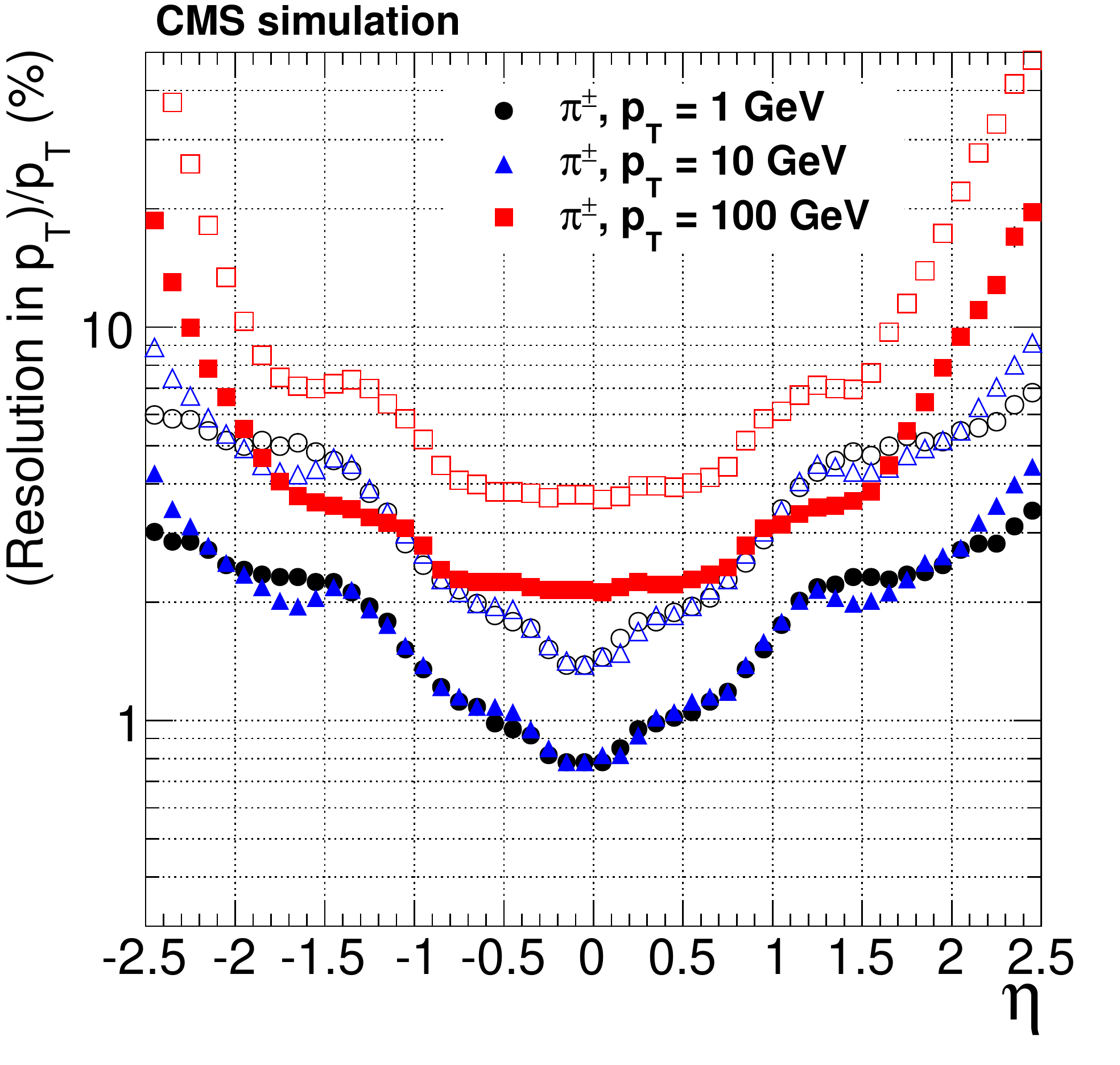}
  \caption {
    Resolution, \textit{as a function of pseudorapidity}, in the five track parameters for \textit{single, isolated pions} with
    transverse momenta of 1, 10, and 100\GeV. From top to bottom and left
    to right: transverse and longitudinal impact parameters, $\phi$, $\cot \theta$
     and \pt.
     For each bin in $\eta$, the solid (open) symbols
     correspond to the half-width for 68\% (90\%) intervals centered on the
     mode of the distribution in residuals, as described in  the text.
 }
    \label{fig:resolutionsVsEtaPiMC}
\end{figure}

Charged pions that do not undergo nuclear interactions behave similarly to muons, as they
are subjected to the same multiple scattering effects and to the same mechanism of energy loss through
ionization. The trajectories of this subset of pions are reconstructed using the CTF algorithm with a
precision that is close to that achieved for muons, and therefore these trajectories populate
the core of the distributions of residuals. The five plots in
Fig.~\ref{fig:resolutionsVsEtaPiMC} show resolutions in the five track parameters as a function
of $\eta$. As expected, the
results are very close to those observed for muons in Fig.~\ref{fig:resolutionsVsEtaMuMC}.
However, the resolutions obtained
for the 90\% interval have a somewhat different pattern for muons than for pion tracks crossing
the barrel-endcap transition region of the tracker. The residuals are generally larger for pions,
as they can interact inelastically, and thereby fail to reach the outer layers of the tracking system.
Their trajectories are measured therefore using smaller lever arms, with degraded resolutions.

Three of the track parameters ($d_0$, $\phi$ and \pt) for electrons have very asymmetric residual distributions
because of bremsstrahlung, and we therefore alter the definition of their resolution. The distribution in track residuals
is split into two regions, separated at the mode of the distribution. Only one of these two regions contains long, non-Gaussian tails due to bremsstrahlung and the resolution is now redefined using only the distribution in this region. It is given by the width of an interval that starts at the
mode of the distribution and is wide enough to include 68\% of the entries in the region. A similar definition
of the resolution corresponding to the width of a 90\% probability interval is used to quantify the size of the non-Gaussian
tails. Note that if the distribution of residuals had been symmetric, then the results obtained with
these new definitions of the resolution would be identical to those that would have been obtained with
the original definitions from the beginning of Section~\ref{subSec:TrackParResolution}.
The other two parameters ($\cot \theta$  and $z_0$) are less affected by bremsstrahlung, and we therefore continue
to use the same definition of resolution as for muons and pions.

\begin{figure}[hbtp]
  \centering
    \includegraphics[width=0.45\textwidth]{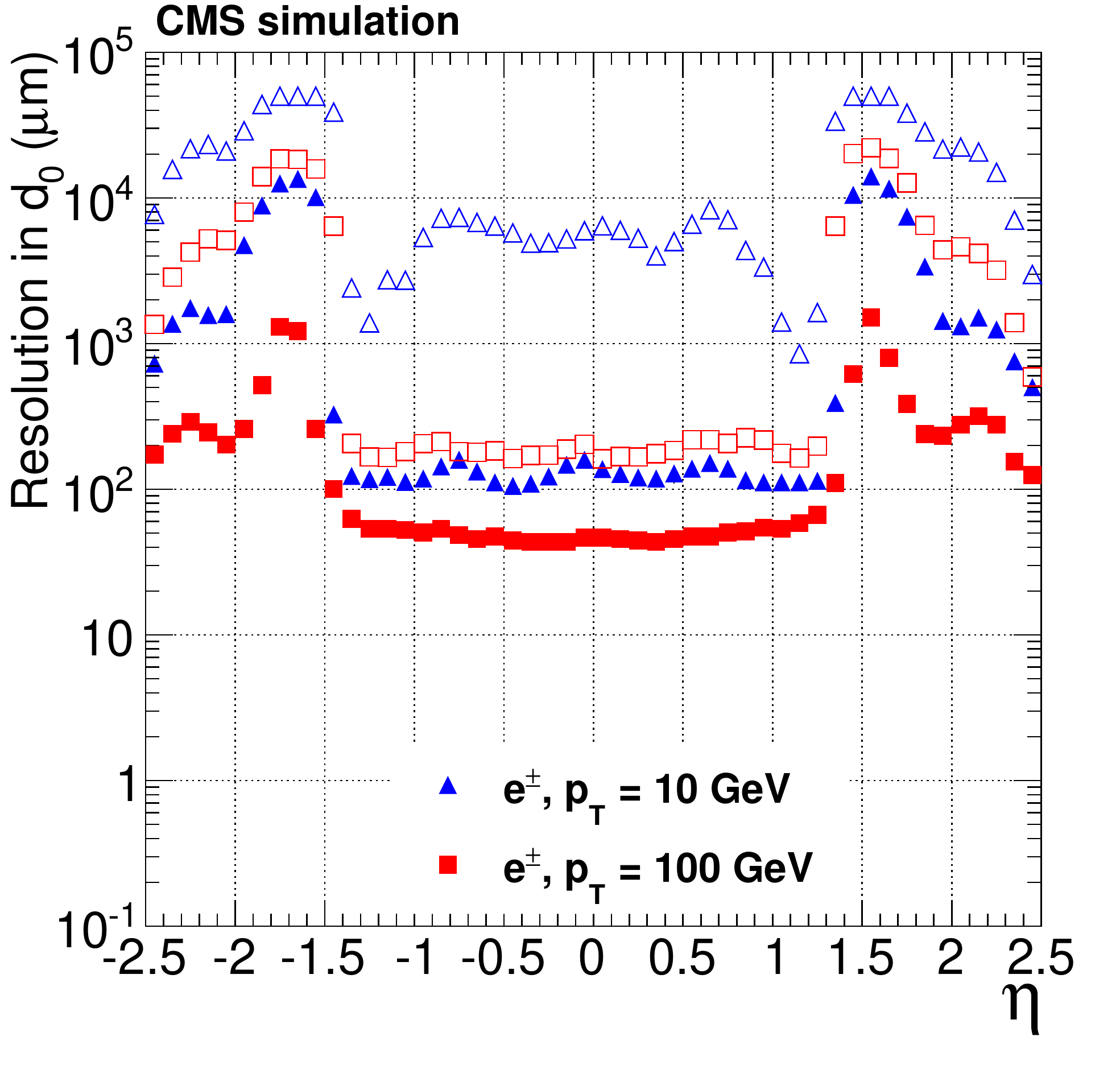}
    \includegraphics[width=0.45\textwidth]{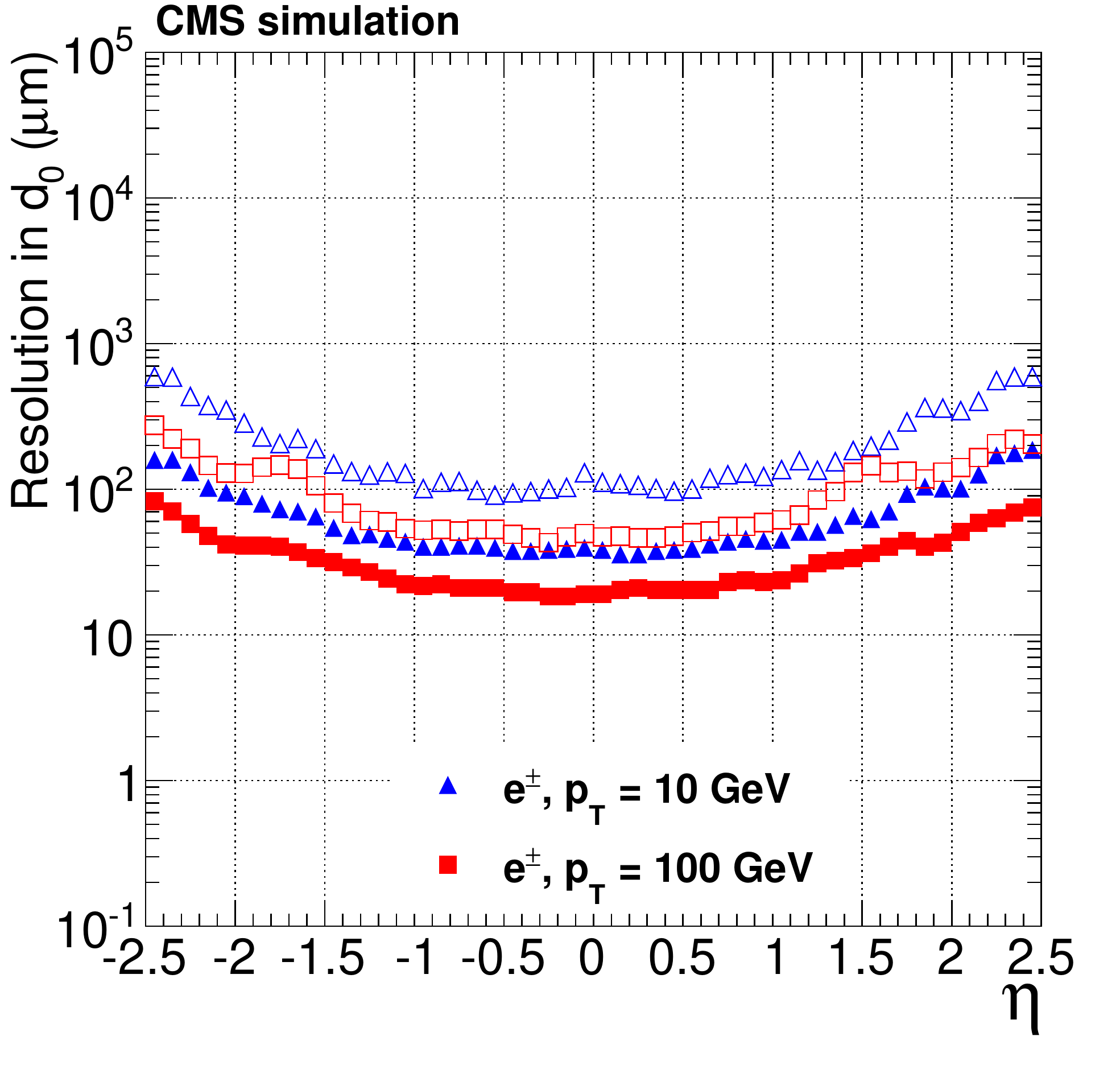}

    \includegraphics[width=0.45\textwidth]{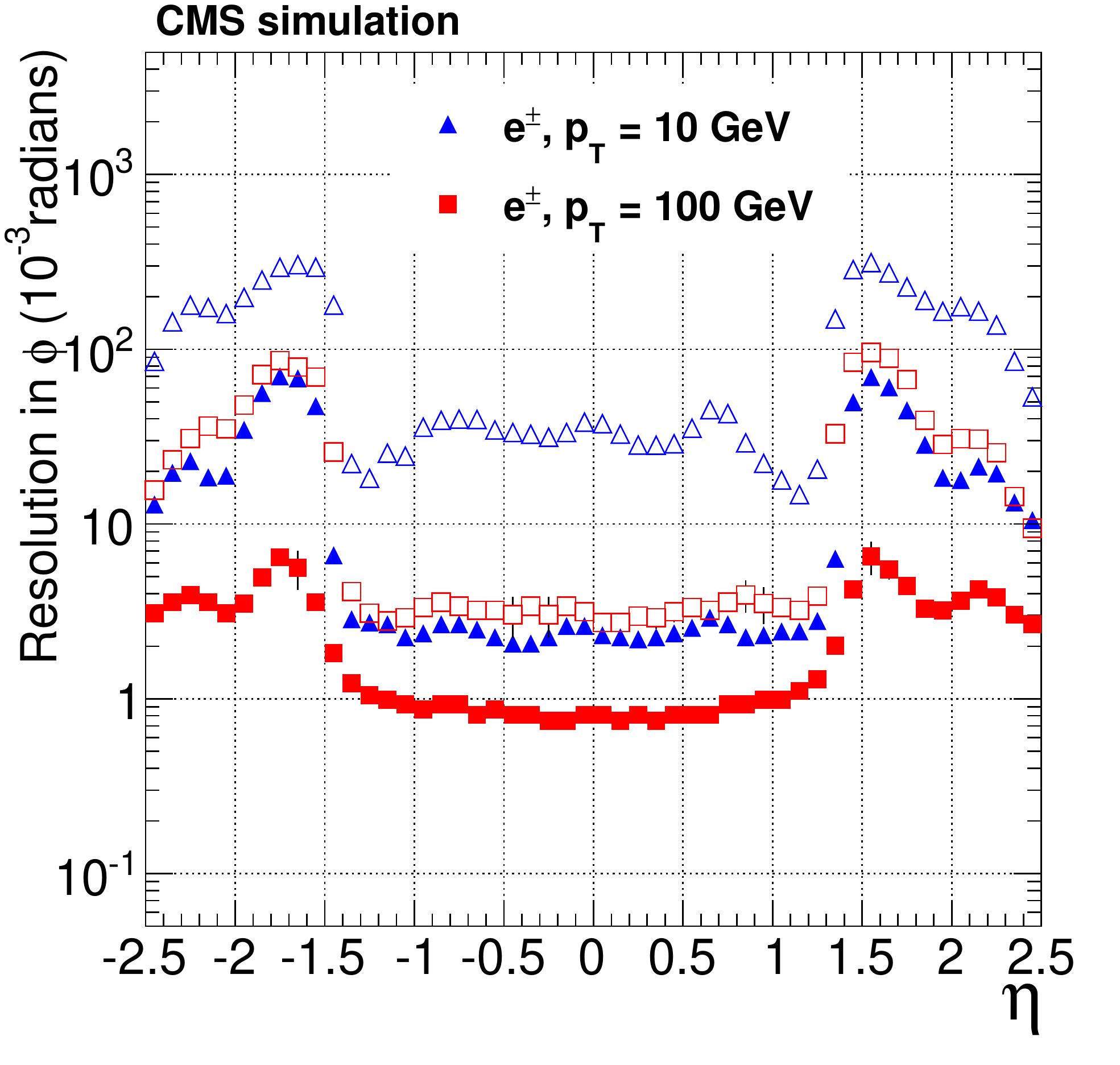}
    \includegraphics[width=0.45\textwidth]{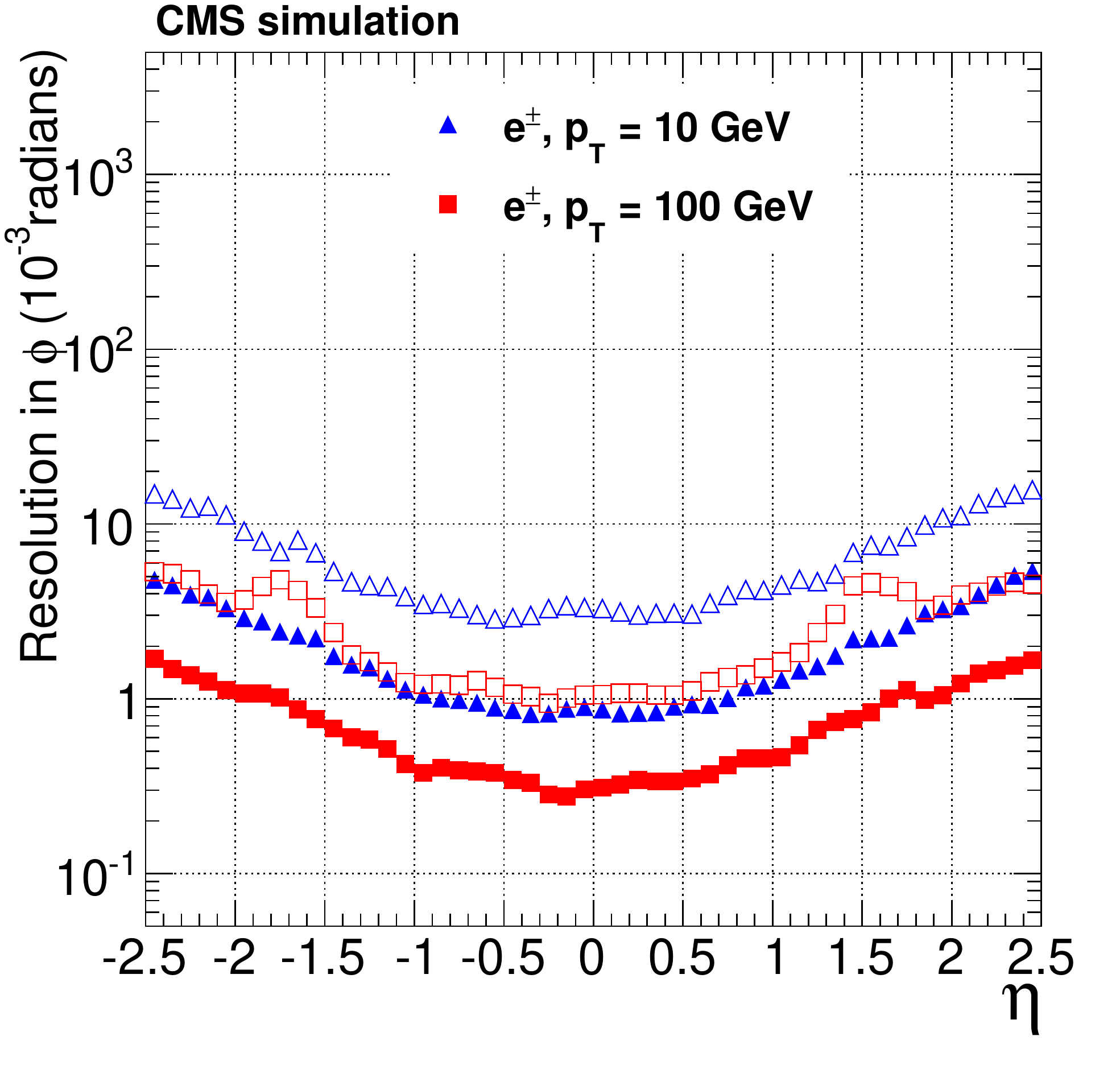}

    \includegraphics[width=0.45\textwidth]{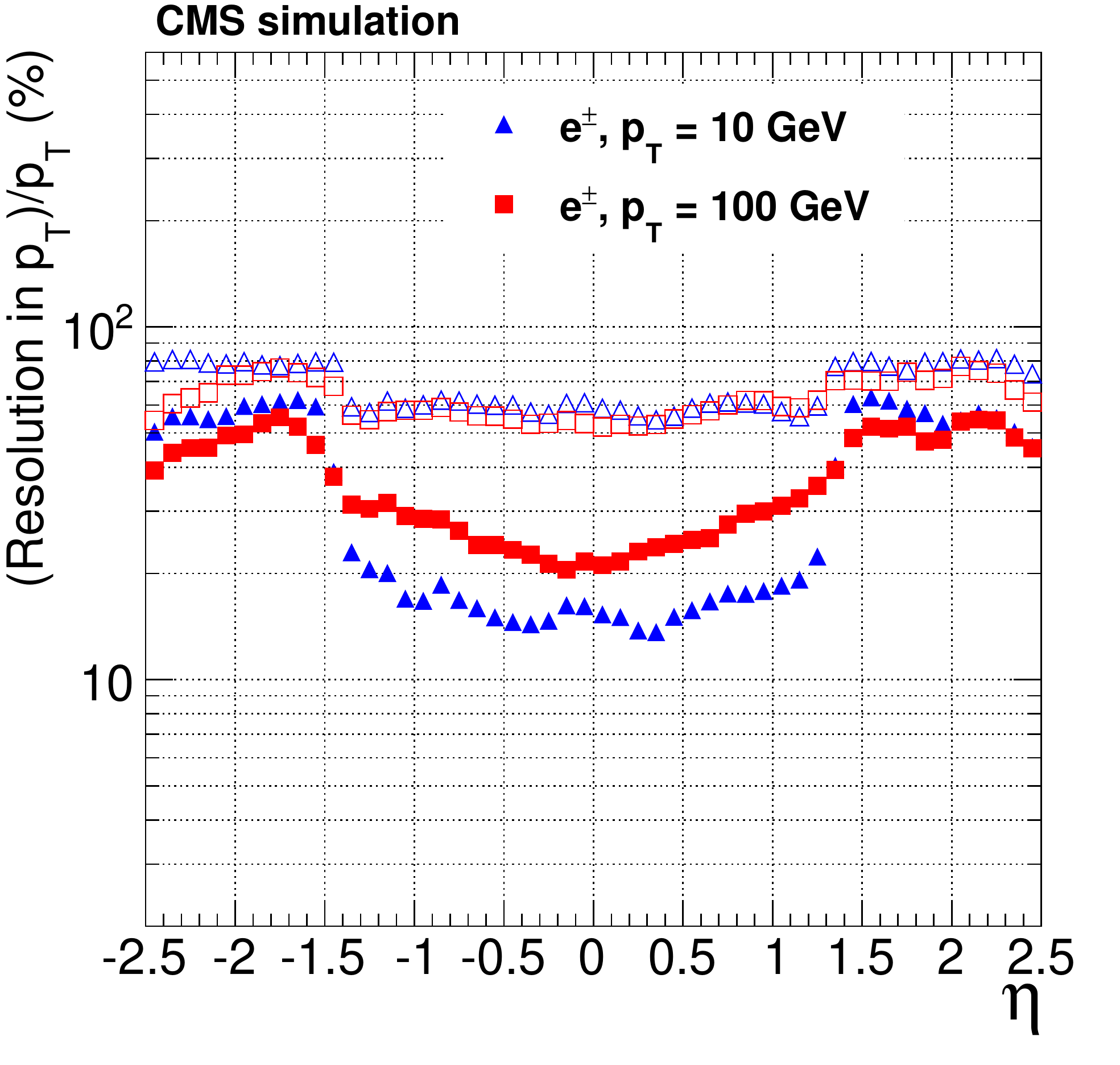}
    \includegraphics[width=0.45\textwidth]{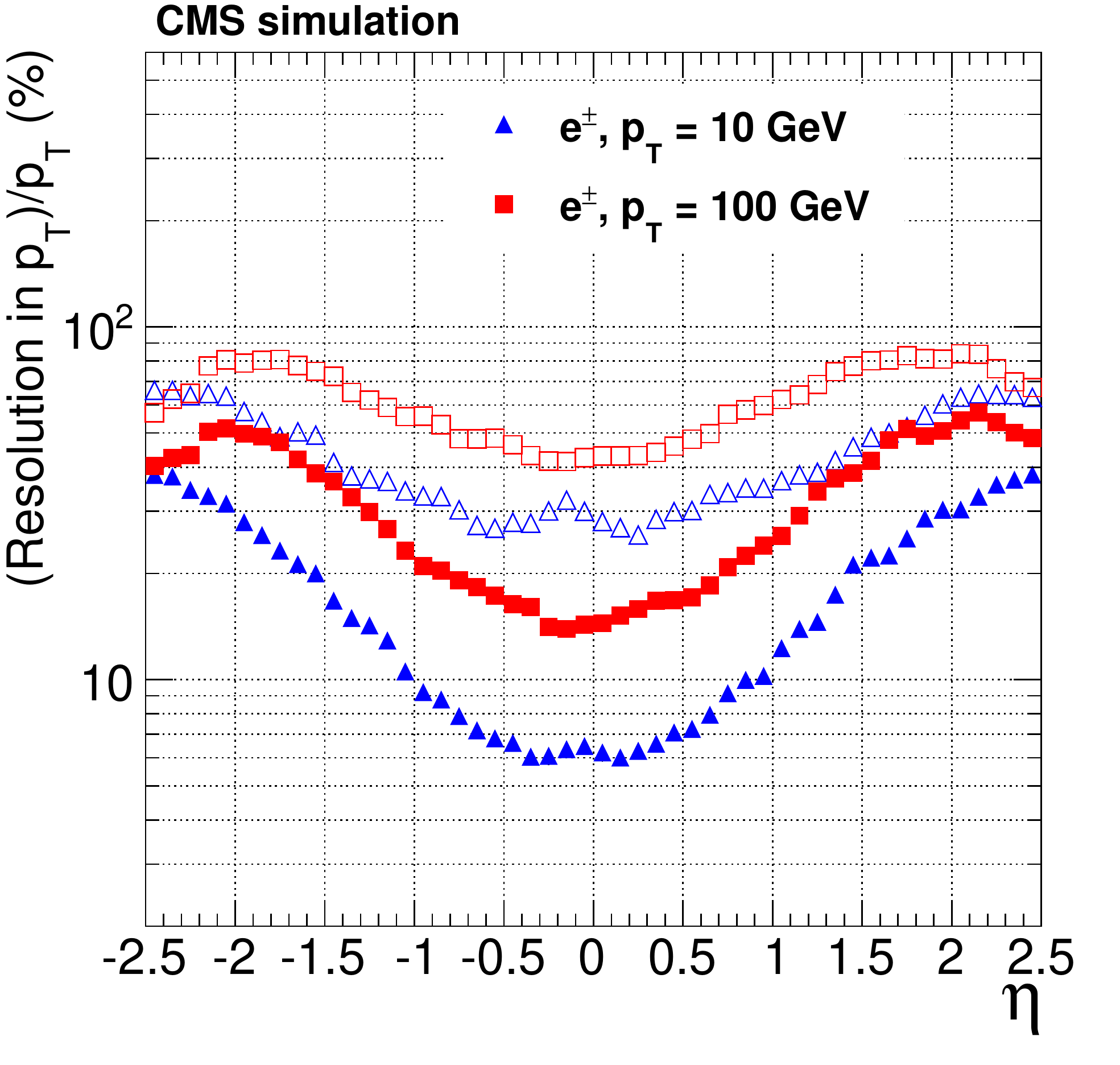}

  \caption {
    Resolution, \textit{as a function of pseudorapidity},
    in the $d_0$, $\phi$ and \pt track parameters for \textit{single, isolated electrons} with
    $\pt = 10$ and 100\GeV.  For each bin in $\eta$, the solid (open) symbols
    correspond to the width of the 68\% (90\%) intervals having its origin on the
    mode of the distribution in residuals, as described in  the text.
    Only the half of the residuals distribution
    that does contain the non-Gaussian tail due to bremsstrahlung is considered in the interval calculation. The left (right) plots are
    of electrons reconstructed with the CTF (GSF) algorithm.
  }
  \label{fig:resolutionsVsEtaEleMCyesBrem}

\end{figure}

\begin{figure}[hbtp]
  \centering
    \includegraphics[width=0.39\textwidth]{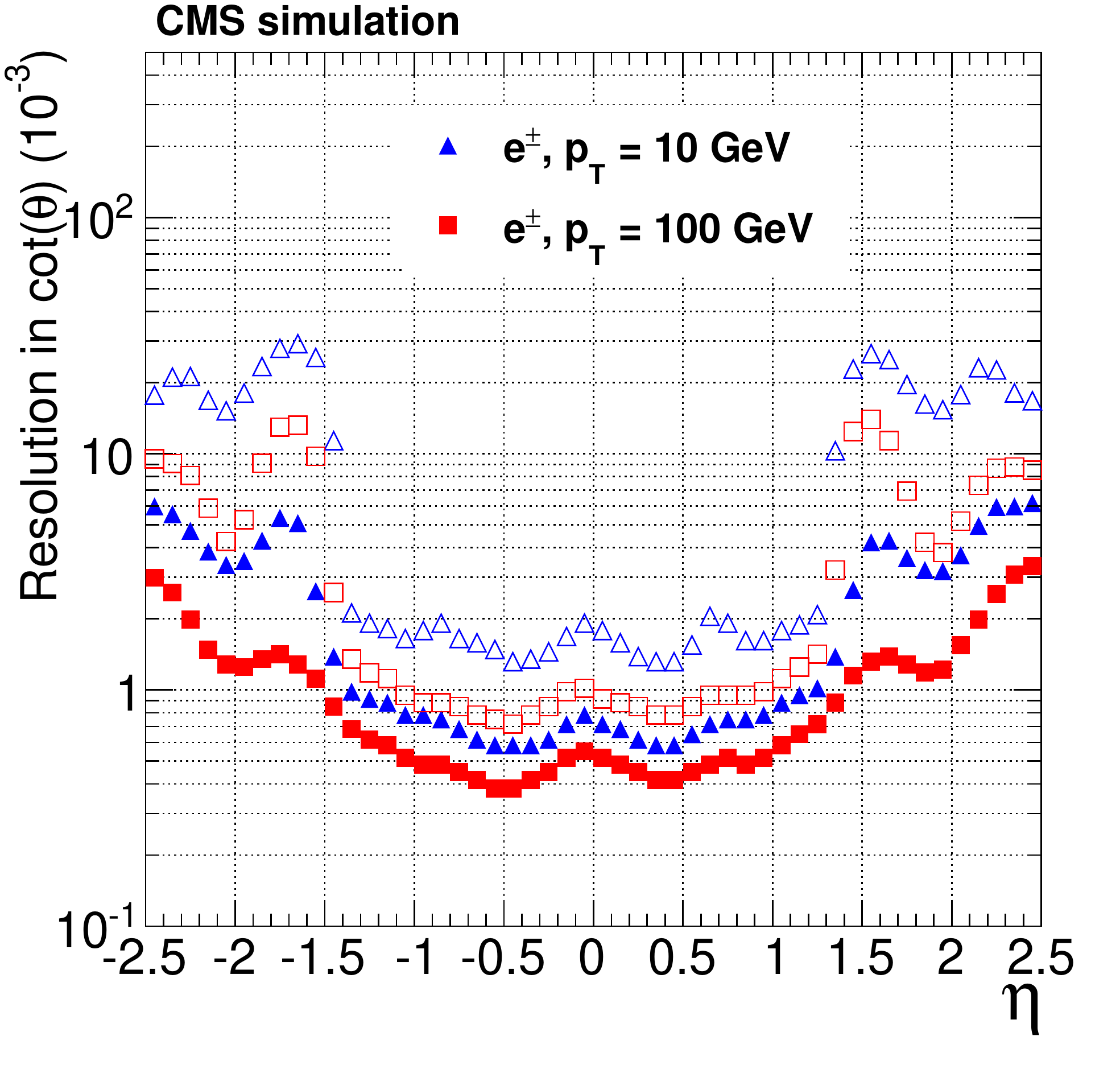}
    \includegraphics[width=0.39\textwidth]{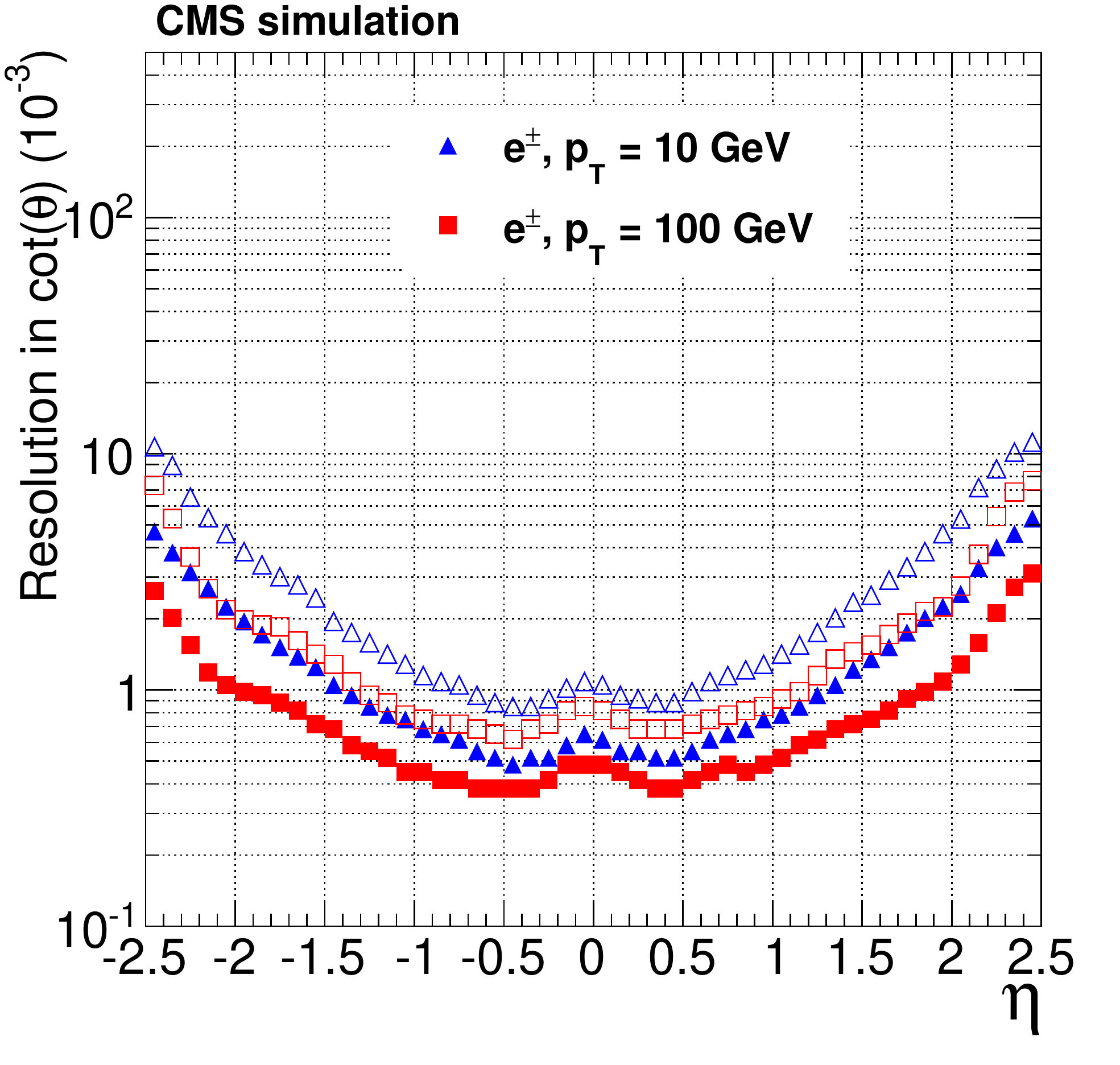}

    \includegraphics[width=0.39\textwidth]{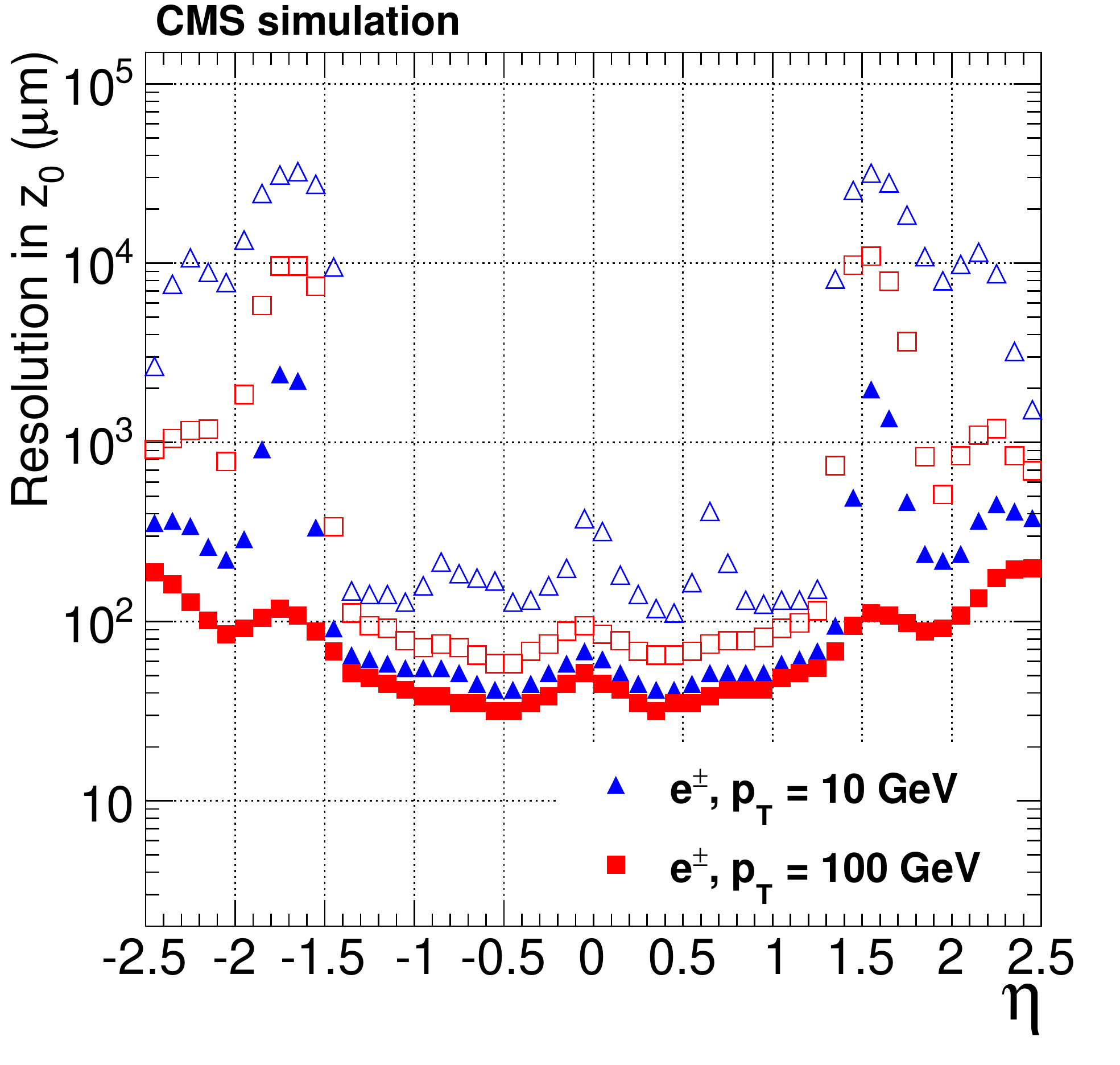}
    \includegraphics[width=0.39\textwidth]{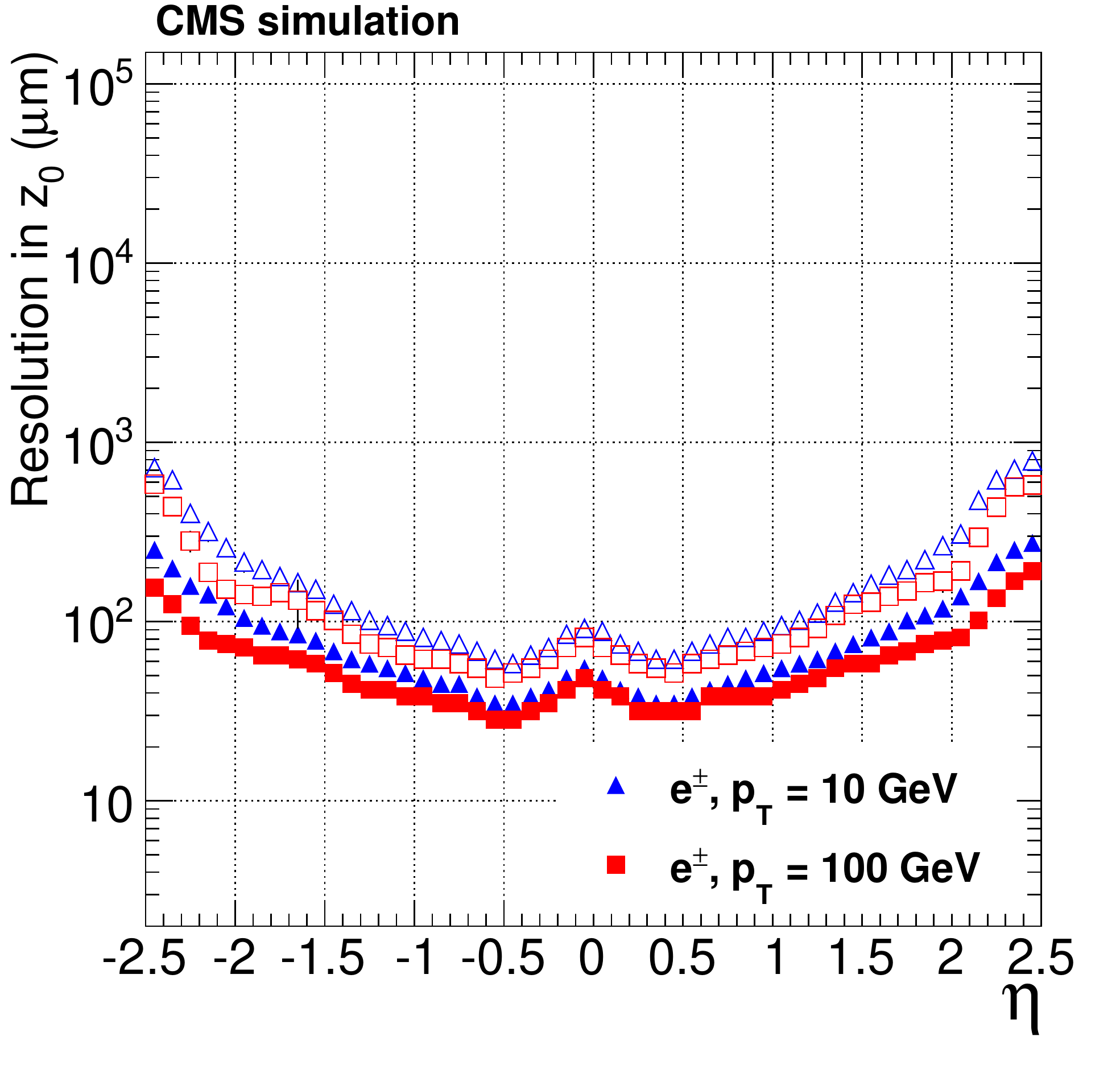}
  \caption {
    Resolution, \textit{as a function of pseudorapidity},
    in the $\cot \theta$  and $z_0$ track parameters for \textit{single, isolated electrons} with
    $\pt = 10$ and 100\GeV.
    For each bin in $\eta$, the solid (open) symbols
    correspond to the half-width of the 68\% (90\%) intervals centered on the
    mode of the distribution in residuals, as described in  the text.
    The left (right) plots are of electrons reconstructed with the CTF (GSF) algorithm.
 }
    \label{fig:resolutionsVsEtaEleMCnoBrem}
\end{figure}

In Fig.~\ref{fig:resolutionsVsEtaEleMCyesBrem}, we show the resolutions in the
$d_0$, $\phi$, and \pt track parameters as a function of $\eta$ for single, isolated electrons
for simulated \pt values of 10 and 100\GeV. These resolutions
are calculated for using both the standard CTF algorithm as well as using the GSF
algorithm, described in Section~\ref{sec:GSF}. However, the GSF requirements described in
Section~\ref{sec:GSF} for consistency of tracks with energy depositions in the ECAL were not applied, as they are
beyond the scope of this discussion. Because the GSF algorithm handles bremsstrahlung in a better way both the 68\% and the
90\% resolutions are significantly improved relative to those obtained with CTF. Similar effects can also be observed
for the resolution in the $\cot \theta$  and $z_0$ parameters, as shown in Fig.~\ref{fig:resolutionsVsEtaEleMCnoBrem}.

\begin{figure}[hbtp]
  \centering
    \includegraphics[width=0.39\textwidth]{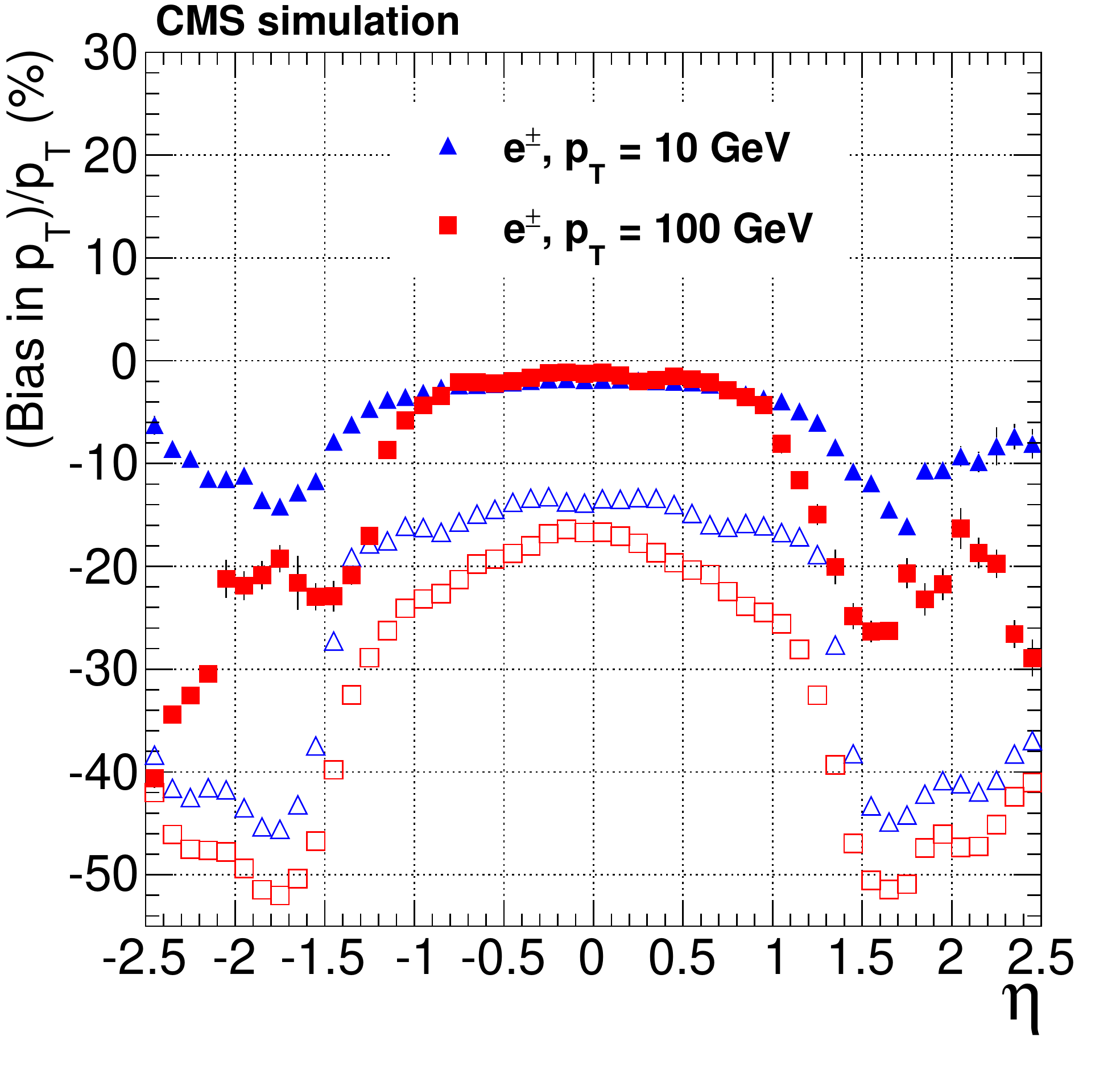}
    \includegraphics[width=0.39\textwidth]{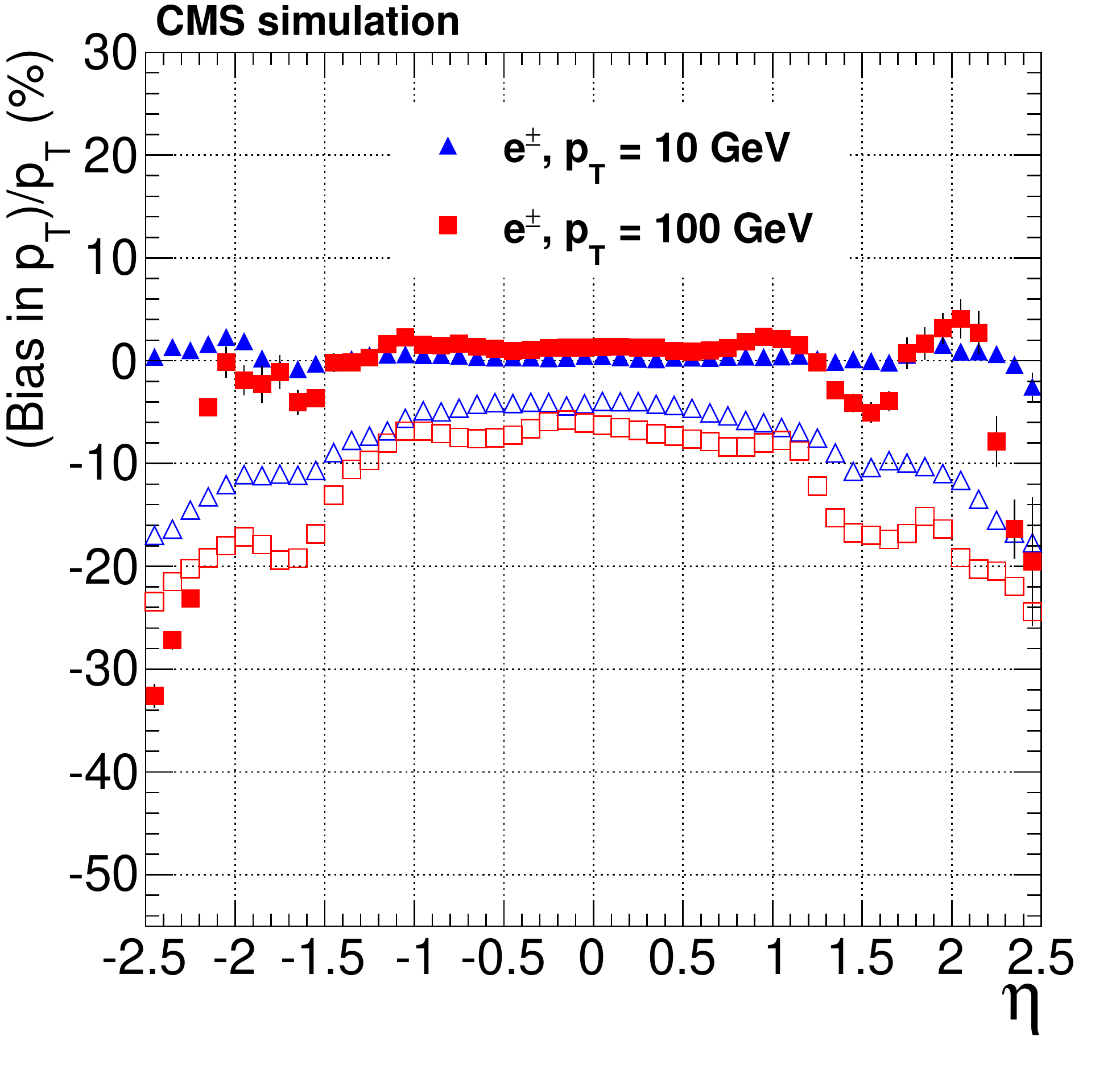}
  \caption {
    \textit{Bias, as a function of pseudorapidity},
    on the \pt track parameter for \textit{single, isolated electrons} with
    $\pt = 10$ and 100\GeV. For each bin in $\eta$, the solid (open) symbols
    correspond to the mode (mean) of the distribution in residuals. The left (right) plots are
    of electrons reconstructed with the CTF (GSF) algorithm.
 }
    \label{fig:biasVsEtaEleMC}
\end{figure}

Figure~\ref{fig:biasVsEtaEleMC} shows the bias in the reconstructed \pt of electrons as a
function of $\eta$. The bias is defined by the mode of the distribution of residuals.
An alternative definition, based on the mean value of residuals is also shown.
The momenta are systematically underestimated by the CTF algorithm for electrons outside the barrel region. However, the bias
is almost completely recovered using the GSF algorithm except for electrons with $\abs{\eta} > 2.0$, where it is affected
more severely by the large amount of material in the pixel endcaps.

\clearpage
\subsubsection{Results from simulated pp collision events}
\label{sec:PerfResolutionsMCcollision}

\begin{figure}[hbtp]
  \centering
    \includegraphics[width=0.45\textwidth]{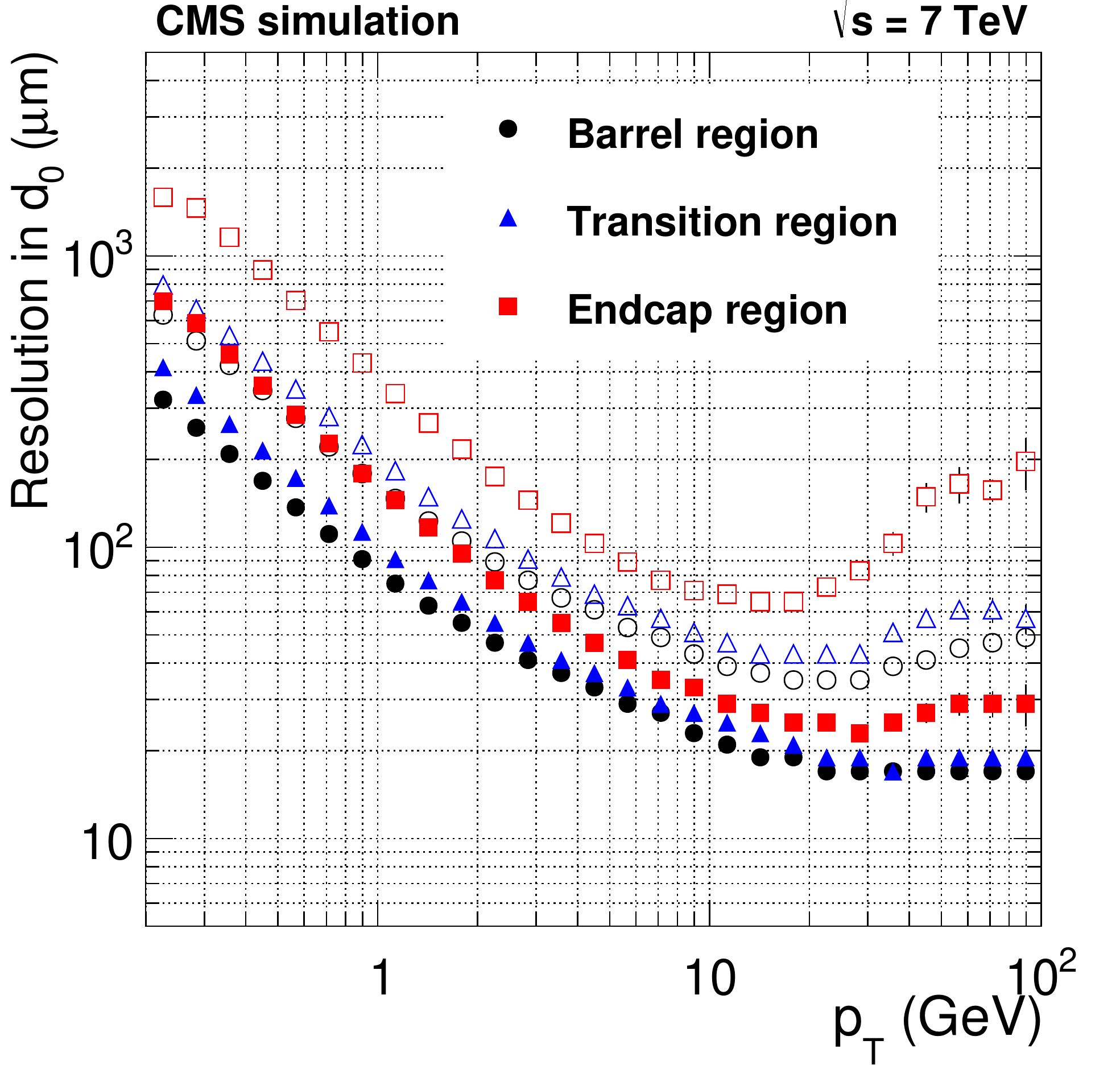}
    \includegraphics[width=0.45\textwidth]{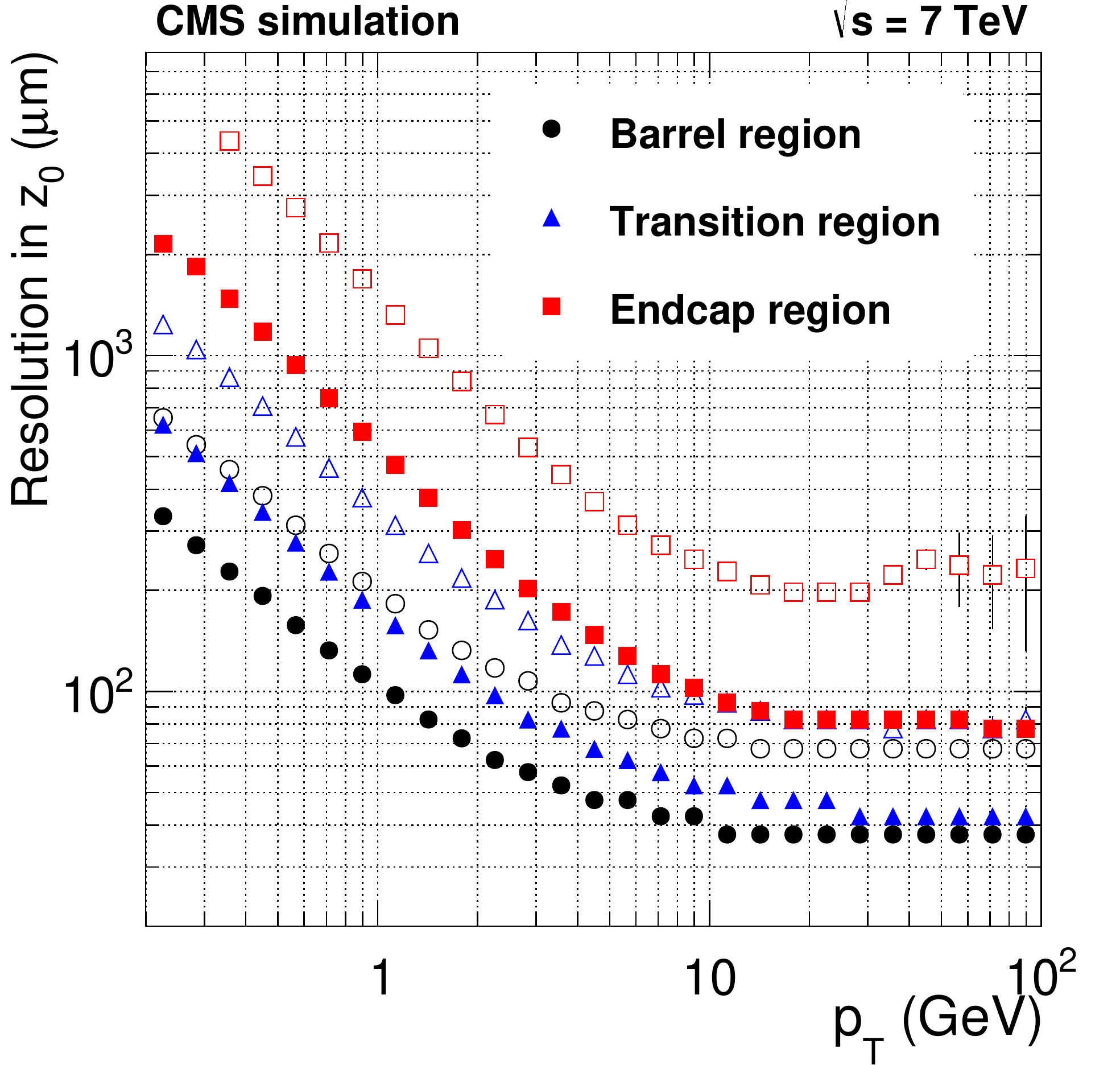} \\
    \includegraphics[width=0.45\textwidth]{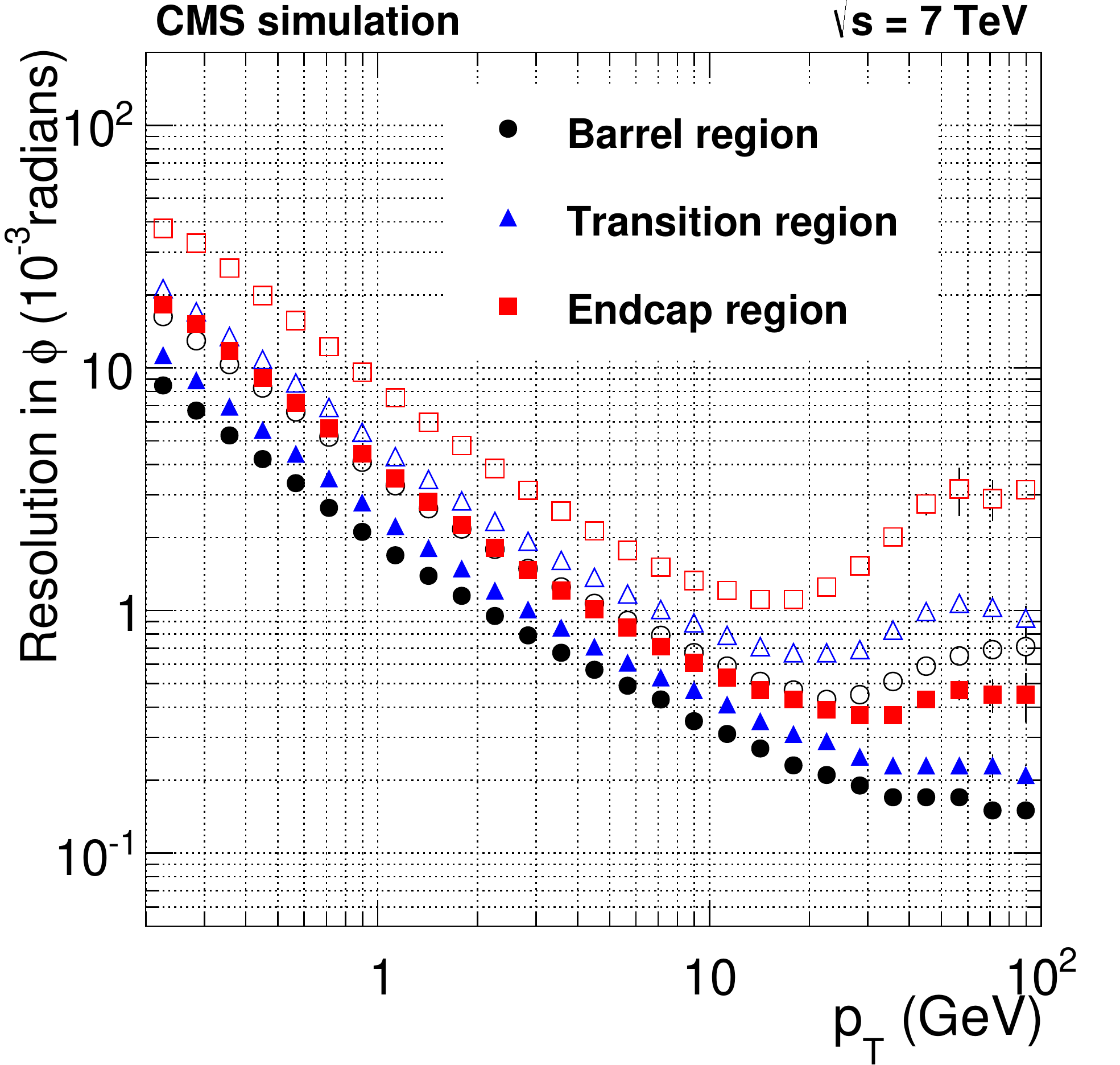}
    \includegraphics[width=0.45\textwidth]{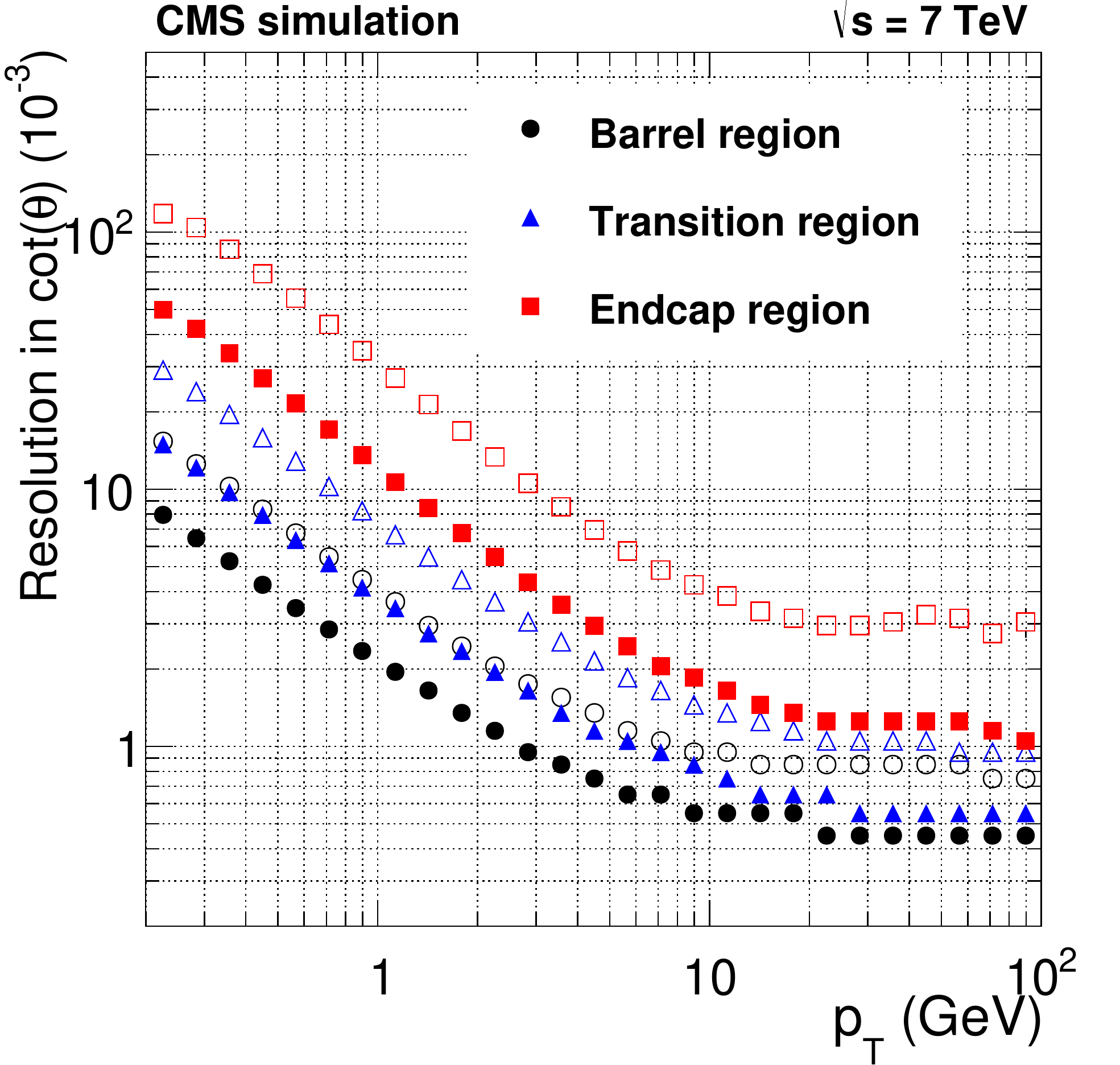} \\
    \includegraphics[width=0.45\textwidth]{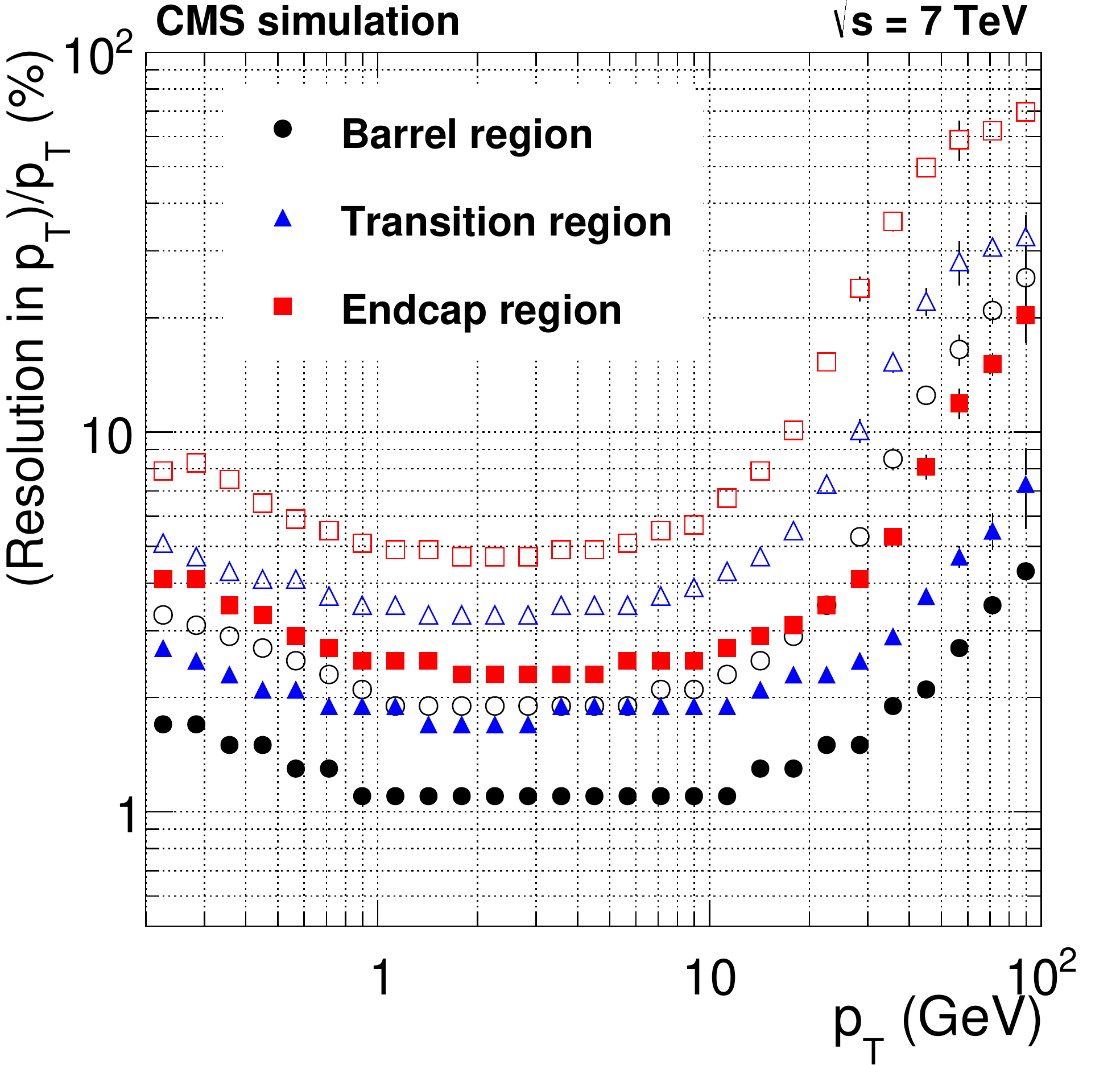}
  \caption {
    Resolution, \textit{as a function of \pt}, in the five track parameters for charged particles in
    simulated \textit{$t\bar t$ events} with pileup.
    The number of pileup interactions superimposed to each simulated event is generated randomly from a Poisson distribution
    with a mean value of 8.
    From top to bottom and left
    to right: transverse and longitudinal impact parameters, $\phi$, $\cot \theta$,
    and \pt.
    For each bin in $\pt$, the solid (open) symbols
    correspond to the half-width of the 68\% (90\%) intervals centered on the
    mode of the distribution in residuals, as described in  the text.
 }
    \label{fig:resolutionVsPtTTbar}
\end{figure}

The resolutions for tracks in \ttbar events, with superimposed pileup interactions, are shown as
a function of track \pt in Fig.~\ref{fig:resolutionVsPtTTbar}. For the five track parameters,
the functional dependence is very similar to that observed for single particles
(Fig.~\ref{fig:resolutionsVsPtMuMC}), except for \pt beyond 20--30\GeV and $\eta$
corresponding to the regions outside the tracker barrel.

The impact of pileup on these resolutions is generally negligible.

\subsection{CPU execution time}
\label{sec:CPU}

\begin{figure}[hbtp]
  \centering
    \includegraphics[width=0.6\textwidth]{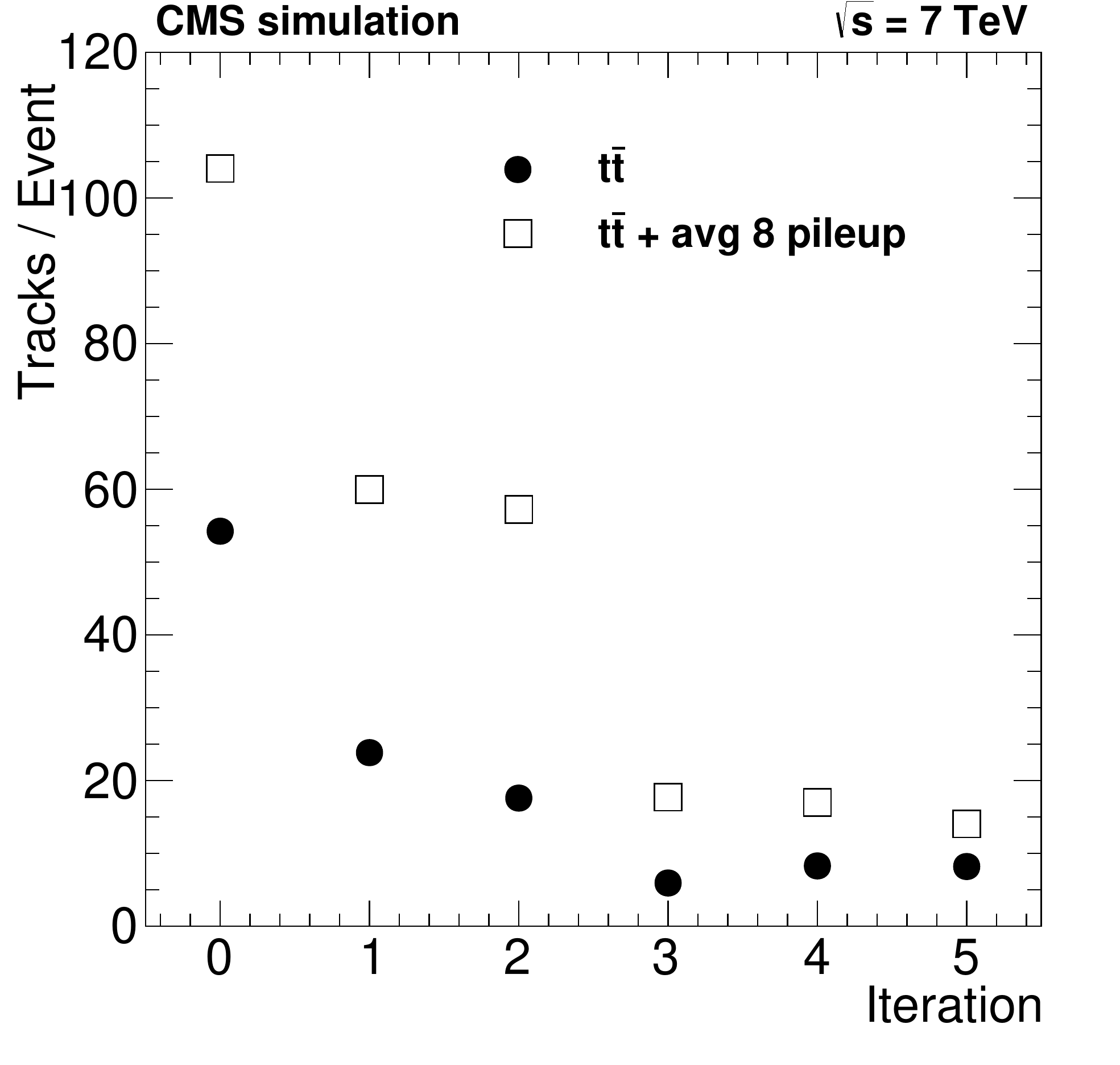}
  \caption {
    The number of additional tracks per event reconstructed after each individual iteration, for \ttbar events
    generated without pileup and with an average of 8 pileup events.
    The distributions include only tracks associated with a simulated charged particle.}
    \label{fig:TracksByAlgo}
\end{figure}

Track reconstruction is, by far, the most computationally challenging part of CMS data reconstruction: for
processing pp events with pileup, it requires almost as much CPU time as all the other reconstruction modules together.
Furthermore, as the number of pileup events increases, the number of tracks increases in proportion, but
the number of hit combinations that can be assembled into seeds and track candidates increases much more quickly,
leading to a far more rapid increase in the required CPU time. The mean
CPU time per event for reconstructing tracks is shown in Table~\ref{tab:CPU}, separated into needs for
tracking iterations and for
computational steps (track seeding, finding, fitting, etc).  The CPU times are given for \ttbar events, simulated
either without pileup or with an average of 8 pileup events.
As the table shows, the presence of pileup significantly increases the total required CPU time, for
example, by a factor 2.4 for Iteration~0 and a factor 8.6 for Iteration~1. Figure~\ref{fig:TracksByAlgo}
shows the number of tracks per event reconstructed in each iteration.  The presence of pileup
increases the number of low-\pt tracks, and as these are mainly reconstructed in Iterations~1--3, pileup
has the biggest effect on these three iterations, increasing thereby both the number of tracks and
the use of CPU time.

\begin{table}[htbp]
\centering
  \topcaption{\label{tab:CPU} The mean CPU time per event attributable to components of the track reconstruction algorithm.
The top table is divided according to the iteration and the bottom table according to the type of task.  The ``other'' category
includes removing clusters between iterations, assignment of track quality, and merging of track collections.  Results are given
for simulated \ttbar events without pileup and with an average of 8 pileup events.  The times are obtained
from an unloaded machine containing an Intel Core i7~CPU~960 running at 3.20\unit{GHz}.}
\begin{tabular}{l|cc}
\hline
Task    & Time for \ttbar (s) & Time for \ttbar + 8 pileup (s) \\
\hline
Pixel tracking \& vertexing & 0.01 & 0.03 \\
Iteration 0 & 0.17 & 0.40 \\
Iteration 1 & 0.13 & 1.12 \\
Iteration 2 & 0.10 & 0.67 \\
Iteration 3 & 0.08 & 0.59 \\
Iteration 4 & 0.11 & 0.48 \\
Iteration 5 & 0.07 & 0.19 \\
Merging of track collections & 0.02 & 0.10 \\
\hline
Total       & 0.68 & 3.58 \\
\hline
\end{tabular}

\vspace{10pt}

\begin{tabular}{l|cc}
\hline
Task    & Time for \ttbar (s) & Time for \ttbar + 8 pileup (s) \\
\hline
Pixel tracking \& vertexing & 0.01 & 0.03 \\
Seeding & 0.06 & 0.55 \\
Track finding & 0.43 & 2.29 \\
Track fitting & 0.14 & 0.56 \\
Other     & 0.04 & 0.15 \\
\hline
Total       & 0.68 & 3.58 \\
\hline
\end{tabular}
\end{table}

\clearpage

\section{Beam spot and primary-vertex reconstruction and its performance}
\label{sec:beamSpotAndPV}

\subsection{Primary-vertex reconstruction}
\label{sec:pvtxreco}
The goal of primary-vertex reconstruction~\cite{CMS_NOTE_2006-032} is to measure the
location, and the associated uncertainty, of all proton-proton interaction vertices in each event,
including the `signal' vertex and any vertices from pileup collisions, using the available
reconstructed tracks. It consists of three steps: (i) selection of the tracks, (ii) clustering
of the tracks that appear to originate from the same interaction vertex, and (iii) fitting for
the position of each vertex using its associated tracks.

Track selection involves choosing tracks consistent with being produced promptly in the primary interaction region,
by imposing requirements on the maximum value of significance of the transverse impact parameter ($<$5)
relative to the centre of the beam spot (which is reconstructed as described in Section~\ref{sec:beamspot}),
the number of strip and pixel hits associated with a track ($\ge$2 pixel layers, pixel+strip  $\ge$5 ),
and the normalized $\chi^2$ from a fit to the trajectory ($<$20).
To ensure high reconstruction efficiency, even for minimum-bias events,
there is no requirement on the \pt of the tracks.

The selected tracks are then clustered on the basis of their $z$-coordinates at their point of closest
approach to the centre of the beam spot.   This clustering allows for the reconstruction of any number of proton-proton
interactions in the same LHC bunch crossing.  The clustering algorithm must balance the
efficiency for resolving nearby vertices in cases of high pileup against the possibility of
accidentally splitting a single, genuine interaction vertex into more than one cluster of tracks.

A simple `gap clustering' algorithm was used in past reconstruction of the CMS data recorded in 2010~\cite{CMS_PAS_TRK-10-005},
with all tracks ordered according to the $z$-coordinate of their point of closest approach to the centre of the beam spot.
When any two neighbouring elements in this ordered set of coordinates had a gap exceeding a distance $z_\text{sep} = 2$\mm,
the gap was used for splitting the tracks on either side into separate vertices.
Interaction vertices separated by  less than $z_\text{sep}$ were merged in this algorithm, making it
a poor choice for high-pileup LHC conditions.

Track clustering is therefore now performed using a \textit{deterministic annealing} (DA) algorithm
\cite{IEEE_DetAnnealing},
finding the global minimum for a problem with many degrees of freedom,
in a way that is analogous to that of a physical system approaching a state of minimal energy through
a series of gradual temperature reductions.
The $z$-coordinates of the points of closest approach of the tracks to the centre of the beam spot are referred to as
\zi, and their associated uncertainties as $\sigma_i^z$.
The tracks must be assigned to some unknown number of vertices at positions \zk.
`Hard' assignments, where a track is assigned to one and only one vertex,
can be represented by values of probability $p_{ik}$ that equal 1 if track $i$ is assigned
to vertex $k$, and 0 otherwise.
In the DA framework, assignments are `soft', meaning tracks can be associated with more than one vertex,
with probability $p_{ik}$ between $0$ and $1$ that
can be interpreted as the probability of the assignment of track $i$ to vertex $k$
in a large ensemble of possible assignments.
Postulating that a priori every possible configuration is equally likely, this
is analogous to calculations in statistical mechanics if the vertex $\chi^2$ represents the
role of the energy.
The most probable vertex positions at ``temperature'' $T$ follow from the
minimization of the analogue of the free energy in statistical mechanics,
\begin{linenomath}
\begin{equation}
F = -T \sum_i^{\text{\# tracks}} p_i \log \sum_k^{\text{\# vertices}}  \rho_k \exp \left[-\frac{1}{T} \frac{(\zi-\zk)^2}{{\sigma_i^z}^2}\right],
\label{eq:DAF}
\end{equation}
\end{linenomath}
relative to the positions of the vertices \zk with vertex weights $\rho_k$.
The sums run over the tracks $i$, and the set of vertices $k$
that reflect the temperature $T$.
Tracks enter with constant weights, $p_i$, reflecting their consistency with originating from
the beam spot.
The number of prototype vertices can be chosen to be arbitrarily large, but
after minimizing $F$ with respect to the $z_k$,
many of the prototype positions coincide.  Then a finite number of effective vertices
emerge at distinct positions, independent of the number of prototypes.
It is computationally more efficient to use those effective vertices with weights $\rho_k$ that correspond to
the fraction of unweighted prototypes that coincide at position $z_k$.
The weights are variable, but the sum is always constrained to unity.
(This version of DA is called ``mass-constrained clustering'' in~\cite{IEEE_DetAnnealing}, because $\sum_k \rho_k = 1$.)

The assignment probabilities are given by
\begin{linenomath}
\begin{equation}
 p_{ik} = \frac{ \rho_k \exp \left[-\frac{1}{T} \frac{\left(\zi-\zk\right)^2}{{\sigma_i^z}^2}\right] }{\sum_{k'} \rho_{k'} \exp \left[-\frac{1}{T} \frac{\left(\zi-\zkprime\right)^2}{{\sigma_i^z}^2}\right] },
\label{eq:DApik}
\end{equation}
\end{linenomath}
where
the resolutions $\sigma_i^z$ are effectively scaled by $\sqrt{T}$.
At very high $T$, all $p_{ik}$ become equal, and all tracks become compatible with a
 single vertex.
For $T\rightarrow 0$ every track becomes compatible with exactly one vertex, resulting in hard assignment.

The DA algorithm is initiated at a very high temperature with a single vertex.
$T$ is gradually decreased,
and $\partial{F}/\partial{\zk}=\partial{F}/\partial{\rho_k}=0$ is implemented iteratively
at each new temperature,
starting with the result of the previous step in temperature.
Because local minima are smeared out by the effective scaling of resolutions as a function of $T$,
this procedure traces the global minimum of $F$ from high to low temperature.

The number of vertices increases as the temperature falls, and rises each time the minimum of $F$ turns
into a saddle point at lower temperatures.  This happens whenever $T$ falls below
the critical temperature of one of the vertices,
\begin{equation}
T_c^k=2\sum_i \frac{p_i  p_{ik}}{{\sigma_i^z}^2} \left(\frac{\zi-\zk}{\sigma_i^z}\right)^2\Big/\sum_i  \frac{p_i  p_{ik}}{{\sigma_i^z}^2}.
\end{equation}
When this happens,
the vertex involved is then replaced by two nearby vertices before the temperature is decreased again.
The sum of the weights $\rho_k$ of the two resultant vertices is initially set equal to the weight of the parent.
The DA process thereby finds not only positions and assignments of tracks to vertices but also the
number of vertices.

The starting temperature for the whole process is chosen to be above the first critical temperature,
evaluated for $\rho_1=p_{i1}=1$. The temperature is decreased at every step by a
cooling factor of $0.6$.
The `annealing' is continued down to a minimum temperature of $T_{\text{min}}=4$, which
represents a compromise between the resolving power and the possibility of
incorrectly splitting true vertices.

Because of the inherently tentative assignment of tracks in the DA framework,
there is a possibility that tracks can be assigned to multiple vertices.
For the final assignment, the annealing is
continued down to $T=1$, but without more splitting.

As described, the DA algorithm is not robust against outliers,
such as secondary or mismeasured tracks.
Above  $T_{\text{min}}$,  outlier rejection competes with splitting, and
is therefore not used.
Below $T_{\text{min}}$, an outlier rejection term
$Z_i=\exp(-\mu^2/T)$
is added to the vertex sums in Eq.~(\ref{eq:DAF}), which acts as a cutoff
for the assignment probabilities in the denominator of Eq.~(\ref{eq:DApik}).
Tracks that are more than $\mu$ standard deviations away from the nearest vertex
are down-weighted, and the algorithm becomes a one-dimensional robust adaptive multi-vertex fit \cite{FruehwirtStrandlie99}.
The default value for the cutoff is $\mu=4$.

Outliers tend to create false vertices when other tracks, typically
worse in resolution, are available nearby.
Candidate vertices are therefore retained only if
at least two of their
tracks are incompatible with originating from other vertices.
The tracks assigned to the rejected candidate vertices are not removed but reassigned to other vertices through another minimization of $F$.
The outlier rejection term at this stage allows individual tracks to have low assignment probability
to all remaining vertex candidates.
A minimal probability of 0.5 is required for making the final assignment
when $T=1$ has been reached.

After identifying candidate vertices based on the DA clustering in $z$,
those candidates containing at least two tracks are then fitted using an \textit{adaptive vertex
fitter}~\cite{CMS_NOTE_2007-008} to compute the best estimate of vertex
parameters, including its $x$, $y$ and $z$ position and covariance matrix, as well as the
indicators for the success of the fit, such as the number of degrees of
freedom for the vertex, and weights of the tracks used in the vertex.
In the adaptive vertex fit, each track in the vertex is assigned a
weight between 0 and 1, which reflects the likelihood that it genuinely belongs to the vertex.
Tracks that are consistent with the position of the reconstructed vertex have
a weight close to 1, whereas tracks that lie more than a few standard deviations from the vertex have small
weights.  The number of degrees of freedom in the fit is defined as
\begin{linenomath}
\begin{equation}
n_\mathrm{dof} = -3 + 2\sum_{i=1}^{\text{\# tracks}}{w_i},
\label{eq:ndofvtx}
\end{equation}
\end{linenomath}
where $w_i$ is the weight of the $i$th track, and the sum runs over all tracks associated with the vertex.
The value of $n_\mathrm{dof}$ is therefore strongly correlated with the number of
tracks compatible with arising from the interaction region.  For this
reason,
$n_\mathrm{dof}$
can be also used to
select true proton-proton interactions.

\subsubsection{Primary-vertex resolution \label{subsec:pvtxresolution}}

The resolution in a reconstructed primary-vertex position depends strongly on the number of tracks
used to fit the vertex and the \pt of those tracks.  In this
section, we introduce a `splitting method' for measuring the resolution
as a function of the number of tracks emanating from a vertex.
The tracks used in any given vertex are split equally into two sets.  During the splitting
procedure, the tracks are first sorted in descending order of \pt,
and then grouped in pairs starting from the track with the largest \pt.
For each pair, tracks are assigned randomly to one or the other set of tracks.
This ensures that the two sets of tracks have, on average, the
same kinematic properties.  These two sets of tracks
are then fitted independently with the adaptive vertex fitter.
To extract the resolution, the distributions in the difference of the fitted vertex positions for a
given number of tracks are fitted using a single Gaussian distribution,
whose fitted \rms width is then divided by $\sqrt{2}$, because the two measurements
of the vertex used in the difference have the same resolution.
The range of the fit is constrained to be within twice
the \rms of the distribution.

\begin{figure}[!ht]
  \centering
    \includegraphics[width=0.45\textwidth]{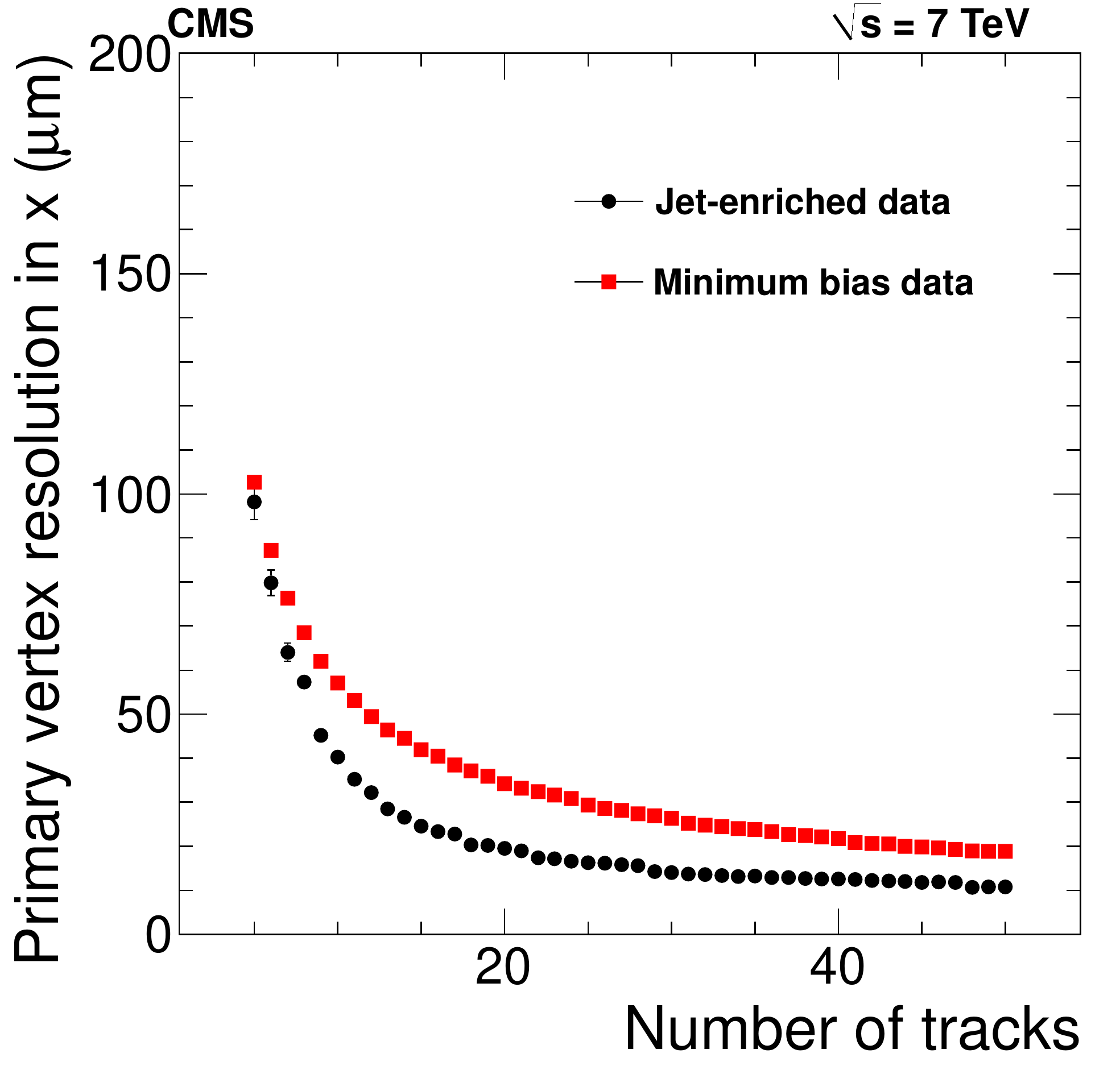}
    \includegraphics[width=0.45\textwidth]{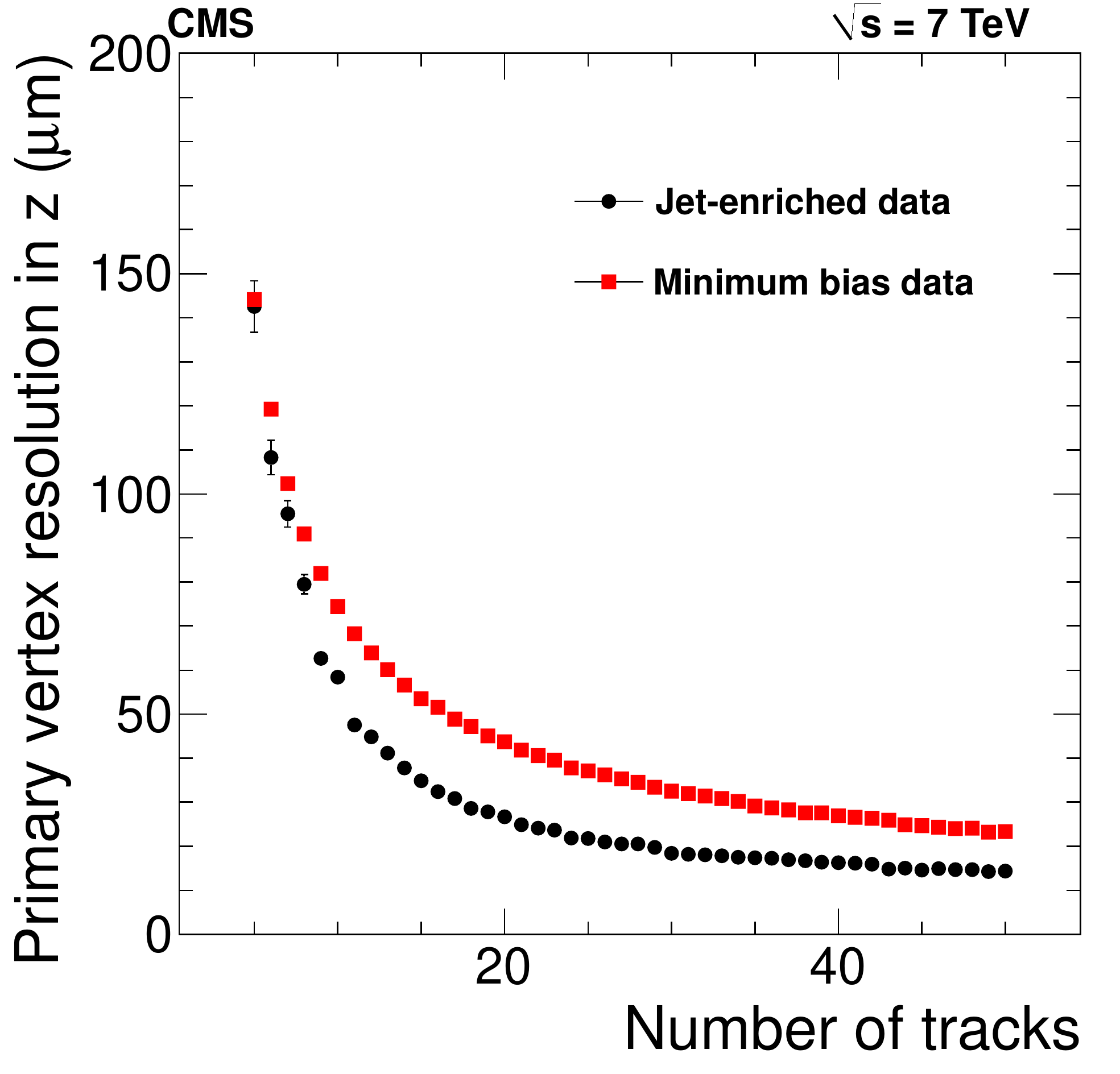}
    \caption{Primary-vertex resolution
    in $x$ (left)
 and
 in $z$ (right) as a function of the number of
 tracks at the fitted vertex, for two kinds of events with different average track \pt values (see text).
 \label{fig:pvtx_respt}}
\end{figure}

Results from a study of the primary-vertex resolution in $x$
and $z$ as a function of the number of tracks associated to the vertex,
using both minimum-bias and jet-enriched data samples,
are shown in Fig.~\ref{fig:pvtx_respt}.
The resolution in $y$ is almost identical to that in $x$, and is therefore omitted.
The minimum-bias sample is collected from a suite of triggers requiring, for example,
only a coincidence of signals from the Beam Scintillator Counters or minimal requirements on the hit or track multiplicity in the pixel detectors.
The jet-enriched samples are produced by requiring each event to have a reconstructed jet
with transverse energy $\et > 20$\GeV.  The tracks in these events have significantly higher
mean \pt, resulting in higher resolution in the track impact parameter and consequently better vertex resolution.
For minimum-bias events, the resolutions in $x$ and $z$ are, respectively, less than
$20\mum$ and $25\mum$, for primary vertices reconstructed using at least 50 tracks.
The resolution is better for the jet-enriched sample across the full
range of the number of tracks used to fit the vertex,
approaching $10\mum$ in $x$ and $12\mum$ in $z$ for primary vertices using at least 50 tracks.
The primary-vertex resolution
for the minimum-bias data from pp collisions has also been compared with simulated minimum-bias events
(\PYTHIA6, Tune Z2~\cite{Field:2011iq}), and found to be in excellent agreement.

The difference between the measured vertex positions,
divided by the sum of the contributions to the uncertainty from the fit, taken in quadrature, is
referred to as the ``pull.''
The standard deviation of the Gaussian function fitted to the pull distribution is roughly
independent of the number of tracks at the vertex and
is found to be approximately 0.93 in data and 0.90 in simulation,
indicating that the position uncertainty from the fit to a vertex is slightly overestimated for both.
This is consistent with the slightly overestimated track uncertainties observed in MC studies.
\ifANnote{The mean of the pull distribution in $x$ and $z$ as a function of the number
of tracks in the vertex for minimum-bias data and simulation
is shown in Fig.~\ref{fig:pvtx_pulls}.}

\subsubsection{Efficiency of primary-vertex reconstruction \label{subsec:pvtxefficiency}}
\begin{figure}[!ht]
  \centering
  \includegraphics[width=0.45\textwidth]{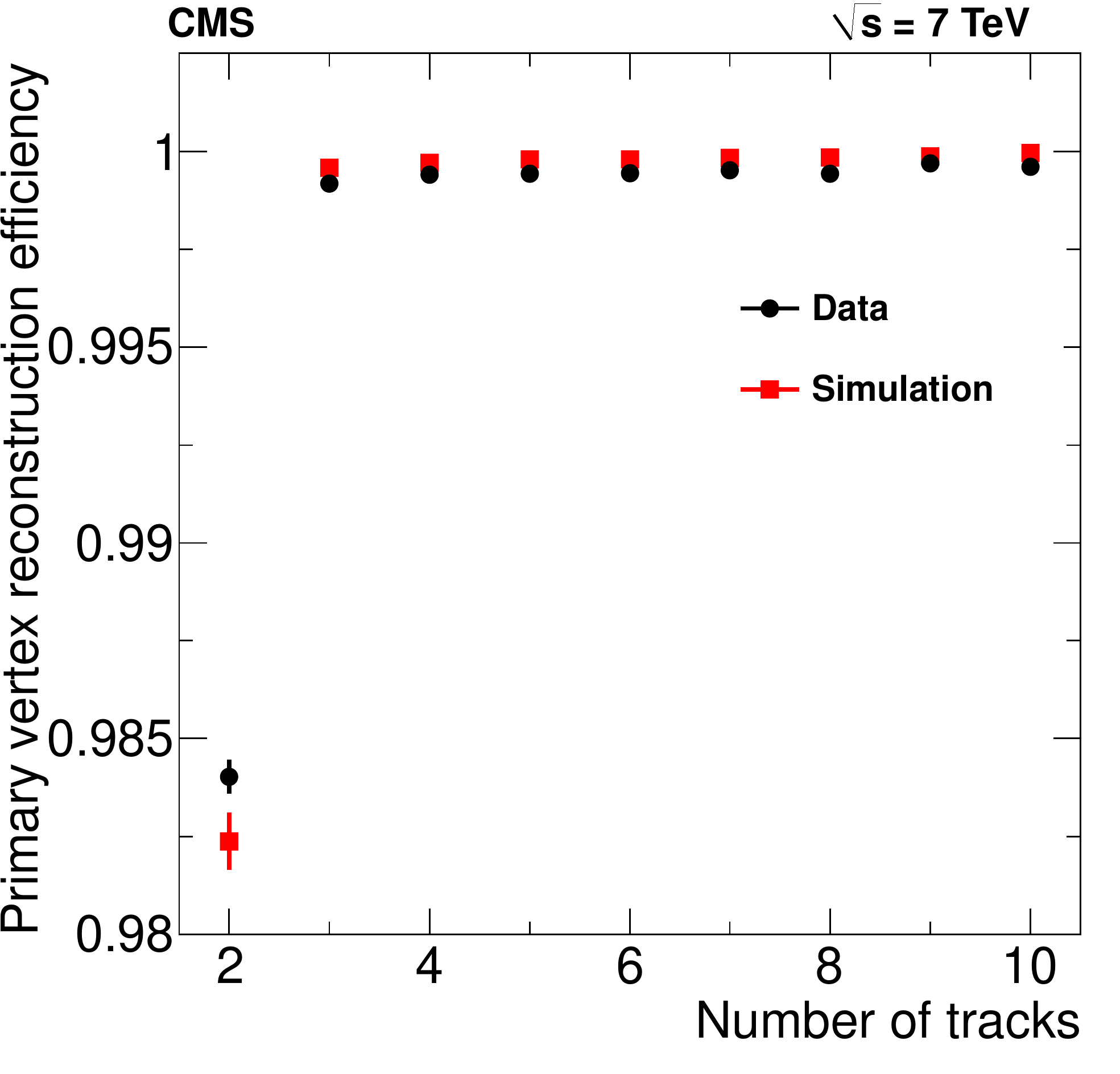}
   \caption[Primary-vertex reconstruction efficiency as a function of the number of
   tracks.]{Primary-vertex reconstruction efficiency as a function of the number of
     tracks in a cluster, measured in minimum-bias data and in MC simulation. \label{fig:pvtx_efficiency}}
\end{figure}
Given an input set of reconstructed tracks, the primary-vertex reconstruction efficiency
is evaluated based on how often a vertex is reconstructed successfully and its
position found consistent with the true value.  Neither
the tracking efficiency nor the probability to produce a minimal
number of charged particles in a minimum-bias interaction is considered
in the extraction of the efficiency for reconstruction of the vertex.

Just as in the measurement of the resolution, the efficiency for
primary-vertex reconstruction depends strongly on the
number of tracks in the cluster.
The same splitting method described in the previous section can be used
to also extract the reconstruction efficiency as a function
of the number of tracks in the vertex cluster.
In this implementation of the method, the tracks used at the vertex
are sorted first in descending order of \pt and then split into two
different sets, such that two-thirds (one-third) of the tracks are randomly assigned as
\textit{tag} (\textit{probe}) tracks.  The asymmetric splitting is used to increase the
number of vertices with a small number of tracks, where the efficiency is
expected to be lowest.  The sets of tag and probe tracks are then
clustered and fitted independently  to extract the vertex
reconstruction efficiency.  While each event is not entirely reclustered,
the contribution to the efficiency from such clustering is not neglected, as the
possibility still remains that a new cluster, using the reduced set of tracks following
splitting, will not be formed.  The effect of pileup on the measurement of the vertexing
efficiency has been checked in simulation, and found to be small.

The efficiency is calculated based on the number of times the probe
vertex is reconstructed and matched
to the original vertex, given that the tag vertex is reconstructed and
matched to the original vertex. A tag or probe vertex is considered
to be matched to the original vertex if the tag or probe vertex position
in $z$ is within $5\sigma$ from the original vertex. The value of $\sigma$ here is
chosen to be the larger of the uncertainty in the fit to a vertex for the tag
or probe tracks and the uncertainty in the original vertex.

Figure~\ref{fig:pvtx_efficiency} shows the efficiency of the primary-vertex reconstruction
as a function of the number of tracks clustered in $z$.
The results are obtained using the splitting method described above,
applied to both minimum-bias data and to MC simulation, and show agreement between the
two samples.  The primary-vertex efficiency is estimated to be close
to 100\% when more than two tracks are used to reconstruct the
vertex.  The effect of pileup on the efficiency is checked using
simulated minimum-bias events, with and without added pileup, and the
loss of efficiency is found to be $ < 0.1$\% for the pileup with a mean value of 8.

\subsection{Track and vertex reconstruction with the pixel detector
\label{sec:pixeltrackvertex}}

CMS has an independent reconstruction of tracks and primary vertices based purely on pixel
hits. The pixel track reconstruction is extremely fast, because only three tracker layers are used, and
the low occupancy and high 3-D spatial resolution of the pixel detector make it ideally suited to track finding.
Such reconstructed pixel tracks and primary vertices can be found extremely fast, hence making them valuable tools for the HLT.

Pixel tracks are formed in the same fashion as the pixel triplets, described in
Section~\ref{sec:SeedGeneration}, requiring $\pt>0.9$\GeV.
Vertex finding using pixel tracks provides a simple
and efficient method for measuring the position of the primary vertex.
The clustering of tracks
is performed using a gap clustering algorithm, with vertex candidates having at least
two tracks fitted using an adaptive vertex fit, as described in Section~\ref{sec:pvtxreco}.

The great speed with which pixel tracks and pixel primary vertices can be reconstructed also makes
them a useful tool for many algorithms in the HLT. For example, counting
the number of pixel tracks near a lepton can help determine if the lepton is isolated.
Similarly, measuring the impact parameter of pixel tracks relative to their
vertex can be used to identify the displaced tracks expected from b-hadron decays.

\subsubsection{Tracking efficiency and fake rate for pixel tracks}

The reconstruction efficiency of pixel tracks is estimated by
comparing the reconstructed tracks with the particles generated in simulation.
Since pixel tracks have only three hits, it is required that all three hits must
be produced by the same simulated particle, for the pixel track and simulated
particle to be associated.
The efficiency for reconstructing a particle as a pixel track is defined
as the fraction of simulated particles that can be associated with a
reconstructed pixel track. The fake rate is defined as the fraction of reconstructed tracks that
are not associated with any simulated particle.

Plots on top left and top right in Fig.~\ref{fig:pixel_track_eff} show the
dependence of the measured pixel track efficiency
on the simulated track $\eta$ and \pt, for inclusive \ttbar events with
and without superimposed pileup (where the number of pileup interactions
is 8, as mentioned in Section~\ref{sec:trackPerformance}). The maximum efficiency for the
pixel tracks is $\sim$85\%. The $\sim$15\% inefficiency arises
mainly from the presence of defective pixel modules ($\sim$2.4\% of the
read out chips in the CMS pixel detector are inoperative) and geometric inefficiency.
The asymmetry between positive and negative $\eta$ reflects the non-uniform distribution
of the affected pixel modules.
In the top-right plot of Fig.~\ref{fig:pixel_track_eff}, the
efficiency drops at low \pt because of the $\pt>0.9$\GeV requirement on pixel tracks.
Figure~\ref{fig:pixel_track_eff} also shows that the addition of
pileup events leads to only a small loss in efficiency.

Plots at the bottom left and bottom right in Fig.~\ref{fig:pixel_track_eff}
show the fake rate as a function of $\eta$ and \pt, both with and
without the presence of pileup.
As observed for the full tracking algorithms in Section~\ref{subSec:EfficiencyFakeRate}, the fake rate increases significantly with $\abs{\eta}$
and \pt.
The effect of pileup is also clearly visible, as the fake
rate increases by 50\% with high pileup.

\begin{figure}[hbtp]
  \centering
    \includegraphics[width=0.45\textwidth]{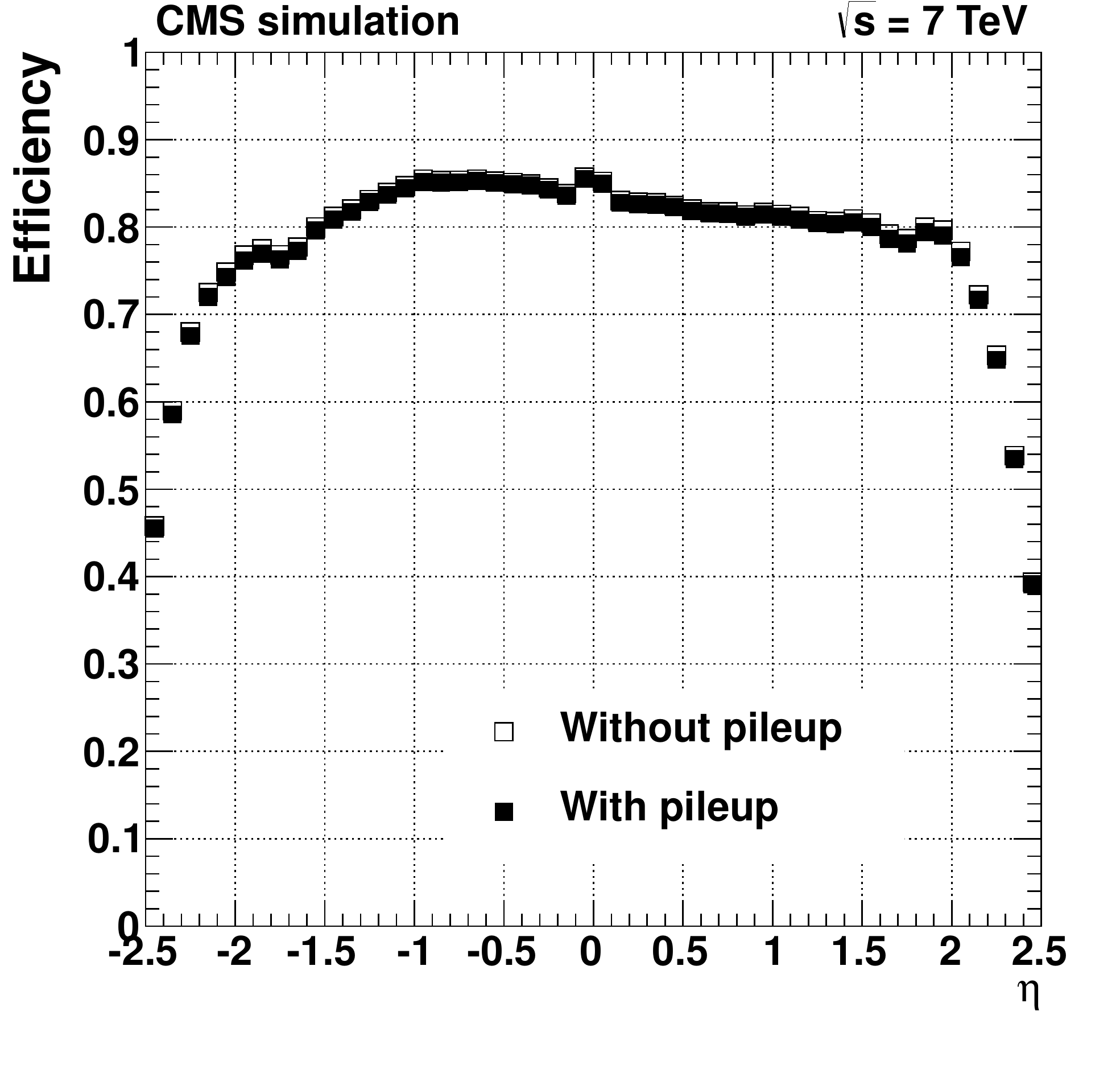}
    \includegraphics[width=0.45\textwidth]{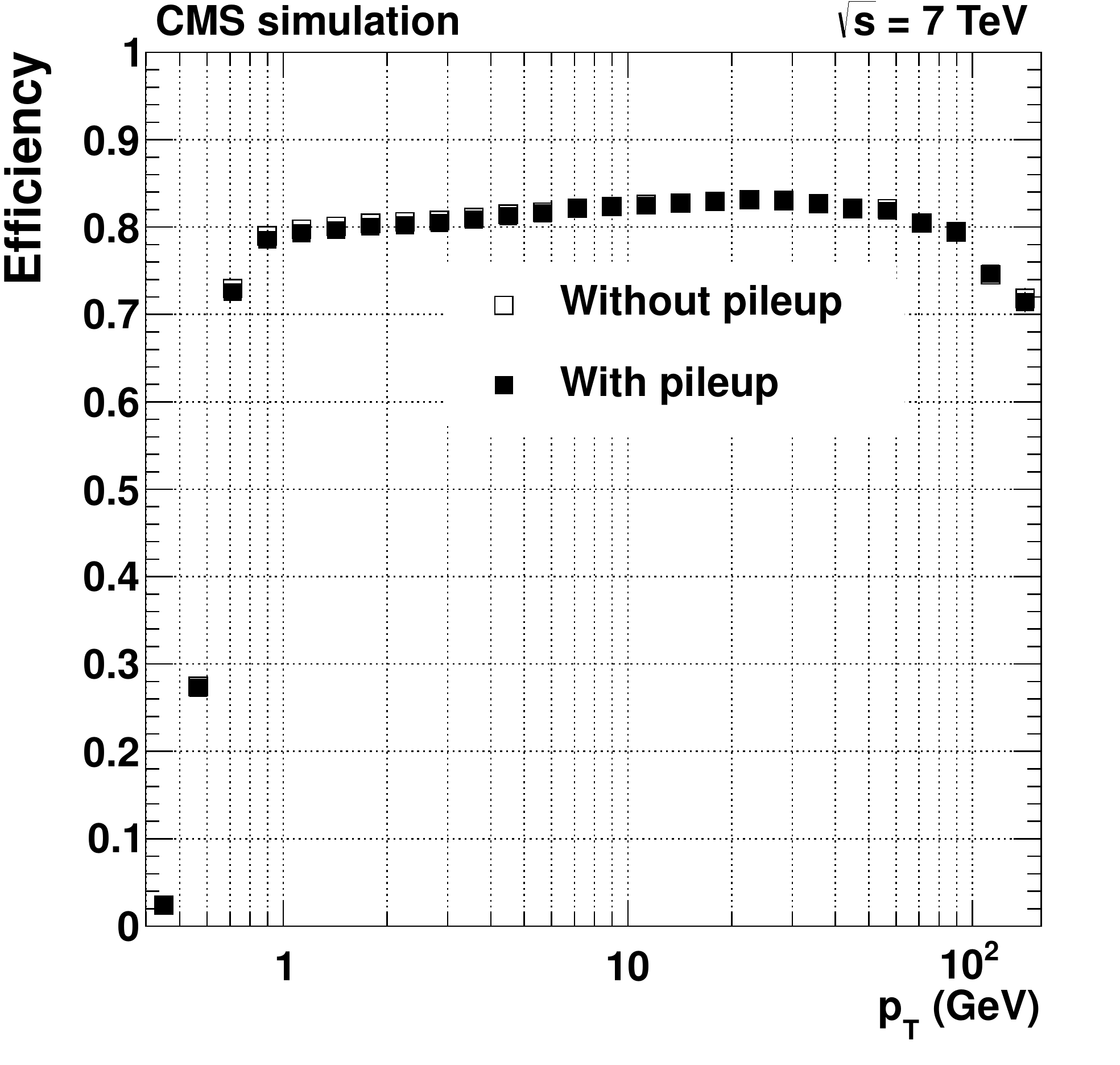} \\
    \includegraphics[width=0.45\textwidth]{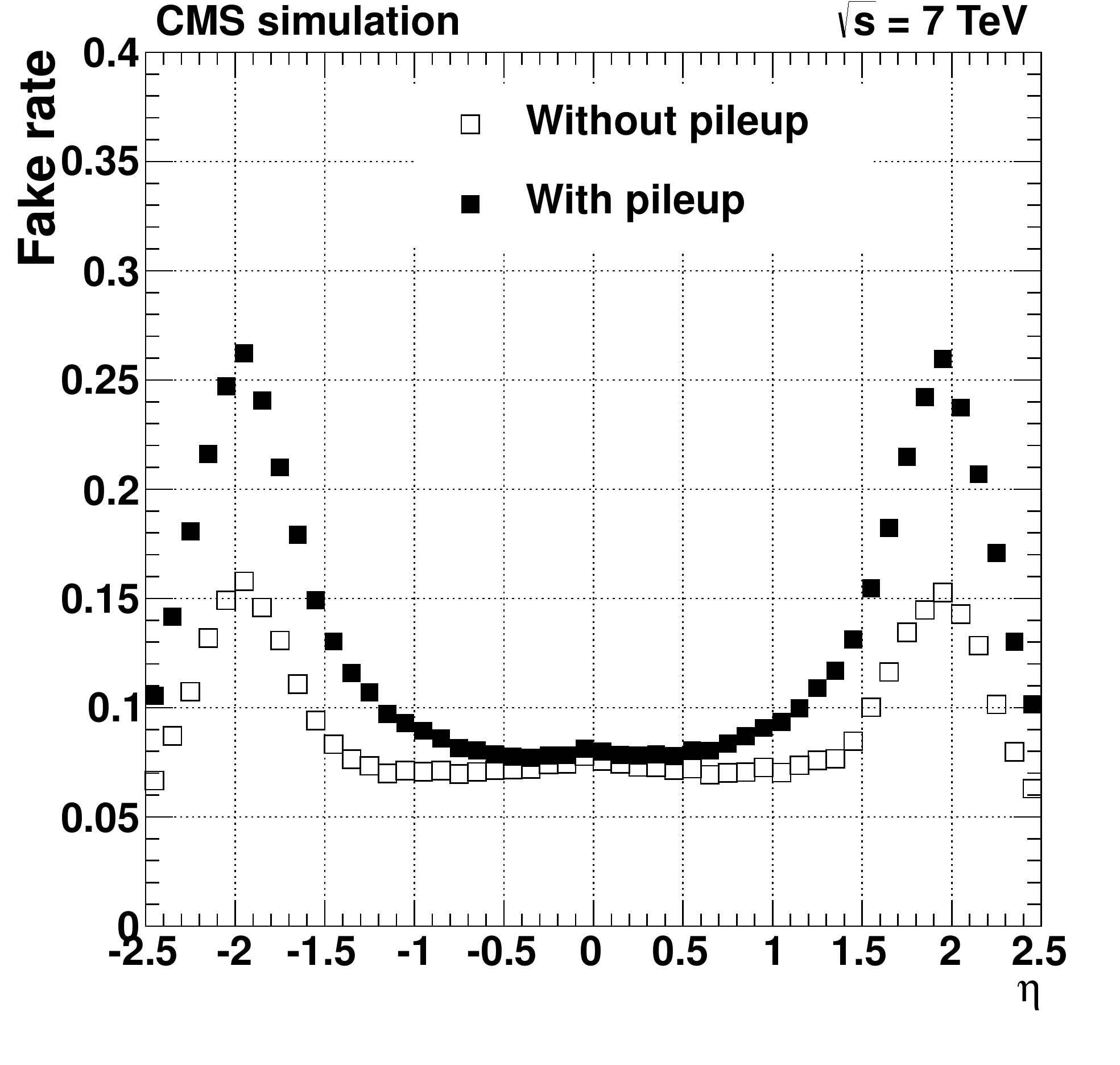}
    \includegraphics[width=0.45\textwidth]{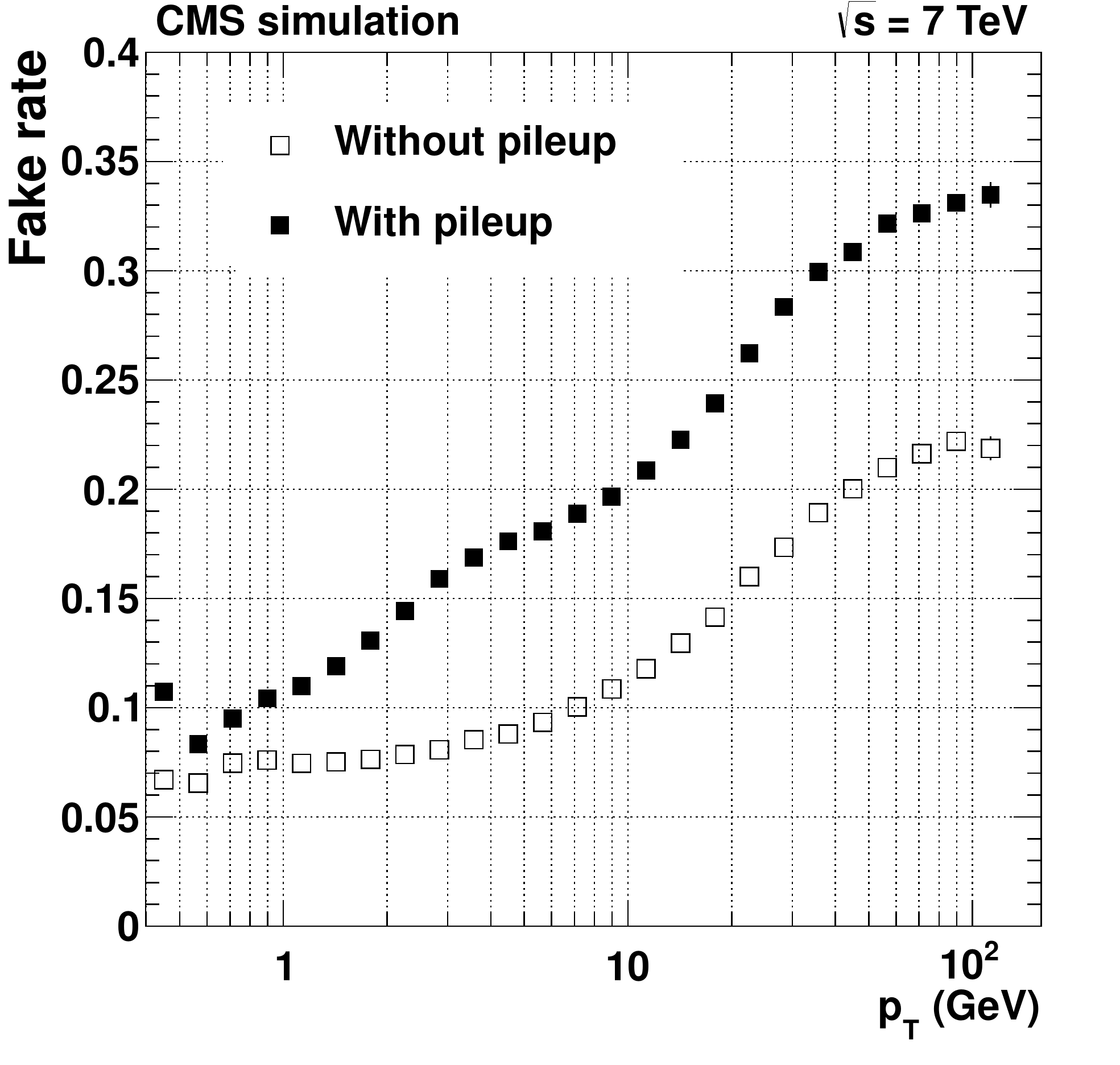} \\
  \caption {
    Pixel tracking \textit{efficiency} (top) and \textit{fake rate} (bottom) for \textit{$t\bar t$ events} simulated with and without superimposed pileup
    collisions.
    The number of pileup interactions superimposed on each simulated event is randomly generated from a Poisson distribution with mean
    equal to 8.
   The two plots of
    efficiency and fake rate as a function of pseudorapidity are produced
    applying a $\pt >0.9$\GeV selection.
  \label{fig:pixel_track_eff}}

\end{figure}

\subsubsection{Resolution in the parameters of pixel tracks}

The resolutions in transverse and longitudinal impact parameters $d_0$ and $d_z$ can be
extracted from simulated events in the same way as in Section~\ref{subSec:TrackParResolution}.
Figure~\ref{fig:pixel_ip_resolution_vs_pt} shows
the resolutions for the five pixel track parameters as a function of
pixel track $\pt$ that includes the effect from pileup.
The distributions are similar in form, but somewhat poorer resolution than those shown for standard tracking
in Fig.~\ref{fig:resolutionVsPtTTbar}. The pixel track resolution in \pt degrades by
over 30\% for track $\pt>10$\GeV.

\begin{figure}[hbtp]
  \centering
    \includegraphics[width=0.45\textwidth]{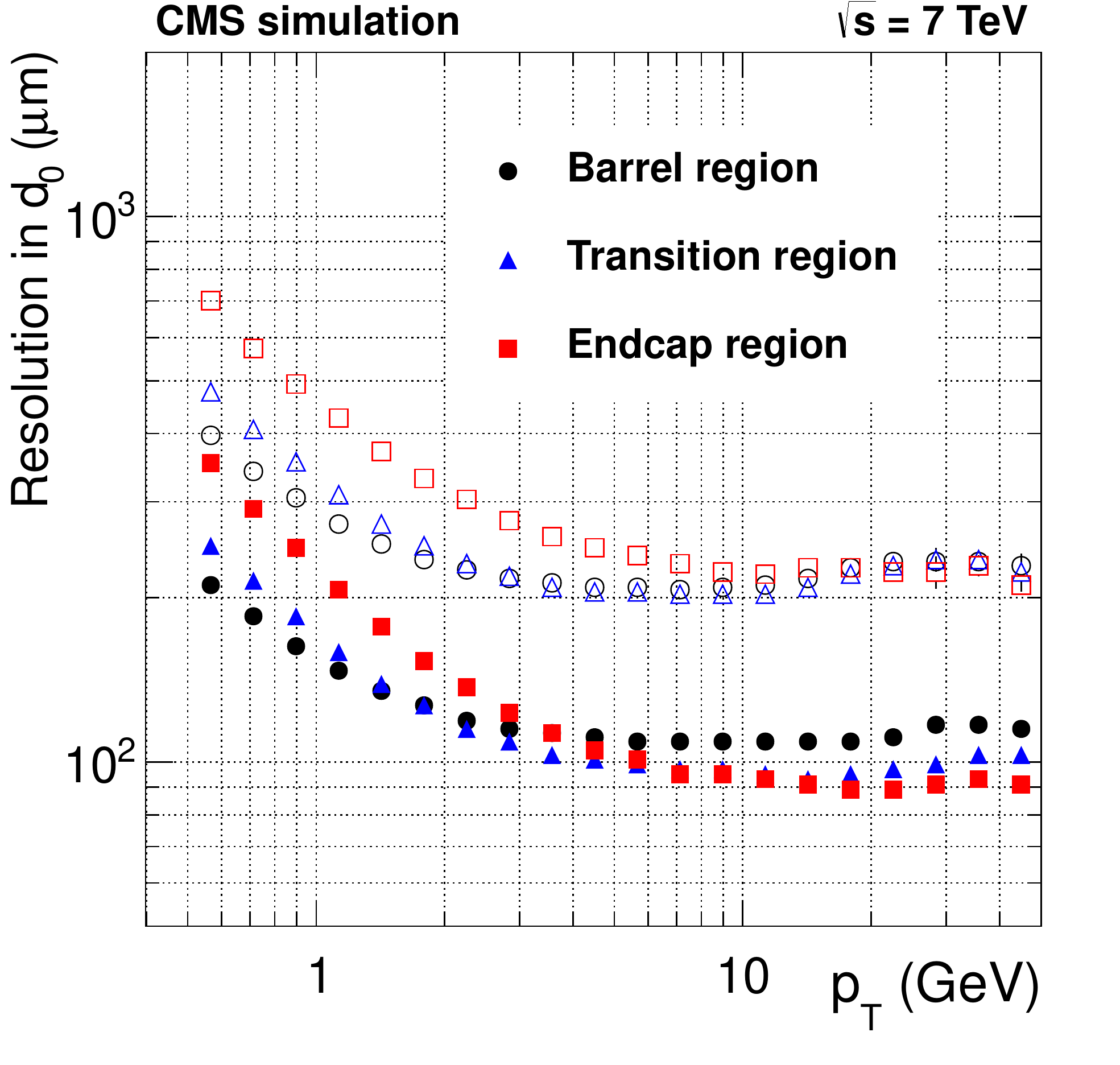}
    \includegraphics[width=0.45\textwidth]{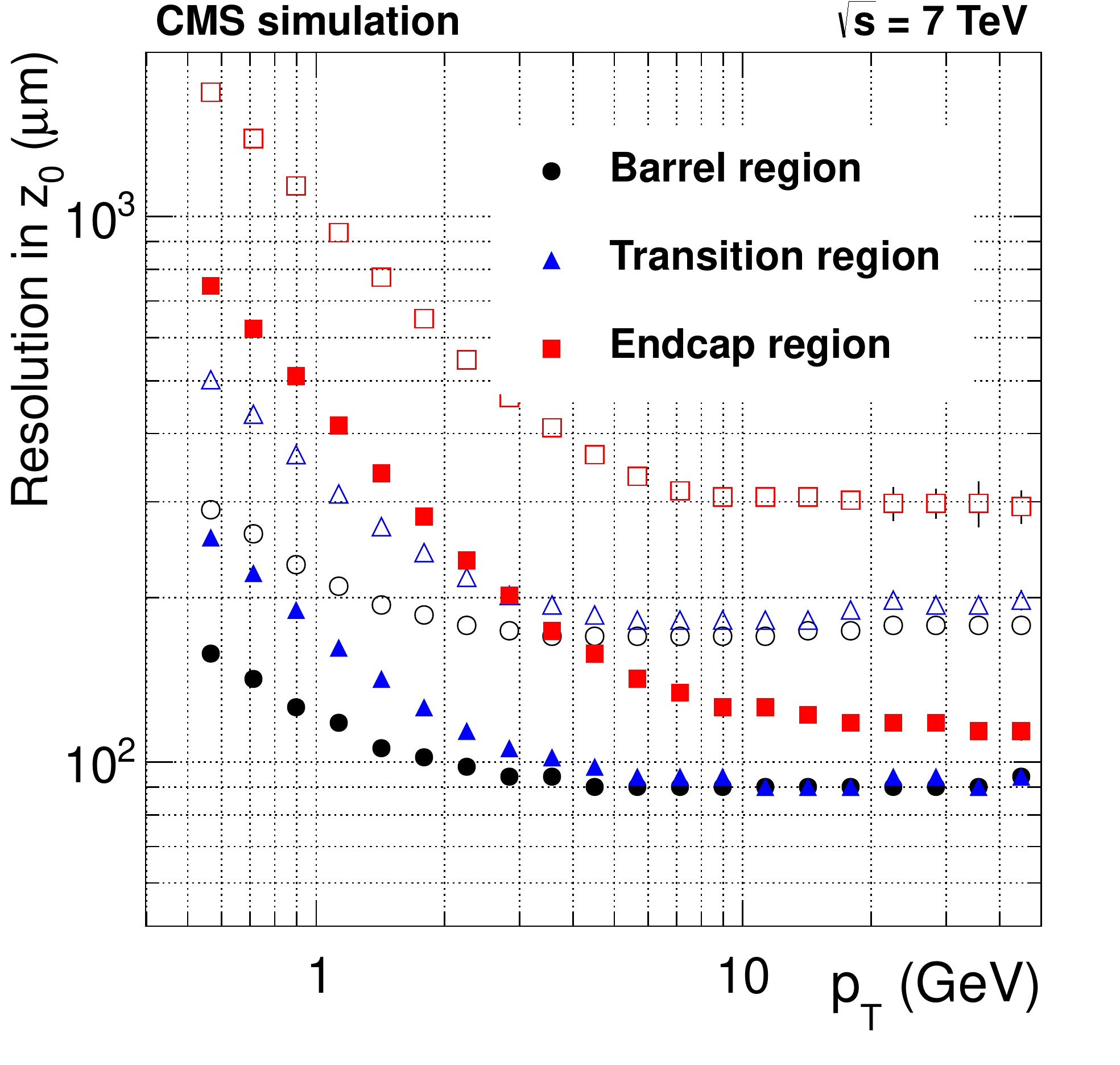} \\
    \includegraphics[width=0.45\textwidth]{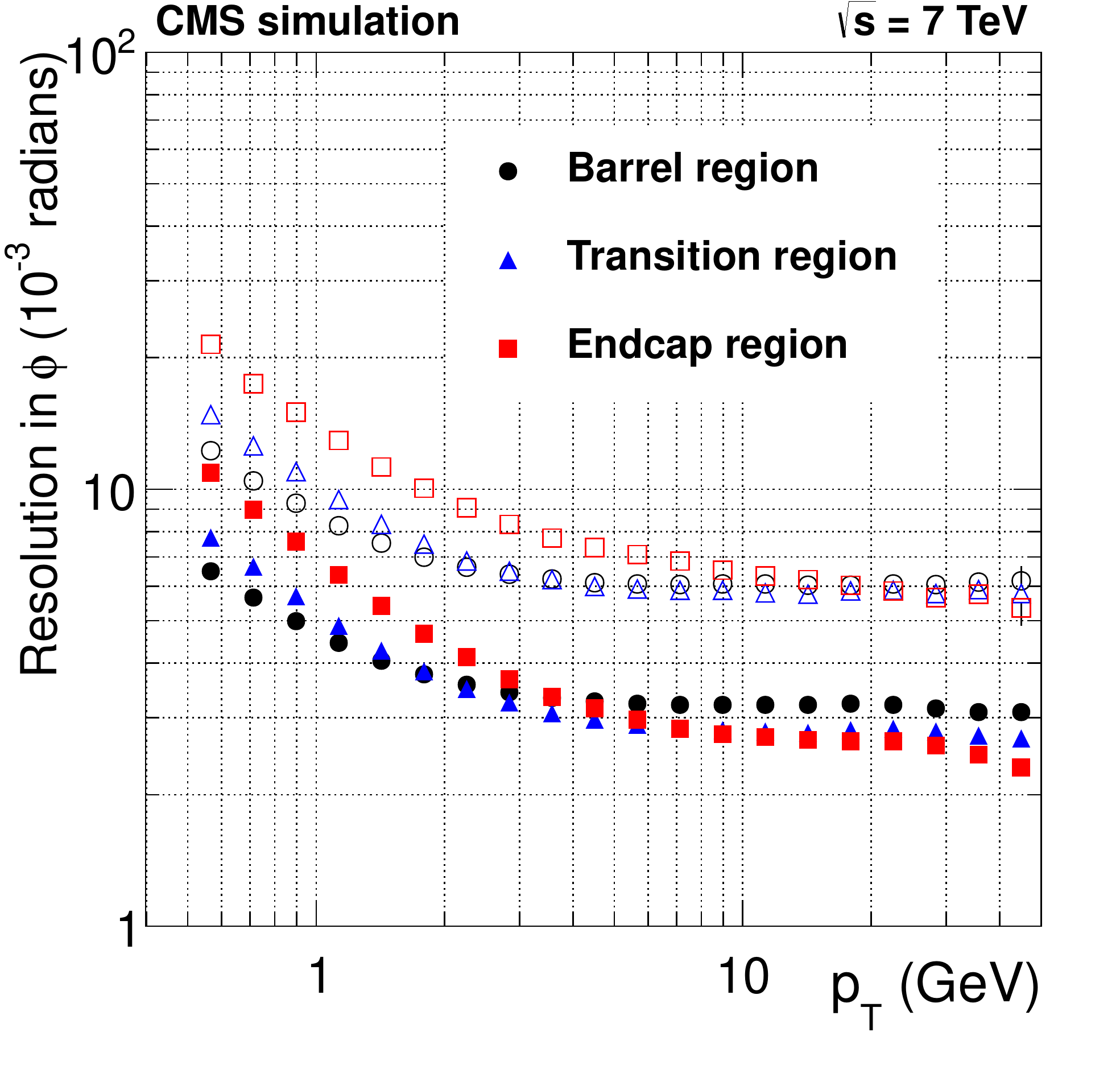}
    \includegraphics[width=0.45\textwidth]{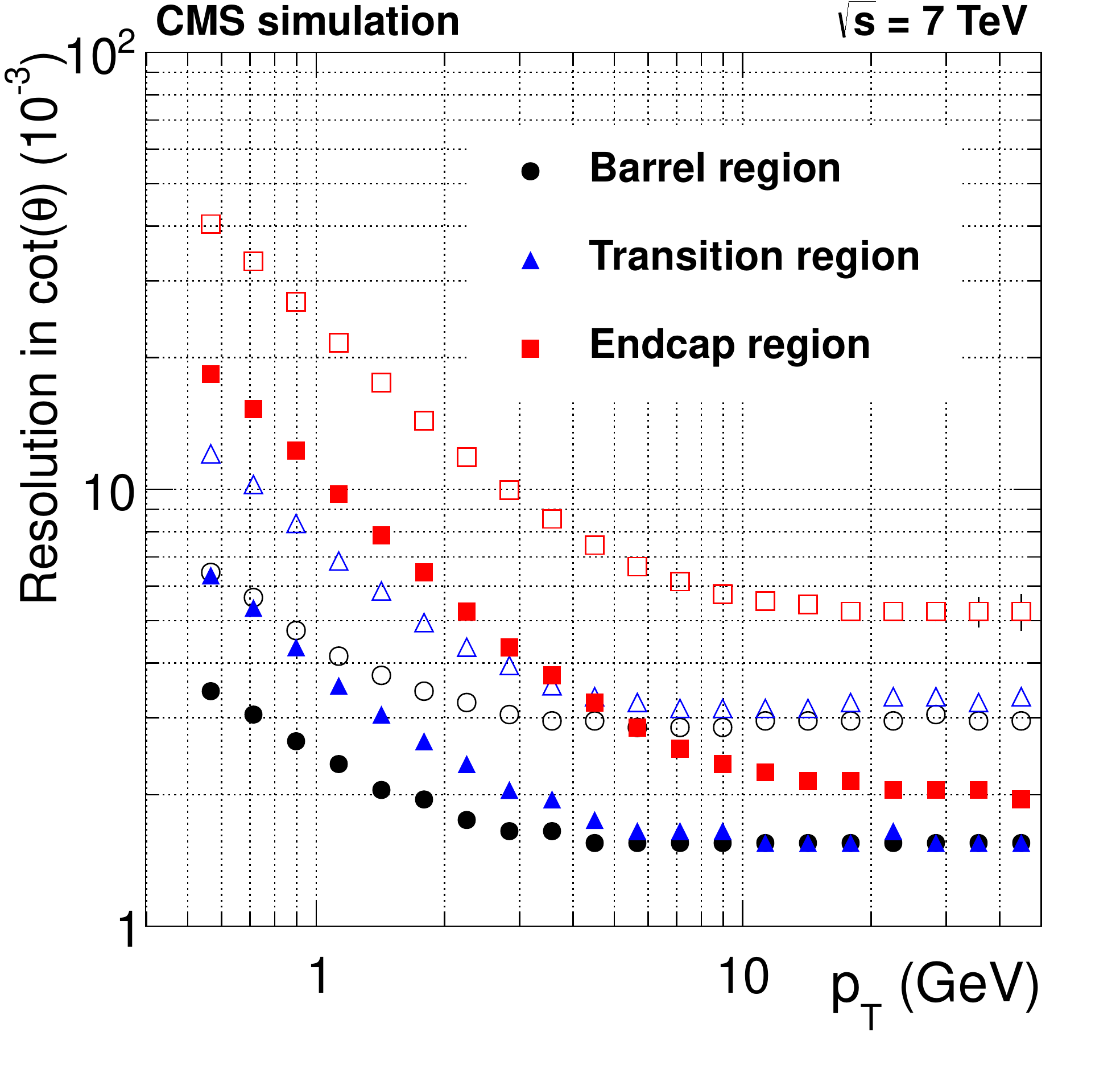} \\
    \includegraphics[width=0.45\textwidth]{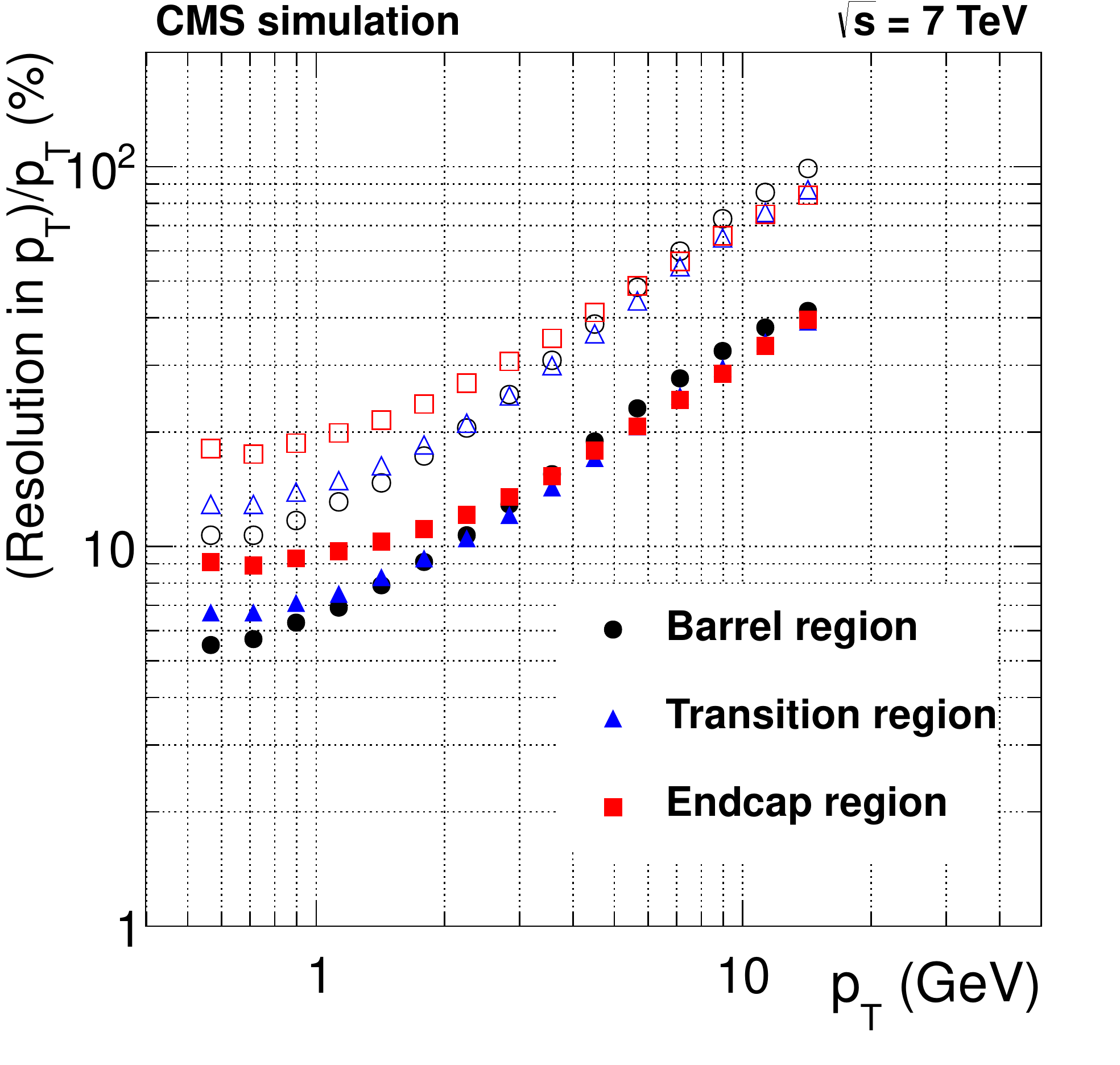}
  \caption {
    \textit{Resolution, as a function of \pt}, in the five track parameters for pixel tracks
in simulated \textit{\ttbar events} with pileup in the barrel, transition and endcap regions,
defined by the pseudorapidity intervals 0--0.9, 0.9--1.4 and 1.4--2.5, respectively. From top
to bottom and left to right: transverse and longitudinal impact parameters, $\varphi$,
$\cot \vartheta$ and transverse momentum. For each bin in $\pt$, the solid (open) symbol
 corresponds to the half-width of the 68\% (90\%) interval centered on the most probable
value of the residuals distributions.
}
    \label{fig:pixel_ip_resolution_vs_pt}
\end{figure}

\subsubsection{Position resolution for pixel based vertices}
\label{sec:pixelPV}

The position resolution for pixel vertices is extracted using the same method
used to measure primary-vertex
resolutions in Section~\ref{subsec:pvtxresolution} (split method).
Figure~\ref{fig:pixelpvtx_resolution_MBias_Jets_data} shows the measured
resolution as a function of the number of tracks in $x$ (left),
and $z$ (right), using both minimum-bias and a
jet-enriched data.
(The resolution in $y$ is almost identical to that in $x$, and hence it is omitted.)
The resolution is better for the jet-enriched sample, across the full
range of associated tracks used to reconstruct the pixel vertex. For example, with 50
tracks, the $x$ resolution of a pixel vertex is $30\mum$ for the minimum-bias
sample, compared to $25\mum$ for the jet-enriched sample. This is due to
the fact that tracks in the jet-enriched data have higher
mean \pt compared to those in the minimum-bias sample. As before,
the pixel vertex resolution in the minimum-bias data has also been compared with
that in simulated minimum-bias events and again found to be in good agreement.

\begin{figure}[hbtp]
	  \centering
\includegraphics[width=0.45\textwidth]{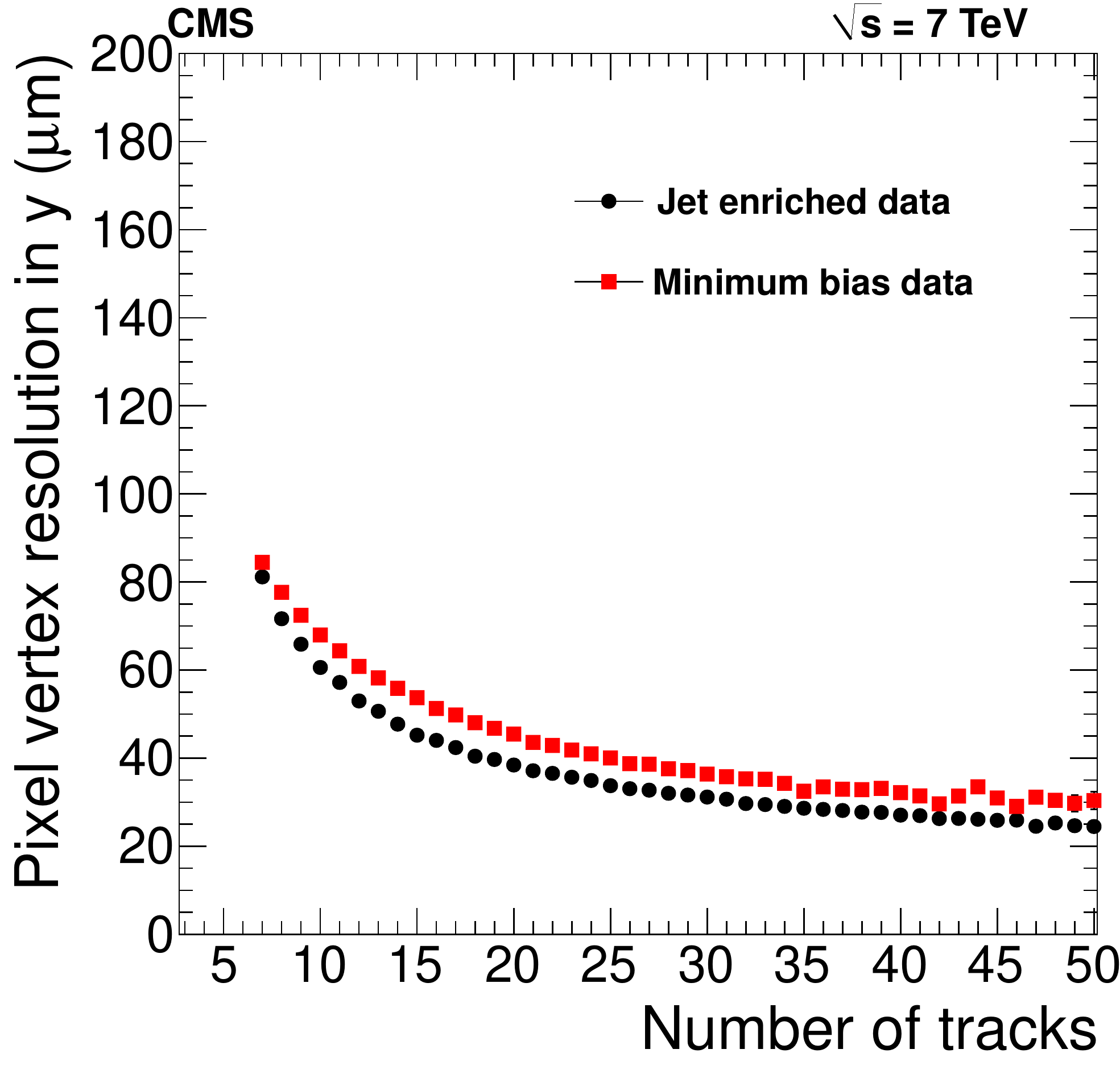}
       \includegraphics[width=0.48\textwidth]{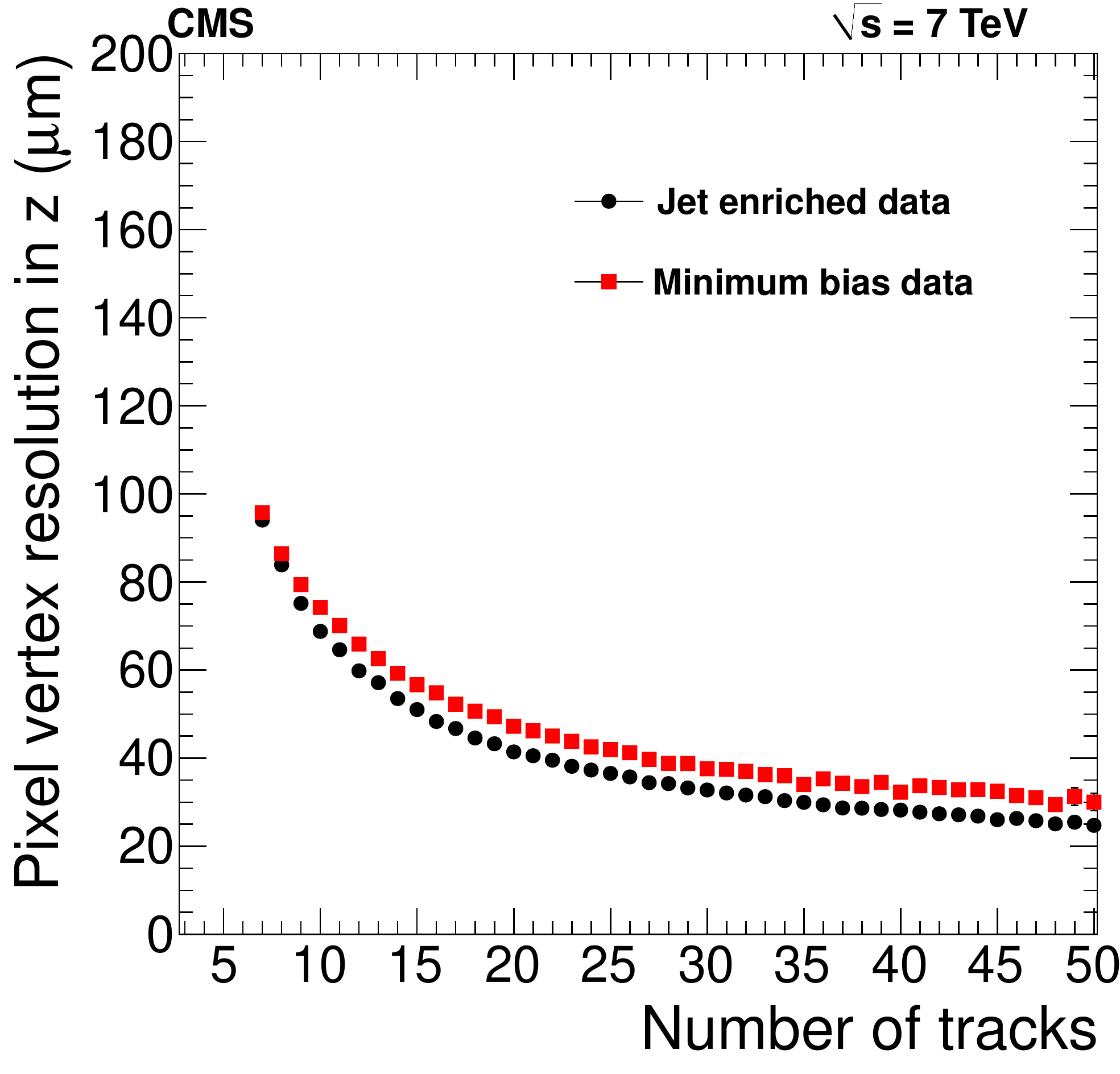}
  \caption{Pixel vertex position resolutions in $x$ (left) and $z$ (right) as a function of the number of
    tracks used in the fitted vertex, for minimum-bias and jet-enriched data.}
    \label{fig:pixelpvtx_resolution_MBias_Jets_data}
\end{figure}

\subsection{Reconstruction of the LHC beam spot
\label{sec:beamspot}}

The beam spot represents a 3-D profile of the luminous
region, where the LHC beams collide in the CMS detector.  The beam spot parameters are
determined from an average over many events, in contrast to the
event-by-event primary vertex that gives the precise position of a
single collision.  Measurements of the centre and dependence of
the luminous region on $r$ and $z$ are important components of event reconstruction.
The position of the centre of the beam spot, corresponding to the centre of the luminous region,
is used, especially in the HLT,
(i) to estimate the position of the interaction point prior
to the reconstruction of the primary vertex; (ii) to provide an additional
constraint in the reconstruction of all the primary vertices of an event;
and (iii) to provide the primary interaction point in the full reconstruction of
low-multiplicity data.

\subsubsection{Determination of the position of the centre of the beam spot}
The position of the centre of the beam spot can be determined in two ways.  The first method is through the
reconstruction of primary vertices (see Section~\ref{sec:pvtxreco}),
which map out the collisions as a function of $x$, $y$, and $z$, and therefore the shape of
the beam spot.  The mean position in $x$, $y$, and $z$,  and the size
of the luminous region can be
determined through a fit of a likelihood to the 3-D distribution of vertex positions.
The second method utilises a correlation between $d_{0}$ and $\phi$ that
appears when the centre of the beam spot is displaced relative to its expected position.
The $d_{0}$ for tracks coming from a primary vertex can
be parametrized as:
\begin{linenomath}
\begin{equation}
   d_{0}(\phi, z_0) = x_\mathrm{BS} \sin\phi + \dd{x}{z} \sin\phi\,  (z_0-z_\mathrm{BS})
                  - y_\mathrm{BS} \cos\phi - \dd{y}{z} \cos\phi\,  (z_0-z_\mathrm{BS}) ,
\end{equation}
\end{linenomath}
where $x_\mathrm{BS}$ and $y_\mathrm{BS}$ are the $x$ and $y$ positions of the beam at
$z=z_\mathrm{BS}$ (the centre of the beam spot along the beam direction), and
$\ddinline{x}{z}$ and $\ddinline{y}{z}$ are the derivatives (slopes) of $x$ and $y$ relative to $z$.
The fit of the beam spot~\cite{CMS_NOTE_2007-021} uses an iterative
$\chi^2$ fit to utilise this correlation between $d_{0}$ and
$\phi$.  With a sample of 1000 tracks, the position can be
determined with a statistical precision of approximately 5\mum.

The two methods have been checked against each other, and provide consistent results.
The precision of the $d_{0}$--$\phi$ fit is better in lower-multiplicity events, however
the width and length of the luminous region can not be obtained with the same algorithm.
Therefore, a combination of the two methods is used
to measure the beam spot used in the full reconstruction of each event.
The $d_{0}$--$\phi$ fit is used to determine the centre of the luminous region
in the transverse plane, ($x_\mathrm{BS}, y_\mathrm{BS}$), and the slopes, $\ddinline{x}{z}$ and $\ddinline{y}{z}$;
while $z_\mathrm{BS}$ and the \rms widths of the luminous region
$\sigma_x$, $\sigma_y$, and $\sigma_z$ are all determined from the
fit to the 3-D vertex distribution.  The beam spot is determined in every
luminosity section (LS), corresponding to the events collected during
a period of 23\unit{seconds}.
When the results from all LS intervals of a run are available, they are combined to extract
the final beam spot values.
A weighted average is performed,
with a check implemented to assure that no significant shift occurred in the parameters that might indicate a movement of the beam spot.
To protect against slow drifts
of the beam, no more than 60 consecutive LS are combined at a time.

\begin{figure}[p]
  \centering
    \includegraphics[width=0.9\textwidth]{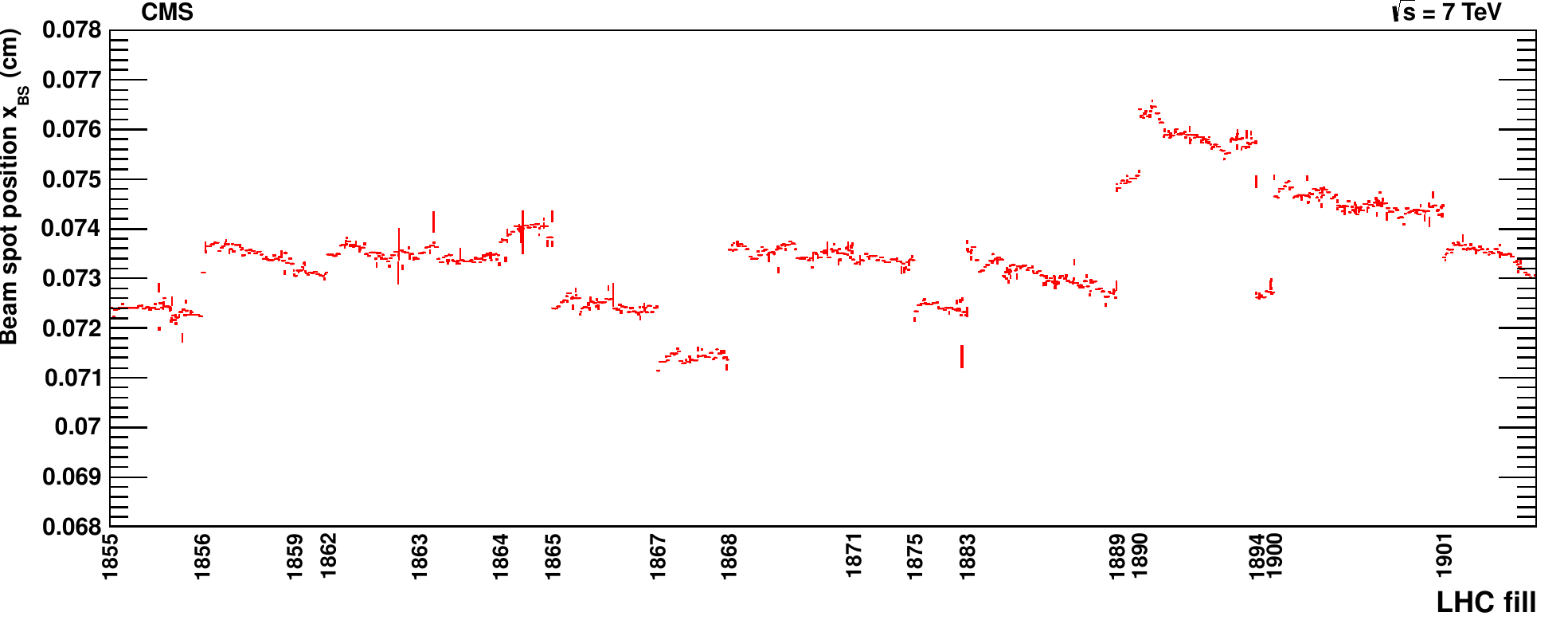}
    \includegraphics[width=0.9\textwidth]{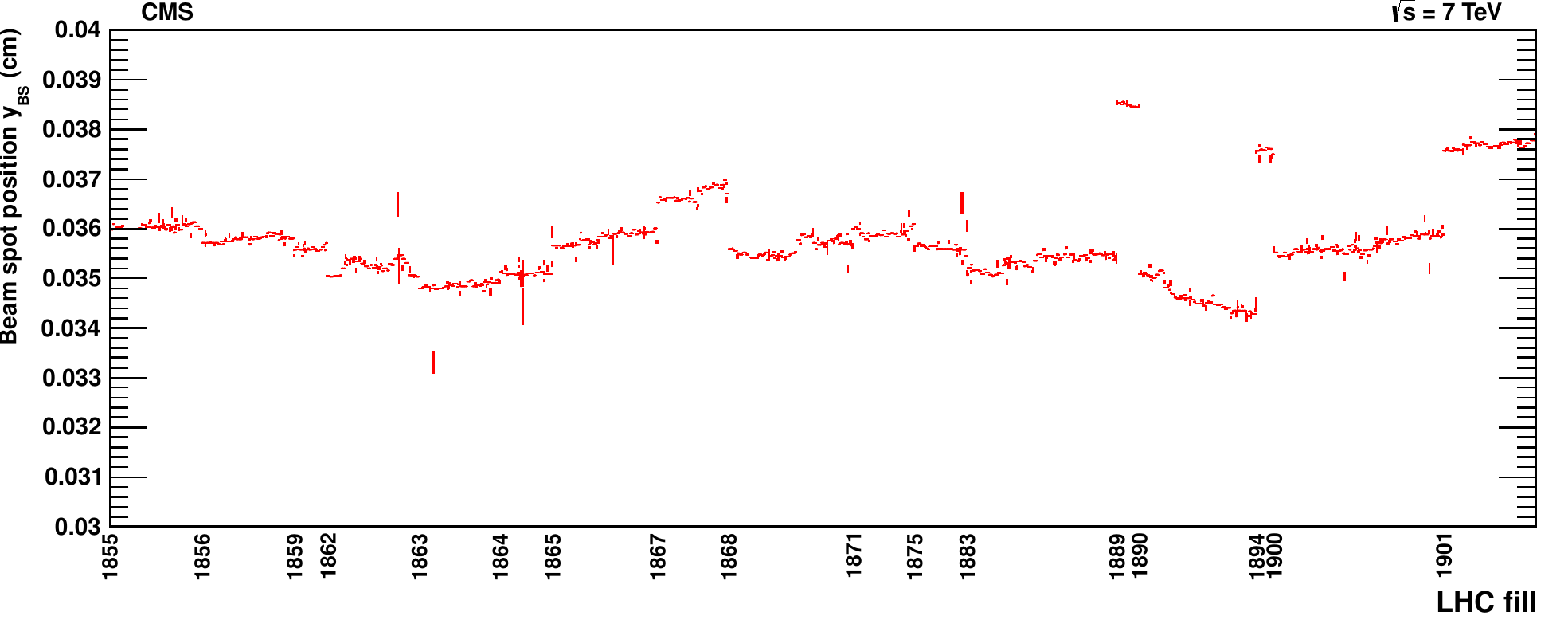}
    \includegraphics[width=0.9\textwidth]{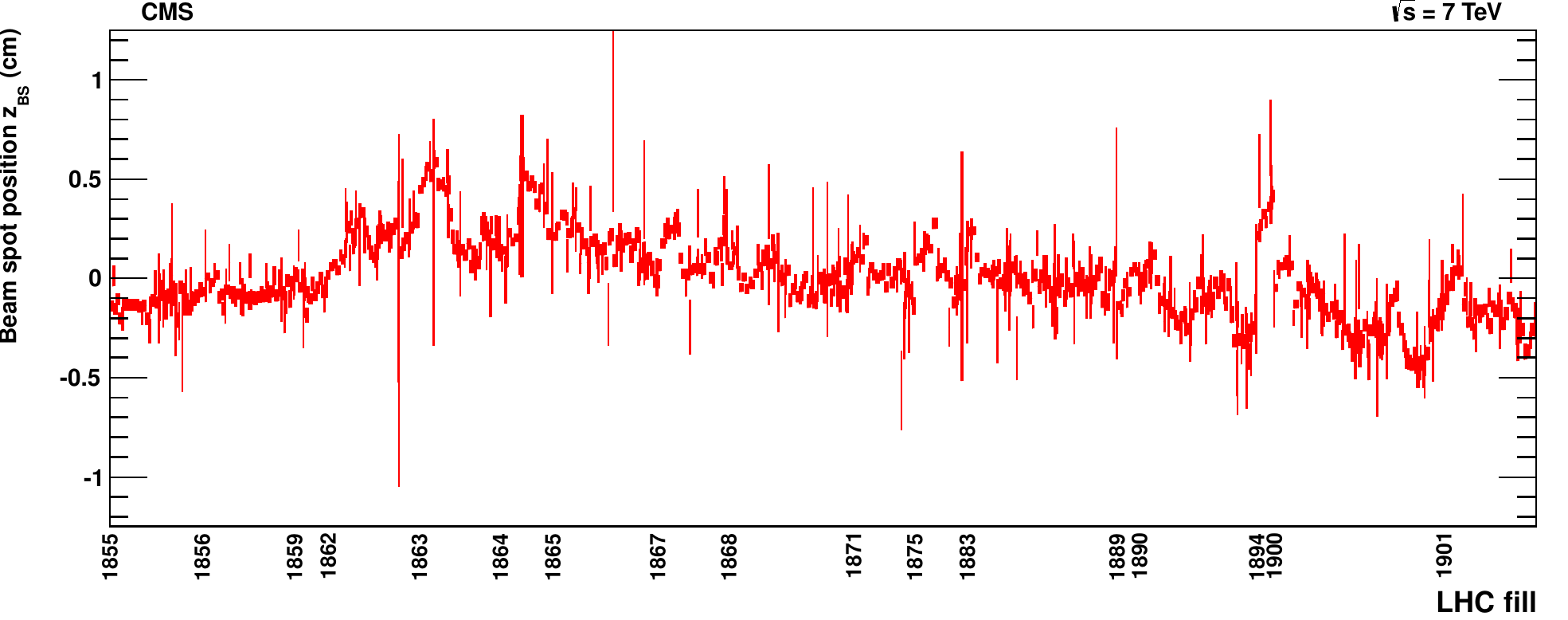}
    \caption[Position of the centre of the beam spot as a function of time.]
     {Fitted $x_\mathrm{BS}$ (top), $y_\mathrm{BS}$ (middle) and $z_\mathrm{BS}$ (bottom)
      positions of the centre of the luminous region as a
      function of time during early 2011 running.  The $x_\mathrm{BS}$ and $y_\mathrm{BS}$ values are extracted from the
      a fit to the $d_{0}$--$\phi$ distribution, and the value of $z_\mathrm{BS}$ is extracted from the fit to the primary-vertex distribution.
      Each point represents one luminosity section of 23\unit{seconds}.
      The error bars reflect the statistical uncertainty from the fit.
\label{fig:beamxyz_bylumi}}
\end{figure}
Figure~\ref{fig:beamxyz_bylumi} shows the fitted positions as a function of time
for LHC fills during early 2011.
The results demonstrate that, within a fill, the position
is quite stable, while occasionally there are larger shifts between fills.

\subsubsection{Determining the size of the beam spot}
The size of the luminous region is also determined with two methods.
The first one is based on
the reconstructed primary-vertex distribution, where the values of $\sigma$ are obtained through
the likelihood fit described above.
The second method, described below, which measures only the transverse size, is based on
event-by-event correlations between the transverse impact
parameters of two tracks originating from the same vertex.

The displacement of the interaction point within the interaction
region introduces a common displacement of trajectories
of all particles from the interaction. This shift of the trajectories
produces a correlation between the transverse impact parameters of
tracks relative to the nominal beam position. The strength of the
correlation reflects the transverse size of the beam.
The correlation between the transverse impact
parameters of two tracks from the vertex of one
interaction (labelled (1) and (2)) can be expressed by the expectation
value
\begin{linenomath}
\begin{equation}
\left< d_{0}^{(1)} \ d_{0}^{(2)} \right> =
\frac{\sigma_x^2+\sigma_y^2}{2} \cos(\phi_1 - \phi_2)
+ \frac{\sigma_y^2-\sigma_x^2}{2} \cos(\phi_1 + \phi_2),
\label{eq:ipcorr}
\end{equation}
\end{linenomath}
where $\phi_1$ and $\phi_2$ are the azimuthal angles of the tracks
measured at the point of closest approach to the beam. A particular
feature of this correlation is that its size is independent of
the resolutions in vertex positions and impact parameters, and therefore
corrections to remove contributions from the resolution are not required.
Assuming no correlation between $\phi_1 - \phi_2$ and $\phi_1 + \phi_2$, the
coefficients in Eq.~(\ref{eq:ipcorr}) can be obtained through the slopes
of straight lines fitted to the respective dependence of $\left<
  d_{0}^{(1)} \ d_{0}^{(2)} \right>$.

Both methods have been used to extract $\sigma_x$ and $\sigma_y$ and
the results averaged over an LHC fill are found to be
consistent to 2-3\mum~\cite{CMS_PAS_TRK-10-005}.
Figure~\ref{fig:beamwidth_bylumi}
shows $\sigma_x$, $\sigma_y$, and $\sigma_z$ as a function of time for LHC fills in early
2011, obtained using the likelihood fit to the primary-vertex distribution.
The size of the beam grows with time during each fill, reflecting the growth of
the beam emittance.
The emittance growth has been directly observed with dedicated instrumentation
and correlated with real-time measurements of the beam size~\cite{White:2012zzc}.
\begin{figure}[p]
  \centering
    \includegraphics[width=0.9\textwidth]{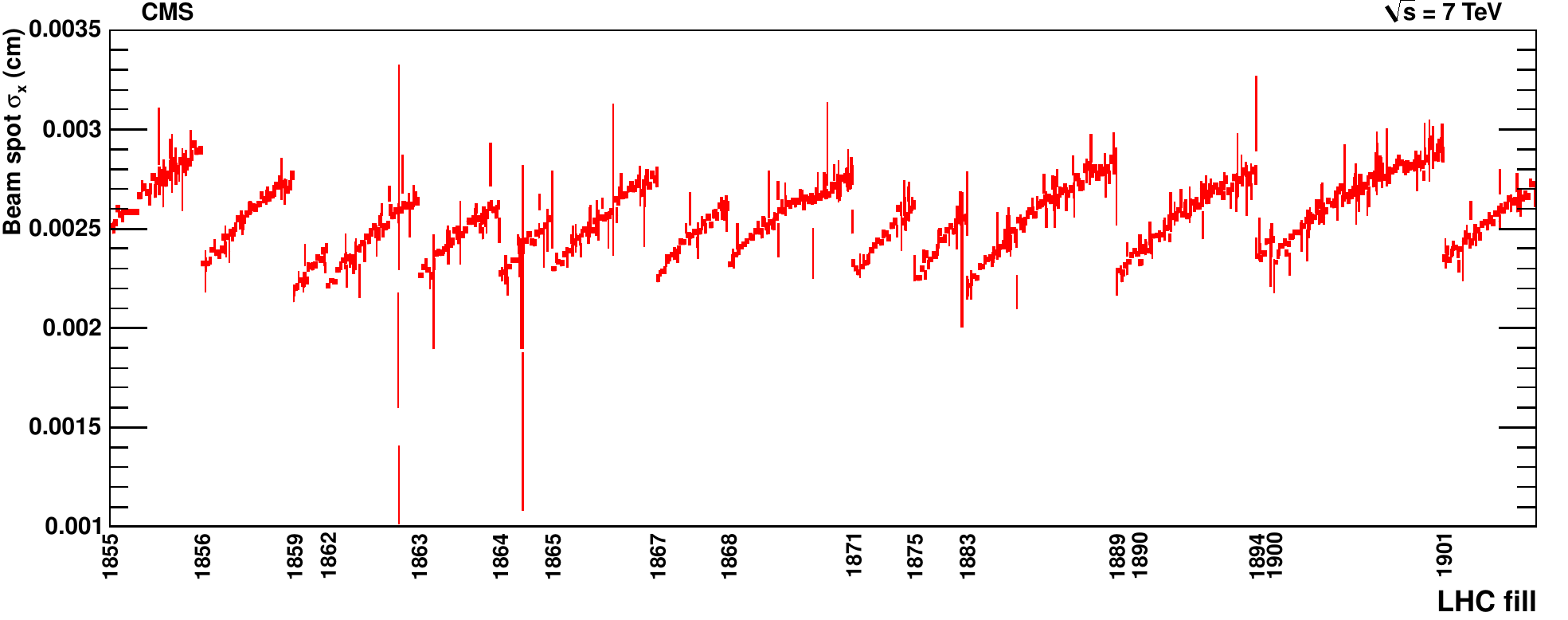}
    \includegraphics[width=0.9\textwidth]{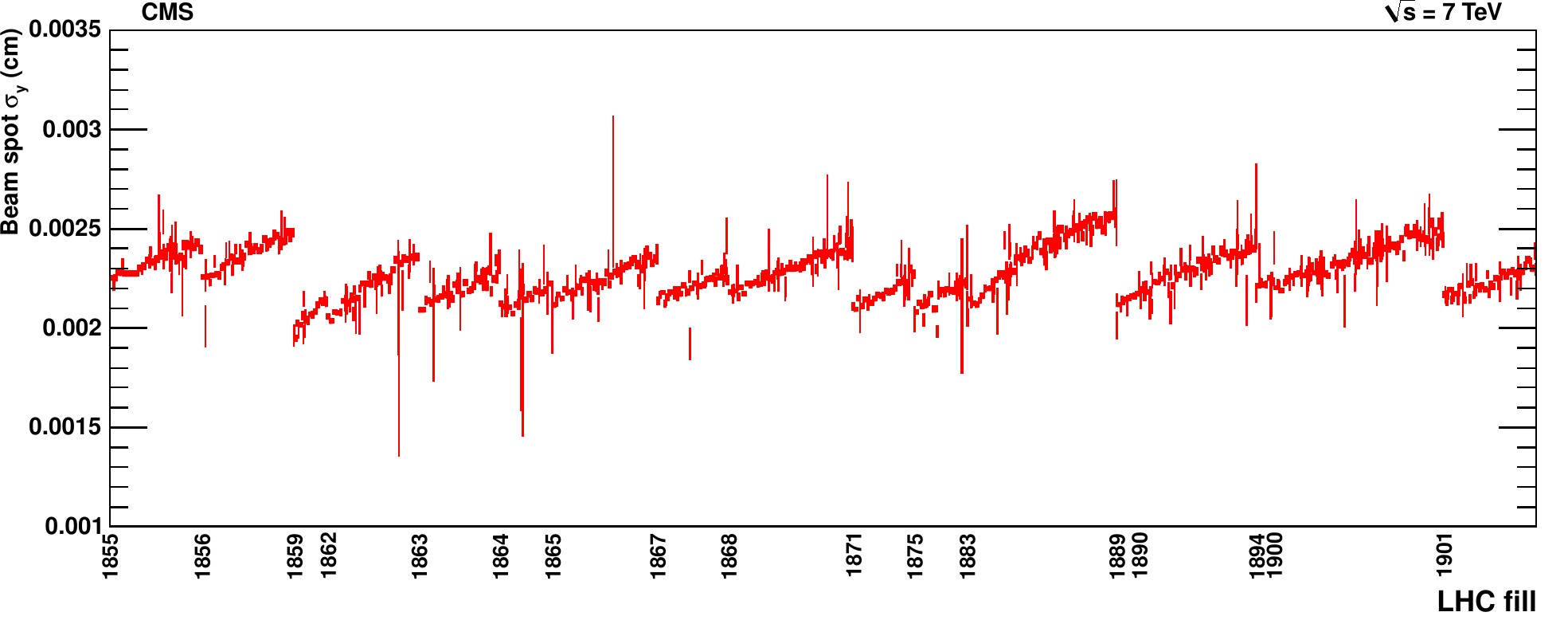}
    \includegraphics[width=0.9\textwidth]{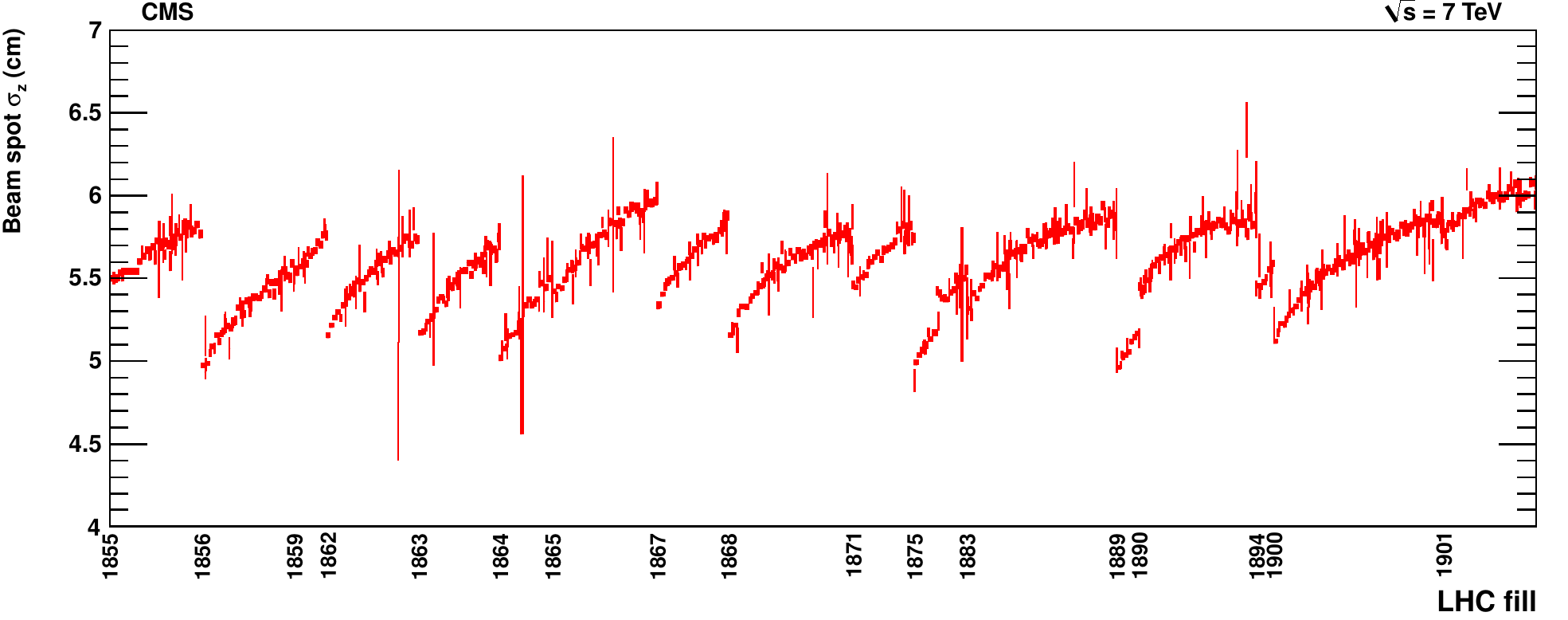}
    \label{fig:beamwidth_z}
    \caption[Beam size as a function of time.]{Fitted widths
      $\sigma_x$ (top) and $\sigma_y$ (middle),
      and length $\sigma_z$ (bottom) of the luminous region as a
      function of time during early 2011 running, extracted from the fit to
      the distribution of reconstructed primary vertices.
      Each point represents one luminosity section of 23~seconds.
      The error bars reflect the statistical uncertainty from the fit.
\label{fig:beamwidth_bylumi}}
\end{figure}
\clearpage

\section{Summary and conclusions}
\label{sec:conclusions}

CMS has developed sophisticated tracking and vertexing software algorithms,
based on
the Kalman filter, the Gaussian sum filter, and the deterministic annealing filter, to reconstruct
the proton-proton collision data provided by the CMS silicon tracker.
The implementation of these algorithms has been optimized for computational efficiency,
required to keep up with the high data rates recorded using the CMS apparatus.
The flexibility of this software is evident from the fact that, with only few changes,
it has been adapted to provide the fast tracking needed for the CMS high-level trigger,
which processes events at rates of up to 100\unit{kHz}. Furthermore, a dedicated version of the software that
accommodates bremsstrahlung energy loss in the tracker material, is used to reconstruct electrons.

The tracking algorithms reconstruct tracks over the full pseudorapidity
range of the tracker $\abs{\eta} < 2.5$, finding charged particles with \pt as low as 0.1\GeV, or produced as
far as 60\cm from the beam line (such as pions from \Kzero decay).
Promptly produced, isolated muons of $\pt > 0.9$\GeV are reconstructed with essentially 100\%
efficiency for $\abs{\eta} < 2.4$. In the central region ($\abs{\eta} < 1.4$), where the resolution is best,
muons of $\pt = 100$\GeV have resolutions of approximately 2.8\% in \pt, and
10 and 30\mum in transverse and longitudinal impact parameter, respectively.

For prompt, charged particles of $\pt > 0.9$\GeV in simulated \ttbar events, under typical 2011 LHC
pileup conditions, the average track-reconstruction efficiency is 94\% in the barrel region ($\abs{\eta} < 0.9$) of the
tracker and 85\% at higher pseudorapidity. Most of the inefficiency is caused by hadrons undergoing
nuclear interactions in the tracker material. In the same \pt range, the fraction of falsely reconstructed
tracks is at the few percent level.
In the central region, tracks with $1 < \pt < 10$\GeV have a resolution in \pt of approximately 1.5\%.
The resolution in their transverse (longitudinal) impact parameters improves from
90\mum (150\mum) at $\pt = 1$\GeV to 25\mum (45\mum) at $\pt = 10$\GeV. In this momentum range,
the resolution in the track parameters is dominated by multiple scattering.

Tracks are used to reconstruct the primary interaction vertices in each event. For vertices
with many tracks, characteristic of interesting events, the achieved vertex position resolution
is 10--12\mum in each of the three spatial dimensions.

When the LHC was first proposed, it was not at all certain that tracking of such high quality could be
achieved. To make this possible, the CMS collaboration elected to build the world's largest all-silicon
tracker, which would provide a relatively small number of high precision hit position measurements, and
immersed it in a powerful coaxial magnetic field. The collaboration then devoted many years
to the development and study of different tracking algorithms, before finally selecting the ones
described in this paper.
For example, it was thought initially that track finding should be seeded using hits in the outer
layers of the tracker, where the channel occupancy is relatively low. Only later was it broadly appreciated that
the pixel tracker is much better for this purpose, with its high granularity giving it
excellent resolution in three dimensions and an even lower channel occupancy, despite the high
track density.
The CMS track and primary-vertex reconstruction software has already achieved or surpassed the
performance levels predicted at the time that the tracker was originally designed~\cite{ttdr}.
Evolution and refinement of tracking and vertexing algorithms will continue in the future, in order to meet
the challenges of ever increasing LHC luminosity.

\section*{Acknowledgements}

\hyphenation{Bundes-ministerium Forschungs-gemeinschaft Forschungs-zentren} We congratulate our colleagues in the CERN accelerator departments for the excellent performance of the LHC and thank the technical and administrative staffs at CERN and at other CMS institutes for their contributions to the success of the CMS effort. In addition, we gratefully acknowledge the computing centres and personnel of the Worldwide LHC Computing Grid for delivering so effectively the computing infrastructure essential to our analyses. Finally, we acknowledge the enduring support for the construction and operation of the LHC and the CMS detector provided by the following funding agencies: the Austrian Federal Ministry of Science, Research and Economy and the Austrian Science Fund; the Belgian Fonds de la Recherche Scientifique, and Fonds voor Wetenschappelijk Onderzoek; the Brazilian Funding Agencies (CNPq, CAPES, FAPERJ, and FAPESP); the Bulgarian Ministry of Education and Science; CERN; the Chinese Academy of Sciences, Ministry of Science and Technology, and National Natural Science Foundation of China; the Colombian Funding Agency (COLCIENCIAS); the Croatian Ministry of Science, Education and Sport, and the Croatian Science Foundation; the Research Promotion Foundation, Cyprus; the Ministry of Education and Research, Estonian Research Council via IUT23-4 and IUT23-6 and European Regional Development Fund, Estonia; the Academy of Finland, Finnish Ministry of Education and Culture, and Helsinki Institute of Physics; the Institut National de Physique Nucl\'eaire et de Physique des Particules~/~CNRS, and Commissariat \`a l'\'Energie Atomique et aux \'Energies Alternatives~/~CEA, France; the Bundesministerium f\"ur Bildung und Forschung, Deutsche Forschungsgemeinschaft, and Helmholtz-Gemeinschaft Deutscher Forschungszentren, Germany; the General Secretariat for Research and Technology, Greece; the National Scientific Research Foundation, and National Innovation Office, Hungary; the Department of Atomic Energy and the Department of Science and Technology, India; the Institute for Studies in Theoretical Physics and Mathematics, Iran; the Science Foundation, Ireland; the Istituto Nazionale di Fisica Nucleare, Italy; the Korean Ministry of Education, Science and Technology and the World Class University program of NRF, Republic of Korea; the Lithuanian Academy of Sciences; the Ministry of Education, and University of Malaya (Malaysia); the Mexican Funding Agencies (CINVESTAV, CONACYT, SEP, and UASLP-FAI); the Ministry of Business, Innovation and Employment, New Zealand; the Pakistan Atomic Energy Commission; the Ministry of Science and Higher Education and the National Science Centre, Poland; the Funda\c{c}\~ao para a Ci\^encia e a Tecnologia, Portugal; JINR, Dubna; the Ministry of Education and Science of the Russian Federation, the Federal Agency of Atomic Energy of the Russian Federation, Russian Academy of Sciences, and the Russian Foundation for Basic Research; the Ministry of Education, Science and Technological Development of Serbia; the Secretar\'{\i}a de Estado de Investigaci\'on, Desarrollo e Innovaci\'on and Programa Consolider-Ingenio 2010, Spain; the Swiss Funding Agencies (ETH Board, ETH Zurich, PSI, SNF, UniZH, Canton Zurich, and SER); the Ministry of Science and Technology, Taipei; the Thailand Center of Excellence in Physics, the Institute for the Promotion of Teaching Science and Technology of Thailand, Special Task Force for Activating Research and the National Science and Technology Development Agency of Thailand; the Scientific and Technical Research Council of Turkey, and Turkish Atomic Energy Authority; the National Academy of Sciences of Ukraine, and State Fund for Fundamental Researches, Ukraine; the Science and Technology Facilities Council, UK; the US Department of Energy, and the US National Science Foundation.

Individuals have received support from the Marie-Curie programme and the European Research Council and EPLANET (European Union); the Leventis Foundation; the A. P. Sloan Foundation; the Alexander von Humboldt Foundation; the Belgian Federal Science Policy Office; the Fonds pour la Formation \`a la Recherche dans l'Industrie et dans l'Agriculture (FRIA-Belgium); the Agentschap voor Innovatie door Wetenschap en Technologie (IWT-Belgium); the Ministry of Education, Youth and Sports (MEYS) of the Czech Republic; the Council of Science and Industrial Research, India; the Compagnia di San Paolo (Torino); the HOMING PLUS programme of Foundation for Polish Science, cofinanced by EU, Regional Development Fund; and the Thalis and Aristeia programmes cofinanced by EU-ESF and the Greek NSRF.

\bibliography{auto_generated}   % will be created by the tdr script.

\providecommand{\href}[2]{#2}\begingroup\raggedright\begin{thebibliography}{10}%
\makeatletter
\providecommand{\hrefCMSnoop }[0]{\@secondoftwo}%
\makeatother
\providecommand{\doi}{\texttt{doi:}\begingroup \urlstyle{tt}\Url}

\bibitem{Bayatian:2006zz}
\href {http://cds.cern.ch/record/922757} {{ CMS} Collaboration, ``{Physics
  technical design report, volume I: Detector performance and software}'',}
  CERN-LHCC {2006-001}, 2006.

\bibitem{Ball:2007zza}
\hrefCMSnoop {} {{ CMS} Collaboration, ``{Physics technical design report,
  volume II: Physics performance}'',} \textit{ J. Phys. G} \textbf{ 34} (2007)
  995,
\href{http://dx.doi.org/10.1088/0954-3899/34/6/S01}{\doi{10.1088/0954-3899/34/6/S01}}.
%%CITATION = JPHGB,G34,995;%%.

\bibitem{CMS_PAS_PFT-09-001}
\href {http://cds.cern.ch/record/1194487/} {{ CMS} Collaboration,
  ``{Particle-Flow Event Reconstruction in CMS and Performance for Jets, Taus
  and $E_T^{miss}$}'',} CMS Physics Analysis Summary CMS-PAS-PFT-09-001, 2009.

\bibitem{Chatrchyan:2012jua}
\hrefCMSnoop {} {{ CMS} Collaboration, ``{Identification of b-quark jets with
  the CMS experiment}'',} \textit{ JINST} \textbf{ 8} (2013) P04013,
  \href{http://dx.doi.org/10.1088/1748-0221/8/04/P04013}{\doi{10.1088/1748-0221/8/04/P04013}},
\href{http://www.arXiv.org/abs/1211.4462}{\texttt{ arXiv:1211.4462}}.
%%CITATION = ARXIV:1211.4462;%%.

\bibitem{:2008zzk}
\hrefCMSnoop {} {{ CMS} Collaboration, ``The {CMS} experiment at the {CERN
  LHC}'',} \textit{ JINST} \textbf{ 3} (2008) S08004,
\href{http://dx.doi.org/10.1088/1748-0221/3/08/S08004}{\doi{10.1088/1748-0221/3/08/S08004}}.
%%CITATION = JINST,3,S08004;%%.

\bibitem{Agostinelli:2002hh}
\hrefCMSnoop {} {{ GEANT4} Collaboration, ``{GEANT4}---a simulation toolkit'',}
  \textit{ Nucl. Instrum. Meth. A} \textbf{ 506} (2003) 250,
\href{http://dx.doi.org/10.1016/S0168-9002(03)01368-8}{\doi{10.1016/S0168-9002(03)01368-8}}.
%%CITATION = NUIMA,A506,250;%%.

\bibitem{craft_paper}
\hrefCMSnoop {} {{ CMS} Collaboration, ``Commissioning of the {CMS} experiment
  and the cosmic run at four Tesla'',} \textit{ JINST} \textbf{ 5} (2010)
  T03001,
  \href{http://dx.doi.org/10.1088/1748-0221/5/03/T03001}{\doi{10.1088/1748-0221/5/03/T03001}},
\href{http://www.arXiv.org/abs/0911.4845}{\texttt{ arXiv:0911.4845}}.
%%CITATION = ARXIV:0911.4845;%%.

\bibitem{craft_pixel_paper}
\hrefCMSnoop {} {{ CMS} Collaboration, ``Commissioning and performance of the
  {CMS} pixel tracker with cosmic ray muons'',} \textit{ JINST} \textbf{ 5}
  (2010) T03007,
  \href{http://dx.doi.org/10.1088/1748-0221/5/03/T03007}{\doi{10.1088/1748-0221/5/03/T03007}},
\href{http://www.arXiv.org/abs/0911.5434}{\texttt{ arXiv:0911.5434}}.
%%CITATION = ARXIV:0911.5434;%%.

\bibitem{craft_strip_paper}
\hrefCMSnoop {} {{ CMS} Collaboration, ``Commissioning and performance of the
  {CMS} silicon strip tracker with cosmic ray muons'',} \textit{ JINST}
  \textbf{ 5} (2010) T03008,
  \href{http://dx.doi.org/10.1088/1748-0221/5/03/T03008}{\doi{10.1088/1748-0221/5/03/T03008}},
\href{http://www.arXiv.org/abs/0911.4996}{\texttt{ arXiv:0911.4996}}.
%%CITATION = ARXIV:0911.4996;%%.

\bibitem{craft_alignment_paper}
\hrefCMSnoop {} {{ CMS} Collaboration, ``Alignment of the {CMS} silicon tracker
  during commissioning with cosmic rays'',} \textit{ JINST} \textbf{ 5} (2010)
  T03009,
  \href{http://dx.doi.org/10.1088/1748-0221/5/03/T03009}{\doi{10.1088/1748-0221/5/03/T03009}},
\href{http://www.arXiv.org/abs/0910.2505}{\texttt{ arXiv:0910.2505}}.
%%CITATION = ARXIV:0910.2505;%%.

\bibitem{craft_bfield_paper}
\hrefCMSnoop {} {{ CMS} Collaboration, ``Precise mapping of the magnetic field
  in the {CMS} barrel yoke using cosmic rays'',} \textit{ JINST} \textbf{ 5}
  (2010) T03021,
  \href{http://dx.doi.org/10.1088/1748-0221/5/03/T03021}{\doi{10.1088/1748-0221/5/03/T03021}},
\href{http://www.arXiv.org/abs/0910.5530}{\texttt{ arXiv:0910.5530}}.
%%CITATION = ARXIV:0910.5530;%%.

\bibitem{TRK10_001}
\hrefCMSnoop {} {{ CMS} Collaboration, ``{CMS} Tracking Performance Results
  from early {LHC} Operation'',} \textit{ Eur. Phys. J. C} \textbf{ 70} (2010)
  1165,
  \href{http://dx.doi.org/10.1140/epjc/s10052-010-1491-3}{\doi{10.1140/epjc/s10052-010-1491-3}},
\href{http://www.arXiv.org/abs/1007.1988}{\texttt{ arXiv:1007.1988}}.
%%CITATION = ARXIV:1007.1988;%%.

\bibitem{CMS_PAS_TRK-10-003}
\href {http://cds.cern.ch/record/1279138} {{ CMS} Collaboration, ``{Studies of
  Tracker Material}'',} CMS Physics Analysis Summary CMS-PAS-TRK-10-003, 2010.

\bibitem{Kotlinski:2006jz}
D.~Kotlinski\hrefCMSnoop {} { {et~al.}, ``The control and readout systems of
  the {CMS} pixel barrel detector'',} \textit{ Nucl. Instrum. Meth. A} \textbf{
  565} (2006) 73,
\href{http://dx.doi.org/10.1016/j.nima.2006.04.065}{\doi{10.1016/j.nima.2006.04.065}}.
%%CITATION = NUIMA,A565,73;%%.

\bibitem{Kastli:2005jj}
H.~C. K\"astli\hrefCMSnoop {} { {et~al.}, ``Design and performance of the {CMS}
  pixel detector readout chip'',} \textit{ Nucl. Instrum. Meth. A} \textbf{
  565} (2006) 188,
  \href{http://dx.doi.org/10.1016/j.nima.2006.05.038}{\doi{10.1016/j.nima.2006.05.038}},
\href{http://www.arXiv.org/abs/physics/0511166}{\texttt{
  arXiv:physics/0511166}}.
%%CITATION = PHYSICS/0511166;%%.

\bibitem{Cucciarelli:687475}
\href {http://cds.cern.ch/record/687475} {S.~Cucciarelli, D.~Kotlinski, and
  T.~Todorov, ``Position Determination of Pixel Hits'',} CMS Note 2002-049,
  2002.

\bibitem{CMS_NOTE_2007_033}
M.~Swartz\href {https://cds.cern.ch/record/1073691} { {et~al.}, ``{A new
  technique for the reconstruction, validation, and simulation of hits in the
  CMS pixel detector}'',} CMS Note 2007-033, 2007.

\bibitem{Swartz:2003ch}
\hrefCMSnoop {} {M.~Swartz, ``{CMS pixel simulations}'',} \textit{ Nucl.
  Instrum. Meth. A} \textbf{ 511} (2003) 88,
  \href{http://dx.doi.org/10.1016/S0168-9002(03)01757-1}{\doi{10.1016/S0168-9002(03)01757-1}}.

\bibitem{Chiochia:2004qh}
\hrefCMSnoop {} {V.~Chiochia {et~al.}, ``Simulation of Heavily Irradiated
  Silicon Pixel Sensors and Comparison with Test Beam Measurements'',} \textit{
  IEEE Trans. Nucl. Sci.} \textbf{ 52} (2005) 1067,
  \href{http://dx.doi.org/10.1109/TNS.2005.852748}{\doi{10.1109/TNS.2005.852748}},
\href{http://www.arXiv.org/abs/physics/0411143}{\texttt{
  arXiv:physics/0411143}}.
%%CITATION = PHYSICS/0411143;%%.

\bibitem{Swartz:2005vp}
\hrefCMSnoop {} {M.~Swartz {et~al.}, ``Observation, modeling, and temperature
  dependence of doubly peaked electric fields in irradiated silicon pixel
  sensors'',} \textit{ Nucl. Instrum. Meth. A} \textbf{ 565} (2006) 212,
  \href{http://dx.doi.org/10.1016/j.nima.2006.05.002}{\doi{10.1016/j.nima.2006.05.002}},
\href{http://www.arXiv.org/abs/physics/0510040}{\texttt{
  arXiv:physics/0510040}}.
%%CITATION = PHYSICS/0510040;%%.

\bibitem{ttdr}
\href {http://cds.cern.ch/record/368412} {{CMS Collaboration}, ``{The CMS
  tracker system project: Technical design report}'',} CERN-LHCC 98-006, 1998.

\bibitem{FED}
\hrefCMSnoop {} {C.~Foudas {et~al.}, ``The {CMS} tracker readout front end
  driver'',} \textit{ IEEE Trans. Nucl. Sci.} \textbf{ 52} (2005) 2836,
  \href{http://dx.doi.org/10.1109/TNS.2005.860173}{\doi{10.1109/TNS.2005.860173}},
\href{http://www.arXiv.org/abs/physics/0510229}{\texttt{
  arXiv:physics/0510229}}.
%%CITATION = PHYSICS/0510229;%%.

\bibitem{French:2001xb}
M.~J. French\hrefCMSnoop {} { {et~al.}, ``{Design and results from the APV25, a
  deep sub-micron CMOS front-end chip for the CMS tracker}'',} \textit{ Nucl.
  Instrum. Meth. A} \textbf{ 466} (2001) 359,
\href{http://dx.doi.org/10.1016/S0168-9002(01)00589-7}{\doi{10.1016/S0168-9002(01)00589-7}}.
%%CITATION = NUIMA,A466,359;%%.

\bibitem{tifPaper}
\hrefCMSnoop {} {{ CMS Tracker} Collaboration, ``{Stand-Alone Cosmic Muon
  Reconstruction before Installation of the CMS Silicon Strip Tracker}'',}
  \textit{ JINST} \textbf{ 4} (2009) P05004,
  \href{http://dx.doi.org/10.1088/1748-0221/4/05/P05004}{\doi{10.1088/1748-0221/4/05/P05004}},
\href{http://www.arXiv.org/abs/0902.1860}{\texttt{ arXiv:0902.1860}}.
%%CITATION = 0902.1860;%%.

\bibitem{Chatrchyan:2014wfa}
\hrefCMSnoop {} {{CMS Collaboration}, ``{Alignment of the CMS tracker with LHC
  and cosmic ray data}'',} (2014).
  \href{http://www.arXiv.org/abs/1403.2286}{\texttt{ arXiv:1403.2286}}.
Submitted to JINST.
%%CITATION = ARXIV:1403.2286;%%.

\bibitem{Regler:1996yw}
\hrefCMSnoop {} {M.~Regler, R.~Fr{\"u}hwirth, and W.~Mitaroff, ``Filter methods
  in track and vertex reconstruction'',} \textit{ Int. J. Mod. Phys. C}
  \textbf{ 7} (1996) 521,
\href{http://dx.doi.org/10.1142/S0129183196000454}{\doi{10.1142/S0129183196000454}}.
%%CITATION = IMPAE,C7,521;%%.

\bibitem{Mankel:2004yv}
\hrefCMSnoop {} {R.~Mankel, ``Pattern recognition and event reconstruction in
  particle physics experiments'',} \textit{ Rept. Prog. Phys.} \textbf{ 67}
  (2004) 553,
  \href{http://dx.doi.org/10.1088/0034-4885/67/4/R03}{\doi{10.1088/0034-4885/67/4/R03}},
\href{http://www.arXiv.org/abs/physics/0402039}{\texttt{
  arXiv:physics/0402039}}.
%%CITATION = PHYSICS/0402039;%%.

\bibitem{Strandlie:2010zz}
\hrefCMSnoop {} {A.~Strandlie and R.~Fr{\"u}hwirth, ``Track and vertex
  reconstruction: From classical to adaptive methods'',} \textit{ Rev. Mod.
  Phys.} \textbf{ 82} (2010) 1419,
\href{http://dx.doi.org/10.1103/RevModPhys.82.1419}{\doi{10.1103/RevModPhys.82.1419}}.
%%CITATION = RMPHA,82,1419;%%.

\bibitem{Billoir:1989mh}
\hrefCMSnoop {} {P.~Billoir, ``{Progressive track recognition with a Kalman
  like fitting procedure}'',} \textit{ Comput. Phys. Commun.} \textbf{ 57}
  (1989) 390,
\href{http://dx.doi.org/10.1016/0010-4655(89)90249-X}{\doi{10.1016/0010-4655(89)90249-X}}.
%%CITATION = CPHCB,57,390;%%.

\bibitem{Billoir:1990we}
\hrefCMSnoop {} {P.~Billoir and S.~Qian, ``{Simultaneous pattern recognition
  and track fitting by the Kalman filtering method}'',} \textit{ Nucl. Instrum.
  Meth. A} \textbf{ 294} (1990) 219,
\href{http://dx.doi.org/10.1016/0168-9002(90)91835-Y}{\doi{10.1016/0168-9002(90)91835-Y}}.
%%CITATION = NUIMA,A294,219;%%.

\bibitem{Mankel:1997dy}
\hrefCMSnoop {} {R.~Mankel, ``A Concurrent track evolution algorithm for
  pattern recognition in the {HERA-B} main tracking system'',} \textit{ Nucl.
  Instrum. Meth. A} \textbf{ 395} (1997) 169,
\href{http://dx.doi.org/10.1016/S0168-9002(97)00705-5}{\doi{10.1016/S0168-9002(97)00705-5}}.
%%CITATION = NUIMA,A395,169;%%.

\bibitem{Fruhwirth:1987fm}
\hrefCMSnoop {} {R.~Fr{\"u}hwirth, ``Application of {Kalman} filtering to track
  and vertex fitting'',} \textit{ Nucl. Instrum. Meth. A} \textbf{ 262} (1987)
  444,
  \href{http://dx.doi.org/10.1016/0168-9002(87)90887-4}{\doi{10.1016/0168-9002(87)90887-4}}.

\bibitem{Strandlie:927379}
\href {http://cds.cern.ch/record/927379} {A.~Strandlie and W.~Wittek,
  ``{Propagation of covariance matrices of track parameters in homogeneous
  magnetic fields in CMS}'',} CMS Note 2006-001, 2006.

\bibitem{Cash:1990:VOR:79505.79507}
\hrefCMSnoop {} {J.~R. Cash and A.~H. Karp, ``{A variable order Runge-Kutta
  method for initial value problems with rapidly varying right-hand sides}'',}
  \textit{ {ACM Trans. Math. Softw.}} \textbf{ {16/3}} ({1990}) 201,
  \href{http://dx.doi.org/10.1145/79505.79507}{\doi{10.1145/79505.79507}}.

\bibitem{Baffioni:2006cd}
S.~Baffioni\hrefCMSnoop {} { {et~al.}, ``Electron reconstruction in {CMS}'',}
  \textit{ Eur. Phys. J. C} \textbf{ 49} (2007) 1099,
  \href{http://dx.doi.org/10.1140/epjc/s10052-006-0175-5}{\doi{10.1140/epjc/s10052-006-0175-5}}.

\bibitem{CMS_PAS_PFT-10-003}
\href {https://cds.cern.ch/record/1279347} {{ CMS} Collaboration,
  ``{Commissioning of particle-flow event reconstruction with leptons from
  \JPsi and W decays at 7~\TeV}'',} CMS Physics Analysis Summary
  CMS-PAS-PFT-10-003, 2010.

\bibitem{Adam:2003kg}
\hrefCMSnoop {} {W.~Adam, R.~Fr{\"u}hwirth, A.~Strandlie, and T.~Todorov,
  ``Reconstruction of electrons with the Gaussian-sum filter in the {CMS}
  tracker at {LHC}'',} \textit{ J. Phys. G} \textbf{ 31} (2005) N9,
  \href{http://dx.doi.org/10.1088/0954-3899/31/9/N01}{\doi{10.1088/0954-3899/31/9/N01}},
\href{http://www.arXiv.org/abs/physics/0306087}{\texttt{
  arXiv:physics/0306087}}.
%%CITATION = PHYSICS/0306087;%%.

\bibitem{CMS_PAS_EGM-10-004}
\href {http://cds.cern.ch/record/1299116} {{ CMS} Collaboration, ``Electron
  reconstruction and identification at $\sqrt{s}=7$~TeV'',} CMS Physics
  Analysis Summary CMS-PAS-EGM-10-004, 2010.

\bibitem{LHCC_2007_021}
\href {http://cds.cern.ch/record/1043242/files/lhcc-2007-021.pdf} {{CMS
  Collaboration}, ``{CMS} high level trigger'',} CERN/LHCC 2007-021, 2007.

\bibitem{Wingham:2008zz}
\href {http://cds.cern.ch/record/1133150} {M.~Wingham, ``Commissioning of the
  {CMS} tracker and preparing for early physics at the {LHC}'',} PhD thesis
  CERN-THESIS-2008-081, 2008.

\bibitem{CMS_PAS_TRK-10-004}
\href {https://cds.cern.ch/record/1279137} {{CMS Collaboration}, ``Measurement
  of the Momentum Scale and Resolution using Low-Mass Resonances and Cosmic-Ray
  Muons'',} CMS Physics Analysis Summary CMS-PAS-TRK-10-004, 2010.

\bibitem{Chatrchyan:2013mxa}
\hrefCMSnoop {} {{ CMS} Collaboration, ``{Measurement of the properties of a
  Higgs boson in the four-lepton final state}'',} \textit{ Phys. Rev. D}
  \textbf{ 89} (2014) 092007,
  \href{http://dx.doi.org/10.1103/PhysRevD.89.092007}{\doi{10.1103/PhysRevD.89.092007}},
\href{http://www.arXiv.org/abs/1312.5353}{\texttt{ arXiv:1312.5353}}.
%%CITATION = ARXIV:1312.5353;%%.

\bibitem{CMS_PAS_TRK-10-005}
\href {http://cds.cern.ch/record/1279383} {{CMS Collaboration}, ``Tracking and
  Primary Vertex Results in First 7 TeV Collisions'',} CMS Physics Analysis
  Summary CMS-PAS-TRK-10-005, 2010.

\bibitem{CMS_PAS_TRK-10-002}
\href {http://cds.cern.ch/record/1279139} {{ CMS} Collaboration, ``{Measurement
  of Tracking Efficiency}'',} CMS Physics Analysis Summary CMS-PAS-TRK-10-002,
  2010.

\bibitem{Pythia6}
\hrefCMSnoop {} {T.~Sj{\"o}strand, S.~Mrenna, and P.~Skands, ``PYTHIA 6.4
  physics and manual'',} \textit{ JHEP} \textbf{ 05} (2006) 026,
  \href{http://dx.doi.org/10.1088/1126-6708/2006/05/026}{\doi{10.1088/1126-6708/2006/05/026}}.

\bibitem{PDG2012}
\hrefCMSnoop {} {{Particle Data Group}, J.~Beringer {et~al.}, ``{Review of
  Particle Physics}'',} \textit{ Phys. Rev. D} \textbf{ 86} (2012) 010001,
  \href{http://dx.doi.org/10.1103/PhysRevD.86.010001}{\doi{10.1103/PhysRevD.86.010001}}.

\bibitem{CMS:2011aa}
\hrefCMSnoop {} {{ CMS} Collaboration, ``Measurement of the inclusive {W} and
  {Z} production cross sections in pp collisions at $\sqrt{s}=7$~TeV with the
  {CMS} experiment'',} \textit{ JHEP} \textbf{ 10} (2011) 132,
  \href{http://dx.doi.org/10.1007/JHEP10(2011)132}{\doi{10.1007/JHEP10(2011)132}},
\href{http://www.arXiv.org/abs/1107.4789}{\texttt{ arXiv:1107.4789}}.
%%CITATION = ARXIV:1107.4789;%%.

\bibitem{Chatrchyan:2012xi}
\hrefCMSnoop {} {{ CMS} Collaboration, ``Performance of {CMS} muon
  reconstruction in pp collision events at $\sqrt{s}=7$~TeV'',} \textit{ JINST}
  \textbf{ 7} (2012) P10002,
  \href{http://dx.doi.org/10.1088/1748-0221/7/10/P10002}{\doi{10.1088/1748-0221/7/10/P10002}},
\href{http://www.arXiv.org/abs/1206.4071}{\texttt{ arXiv:1206.4071}}.
%%CITATION = ARXIV:1206.4071;%%.

\bibitem{Oreglia:1980cs}
\href {http://www.slac.stanford.edu/cgi-wrap/getdoc/slac-r-236.pdf} {M.~J.
  Oreglia, ``A Study of the Reactions $\psi^\prime \to \gamma \gamma \psi$'',}
  PhD thesis SLAC-0236 UMI-81-08973, 1980.

\bibitem{CMS_NOTE_2006-032}
T.~Speer\href {https://cds.cern.ch/record/927395} { {et~al.}, ``Vertex Fitting
  in the CMS Tracker'',} CMS Note 2006-032, 2006.

\bibitem{IEEE_DetAnnealing}
\hrefCMSnoop {} {K.~Rose, ``{Deterministic Annealing for Clustering,
  Compression, Classification, Regression and related Optimisation
  Problems}'',} \textit{ Proceedings of the IEEE} \textbf{ 86} (1998)
  \href{http://dx.doi.org/10.1109/5.726788}{\doi{10.1109/5.726788}}.

\bibitem{FruehwirtStrandlie99}
\hrefCMSnoop {} {R.~Fr{\"u}hwirth and A.~Strandlie, ``Track fitting with
  ambiguities and noise: a study of elastic tracking and nonlinear filters'',}
  \textit{ Comput. Phys. Commun.} \textbf{ 120} (1999) 197,
  \href{http://dx.doi.org/10.1016/S0010-4655(99)00231-3}{\doi{10.1016/S0010-4655(99)00231-3}}.

\bibitem{CMS_NOTE_2007-008}
\href {https://cds.cern.ch/record/1027031} {R.~Fr{\"u}hwirth, W.~Waltenberger,
  and P.~Vanlaer, ``Adaptive Vertex Fitting'',} CMS Note 2007-008, 2007.

\bibitem{Field:2011iq}
\hrefCMSnoop {} {R.~Field, ``{Min-Bias and the Underlying Event at the LHC}'',}
  \textit{ Acta Phys. Polon. B} \textbf{ 42} (2011) 2631,
  \href{http://dx.doi.org/10.5506/APhysPolB.42.2631}{\doi{10.5506/APhysPolB.42.2631}},
\href{http://www.arXiv.org/abs/1110.5530}{\texttt{ arXiv:1110.5530}}.
%%CITATION = ARXIV:1110.5530;%%.

\bibitem{CMS_NOTE_2007-021}
\href {https://cds.cern.ch/record/1061285} {T.~Miao, N.~Leioatts, H.~Wenzel,
  and F.~Yumiceva, ``Beam Position Determination using Tracks'',} CMS Note
  2007-021, 2007.

\bibitem{White:2012zzc}
S.~M. White\href
  {http://accelconf.web.cern.ch/AccelConf/IPAC2012/papers/MOPPR076.PDF} {
  {et~al.}, ``{Using the BRAN Luminosity Detectors for Beam Emittance
  Monitoring During LHC Physics Runs}'',} in \textit{ Proceedings, 3rd Int.
  Particle Accelerator Conf. (IPAC 2012)}, p.~966.
\newblock
2012.
\newblock
%%CITATION = CONFP,C1205201,966;%%.

\end{thebibliography}\endgroup

\cleardoublepage \appendix\section{The CMS Collaboration \label{app:collab}}\begin{sloppypar}\hyphenpenalty=5000\widowpenalty=500\clubpenalty=5000\textbf{Yerevan Physics Institute,  Yerevan,  Armenia}\\*[0pt]
S.~Chatrchyan, V.~Khachatryan, A.M.~Sirunyan, A.~Tumasyan
\vskip\cmsinstskip
\textbf{Institut f\"{u}r Hochenergiephysik der OeAW,  Wien,  Austria}\\*[0pt]
W.~Adam, T.~Bergauer, M.~Dragicevic, J.~Er\"{o}, C.~Fabjan\cmsAuthorMark{1}, M.~Friedl, R.~Fr\"{u}hwirth\cmsAuthorMark{1}, V.M.~Ghete, C.~Hartl, N.~H\"{o}rmann, J.~Hrubec, M.~Jeitler\cmsAuthorMark{1}, W.~Kiesenhofer, V.~Kn\"{u}nz, M.~Krammer\cmsAuthorMark{1}, I.~Kr\"{a}tschmer, D.~Liko, I.~Mikulec, D.~Rabady\cmsAuthorMark{2}, B.~Rahbaran, H.~Rohringer, R.~Sch\"{o}fbeck, J.~Strauss, A.~Taurok, W.~Treberer-Treberspurg, W.~Waltenberger, C.-E.~Wulz\cmsAuthorMark{1}
% ...
\vskip\cmsinstskip
% ...
1:~~Also at Vienna University of Technology, Vienna, Austria\\
2:~~Also at CERN, European Organization for Nuclear Research, Geneva, Switzerland\\
% ...

\end{sloppypar}
\end{document}